\documentclass{aa}

\usepackage{graphicx}
\usepackage{xcolor}
\usepackage{txfonts}
\usepackage{subfigure}
\usepackage{float}
\usepackage{natbib,twoopt}
\usepackage{xspace}
\usepackage{balance}
\usepackage{upgreek}
\usepackage{multirow}
\usepackage[normalem]{ulem}
\usepackage{soul}
\usepackage[colorlinks=true, allcolors=blue]{hyperref}
\usepackage{outlines}

\usepackage[flushleft]{threeparttable}

\def\apjl{ApJ Lett.}
\def\apjs{ApJ Suppl.}

\def\aap{A\&A}

\bibpunct{(}{)}{;}{a}{}{,}
\DeclareGraphicsExtensions{.eps,.ps,.pdf}

\defcitealias{Bellagamba19}{B19}
\defcitealias{Lesci_colours}{L24}
\defcitealias{AMICOKiDS-1000}{M25}
\defcitealias{Planck18}{Planck18}

\newcommand\Ocdm{0.169}
\newcommand\OcdmUp{0.024}
\newcommand\OcdmLow{0.022}
\newcommand\As{3.75}
\newcommand\AsUp{1.07}
\newcommand\AsLow{0.80}

\newcommand\Om{0.218}
\newcommand\OmUp{0.024}
\newcommand\OmLow{0.021}
\newcommand\seight{0.86}
\newcommand\seightUp{0.03}
\newcommand\seightLow{0.03}
\newcommand\Sotto{0.74}
\newcommand\SottoUp{0.03}
\newcommand\SottoLow{0.03}

\newcommand\cZero{0.42}
\newcommand\cZerolow{0.13}
\newcommand\cZeroup{0.13}

\newcommand\soffZero{0.24}
\newcommand\soffZerolow{0.11}
\newcommand\soffZeroup{0.11}

\newcommand\alphaScaling{-0.170}
\newcommand\alphaScalinglow{0.020}
\newcommand\alphaScalingup{0.017}
\newcommand\betaScaling{0.553}
\newcommand\betaScalinglow{0.021}
\newcommand\betaScalingup{0.021}
\newcommand\gammaScaling{0.32}
\newcommand\gammaScalinglow{0.26}
\newcommand\gammaScalingup{0.28}

\newcommand\scatterZero{0.052}
\newcommand\scatterZerolow{0.015}
\newcommand\scatterZeroup{0.023}

\begin{document}

\title{AMICO galaxy clusters in KiDS-1000: Cosmological constraints and mass calibration from counts and weak lensing}

\author
{G. F. Lesci\inst{\ref{1},\ref{2}} 
\and F. Marulli\inst{\ref{1},\ref{2},\ref{3}}
\and L. Moscardini\inst{\ref{1},\ref{2},\ref{3}}
\and M. Maturi\inst{\ref{5},\ref{8}}
\and M. Sereno\inst{\ref{2},\ref{3}}
\and M. Radovich\inst{\ref{7}}
\and M. Romanello\inst{\ref{1},\ref{2}}
\and \\C. Giocoli\inst{\ref{1},\ref{2},\ref{3}}
\and A. H. Wright\inst{\ref{bochum}}
\and S. Bardelli\inst{\ref{2}}
\and M. Bilicki\inst{\ref{bilicki}}
\and G. Castignani\inst{\ref{2}}
\and H. Hildebrandt\inst{\ref{bochum}}
\and L. Ingoglia\inst{\ref{1},\ref{10}}
\and \\S. Joudaki\inst{\ref{madrid},\ref{portsmouth}}
\and A. Kannawadi\inst{\ref{princeton}}
\and E. Puddu\inst{\ref{9}}
}

\offprints{G. F. Lesci \\ \email{giorgio.lesci2@unibo.it}}

\institute{
  Dipartimento di Fisica e Astronomia ``Augusto Righi'' - Alma Mater Studiorum
  Universit\`{a} di Bologna, via Piero Gobetti 93/2, I-40129 Bologna,
  Italy\label{1}
  \and INAF - Osservatorio di Astrofisica e Scienza dello Spazio di
  Bologna, via Piero Gobetti 93/3, I-40129 Bologna, Italy\label{2}
  \and INFN - Sezione di Bologna, viale Berti Pichat 6/2, I-40127
  Bologna, Italy\label{3}
  \and Zentrum f\"ur Astronomie, Universit\"at Heidelberg, Philosophenweg 12, D-69120 Heidelberg, Germany\label{5}
  \and ITP, Universit\"at Heidelberg, Philosophenweg 16, 69120 Heidelberg, Germany\label{8}  
  \and INAF - Osservatorio Astronomico di Padova, vicolo dell'Osservatorio 5, I-35122 Padova, Italy\label{7}
  \and Ruhr University Bochum, Faculty of Physics and Astronomy, Astronomical Institute (AIRUB), German Centre for Cosmological Lensing, 44780 Bochum, Germany\label{bochum}
  \and Center for Theoretical Physics, Polish Academy of Sciences, al. Lotników 32/46, 02-668 Warsaw, Poland\label{bilicki}
  \and Department of Astrophysical Sciences, Princeton University, 4 Ivy Lane, Princeton, NJ 08544, USA\label{princeton}
  \and INAF - Istituto di Radioastronomia, Via Piero Gobetti 101, 40129 Bologna, Italy\label{10}
  \and Centro de Investigaciones Energéticas, Medioambientales y Tecnológicas (CIEMAT), Av. Complutense 40, E-28040 Madrid, Spain\label{madrid}
  \and Institute of Cosmology \& Gravitation, Dennis Sciama Building, University of Portsmouth, Portsmouth, PO1 3FX, United Kingdom\label{portsmouth}
  \and INAF - Osservatorio Astronomico di Capodimonte, Salita Moiariello 16, Napoli 80131, Italy\label{9}
}

\date{Received --; accepted --}

\abstract
   {}
   {We present the joint modelling of weak-lensing and count measurements of the galaxy clusters detected with the Adaptive Matched Identifier of Clustered Objects (AMICO) code, in the fourth data release of the Kilo Degree Survey (KiDS-1000). The analysed sample comprises approximately 8000 clusters that cover an effective area of 839~deg$^{2}$ and extend up to a redshift of $z = 0.8$. This modelling provides the first mass calibration of this cluster sample, as well as the first cosmological constraints derived from it.}
   {We derived stacked cluster weak-lensing and count measurements in bins of redshift and intrinsic richness, $\lambda^*$. To define the background galaxy samples for the stacked profiles, we used a combination of selections based on photometric redshifts (photo-$z$s) and colours. Then, based on self-organising maps, we reconstructed the true redshift distributions of the background galaxy samples. In the joint modelling of weak lensing and counts, we accounted for the systematic uncertainties arising from impurities in the background and cluster samples, biases in the cluster $z$ and $\lambda^*$, projection effects, halo orientation and miscentring, truncation of cluster halo mass distributions, matter correlated with cluster haloes, multiplicative shear bias, baryonic matter, geometric distortions in the lensing profiles, uncertainties in the theoretical halo mass function, and super-sample covariance. In addition, we employed a blinding strategy based on perturbing the cluster sample completeness.}
   {The improved statistics and photometry, along with the refined analysis compared to the previous KiDS data release, KiDS-DR3, led to a halving of the uncertainties on $\Omega_{\rm m}$ and $\sigma_8$, as we obtained $\Omega_{\rm m}=\Om^{+\OmUp}_{-\OmLow}$ and $\sigma_8=\seight^{+\seightUp}_{-\seightLow}$, despite a more extensive modelling of systematic uncertainties. The constraint on $S_8 \equiv \sigma_8(\Omega_{\rm m}/0.3)^{0.5}$, $S_8=\Sotto^{+\SottoUp}_{-\SottoLow}$, is in excellent agreement with recent cluster count and KiDS-1000 cosmic shear analyses, while it shows a $2.8\sigma$ tension with Planck cosmic microwave background results. The constraints on the $\log\lambda^*-\log M_{200}$ relation imply a mass precision of 8\%, on average, which is an improvement of three percentage points compared to KiDS-DR3. In addition, the result on the intrinsic scatter of the $\log\lambda^*-\log M_{200}$ relation, $\sigma_{\rm intr}=\scatterZero^{+\scatterZeroup}_{-\scatterZerolow}$, confirms that $\lambda^*$ is an excellent mass proxy.}
   {}

\keywords{clusters -- Cosmology: observations -- large-scale structure
  of Universe -- cosmological parameters}

\authorrunning{G. F. Lesci et al.}

\titlerunning{AMICO galaxy clusters in KiDS-1000: cosmology and masses from counts and weak lensing}

\maketitle

\section{Introduction}
Galaxy clusters are excellent probes for both cosmological \citep{Mantz15,Planck_counts,Costanzi19,Marulli21,Lesci22_counts,Romanello24,Seppi24,Ghirardini24} and astrophysical \citep{Vazza17,CHEX-MATE,Zhu21,Sereno21} studies. Counts and clustering represent the most powerful cosmological probes based on galaxy clusters \citep{Sartoris16,Fumagalli24}, followed by for example the two-halo term of cluster profiles \citep{Giocoli21,Ingoglia22} and sparsity \citep{Corasaniti21}. Accurate models describing such summary statistics can be attained through $N$-body dark matter simulations \citep{Borgani11, Angulo12, Giocoli12}, that allow for the derivation of halo mass function \citep{Sheth1999,Tinker08,Despali16,Castro23_HMF} and halo bias \citep{Sheth01,Tinker10,Castro24} models. As baryonic physics is expected to become one of the leading sources of systematic uncertainties in forthcoming cosmological analyses, several theoretical prescriptions describing its impact on cluster statistics have recently been proposed \citep{Cui12, Velliscig14, Bocquet16, Castro21,Castro23}. \\
\indent Cluster mass measurements are the foundation of cosmological investigations based on such tracers, since mass is a fundamental cosmological quantity and several different cluster observational properties are good mass proxies \citep{Teyssier11,Pratt19}. In optical and near-infrared surveys, such mass proxies can be either the richness, which is the number of cluster member galaxies \citep{Rykoff14,Bellagamba18,Maturi19,Wen24}, the signal amplitude, in the case of matched-filter cluster detection algorithms \citep{Bellagamba18,Maturi19}, or the stellar mass of member galaxies \citep{Palmese20,Pereira20,Doubrawa23}. Among the most reliable probes for measuring cluster masses we have weak gravitational lensing, which consists of the deflection of light rays from background sources due to the intervening cluster gravitational potential \citep[see e.g.][]{Bardeau07,Okabe10,Hoekstra12,Melchior15,Sereno17,Schrabback18,Stern19,Giocoli21,Ingoglia22,Sereno24,Grandis24,Kleinebreil24,Payerne25}. In fact, weak-lensing analyses do not rely on any assumptions about the dynamical state of the cluster, which is different from other methods used to estimate cluster masses based on the properties of the gas and member galaxies, such as cluster X-ray emission \citep{Ettori13,Eckert20,Scheck23}, galaxy velocity dispersion \citep{Kodi20,Ho22,Biviano23,Sereno24_2}, or the Sunyaev-Zeldovich (SZ) effect on the cosmic microwave background \citep[CMB,][]{Arnaud10,PlanckXX2013,Hilton21}. Given cluster weak-lensing measurements, mass estimates can be derived using robust models that describe the density profiles of the dark matter haloes that host galaxy clusters \citep*{NFW,BMO,DK14}, calibrated on $N$-body simulations. \\
\indent In this work, we present the cosmological analysis based on the galaxy cluster catalogue by \citet[][referred to as M25 hereafter]{AMICOKiDS-1000}, which was built up through the use of the Adaptive Matched Identifier of Clustered Objects \citep[AMICO,][]{Bellagamba18,Maturi19} in the fourth data release of the Kilo-Degree Survey \citep[referred to as KiDS-1000,][]{Kuijken19}. To this end, we measured the stacked weak-lensing signal of 8730 clusters in bins of redshift and mass proxy, up to redshift $z=0.8$ and based on the gold sample of galaxy shape measurements by \citet{Giblin21}. Then, we jointly modelled these measurements and cluster counts to derive constraints on fundamental cosmological parameters, such as the matter density parameter at $z=0$, $\Omega_{\rm m}$, and the square root of the mass variance computed on a scale of 8 $h^{-1}$Mpc at $z=0$, $\sigma_8$. We note that this work presents an update of the mass calibration performed by \citet[][referred to as B19 hereafter]{Bellagamba19}, which was based on the AMICO cluster catalogue derived by \citet{Maturi19} from the third KiDS data release \citep[KiDS-DR3,][]{deJong17}. The need for a new mass calibration stems from the enhanced survey coverage and improved galaxy photometric redshift (photo-$z$) estimates. The latter, in fact, yield significant variations in the cluster mass proxy estimates between KiDS-DR3 and KiDS-1000 \citepalias{AMICOKiDS-1000}. \\
\indent The present work is part of a series of investigations that exploit the AMICO galaxy clusters in KiDS for both cosmological \citep{Giocoli21,Ingoglia22,Lesci22_counts,Lesci22b,Busillo23,Romanello24} and astrophysical \citep{Bellagamba19,Sereno20,Radovich20,Puddu21,Castignani22,Castignani23,Tramonte23,Giocoli24} studies. The paper is organised as follows. In Sect.\ \ref{sec:DR4_data} we present the galaxy cluster and shear samples, while in Sect.\ \ref{sec:selection_function} the selection function of the cluster sample is discussed. The measurement and stacking of the cluster weak-lensing profiles, along with the calibration of the background redshift distributions, are detailed in Sect.\ \ref{sec:DR4_measure}. Section \ref{sec_modelling} introduces the theoretical models adopted in the analysis, along with the likelihood function and parameter priors. In Sect.\ \ref{sec:DR4_results} we discuss our results and in Sect.\ \ref{sec:DR4_conclusions} we draw our conclusions. \\
\indent Throughout this paper, we assume a concordance flat $\Lambda$ cold dark matter ($\Lambda$CDM) cosmological model. We refer to $M_{200}$ as the mass enclosed within the critical radius $R_{200}$, which is the distance from the cluster centre within which the mean density is $200$ times the critical density of the Universe at the cluster redshift. The AB magnitude system is employed throughout this paper. The base 10 logarithm is referred to as $\log$, while $\ln$ represents the natural logarithm. The statistical analyses presented in this paper are performed using the \texttt{CosmoBolognaLib}\footnote{\url{https://gitlab.com/federicomarulli/CosmoBolognaLib/}} \citep[CBL,][]{cbl}, a set of \textit{free software} C++/Python numerical libraries for cosmological calculations. The linear matter power spectrum is computed with \texttt{CAMB}\footnote{\url{https://camb.info/}} \citep{CAMB}.

\section{Dataset}\label{sec:DR4_data}
This work is based on KiDS-1000 \citep[][]{Kuijken19}, which relies on observations carried out with the OmegaCAM wide-field imager \citep{OmegaCAM}, installed on the European Southern Observatory (ESO) VLT Survey Telescope \citep[VST,][]{VST}. The KiDS-1000 release covers 1006 tiles of about 1 deg$^2$ each. The 2 arcsec aperture photometry in the $u$, $g$, $r$, $i$ bands is provided, with 5$\sigma$ limiting magnitudes of 24.23, 25.12, 25.02 and 23.68 for the four aforementioned bands, respectively. In addition, aperture-matched $Z$, $Y$, $J$, $H$, $K_{\rm s}$ near-infrared photometry from the fully overlapping VISTA Kilo degree INfrared Galaxy survey \citep[VIKING,][]{VIKINGS,VISTA} is included. Based on these photometric bands, the KiDS-1000 galaxy sample includes photo-$z$ estimates, derived with the \texttt{Bayesian Photometric Redshift} \citep[BPZ,][]{Benitez00} code, for more than 100 million objects, with a typical uncertainty of $\sigma^{\rm gal}_z/(1+z)=0.072$ at full depth \citep{Kuijken19}. As discussed in Sect.\ \ref{sec:shear_sample}, the galaxy photo-$z$s adopted in this work differ from those delivered by \citet{Kuijken19}, though they maintain the same precision.

\subsection{Galaxy cluster sample and mass proxy}\label{sec:ClusterSample}
\begin{figure}[t!]
\centering\includegraphics[width = \hsize-1.6cm, height = 9.5cm] {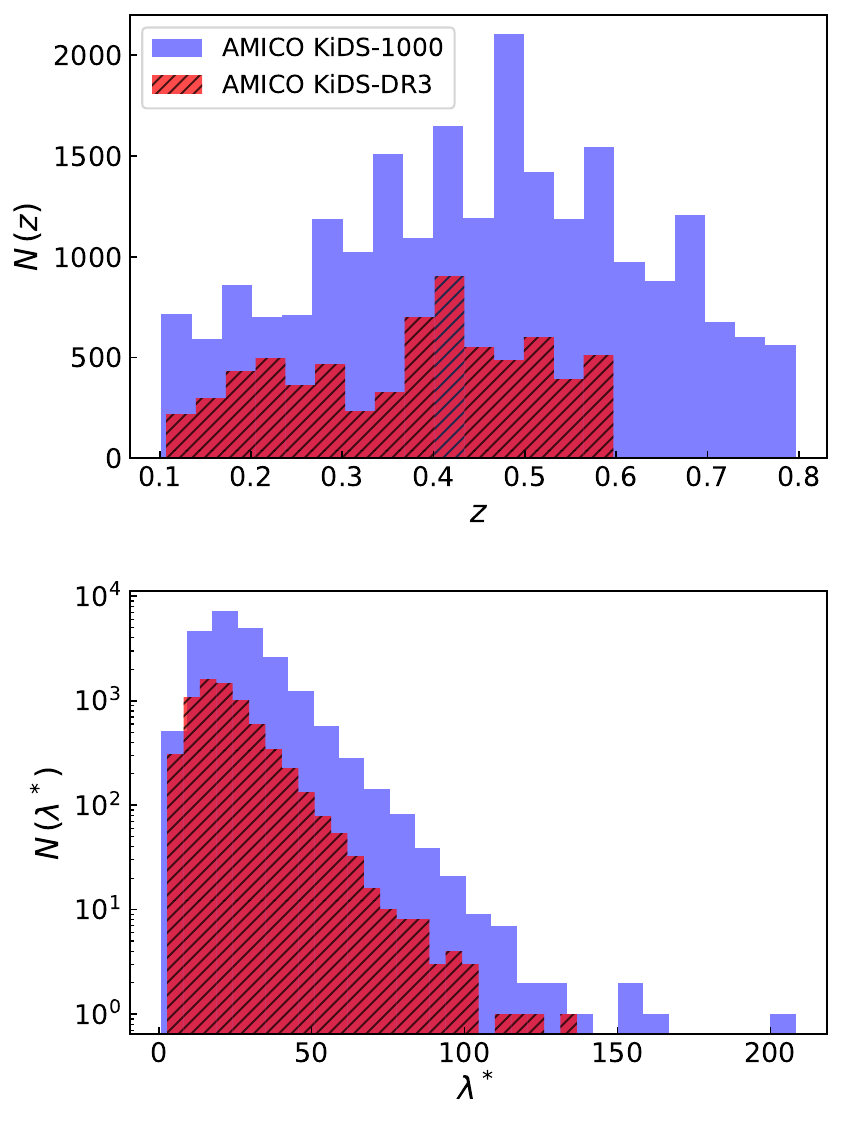}
\caption{Redshift (top panel) and intrinsic richness (bottom panel) distributions of the AMICO galaxy clusters in KiDS-1000 (blue histograms), with ${\rm S/N}>3.5$ and $z\in[0.1,0.8]$, and in KiDS-DR3 (red hatched histograms), following the ${\rm S/N}$ and redshift selections employed in \citetalias{Bellagamba19}. In both panels, no $\lambda^*$ selections have been applied.}
\label{fig:distributions}
\end{figure}
The cluster catalogue on which this work is based, named AMICO KiDS-1000, was built applying the AMICO algorithm \citep{Bellagamba18,Maturi19} to the KiDS-1000 photometric data \citepalias[for a more detailed description, see][]{AMICOKiDS-1000}. AMICO is based on an optimal matched filter and it is one of the two algorithms for cluster identification officially adopted by the Euclid mission \citep{Adam19}, providing highly pure and complete samples of galaxy clusters based on photometric observations. To construct the AMICO KiDS-1000 cluster catalogue, all galaxies located in the regions affected by image artefacts, such as spikes and haloes around bright stars, have been removed. In addition, in this work we exclude the clusters that fall in the regions affected by photometric issues by imposing $\texttt{ARTIFACTS\_FLAG}<2$, and we neglect the tile with central coordinates $\text{R.A.}=196$ and $\text{Dec}=1.5$ due to its low data quality \citepalias[see][]{AMICOKiDS-1000}. This yields a final effective area of 839 deg$^2$, containing cluster detections up to $z=0.9$. Furthermore, we consider clusters having ${\rm S/N}>3.5$ and in the redshift range $z\in[0.1,0.8]$, resulting in a sample of 22\,396 objects. We conservatively exclude cluster detections at $z>0.8$ due to low sample purity and completeness, which also results in poor matching with clusters in existing literature catalogues. Throughout this paper, we employ additional cuts based on the cluster mass proxy that yield 8730 and 7789 clusters in the weak-lensing and count analyses, respectively. The redshift selection is based on the cluster redshifts corrected for the bias derived by \citetalias{AMICOKiDS-1000}, relying on a comparison with a collection of spectroscopic redshifts available within KiDS \citep[referred to as KiDZ,][]{vandenBusch20,vanDenBusch22}, mainly based on the following surveys: Galaxy and Mass Assembly survey data release 4 \citep[GAMA-DR4,][]{Driver11,Liske15,Driver22}, Sloan Digital Sky Survey \citep[SDSS,][]{Alam15}, and 2-degree Field Lensing Survey \citep[2dFLens,][]{Blake16}. The origin of this bias lies in the systematic uncertainties associated with galaxy photo-$z$s. Throughout this paper, we assign these bias-corrected redshifts to each cluster. The inclusion of the statistical uncertainty on this correction leads to a final cluster redshift precision of $\sigma_z/(1+z)=0.014$. In the top panel of Fig.\ \ref{fig:distributions} we show the distribution of the corrected cluster redshifts, along with the redshift distribution of the AMICO KiDS-DR3 cluster sample adopted by \citetalias{Bellagamba19} for weak-lensing mass calibration. In the latter sample, covering an effective area of 377 deg$^2$, detections with ${\rm S/N}>3.5$ and $z\in[0.1,0.6]$ were included, for a total of 6961 clusters. Thus, the KiDS-1000 sample covers a larger redshift range and contains more cluster detections thanks to the larger survey area. Furthermore, thanks to the improved photometry and the inclusion of VIKING bands, the significant incompleteness at $z\sim0.35$ reported by \citet{Maturi19}, caused by the 4000 $\AA$ break between $g$ and $r$ bands, is absent in the current sample. Indeed, the addition of such near-infrared bands improves the accuracy of galaxy photo-$z$s, reducing their dependency on the 4000 $\AA$ break. The cosmological sample used for the AMICO KiDS-DR3 cluster count analysis by \citet{Lesci22_counts} contained fewer objects than the one used by \citetalias{Bellagamba19}, namely 3652. Thus, the number of clusters contributing to the counts in this work is more than twice that of KiDS-DR3. \\
\indent AMICO provides the estimate of two mass proxies for each cluster detection, namely the signal amplitude, $A$, and the intrinsic richness, $\lambda^*$ \citep{Bellagamba18}. The amplitude is obtained by convolving the observed galaxy 3D distribution and the cluster model adopted for the AMICO filter, for which the convolution of a Schechter luminosity function \citep{Schechter1976} and a \citet*[][NFW]{NFW} profile is assumed. Following \citet{Hennig17}, a concentration parameter of $c=3.59$ is adopted in the NFW model \citepalias[for more details, see][]{AMICOKiDS-1000}. The contribution by the Brightest Cluster Galaxy (BCG) is not included in the AMICO filter, in order to prevent biases due to photo-$z$ uncertainties. Indeed, the BCG signal often dominates over other cluster galaxies. The intrinsic richness is defined as follows:
\begin{equation}\label{lambda}
\lambda^*_j=\sum\limits_{i} P_i(j)\;\;\;\;\text{with}\;\;\;\;
\begin{cases}
m_i<m^*(z_j)+1.5 \\ R_i(j)<{R_{\rm max}(z_j)}
\end{cases}
,
\end{equation}
where $P_i(j)$ is the probability assigned by AMICO to the $i$th galaxy of being a member of a given detection $j$, while $z_j$ is the redshift of the $j$th detected cluster, $m_i$ is the magnitude of the $i$th galaxy, $R_i$ corresponds to the distance of the $i$th galaxy from the centre of the cluster, $R_{\rm max}(z_j)$ is the radius enclosing a mass of $M_{200}=10^{14}h^{-1}M_\odot$, which is the typical mass of a galaxy cluster, and $m^*$ corresponds to the $r$-band luminosity at the knee of the Schechter function used in the cluster model assumed for the AMICO filter. Thus, $\lambda^*$ represents the sum of the galaxy membership probabilities, constrained by the conditions given in Eq.\ \eqref{lambda}. As shown by \citet{Bellagamba18,Maturi23,Toni24,Toni25}, the sum of the membership probabilities is an excellent estimator of the true number of member galaxies. We note that the enclosed mass $M_{200}$ is kept fixed in the AMICO detection process to reduce the noise in $\lambda^*$ measurements. \\
\indent In this work, we focused on the calibration of the $\log\lambda^*-\log M_{200}$ relation, using it to model cluster abundance and derive constraints on cosmological parameters. Compared to the amplitude $A$, $\lambda^*$ is less dependent on the cluster model assumed for AMICO detection \citep{Maturi19}. Furthermore, the weight assigned to each galaxy in Eq.\ \eqref{lambda} is by construction lesser than or equal to 1, given the definition of probability. This implies that while some galaxies with especially high weights might increase the value of $A$, their impact on $\lambda^*$ would be milder. Nevertheless, given that $A$ proved to be an effective proxy of the number of cluster galaxies in simulations \citep{Bellagamba18}, we will thoroughly investigate the performance of this mass proxy in future studies based on KiDS data. In the bottom panel of Fig.\ \ref{fig:distributions} we show the $\lambda^*$ distribution of the cluster sample. As detailed in \citetalias{AMICOKiDS-1000}, the $\lambda^*$ distribution of KiDS-DR3 clusters is remarkably different from the one in KiDS-1000, due to the improvement of galaxy photo-$z$s in the latter survey thanks to the inclusion of VIKING photometry. Specifically, the better galaxy photo-$z$s led to the increase of $\lambda^*$ for all KiDS-1000 galaxy clusters. We also note that, as a consequence of the ${\rm S/N}$ cut applied in this analysis, the clusters with the lowest $\lambda^*$ values shown in Fig.\ \ref{fig:distributions} are predominantly located at low redshifts \citepalias{AMICOKiDS-1000}. As discussed in the following, these objects are not included in the analysis due to the low sample purity at small $\lambda^*$.

\subsection{Shear sample}\label{sec:shear_sample}
\begin{figure}[t!]
\centering\includegraphics[width = \hsize-1cm, height = 12cm] {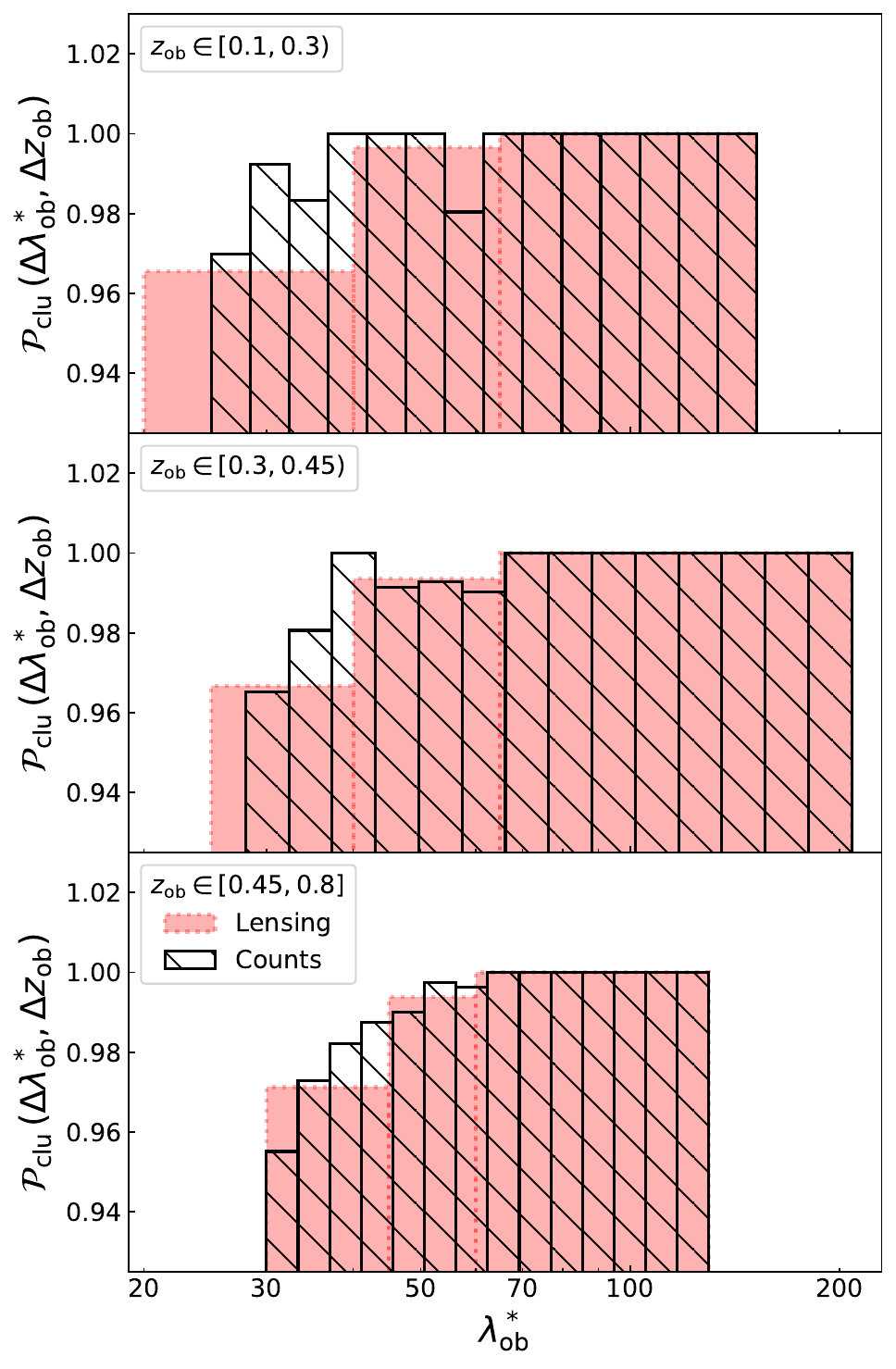}
\caption{Purity of the cluster sample as a function of $\lambda^*_{\rm ob}$, in the redshift bins $z\in[0.1,0.3)$ (top panel), $z\in[0.3,0.45)$ (middle panel), and $z\in[0.45,0.8]$ (bottom panel). The pink histograms show the purity derived in the $\lambda^*_{\rm ob}$ bins used for cluster weak-lensing measurements, while the black hatched histograms display the purity in the bins used for cluster counts.}
\label{fig:P_clu}
\end{figure}
To estimate the weak-lensing signal of the AMICO KiDS-1000 galaxy clusters, we based our analysis on the KiDS-1000 gold shear catalogue \citep{Wright20,Hildebrandt21,Giblin21}. This sample comprises about 21 million galaxies, covering an effective area of $777.4$ deg$^2$, with a weighted number density of $n^{\rm gold}_{\rm eff}= 6.17$ arcmin$^{-2}$. It includes weak-lensing shear measurements from the deep KiDS $r$-band observations, as these are the ones with better seeing properties and yield the highest source density. The estimator used in this analysis is \textit{lensftit} \citep{Miller07,Fenech17}, a likelihood-based model fit method that has been used successfully in the analyses of other datasets, such as the Canada France Hawaii Telescope Lensing Survey \citep[CFHTLenS,][]{Miller13} and the Red Cluster Survey \citep{Hildebrandt16}. 
The photo-$z$ calibration performed for cosmic shear studies \citep{Asgari21}, based on deep spectroscopic reference catalogues that are weighted with the help of self-organising maps \citep[SOMs,][]{Kohonen1982,Wright20}, has been performed by \citet{Hildebrandt21} in the redshift range $z\in[0.1,1.2]$.\\
\indent As in \citet{Kuijken19}, \citetalias{AMICOKiDS-1000} used BPZ \citep{Benitez00} to derive the galaxy photo-$z$s for cluster detection. However, the prior redshift probability distribution used in BPZ by \citet{Kuijken19} appears to generate a remarkable redshift bias for bright, low-redshift galaxies, despite the fact that it reduces uncertainties and catastrophic failures for faint galaxies at higher redshifts. For this reason, \citetalias{AMICOKiDS-1000} chose the same prior on redshift probabilities adopted in the KiDS-DR3 analysis \citep{deJong17}. This leads to a galaxy photo-$z$ uncertainty of $\sigma^{\rm gal}_z/(1+z)=0.074$ given the $r$ magnitude limit $r<23$, in agreement with \citet{Kuijken19}. In this analysis, we adopted the galaxy photo-$z$s derived in this way. As we shall discuss later in Sect.\ \ref{Sec_sys_2}, we calibrated the galaxy redshift distribution using the SOM algorithm provided by \citet{Wright20}.

\section{Cluster selection function and blinding}\label{sec:selection_function}
\begin{figure}[t!]
\centering\includegraphics[width = \hsize, height = 9.5cm] {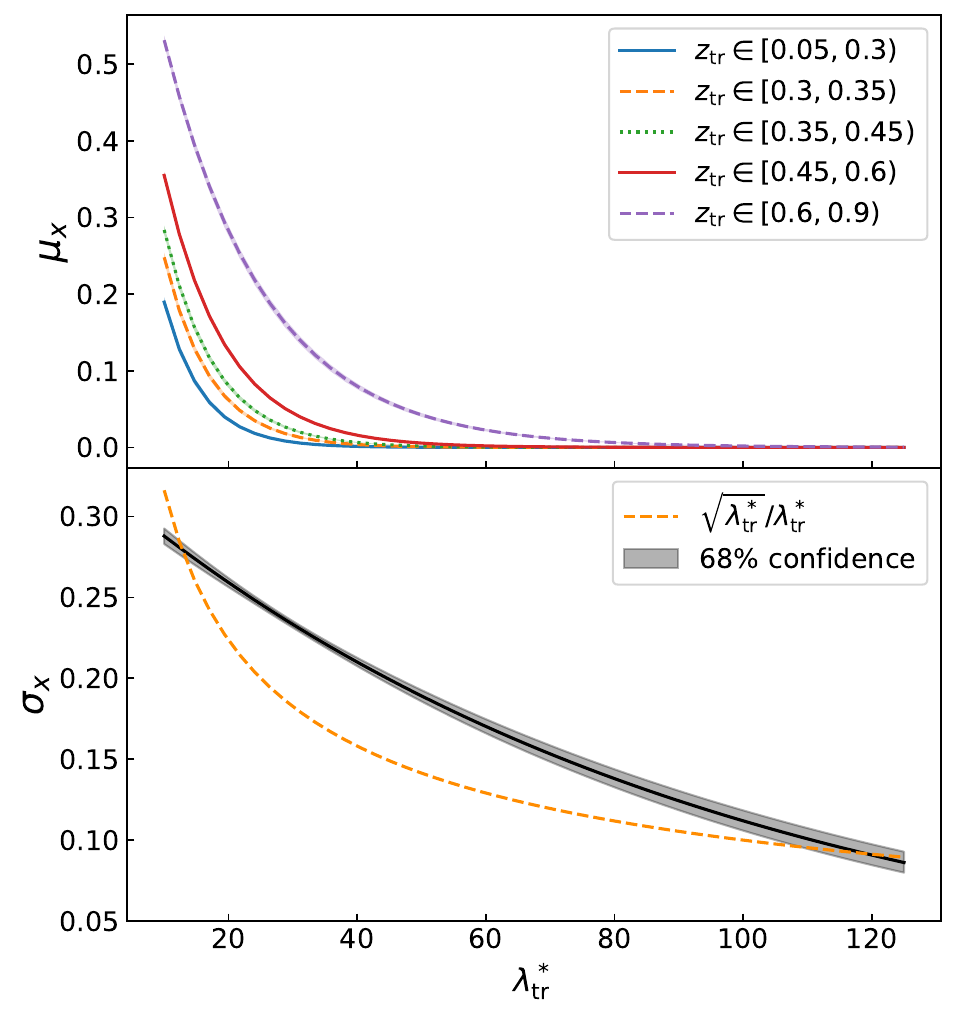}
\caption{Top panel: Mean of $P(x\,|\,\lambda^*_{\rm tr}, z_{\rm tr})$, namely $\mu_x$, as a function of $\lambda^*_{\rm tr}$ and in different bins of $z_{\rm tr}$ (see legend). Bottom panel: Grey band represents the 68\% confidence level of the $\sigma_x$ model, while the dashed orange line shows the Poisson relative uncertainty on $\lambda^*_{\rm tr}$.}
\label{fig:Plambda}
\end{figure}

\begin{figure}[t!]
\centering\includegraphics[width = \hsize-1cm, height = 5.5cm] {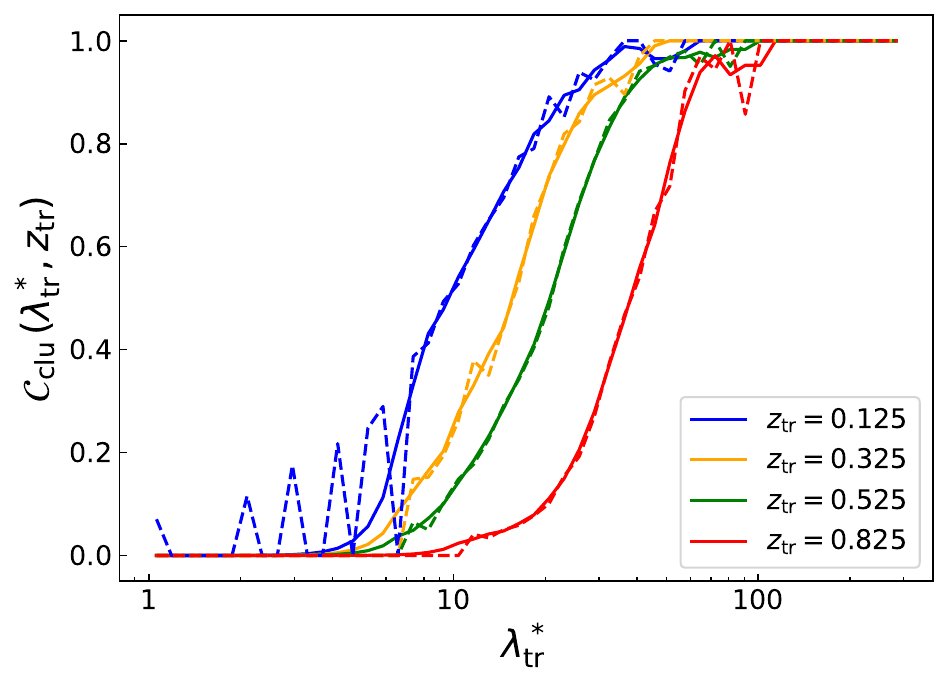}
\caption{Completeness of the cluster sample as a function of $\lambda^*_{\rm tr}$, for some values of $z_{\rm tr}$ (see the legend). Dashed lines display the completeness measurements, while solid lines show the completeness smoothed with Chebyshev polynomials.}
\label{fig:C_clu}
\end{figure}

In this section, we delve into the description of the cluster selection function estimates and summarise the blinding of the statistical analysis presented in this paper. To derive the selection function of the galaxy cluster sample, namely its completeness and the uncertainties associated with cluster observables, along with its purity, we used the data-driven mock cluster catalogue produced by \citetalias{AMICOKiDS-1000}. This mock sample was constructed using the Selection Function extrActor (SinFoniA) code \citep{Maturi19,AMICOKiDS-1000}. SinFoniA created mock galaxy catalogues based on the KiDS-1000 photometric data, randomly extracting field galaxies and cluster members according to their measured membership probabilities using a Monte Carlo approach. These mock catalogues maintain the autocorrelations of galaxies and clusters, the cross-correlation between clusters and galaxies, the survey footprint, and masks, along with the local variations in data properties caused by the varying depth of the survey. The mock catalogues are then processed using AMICO, allowing for a comparison between the detections and the clusters injected into the mock catalogues, ultimately deriving the final statistical properties of the sample.\\
\indent We selected all AMICO detections in this mock catalogue that have a ${\rm S/N}>3.5$, which corresponds to the ${\rm S/N}$ threshold used to define the cluster sample described in Sect.\ \ref{sec:ClusterSample}. Then, we measured the cluster purity, $\mathcal{P}_{\rm clu}$, defined as the number of AMICO detections having a true cluster counterpart over the total number of AMICO detections, as a function of observed intrinisc richness and redshift, namely $\lambda^*_{\rm ob}$ and $z_{\rm ob}$, respectively. Figure \ref{fig:P_clu} displays the values of $\mathcal{P}_{\rm clu}$ related to each bin considered for the cluster count and weak-lensing measurements detailed in the following. We remind that the impurities of the cluster sample yield abundance measurements that are biased high, while the stacked weak-lensing measurements are biased low. Indeed, weak-lensing profiles derived around random positions in the sky are expected to be statistically consistent with zero. We account for these effects in the models described in Sect.\ \ref{sec_modelling}. \\
\indent To estimate the uncertainties on the observed intrinisc richness, $\lambda^*_{\rm ob}$, in the mock catalogue we measure the probability density function $P(\Delta x\,|\,\Delta\lambda^*_{\rm tr},\Delta z_{\rm tr})$, where $x=(\lambda^*_{\rm ob}-\lambda^*_{\rm tr})/\lambda^*_{\rm tr}$. Here, $\lambda^*_{\rm tr}$ is the true intrinisc richness, while $\Delta\lambda^*_{\rm tr}$ and $\Delta z_{\rm tr}$ represent bins of true intrinsic richness and true redshift, respectively. We find that $P(\Delta x\,|\,\Delta\lambda^*_{\rm tr},\Delta z_{\rm tr})$ can be accurately modelled as a Gaussian for any $\Delta z_{\rm tr}$ and $\Delta \lambda^*_{\rm tr}$. Through a joint modelling of this distribution in different bins of $z_{\rm tr}$ and $\lambda^*_{\rm tr}$, performed by means of a Bayesian analysis based on a Markov chain Monte Carlo (MCMC), assuming a Gaussian likelihood, Poisson uncertainties, and large uniform priors on the free parameters presented in the following, we find that the mean of $P(\Delta x\,|\,\Delta\lambda^*_{\rm tr},\Delta z_{\rm tr})$ can be expressed as
\begin{equation}\label{eq:Plambda_mean}
    \mu_x = \exp[- \lambda^*_{\rm tr} \, (A_\mu + B_\mu \, z_{\rm tr})]\,,
\end{equation}
where $A_\mu=0.198\pm0.005$ and $B_\mu = -0.179\pm0.007$, while its standard deviation has the following form
\begin{equation}\label{eq:Plambda_std}
    \sigma_x = A_\sigma \exp(- B_\sigma \lambda^*_{\rm tr})\,,
\end{equation}
where $A_\sigma = 0.320 \pm 0.008$ and $B_\sigma = 0.011\pm0.001$. Equations \eqref{eq:Plambda_mean} and \eqref{eq:Plambda_std} are represented in Fig.\ \ref{fig:Plambda}. We notice that, differently from $\mu_x$, $\sigma_x$ depends only on $\lambda^*_{\rm tr}$. Indeed, we do not find any significant trend with $z_{\rm tr}$ for this quantity. Due to masking, blending of cluster detections, and projection effects, $\sigma_x$ is up to $35\%$ larger than the Poisson relative uncertainty on $\lambda^*_{\rm tr}$. In addition, we note that $\mu_x$ decreases with $\lambda^*_{\rm tr}$ and increases with $z_{\rm tr}$ mainly due to projection effects \citep[see e.g.][]{Myles21,Cao25}. We also observe that, as optical cluster finders preferentially select haloes embedded within filaments aligned with the line of sight, the $\lambda^*$ bias due to projection effects is positively correlated with a cluster shear bias, amounting to up to 25\% in the two-halo regime  \citep{Sunayama20,Wu22,Sunayama23,Zhou23}. Nonetheless, as will be discussed in Sect.\ \ref{sec:DR4_results:sys}, this weak-lensing bias has a negligible impact on our results thanks to our choice of the cluster-centric radial range. Following an approach analogous to the one used to estimate $P(x\,|\,\lambda^*_{\rm tr}, z_{\rm tr})$, we find that $P(z_{\rm ob}\,|\,\lambda^*_{\rm tr}, z_{\rm tr})$ is Gaussian and its mean and standard deviation do not significantly depend on $\lambda^*_{\rm tr}$. We also remark that, as mentioned in Sect.\ \ref{sec:ClusterSample}, throughout this paper we use the cluster redshift uncertainties derived by \citetalias{AMICOKiDS-1000} from the comparison of the AMICO sample with the KiDZ spectroscopic redshifts. \\
\indent While $P(x\,|\,\lambda^*_{\rm tr}, z_{\rm tr})$ is a probability density that allows one to quantify how many clusters fall in a nearby $\lambda^*_{\rm ob}$ bin due to observational uncertainties, the amount of clusters recovered from the underlying true distribution is given by the completeness of the sample $\mathcal{C}_{\rm clu}(\lambda^*_{\rm tr},z_{\rm tr})$. It is defined as the number of cluster detections with ${\rm S/N}>3.5$ having a true counterpart divided by the number of true clusters, as a function of $\lambda^*_{\rm tr}$ and $z_{\rm tr}$. Figure \ref{fig:C_clu} displays the measured completeness and its smoothed version based on Chebyshev polynomials, as discussed in \citetalias{AMICOKiDS-1000}, for some values of $z_{\rm tr}$. Although we evaluated $\mathcal{C}_{\rm clu}$ in seven redshift intervals ranging from 0.1 to 0.9, only four are displayed to avoid overcrowding the plot. These smoothed measurements are then interpolated as a function of $z_{\rm tr}$ and $\lambda^*_{\rm tr}$, in order to include $\mathcal{C}_{\rm clu}$ in the theoretical models described in Sect.\ \ref{sec_modelling}. We note that $\mathcal{C}_{\rm clu}$ is estimated up to $z_{\rm tr}=0.9$ to account for uncertainties in the observed redshift, $z_{\rm ob}$, which is capped at $z_{\rm ob}=0.8$ (see the likelihood models in Sect.\ \ref{sec_modelling}). \\
\indent As detailed in \citetalias{AMICOKiDS-1000}, cluster completeness estimates remained blind until the complete definition and run of the pipeline presented in this paper. Specifically, GFL received three versions of $\mathcal{C}_{\rm clu}$, two of which were intentionally biased in relation to the true one by MM, and built up the pipeline using only one of them. Except for MM, none of the authors knew the true value of $\mathcal{C}_{\rm clu}$. More details on this blinding strategy and on its outcome are provided in Appendix \ref{appendix:blinding} and \citetalias{AMICOKiDS-1000}. \\
\indent We note that, due to the complexity of the cosmological analysis presented in this work, driven by both observational and theoretical systematic uncertainties, we do not select the cluster sample based on purity and completeness estimates. Instead, we adopt a data-driven approach, modelling the sample assuming alternative $\lambda^*_{\rm ob}$ and $z_{\rm ob}$ cuts to identify the combination that maximises the quality of the fit and the number of clusters included in the analysis. The impact of this sample selection is assessed in Sect. \ref{sec:DR4_results}.

\section{Stacked weak-lensing profiles measurement}\label{sec:DR4_measure}
\begin{figure*}[t!]
\centering\includegraphics[width = \hsize-1.5cm, height = 5.3cm] {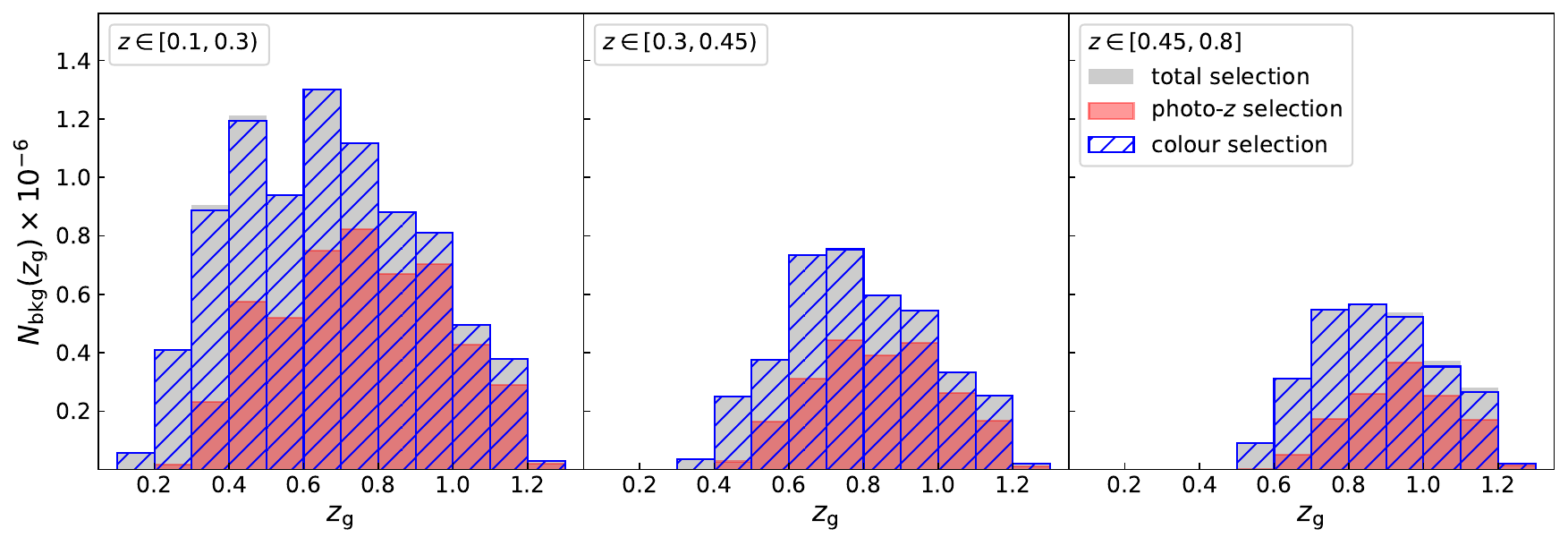}
\caption{Number of galaxies that satisfy the photo-$z$ selection (red histograms) and the colour selection (blue hatched histograms) as a function of $z_{\rm g}$, in different cluster redshift bins, namely $z\in[0.1,0.3)$ (left panel), $z\in[0.3,0.45)$ (middle panel), and $z\in[0.45,0.8]$ (right panel). The grey histograms show the galaxies selected through the total selection criterion, defined in Eq.\ \eqref{eq:tot_selection}.}
\label{fig:n_eff}
\end{figure*}
In this section, we present the pipeline used to measure the weak-lensing profiles of the AMICO KiDS-1000 galaxy clusters. We detail the galaxy selections employed to define the background samples associated with the clusters in our sample. Then, we introduce a density profile estimator based on observed galaxy ellipticities and discuss the stacking of our measurements. Lastly, we assess the purity of the background galaxy samples and calibrate their redshift distributions. To perform the cluster weak-lensing measurements, we assume a flat Universe with $H_0=70$ km s$^{-1}$ Mpc$^{-1}$ and $\Omega_{\rm m}=0.3$, where $H_0$ is the Hubble constant and $\Omega_{\rm m}$ is the matter density parameter. We note that these cosmological assumptions do not affect the final results. Indeed, they induce geometric distortions in the weak-lensing measurements that are properly modelled in the cosmological analysis (see Sect.\ \ref{sec:modelling:stacked_g}).

\subsection{Selection of background sources}\label{Sec_background}
If galaxies that belong to the clusters or in the foreground are mistakenly considered as background, the measured lensing signal can be significantly diluted \citep[see e.g.][]{Broadhurst05,Medezinski07,Dietrich19,Rau24}. To minimise such contamination, we employed stringent selections based on the galaxy photo-$z$ probability density function, denoted as $p(z_{\rm g})$, where $z_{\rm g}$ represents the galaxy redshift, and on galaxy colours. This approach is in line with previous works, such as \citet{Sereno17} and \citetalias{Bellagamba19}. \\
\indent Specifically, to exclude galaxies with a significant probability of being at a redshift equal to or lower than the cluster redshift, we adopted the following photo-$z$ selection
\begin{equation}\label{eq:photoz_sel}
z_{\rm g,min} > z+0.05\,,
\end{equation}
where $z_{\rm g,min}$ is the minimum of the interval containing 95\% of the probability around the first mode of $p(z_{\rm g})$, namely $\overline{z}_{\rm g}$, while $z$ is the mean redshift of the cluster. In Eq.\ \eqref{eq:photoz_sel}, the 0.05 buffer is larger than the uncertainty associated with cluster redshifts, that is $\sigma_z/(1+z)=0.014$. A larger buffer, accounting for the photo-$z$ quality of the background sources, is not expected to significantly impact our analysis. Indeed, as we will discuss in Sect.\ \ref{sec:WL_stacking_2}, the selection in Eq.\ \ref{eq:photoz_sel} plays a marginal role in defining the background galaxy samples. \\
\indent In addition to Eq.\ \eqref{eq:photoz_sel}, we considered a colour selection in order to enhance the background sample completeness, similarly to \citet[][]{Oguri12,Medezinski18,Dietrich19,Schrabback21}. Specifically, we relied on the colour selection calibrated by \citet[][referred to as L24 hereafter]{Lesci_colours}, based on $griz$ photometry. The $gri$ photometry used to calibrate this selection, corresponding to that employed in SDSS \citep[][]{Aihara11}, closely matches that used in KiDS \citep{Kuijken19}, while the $z$ band is redder in KiDS. Any biases deriving from this difference are accounted for in the self-organising map calibration presented in Sect.\ \ref{Sec_sys_2}. When applied to reference photometric samples, such as COSMOS \citep{Laigle16,Weaver22}, this selection provides a background completeness ranging from 30\% to 84\% and a purity larger than 97\% in the lens redshift range $z\in[0.2,0.8]$. To account for cluster redshift uncertainties, we conservatively applied the colour selection at $\tilde{z}=z+0.05$, where 0.05 corresponds to the buffer used also in Eq.\ \eqref{eq:photoz_sel}. For lenses at $\tilde{z}<0.2$, namely below the minimum of the domain covered by the \citetalias{Lesci_colours} selection, we considered the selection valid for $\tilde{z}=0.2$. We also imposed that candidate background galaxies can pass the colour selection if the following condition is satisfied:
\begin{equation}\label{eq:photoz_peak_selection}
\overline{z}_{\rm g}>z+0.05\,,
\end{equation}
where the 0.05 buffer is the same as in Eq.\ \eqref{eq:photoz_sel}. The complete background selection criterion is defined as follows
\begin{equation}\label{eq:tot_selection}
\text{ (photo-$z$ selection) or (colour selection) }\,.\end{equation}
To assess the performance of photo-$z$ and colour selections, we counted the number of galaxies that meet such selection criteria. We considered the same cluster redshift bins adopted for the weak-lensing measurements presented in the following analysis, namely $z\in[0.1,0.3)$, $z\in[0.3,0.45)$, and $z\in[0.45,0.8]$. Figure \ref{fig:n_eff} shows that the colour selection provides the largest number of background sources, namely $N_{\rm bkg}(z_{\rm g})$, at low $z_{\rm g}$ in all cluster redshift bins. This is expected, as the photo-$z$ selection in Eq.\ \eqref{eq:photoz_sel} excludes a greater number of galaxies close to the clusters. On average, the photo-$z$ selection appears to be more conservative compared to the colour selection, enhancing the number of background galaxies by at most 2\% compared to the case of the colour selection alone. The purity of the background samples and the calibration of their redshift distributions are discussed in Sect.\ \ref{Sec_sys_2}.

\subsection{Measurement of the cluster profiles}\label{sec:WL_stacking}
\begin{figure*}[t!]
\centering\includegraphics[width = \hsize-1.6cm, height = 9cm] {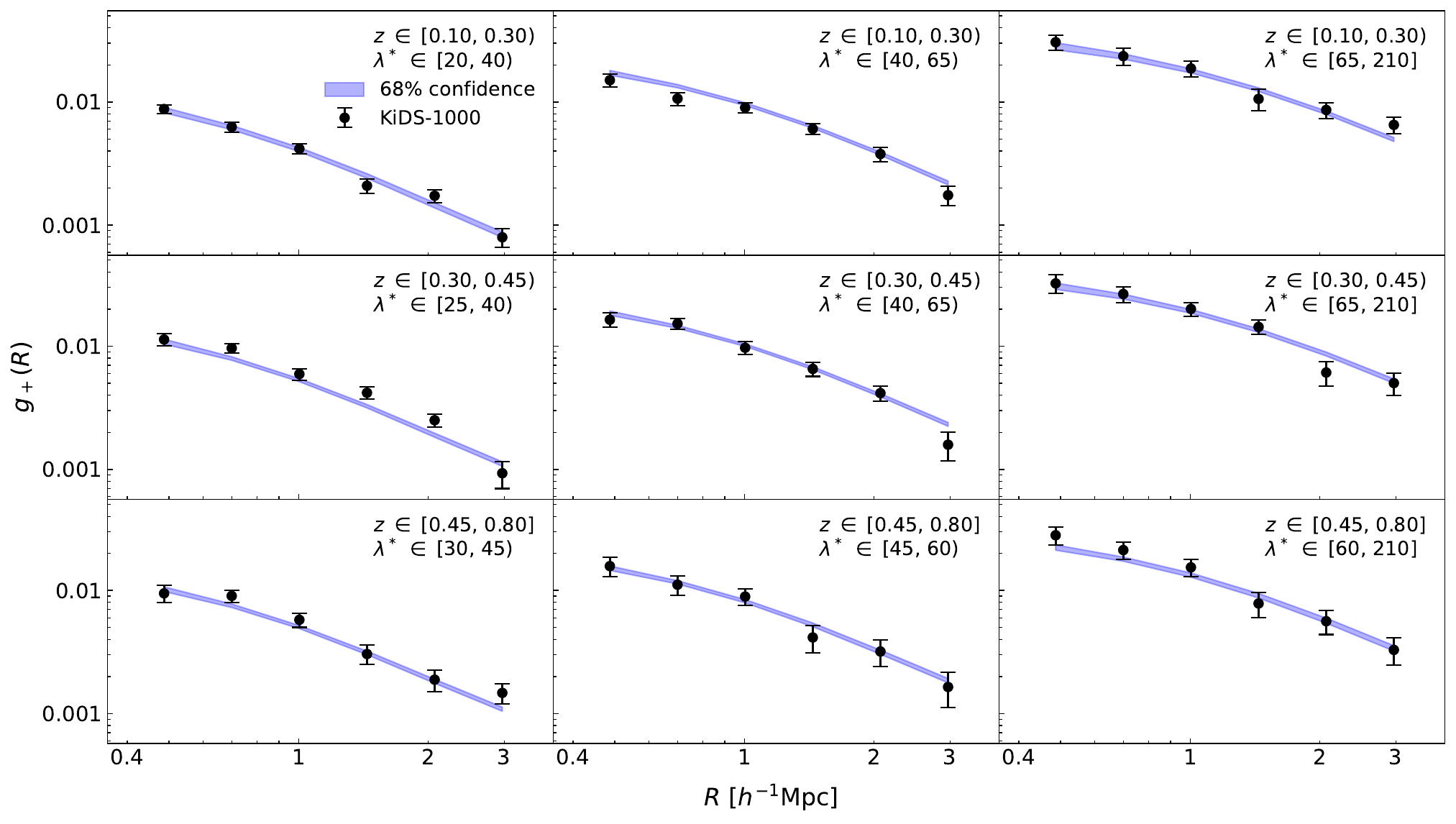}
\caption{Stacked $g_+(R)$ profiles of the AMICO KiDS-1000 galaxy clusters in bins of $z$ (increasing from top to bottom) and $\lambda^*$ (increasing from left to right). The black dots show the measures, while the error bars are the sum of bootstrap errors and statistical uncertainties coming from systematic errors (see Sect.\ \ref{sec:modelling:likelihood}). The blue bands represent the 68\% confidence levels derived from the multivariate posterior of all the free parameters considered in the joint analysis of counts and weak lensing.}
\label{fig:g_t}
\end{figure*}
The tangential shear, $\gamma_+$, is linked to the excess surface density profile of the lens, $\Delta\Sigma_+$, via \citep[see e.g.][]{Sheldon04}
\begin{equation}
\Delta\Sigma_+(R)=\overline{\Sigma}(<R)-\Sigma(R)=\Sigma_{\rm crit}\gamma_+(R),
\end{equation}
where $\Sigma(R)$ is the mass surface density, $\overline{\Sigma}(<R)$ is its mean within the radius $R$, and $\Sigma_{\rm crit}$ is the critical surface density, expressed as \citep{Bartelmann01}
\begin{equation}\label{eq:sigma_crit}
\Sigma_{\rm crit} \equiv \frac{c^2}{4\pi G}\frac{D_{\rm s}}{D_{\rm l}D_{\rm ls}}\,,
\end{equation}
where $D_{\rm s}$, $D_{\rm l}$, and $D_{\rm ls}$ are the observer-source, observer-lens, and lens-source angular diameter distances, respectively. \\
\indent The weak-lensing quantity directly related to galaxy ellipticities is the reduced shear, namely $g=\gamma/(1-\kappa)$, where $\kappa\equiv\Sigma/\Sigma_{\rm crit}$ is the convergence \citep{Schneider1995}. The observed galaxy ellipticity can be conveniently separated into a tangential component, $e_+$, and a cross component, $e_\times$, as follows \citep[see e.g.][]{Viola15}:
\begin{equation}
    \begin{pmatrix}
        e_+ \\
        e_\times
    \end{pmatrix}
    =
        \begin{pmatrix}
        -\cos(2\varphi) & -\sin(2\varphi) \\
        \sin(2\varphi) & -\cos(2\varphi)
    \end{pmatrix}
    \begin{pmatrix}
        e_1 \\
        e_2
    \end{pmatrix}
    \,,
\end{equation}
where $\varphi$ is the position angle of the source with respect to the lens centre, while $e_1$ and $e_2$ are the observed ellipticity components. Therefore, the reduced tangential shear profile, $g_+$, of a cluster labelled as $k$ can be estimated as follows:
\begin{equation}\label{eq:DeltaSigma_measure}
g_{+,k}(R_j)=\left( \frac{\sum_{i\in j}\;w_i\,e_{+,i}}{\sum_{i\in j}\;w_i} \right)\frac{1}{1+\mathcal{M}_j},
\end{equation}
where $j$ is the radial annulus index, with an associated average projected radius $R_j$, corresponding to the central value of the radial bin. The latter is a good approximation of the so-called effective radius \citep[see the Appendix in][]{Sereno17} and we adopted it because its measurement does not require any assumption on the lens properties. In addition, $w_i$ is the statistical weight assigned to the measure of the source ellipticity of the $i$th galaxy \citep{Sheldon04}, satisfying the background selection (Eq.\ \ref{eq:tot_selection}), falling in the $j$th radial annulus of the $k$th cluster. The term $\mathcal{M}_j$ is the average correction due to the multiplicative noise bias in the shear estimate, defined as
\begin{equation}\label{eq:M_bias}
\mathcal{M}_j=\frac{\sum_{i\in j}\;w_i\,m_i}{\sum_{i\in j}\;w_i},
\end{equation}
where $m_i$ is the multiplicative shear bias of the $i$th galaxy. For KiDS-1000, \citet{Giblin21} derived $m$ estimates in five redshift bins, which are consistent within $1\sigma$ with zero and have standard deviation ranging from $10^{-2}$ to $2\times10^{-2}$, decreasing with redshift. Since the galaxy selection and photo-$z$ calibration considered in this work differ from those adopted by \citet{Giblin21}, we extracted for each galaxy a value of $m$ from the uniform distribution $[-0.05,0.05]$, covering the 2$\sigma$ intervals of the $m$ distributions derived by \citet{Giblin21}. This prior leads to $\mathcal{M}_j\simeq0$. Thus, in practice, $\mathcal{M}_j$ does not depend on the radial annulus. To propagate the statistical uncertainty on $m$, namely $\sigma_m$, into the final results, we included $\sigma_m$ into the covariance matrix (see Sect.\ \ref{sec:modelling:likelihood}). Indeed, $\sigma_m$ is also a relative uncertainty on $g_+$. We conservatively assumed $\sigma_m=0.02$, corresponding to the largest $m$ uncertainty value derived for the tomographic bins considered by \citet{Giblin21}. In principle, a term accounting for the shear additive bias, originated by detector-level effects such as inefficiencies in the charge transfer \citep{Fenech17}, should be included in Eq.\ \eqref{eq:DeltaSigma_measure}. \citet{Giblin21} showed that this bias is consistent with zero and has an associated statistical uncertainty of the order of $10^{-4}$, which is negligible in our analysis because it is one order of magnitude lower than the average statistical uncertainty associated with our stacked $g_+$ measurements.

\subsection{Stacked profiles}\label{sec:WL_stacking_2}
\begin{figure}[t!]
\centering\includegraphics[width = \hsize-1cm, height = 6.5cm] {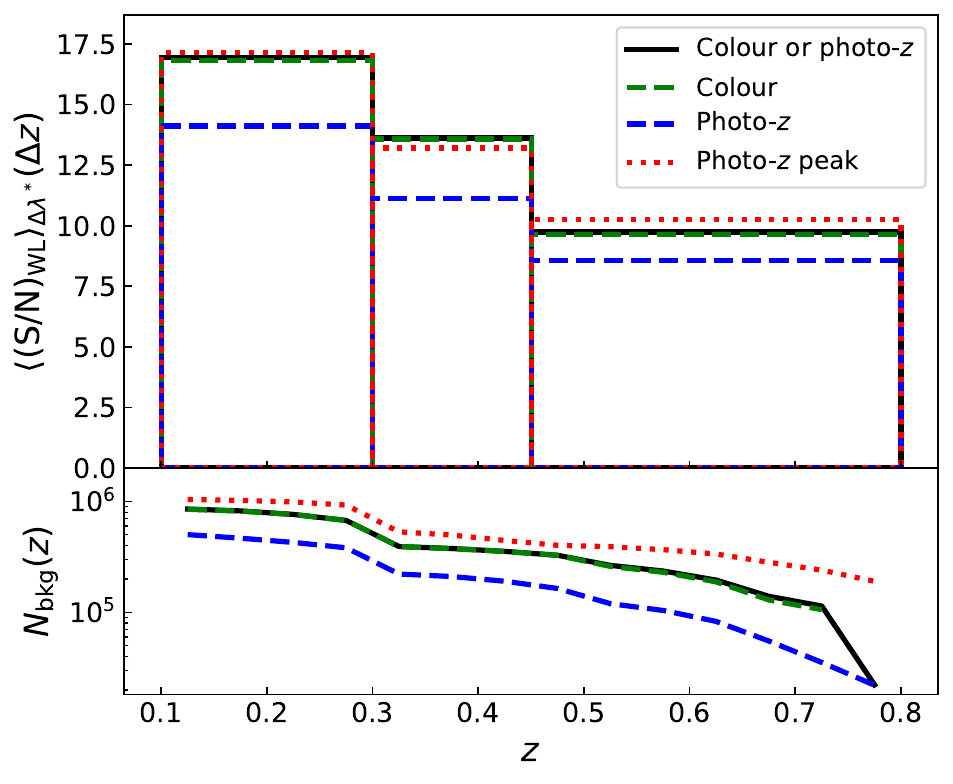}
\caption{Top panel: $\langle({\rm S/N})_{\rm WL}\rangle_{\Delta\lambda^*}$ defined in the cluster redshift bins adopted for the stacking, namely $z\in[0.1,0.3)$, $z\in[0.3,0.45)$, and $z\in[0.45,0.8]$. Bottom panel: Number of background sources as a function of $z$. In both panels, quantities obtained from the combination of colour and photo-$z$ selections (solid black lines), colour selection only (dashed green lines), photo-$z$ selection only (dashed blue lines), and photo-$z$ peak selection are shown. The black and green curves are almost overlapping.}
\label{fig:background_alternative}
\end{figure}

\begin{table}[t]
\small
\caption{\label{tab:sample}Properties of the cluster subsamples used to measure the stacked weak-lensing signal.}
  \centering
    \begin{tabular}{c c c c c} 
      $z$ range & $\lambda^*$ range & $N$ & Median $z$ & Median $\lambda^*$ \\
      \hline
      \rule{0pt}{4ex}
      $[0.1,\,0.3)$ & $[20,\,40)$ & 1411 & 0.24 & 25 \\\rule{0pt}{1.5ex}
      $[0.1,\,0.3)$ & $[40,\,65)$ & 255 & 0.25 & 46 \\\rule{0pt}{1.5ex}
      $[0.1,\,0.3)$ & $[65,\,210]$ & 47 & 0.25 & 78 \\\rule{0pt}{2.5ex}
      $[0.3,\,0.45)$ & $[25,\,40)$ & 1536 & 0.38 & 30 \\\rule{0pt}{1.5ex}
      $[0.3,\,0.45)$ & $[40,\,65)$ & 525 & 0.38 & 46 \\\rule{0pt}{1.5ex}
      $[0.3,\,0.45)$ & $[65,\,210]$ & 90 & 0.38 & 76 \\\rule{0pt}{2.5ex}
      $[0.45,\,0.8]$ & $[30,\,45)$ & 3616 & 0.64 & 35 \\\rule{0pt}{1.5ex}
      $[0.45,\,0.8]$ & $[45,\,60)$ & 930 & 0.63 & 50 \\\rule{0pt}{1.5ex}
      $[0.45,\,0.8]$ & $[60,\,210]$ & 320 & 0.61 & 68
    \end{tabular}
    \tablefoot{The bins of $z$ and $\lambda^*$ are shown in the first and second columns, respectively. The third column lists the number of clusters, $N$, contributing to the stacked signal in the given $z$ and $\lambda^*$ bin. In the last two columns, the median cluster redshift and richness are reported, respectively.}
\end{table}
For most of the clusters in the sample, the weak-lensing signal is too low to precisely measure their density profiles. For this reason, we derived average $g_+$ estimates by stacking the signal from ensembles of clusters, selected according to their mass proxy and redshift. Specifically, the reduced shear profile in the $p$th bin of $\lambda^*$ and the $q$th bin of $z$, namely $\Delta\lambda^*_p$ and $\Delta z_q$, respectively, is expressed as
\begin{equation}\label{eq:DeltaSigma_measure_stack}
g_+(R_j,\Delta\lambda^*_p,\Delta z_q) = \frac{\sum_{k\in \Delta\lambda^*_p,\Delta z_q}W_{k,j}\,g_{+,k}(R_j)}{\sum_{k\in \Delta\lambda^*_p,\Delta z_q}W_{k,j}},
\end{equation}
where $g_{+,k}$ is given by Eq.\ \eqref{eq:DeltaSigma_measure}, $k$ runs over all clusters falling in the bins of $\lambda^*$ and $z$, while $W_{k,j}$ is the total weight
for the $j$th radial bin of the $k$th cluster, estimated as
\begin{equation}\label{eq:W_lensing}
W_{k,j} = \sum_{i\in j} w_i\,,
\end{equation}
where $i$ runs over the background galaxies in the $j$th radial bin. As an alternative to $R_j$ in Eq.\ \eqref{eq:DeltaSigma_measure_stack}, namely the central value of the $j$th radial bin, we computed also an effective radius, $R_{j}^{\rm eff}$, as follows \citep{Umetsu14,Sereno17}:
\begin{equation}
R_{j}^{\rm eff} = \frac{\sum_{k\in \Delta\lambda^*_p,\Delta z_q}\,W_{k,j}R^{\rm eff}_{k,j}}{\sum_{k\in \Delta\lambda^*_p,\Delta z_q}\,W_{k,j}}\,,
\end{equation}
where $R^{\rm eff}_{k,j}$ is the $j$th effective radius of the $k$th cluster, which is expressed as
\begin{equation}
R_{k,j}^{\rm eff} = \frac{\sum_{i\in j}\,w_i\,\tilde{R}_i}{\sum_{i\in j}\,w_i}\,,
\end{equation}
where $\tilde{R}_i$ is the projected distance of the $i$th galaxy from the centre of the $k$th cluster. We verified that $R_{j}^{\rm eff}\simeq R_j$, with differences of approximately 0.1\%, and thus the use of $R^{\rm eff}_{j}$ as a radial estimate does not have a significant impact on the final results. \\
\indent We measured the stacked reduced shear in 6 logarithmically-equispaced radial bins in the cluster-centric radial range $R\in[0.4,3.5]$ $h^{-1}$Mpc. By excluding the central 400 $h^{-1}$kpc, the contamination due to cluster members is significantly reduced \citep[see e.g.][]{hsc_med+al18b,Bellagamba19}. The analysis of the shear signal in the immediate vicinity of the cluster centre might also be affected by inaccuracies in the weak-lensing approximation and by the influence of the BCG on the matter distribution. Furthermore, the exclusion of scales larger than 3.5 $h^{-1}$Mpc reduces dependence of mass calibration results on cosmological parameters, mainly originating from matter correlated with clusters \citep[see e.g.][]{OguriHamana11}. In addition, the impact of possible anisotropic boosts, which affect the correlation functions on large scales due to projection effects, is mitigated \citep{Sunayama20,Wu22,Sunayama23,Park23,Zhou23}. \\
\indent In Fig.\ \ref{fig:g_t} we show the stacked $g_+(R)$ measurements in the bins of redshift and mass proxy listed in Table \ref{tab:sample}. Specifically, we considered the clusters with $\lambda^*>20$ for $z\in[0.1,0.3)$, $\lambda^*>25$ for $z\in[0.3,0.45)$, and $\lambda^*>30$ for $z\in[0.45,0.8]$. This selection yields a total sample of 9049 clusters. Nonetheless, the number of clusters contributing to the stacked weak-lensing measurements is slightly lower, amounting to 8730, due to the smaller area covered by the shear sample compared to the cluster sample. These $\lambda^*$ cuts were validated to ensure adequate model fit quality while maximising the number of clusters included in the analysis. The statistical part of the covariance matrix for each stack of $g_+(R)$ was estimated with a bootstrap procedure with replacement, by performing 10\,000 resamplings of the single cluster profiles contributing to the given stack. Given the large sample sizes, we do not expect significant differences compared to jackknife estimates. We used the bootstrap method to ensure robust off-diagonal covariance estimates, particularly to account for large-scale structure (LSS) contributions. Since we bootstrap the cluster profiles, the covariance arising from source galaxies shared by multiple clusters is not accounted for. However, our method effectively captures the intrinsic scatter in the observable-mass and concentration-mass relations, as well as miscentring effects. In addition, we note that in Stage-III surveys, the contribution from sources shared by multiple clusters is small compared to those from shape noise and LSS \citep{McClintock19}. The inverted covariance matrix is corrected following \citet{Hartlap07}. Lastly, we did not consider the covariance between radial bins in different redshift and mass proxy bins, as its impact on the final results is known to be negligible \citep{McClintock19,Ingoglia22}. \\
\indent In Appendix \ref{appendix:null_tests}, we show that null tests exclude the presence of additive systematic uncertainties affecting weak-lensing measurements. Furthermore, we note that for cosmological purposes, $g_+$ measurements as a function of the angular projected separation from cluster centres are ideal, as they do not require any assumptions on the cosmological model. Nonetheless, in the case of $g_+$ measurements averaged over large redshift bins, a background galaxy selection based on physical projected separations, as the one adopted here, more effectively ensures the exclusion of undesired cluster regions. This selection requires the assumption of a fiducial cosmological model, with this leading to geometric distortions in the measured $g_+$. We accounted for these distortions in the weak-lensing modelling outlined in Sect.\ \ref{sec:modelling:stacked_g}. \\
\indent To assess the impact of alternative background selection criteria on the stacked measurements, we computed the average weak-lensing ${\rm S/N}$ in the $q$th cluster redshift bin as follows \citep{Umetsu20b,Umetsu20}:
\small
\begin{equation}\label{eq:SNR}
    \langle({\rm S/N})_{\rm WL}\rangle_{\Delta\lambda^*}(\Delta z_q) = \frac{1}{N_p}\sum_{p=1}^{N_p} 
    \frac{\sum\limits_{j=1}^{N_R} g_+(R_j,\Delta\lambda^*_p,\Delta z_q) \, w_{g_+}(R_j,\Delta\lambda^*_p,\Delta z_q)}
    {\sqrt{\sum\limits_{j=1}^{N_R} w_{g_+}(R_j,\Delta\lambda^*_p,\Delta z_q)}} \,.
\end{equation} 
\normalsize
Here, $w_{g_+}=\sigma_{g_+}^{-2}$, where $\sigma_{g_+}$ is the uncertainty on $g_+$, derived through bootstrap resampling. In addition, $p$ runs over the number of $\lambda^*$ bins falling within $\Delta z_q$, namely $N_p$, $j$ runs over the number of radial bins within a stack, that is $N_R$. Figure \ref{fig:background_alternative} displays how $\langle({\rm S/N})_{\rm WL}\rangle_{\Delta\lambda^*}$ and the number of background galaxies, $N_{\rm bkg}$, are affected by the background selection choices. For this test, we also included a selection based on Eq.\ \eqref{eq:photoz_peak_selection} only, referred to as a photo-$z$ peak selection. The baseline background selection, namely the combination of colour and photo-$z$ selections (Eq.\ \ref{eq:tot_selection}), leads to $N_{\rm bkg}$ and $\langle({\rm S/N})_{\rm WL}\rangle_{\Delta\lambda^*}$ values that are compatible with those derived from the colour selection alone. This agrees with what discussed in Sect.\ \ref{Sec_background} (see also Fig.\ \ref{fig:n_eff}). Consequently, as shown in Fig.\ \ref{fig:background_alternative}, the photo-$z$ selection alone provides a $\langle({\rm S/N})_{\rm WL}\rangle_{\Delta\lambda^*}$ which is remarkably lower compared to what is derived from the baseline selection. A considerable enhancement of $N_{\rm bkg}$ is achieved at high $z$ by adopting the photo-$z$ peak selection. Nevertheless, due to the low purity of this selection (see Sect.\ \ref{Sec_sys_2}), the $\langle({\rm S/N})_{\rm WL}\rangle_{\Delta\lambda^*}$ is not appreciably larger than that obtained with the baseline selection. Thus, $\langle({\rm S/N})_{\rm WL}\rangle_{\Delta\lambda^*}$ is not a good estimate of the relative contamination between alternative background selection schemes, because contaminants lower both the signal and the noise. In addition, we remark that Eq.\ \eqref{eq:SNR} is an approximation of the S/N, as it neglects the cross-correlation between radial bins.

\begin{figure*}[t!]
\centering
    \includegraphics[width=\linewidth - 4cm, height=6.8cm]{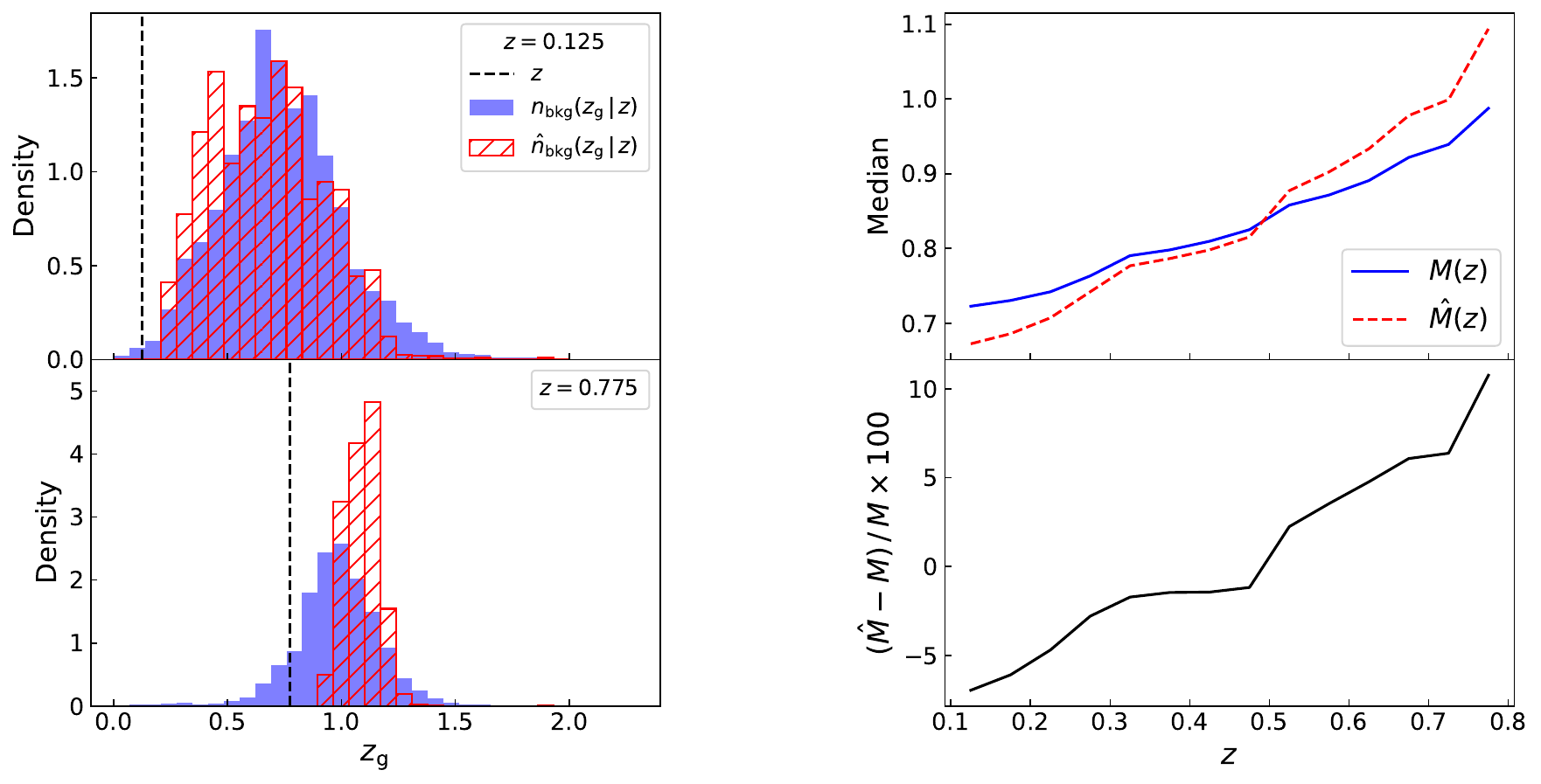}
    \caption{Differences between the uncalibrated, $\hat{n}(z_{\rm g}\,|\,z)$, and the SOM-weighted, $n(z_{\rm g}\,|\,z)$, background redshift distributions. 
    Left panels: $n(z_{\rm g}\,|\,z)$ (blue histograms) and $\hat{n}(z_{\rm g}\,|\,z)$ (red hatched histograms) for $z=0.125$ (top) and $z=0.775$ (bottom). 
    The vertical dashed black lines represent the cluster redshifts. 
    Right panels: In the top subpanel, the median values of $n(z_{\rm g}\,|\,z)$ (solid blue curve) and $\hat{n}(z_{\rm g}\,|\,z)$ (dashed red curve), namely $M$ and $\hat{M}$, respectively, are shown. 
    The percentage differences between them are displayed in the bottom subpanel.}
    \label{fig:SOM_results}
\end{figure*}
\subsection{SOM-reconstructed background distributions}\label{Sec_sys_2}
\begin{figure}[t!]
\centering\includegraphics[width = \hsize-1cm, height = 8cm] {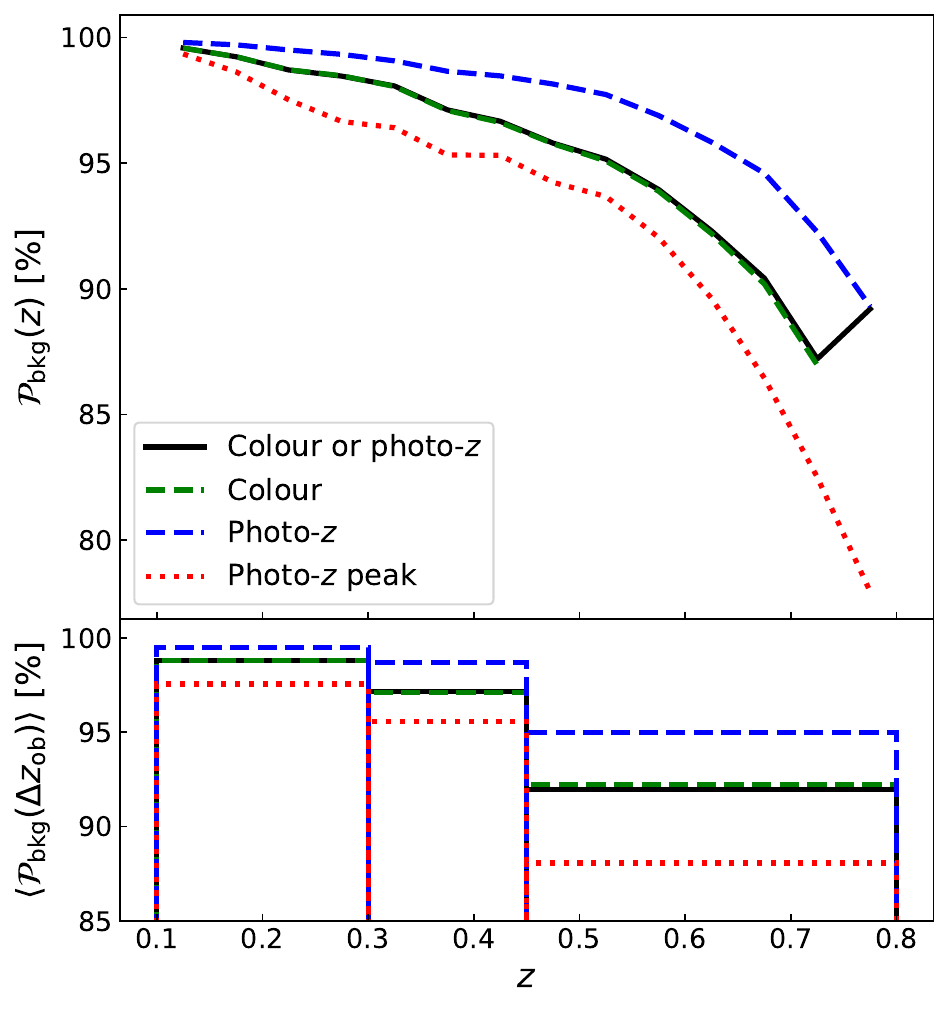}
\caption{Top panel: Background selection purity as a function of the cluster redshift. Bottom panel: Effective selection purity averaged over the cluster redshift distribution, in the cluster redshift bins $z\in[0.1,0.3)$, $z\in[0.3,0.45)$, and $z\in[0.45,0.8]$. In both panels, the symbols are the same as those in Fig.\ \ref{fig:background_alternative}. The black and green curves are almost overlapping.}
\label{fig:purity_background}
\end{figure}

Foreground and cluster galaxies dilute the weak-lensing signal, as their shapes are uncorrelated with the matter distribution of galaxy clusters. This holds strictly true only in the absence of intrinsic alignments. To assess the fraction of such contaminants, we rely on the spectroscopic galaxy sample developed by \citet{vanDenBusch22} and extended by \citet{Wright24}, which includes 93\,224 objects. Due to different selection effects, the redshift distribution of this spectroscopic sample differs remarkably from those of the photometric samples defined via the background selection (Eq.\ \ref{eq:tot_selection}). Thus, to reconstruct the true redshift distributions of the background galaxy samples, the magnitude and colour distributions of the spectroscopic sample have been weighted through the use of the SOM algorithm developed by \citet{Wright20} and adopted by \citet{Hildebrandt21} for the calibration of KiDS-1000 galaxy redshift distributions, based on the \texttt{kohonen} package \citep{Wehrens07,Wehrens18}. We remark that the background selection presented in Sect.\ \ref{Sec_background} is applied galaxy-by-galaxy, while the SOM analysis provides a redshift calibration for a population of objects. For this reason, SOM results cannot be used in the background selection adopted for our $g_+$ measurements. \\
\indent Based on our baseline background selection (Eq.\ \ref{eq:tot_selection}), we derived uncalibrated background redshift distributions given a cluster redshift, namely $\hat{n}_{\rm bkg}(z_{\rm g}\,|\,z)$, in $z$ bins having a width of $\delta z=0.05$ and spanning the whole redshift range of the cluster sample. We remark that $\hat{n}_{\rm bkg}(z_{\rm g}\,|\,z)$ is derived from the background galaxies associated with the clusters in the sample and falling within the cluster-centric projected radius range defined in Sect.\ \ref{sec:WL_stacking_2}. Then, we reconstructed the true background redshift distributions, $n_{\rm bkg}(z_{\rm g}\,|\,z)$, following the methods described in \citet{Wright20}. Specifically, we provided the SOM code with the information on the 9 KiDS and VIKING photometric bands, along with all their colour combinations. We used SOMs with toroidal topology and 50$\times$50 hexagonal cells, verifying that increasing the number of cells does not significantly affect the final results. As an illustration, the left panels in Fig.\ \ref{fig:SOM_results} show $n_{\rm bkg}(z_{\rm g}\,|\,z)$ and $\hat{n}_{\rm bkg}(z_{\rm g}\,|\,z)$ for $z=0.125$ and $z=0.775$. The median values of $n_{\rm bkg}(z_{\rm g}\,|\,z)$ and $\hat{n}_{\rm bkg}(z_{\rm g}\,|\,z)$, namely $M(z)$ and $\hat{M}(z)$, respectively, along with their percentage differences, are shown in the right panels of Fig.\ \ref{fig:SOM_results}. As we can see, $M$ is 8\% larger than $\hat{M}$ at low $z$, with this difference progressively reducing up to $z=0.5$. For $z>0.5$, $\hat{M}$ becomes larger than $M$, with differences reaching about 10\% at large $z$. \\
\indent From $n_{\rm bkg}(z_{\rm g}\,|\,z)$, we derived the purity of the background samples, namely $\mathcal{P}_{\rm bkg}(z)$, defined as the ratio of the number of galaxies with $z_{\rm g}>z$ over the total number of selected galaxies. Figure \ref{fig:purity_background} shows that $\mathcal{P}_{\rm bkg}>96\%$ for $z<0.5$, in agreement with \citetalias{Lesci_colours}. For $z>0.5$, for which $\hat{M}>M$, $\mathcal{P}_{\rm bkg}$ decreases down to 90\% at $z=0.75$. Nevertheless, \citetalias{Lesci_colours} predicted $\mathcal{P}_{\rm bkg}>97\%$ for these values of $z$. This is mainly due to the uncertainties on KiDS photometry, which are larger compared to those affecting the galaxy samples considered by \citetalias{Lesci_colours}. To evaluate the overall impact of such impurities on the stacked cluster weak-lensing measurements, in Fig.\ \ref{fig:purity_background} we also show $\langle\mathcal{P}_{\rm bkg}(\Delta z_{\rm ob})\rangle$, that is the effective purity weighted over the cluster redshift distribution in a given $\Delta z$. We find that $\langle\mathcal{P}_{\rm bkg}(\Delta z_{\rm ob})\rangle=99\%$ for $z\in[0.1,0.3)$, $\langle\mathcal{P}_{\rm bkg}(\Delta z_{\rm ob})\rangle=97\%$ for $z\in[0.3,0.45)$, and $\langle\mathcal{P}_{\rm bkg}(\Delta z_{\rm ob})\rangle=92\%$ for $z\in[0.45,0.8]$. \\
\indent As displayed in Fig.\ \ref{fig:purity_background}, $\mathcal{P}_{\rm bkg}$ can be improved by up to four percentage points at high $z$ by using the photo-$z$ selection alone. Despite the fact that this could make the photo-$z$ selection preferable for the last $\Delta z$ bin, we do not expect this to have a significant impact on the final results. Indeed, as we shall discuss in the following, the final results on cosmology and cluster masses are dominated by low and intermediate redshift measurements (see also $\langle({\rm S/N})_{\rm WL}\rangle_{\Delta\lambda^*}$ in Fig.\ \ref{fig:background_alternative}). Furthermore, we note that the colour selection does not yield any background galaxies for cluster redshifts larger than $z>0.75$, due to the buffer of 0.05 in the lens redshifts described in Sect.\ \ref{Sec_background}. Indeed, we remark that the colour selection by \citetalias{Lesci_colours} used in this work is valid up to $z=0.8$. Consequently, the combination of colour and photo-$z$ selections yields the same results as the photo-$z$ selection alone for $z>0.75$, as shown in Fig.\ \ref{fig:purity_background}. Figure \ref{fig:purity_background} also shows that the photo-$z$ peak selection yields the lowest background purity at any $z$, with $\mathcal{P}_{\rm bkg}\sim80\%$ at $z>0.7$. Together with the fact that this selection does not improve the weak-lensing S/N, as discussed in Sect.\ \ref{sec:WL_stacking_2}, we conclude that the combination of colour and photo-$z$ selection is more robust for our analysis. \\
\indent Furthermore, by reconstructing the background redshift distribution at a given $z$ and cluster-centric projected distance, that is $n_{\rm bkg}(z_{\rm g}\,|\,z,R)$, we find no significant dependence of $\mathcal{P}_{\rm bkg}$ on $R$. This is expected, as the inner regions of the clusters, which may degrade the purity due to the contamination by cluster members, are excluded from the analysis. Consequently, we do not expect any significant dependence of $\mathcal{P}_{\rm bkg}$ on $\lambda^*$. While previous studies \citep[e.g.][]{McClintock19} have demonstrated significant cluster member contamination when selecting source galaxies based on their average redshift estimates, our background selection method (Sect.\ \ref{Sec_background}) minimises this contamination through two key features: a robust colour selection, and stricter galaxy photometric redshift requirements. The reduced shear unaffected by impurities, namely $g_+^{\rm true}$, can be expressed as \citep[see e.g.][]{Dietrich19}
\begin{equation}\label{eq:correction_for_purity}
    g_+^{\rm true}(z) = \frac{g_+(z)}{\mathcal{P}_{\rm bkg}(z)}\,,
\end{equation}
where $g_+(z)$ is the measured reduced shear. Thus, in the following analysis we multiplied the theoretical model by $\mathcal{P}_{\rm bkg}(z)$. \\
\indent Using simulations, \citet{Wright20} showed that the bias on the mean of the SOM-reconstructed redshift distributions is below 1\%, on average, in the different tomographic bins considered for the KiDS-1000 cosmic shear analysis \citep[see also][]{Hildebrandt21,Giblin21}. The statistical uncertainty on the mean redshift, accounting for photometric noise, shot-noise due to the limited sample size, spectroscopic selection effects and incompleteness, sample variance due to LSS, and inherent approximations in the simulations, amounts to at most 2\% \citep{Hildebrandt21}. Despite the differences between the galaxy photo-$z$ estimates used in this work and those adopted for the KiDS-1000 cosmic shear analysis, stemming from the different priors used in BPZ by \citetalias{AMICOKiDS-1000} (see Sect.\ \ref{sec:shear_sample}), we expect that the uncertainties discussed above hold also for our sample. Since this uncertainty impacts the estimation of the ratio $\mathcal{R}=D_{\rm ls}/D_{\rm s}$ (see Eq.\ \ref{eq:sigma_crit}), we derived the mean $\mathcal{R}$, weighted by $n_{\rm bkg}(z_{\rm g}\,|\,z)$, at the central values of the cluster redshift bins used in the analysis. We repeated this process by shifting $n_{\rm bkg}(z_{\rm g}\,|\,z)$ by $\pm2\%$ of its median. This results in a relative uncertainty on $g_+$ of 1\% for the first two redshift bins and 4\% for the last one. As discussed in Sect.\ \ref{sec:modelling:likelihood}, this uncertainty is propagated into the final posteriors.

\section{Modelling}\label{sec_modelling}
In this section, we detail the modelling used to simultaneously constrain the $\log\lambda^*-\log M_{200}$ scaling relation and cosmological parameters. Section \ref{modelling:counts} outlines the cluster count model. After introducing, in Sect.\ \ref{sec:modelling:basic_lens}, the model describing a single galaxy cluster mass profile, we delve into the description of the theoretical expected values of the $g_+$ stacked measurements, in Sect.\ \ref{sec:modelling:stacked_g}. Then, in Sect.\ \ref{sec:modelling:likelihood}, we present the joint Bayesian modelling of all the weak-lensing stacks derived in Sect.\ \ref{sec:DR4_measure} and cluster abundance measurements.

\subsection{Cluster count model}\label{modelling:counts}
The expected number of clusters in a bin of observed intrinsic richness, $\Delta\lambda^*_{\rm ob}$, and redshift, $\Delta z_{\rm ob}$, is expressed as follows:
\begin{align}\label{eq:counts_model}
\langle N(\Delta \lambda^*_{\rm ob},\Delta &z_{\rm ob})\rangle =
\frac{\Omega}{\mathcal{P}_{\rm clu}(\Delta \lambda^*_{\rm ob},\Delta z_{\rm ob})}\int_0^\infty{\rm d}z_{\rm tr}\,\frac{{\rm d}^2 V}{{\rm d} z_{\rm tr}{\rm d}\Omega}\times\nonumber\\
&\times\int_0^\infty{\rm d}M\,\frac{{\rm d} n(M,z_{\rm tr})}{{\rm d} M}\,\mathcal{B}_{\rm\scriptscriptstyle HMF}(M)\,\times\nonumber\\
&\times\int_0^\infty{\rm d}\lambda^*_{\rm tr}\,\mathcal{C}_{\rm clu}(\lambda^*_{\rm tr},z_{\rm tr})\,P(\lambda^*_{\rm tr}|M,z_{\rm tr})\,\times\nonumber\\
&\times\int_{\Delta\lambda^*_{\rm ob}}{\rm d} \lambda^*_{\rm ob} \,\,P(\lambda^*_{\rm ob}|\lambda^*_{\rm tr},z_{\rm tr})\int_{\Delta z_{\rm ob}}{\rm d} z_{\rm ob} \,\,P(z_{\rm ob}|z_{\rm tr})\,,
\end{align}
where $z_{\rm tr}$ is the true redshift, $V$ is the co-moving volume, $\Omega$ is the effective area covered by the cluster sample, $M$ is the true mass, and ${\rm d} n(M,z_{\rm tr})/{\rm d} M$ is the halo mass function, for which we adopted the model by \citet{Tinker08}. In Eq.\ \eqref{eq:counts_model}, $\mathcal{B}_{\rm\scriptscriptstyle HMF}(M)$ is the halo mass function bias. Following \citet{Costanzi19}, $\mathcal{B}_{\rm\scriptscriptstyle HMF}(M)$ is expressed as follows:
\begin{equation}\label{eq:B_HMF}
\mathcal{B}_{\rm\scriptscriptstyle HMF}(M) = s\log \frac{M_{200\rm m}(M)}{M^*}+q\,,
\end{equation}
where $M^*$ is expressed in $h^{-1}$M$_\odot$ and $\log M^*=13.8$, $q$ and $s$ are free parameters of the model (see Sect.\ \ref{sec:modelling:likelihood}), and $M_{200\rm m}(M)$ is the mass within a radius enclosing a mean density 200 times larger than the mean density of the Universe, at the halo redshift, given a generically defined mass $M$. Indeed, \citet{Costanzi19} calibrated Eq.\ \eqref{eq:B_HMF} assuming $M_{200\rm m}$, while we adopted the critical overdensity to define masses throughout this paper. Thus, we estimated $M_{200\rm m}(M)$ using the mass conversion model by \citet[][Eq.\ 13]{Ragagnin21}, testing our implementation against the \texttt{HYDRO\_MC} code.\footnote{\url{https://github.com/aragagnin/hydro_mc}} As we will discuss in Sect.\ \ref{sec:modelling:likelihood}, we propagated the uncertainties on the parameters entering Eq.\ \eqref{eq:B_HMF} into the final results. This ensures that our cosmological constraints are robust against the assumption of alternative halo mass function models. \\
\indent In Eq.\ \eqref{eq:counts_model}, $P(\lambda^*_{\rm tr}|M,z_{\rm tr})$ is a log-normal probability density function, whose mean is given by the $\log\lambda^*-\log M_{200}$ scaling relation and its standard deviation is given by the intrinsic scatter, $\sigma_{\rm intr}$: 
\begin{align}\label{eq:scalingrelation_PDF}
P(\lambda^*_{\rm tr}|M,z_{\rm tr})=&\frac{1}{\ln(10)\lambda^*_{\rm tr}\sqrt{2\pi}\sigma_{\rm intr}}\exp\left(-\frac{[\log \lambda^*_{\rm tr} - \mu(M,z_{\rm tr})]^2}{2\sigma^2_{\rm intr}}\right)\,,
\end{align}
where $\mu(M,z_{\rm tr})$ is the mean of the distribution, expressed as
\begin{align}\label{eq:scalingrelation}
\mu(M,z_{\rm tr})=\alpha+\beta\log\frac{M}{M_{\rm piv}}+\gamma\log\frac{H(z_{\rm tr})}{H(z_{\rm piv})} + \log\lambda^*_{\rm piv}\,,
\end{align}
where $M_{\rm piv}=10^{14}h^{-1}M_\odot$, $z_{\rm piv}=0.4$, and $\lambda^*_{\rm piv}=50$ are the redshift, mass, and intrinsic richness pivots, respectively. The second last term in Eq.\ \eqref{eq:scalingrelation}, including the Hubble function $H(z)$, accounts for deviations in the redshift evolution from what is predicted in the self-similar growth scenario \citep{Sereno15}. As detailed in Sect.\ \ref{sec:modelling:likelihood}, $\alpha$, $\beta$, $\gamma$, and $\sigma_{\rm intr}$ in Eqs.\ \eqref{eq:scalingrelation_PDF} and \eqref{eq:scalingrelation} are free parameters in the modelling. Furthermore, $\mathcal{P}_{\rm clu}$ and $\mathcal{C}_{\rm clu}$ in Eq.\ \eqref{eq:counts_model} are the purity and completeness of the cluster sample, respectively, described in Sect.\ \ref{sec:selection_function}, while $P(z_{\rm ob}|z_{\rm tr})$ is a Gaussian probability density function with mean corresponding to $z_{\rm tr}$ and a standard deviation of $0.014(1+z_{\rm tr})$ (see Sect.\ \ref{sec:ClusterSample}). Lastly, $P(\lambda^*_{\rm ob}|\lambda^*_{\rm tr},z_{\rm tr})$ is a Gaussian derived from the $P(x|\lambda^*_{\rm tr},z_{\rm tr})$ distribution described in Sect.\ \ref{sec:selection_function}. Specifically, following Eq.\ \eqref{eq:Plambda_mean}, the mean of $P(\lambda^*_{\rm ob}|\lambda^*_{\rm tr},z_{\rm tr})$ is expressed as 
\begin{equation}\label{eq:Plambda_mean_final}
    \mu_{\lambda^*} = \lambda^*_{\rm tr} + \lambda^*_{\rm tr} \exp[- \lambda^*_{\rm tr} \, (\bar{A}_\mu + \bar{B}_\mu \, z_{\rm tr})]\,,
\end{equation}
while its standard deviation is based on Eq.\ \eqref{eq:Plambda_std} and is expressed as follows:
\begin{equation}\label{eq:Plambda_std_final}
    \sigma_{\lambda^*} = \bar{A}_\sigma \lambda^*_{\rm tr} \exp(- \bar{B}_\sigma \lambda^*_{\rm tr})\,.
\end{equation}
In Eqs.\ \eqref{eq:Plambda_mean_final} and \eqref{eq:Plambda_std_final}, $\bar{A}_\mu=0.198$, $\bar{B}_\mu = -0.179$, $\bar{A}_\sigma = 0.320$, and $\bar{B}_\sigma = 0.011$.

\subsection{Basic lens model and miscentring}\label{sec:modelling:basic_lens}
Here we present the models used to describe the mass profiles of single galaxy clusters. The expected value of a stacked weak-lensing profile is detailed in Sect.\ \ref{sec:modelling:stacked_g}. Specifically, the 3D mass density profiles of single galaxy clusters are described by a truncated NFW profile \citep*[BMO,][]{BMO},
\begin{equation}\label{rho_BMO}
\rho(r)=\frac{\rho_{\rm s}}{(r/r_{\rm s})\,(1+r/r_{\rm s})^2} \left(\frac{r^2_{\rm t}}{r^2+r^2_{\rm t}}\right)^2,
\end{equation}
where $\rho_{\rm s}$ is the characteristic density, $r_{\rm s}$ is the scale radius, and $r_{\rm t}$ is the truncation radius. The scale radius is expressed as $r_{\rm s} = r_{200}/c_{200}$, where $r_{200}$ is the radius enclosing a mass such that the corresponding mean density is 200 times the critical density of the Universe at that redshift, and $c_{200}$ is the concentration. As discussed in Sect.\ \ref{sec:modelling:stacked_g}, $c_{200}$ is given by a log-linear relation with $M_{200}$ whose amplitude is constrained by our data. The truncation radius is defined as $r_{\rm t}=F_{\rm t}r_{200}$, where $F_{\rm t}$ is the truncation factor. The latter is a free parameter in the analysis, as detailed in Sect.\ \ref{sec:modelling:likelihood}. The BMO profile (Eq. \ref{rho_BMO}) provides more accurate mass estimates than a simple NFW parametrisation. \citet{OguriHamana11} demonstrated that NFW-based fits to shear signals can introduce systematic biases on masses of about 10-15\%. \\
\indent We included a two-halo term in the model, whose contribution to the surface mass density is expressed as \citep{Oguri11}
\begin{equation}\label{eq:2halo}
\Sigma_{\rm 2h}(\theta,M,z) = \frac { \rho_{\text{m}} (z)\, b_{\text{h}}(M,z)} {(1+z)^3\, D^2_{\text{l}} (z)} \int \frac {l {\rm d}l} {2 \pi} J_0(l\theta) P(k_{\text{l}},z)\,\,,
\end{equation}
where $\theta$ is the angular radius, $J_0$ is the 0th order Bessel function, $k_{\text{l}} = l/(1+z)/D_{\text{l}}(z)$, $D_{\rm l}$ is the lens angular diameter distance, $b_{\rm h}$ is the halo bias, for which the model by \citet{Tinker10} was assumed, and $P(k_{\text{l}},z)$ is the linear matter power spectrum. Then the centred surface mass density is expressed as
\begin{equation}\label{Sigma_cen}
\Sigma_{\rm cen}(R) = \Sigma_{\rm 1h}(R) +\Sigma_{\rm 2h}(R),
\end{equation}
where $\Sigma_{\rm 1h}(R)$, for which \citet{OguriHamana11} proposed an analytic form, is derived from Eq.\ \eqref{rho_BMO}. The centred excess surface mass density has the following functional form:
\begin{equation}\label{DeltaSigma_cen}
\Delta\Sigma_{+,\rm cen}(R) = \frac{2}{R^2}\int_0^R \mathrm{d} r\,\, r\, \Sigma_{\rm cen}(r) -\Sigma_{\rm cen}(R)\,.
\end{equation}
From Eqs.\ \eqref{Sigma_cen} and \eqref{DeltaSigma_cen}, we express the tangential reduced shear component of a centred halo, $g_{+,\rm cen}$, as in \citet{Seitz97}:
\begin{equation}\label{eq:g_cen}
g_{+,\rm cen}(R,M,z) = \frac{\Delta\Sigma_{+,\rm cen}(R,M,z)\,\langle\Sigma_{\rm crit}^{-1}(z)\rangle}{1 - \Sigma_{\rm cen}(R,M,z)\,\langle\Sigma_{\rm crit}^{-1}(z)\rangle^{-1}\,\langle\Sigma_{\rm crit}^{-2}(z)\rangle}\,.
\end{equation}
In Eq.\ \eqref{eq:g_cen}, $\langle\Sigma_{\rm crit}^{-\eta}\rangle$ has the following expression
\begin{equation}
\langle\Sigma_{\rm crit}^{-\eta}(z)\rangle = \frac{\int_{z_{\rm g}>z}{\rm d}z_{\rm g}\, \Sigma_{\rm crit}^{-\eta}(z_{\rm g},z)\,n(z_{\rm g}\,|\,z)}{\int_{z_{\rm g}>z}{\rm d}z_{\rm g}\, n(z_{\rm g}\,|\,z)}\,,
\end{equation}
where $\Sigma_{\rm crit}$ is given by Eq.\ \eqref{eq:sigma_crit}, while $n(z_{\rm g}\,|\,z)$ is the SOM-reconstructed background redshift distribution described in Sect.\ \ref{Sec_sys_2}.\\
\indent We also modelled the contribution to $g_+$ due to the miscentred population of clusters, as is done, for example, in \citet{Johnston07} and \citet{Viola15}. We assumed that the probability of a lens being at a projected distance $R_{\rm s}$ from the chosen centre, namely $P(R_{\text{s}})$, follows a Rayleigh distribution, that is 
\begin{equation}\label{eq:Poff}
P(R_{\text{s}}) = \frac {R_{\text{s}}} {\sigma_{\text{off}}^2} \exp \bigg[-\frac 1 2 \bigg( \frac {R_{\text{s}}} {\sigma_{\text{off}}} \bigg)^2 \bigg]\,,
\end{equation}
where $\sigma_{\rm off}$ is the standard deviation of the halo misplacement distribution on the plane of the sky, expressed in units of $h^{-1}$Mpc. The corresponding azimuthally averaged profile is given by \citep{Yang06}
\begin{equation}
\Sigma (R | R_{\text{s}}) = \frac 1 {2 \pi} \int_0^{2 \pi} \Sigma_{\text{cen}} \, \bigg( \sqrt {R^2 + R_{\text{s}}^2 - 2 R R_{\text{s}} \cos \theta} \bigg) \, {\rm d} \theta\,,
\end{equation}
and the surface mass density distribution of a miscentred halo is expressed as
\begin{equation}\label{Sigma_off}
\Sigma_{\text{off}}(R) = \int P(R_{\text{s}})\, \Sigma (R | R_{\text{s}})\, {\rm d} R_{\text{s}} \,.
\end{equation}
Analogously to Eq.\ \eqref{eq:g_cen}, the tangential reduced shear component of a miscentred halo, $g_{+,\rm off}$, has the following expression:
\begin{equation}\label{eq:g_off}
g_{+,\rm off}(R,M,z) = \frac{\Delta\Sigma_{+,\rm off}(R,M,z)\,\langle\Sigma_{\rm crit}^{-1}(z)\rangle}{1 - \Sigma_{\rm off}(R,M,z)\,\langle\Sigma_{\rm crit}^{-1}(z)\rangle^{-1}\,\langle\Sigma_{\rm crit}^{-2}(z)\rangle}\,,
\end{equation}
where $\Sigma_{\rm off}$ is given by Eq.\ \eqref{Sigma_off}, while $\Delta\Sigma_{+,\rm off}$ is derived by replacing $\Sigma_{\rm cen}$ with $\Sigma_{\rm off}$ in Eq.\ \eqref{DeltaSigma_cen}.

\subsection{Stacked weak-lensing model}\label{sec:modelling:stacked_g}
\begin{table*}[t]
\caption{\label{tab:priors_and_posteriors}Parameters considered in the joint analysis of stacked weak lensing and counts.}
  \centering
    \begin{tabular}{l c c r} 
      Parameter & Description & Prior & Posterior \\ 
      \hline
      \rule{0pt}{4ex}
      $\Omega_{\rm\scriptscriptstyle CDM}$ & Cold dark matter density parameter at $z=0$ & [0.1, 0.4] & $\Ocdm^{+\OcdmUp}_{-\OcdmLow}$\\ \rule{0pt}{2.5ex}
      $10^9A_{\rm s}$ & Amplitude of the primordial matter power spectrum & [0.8, 8] & $\As^{+\AsUp}_{-\AsLow}$\\ \rule{0pt}{2.5ex}
      $\Omega_{\rm m}$ & Total matter density parameter at $z=0$ & --- & $\Om^{+\OmUp}_{-\OmLow}$\\ \rule{0pt}{2.5ex}
      $\sigma_8$ & Amplitude of the matter power spectrum at $z=0$ & --- & $\seight^{+\seightUp}_{-\seightLow}$\\ \rule{0pt}{2.5ex}
      $S_8\equiv\sigma_8(\Omega_m/0.3)^{0.5}$ & Clustering amplitude & --- & $\Sotto^{+\SottoUp}_{-\SottoLow}$\\ \rule{0pt}{2.5ex}
      $\Omega_{\rm b}$ & Baryon density parameter at $z=0$ & $\mathcal{N}(0.0493, 0.0016)$ & ---\\ \rule{0pt}{2.5ex}
      $n_{\rm s}$ & Primordial power spectrum spectral index & $\mathcal{N}(0.9649, 0.0210)$ & --- \\ \rule{0pt}{2.5ex}
      $h\equiv H_0/(100$ km/s/Mpc) & Normalised Hubble constant & $\mathcal{N}(0.7, 0.03)$ & --- \\ \rule{0pt}{2.5ex}
      $\alpha$ & Amplitude of the $\log\lambda^*-\log M_{200}$ relation & [-2, 2] & $\alphaScaling^{+\alphaScalingup}_{-\alphaScalinglow}$\\ \rule{0pt}{2.5ex}
      $\beta$ & Slope of the $\log\lambda^*-\log M_{200}$ relation & [0, 3] & $\betaScaling^{+\betaScalingup}_{-\betaScalinglow}$ \\ \rule{0pt}{2.5ex}
      $\gamma$ & Redshift evolution of the $\log\lambda^*-\log M_{200}$ relation & [-3, 3] & $\gammaScaling^{+\gammaScalingup}_{-\gammaScalinglow}$\\ \rule{0pt}{2.5ex}
      $\sigma_{\rm intr}$ & Intrinsic scatter of the $\log\lambda^*-\log M_{200}$ relation & [0.01, 0.5] & $\scatterZero^{+\scatterZeroup}_{-\scatterZerolow}$\\ \rule{0pt}{2.5ex}
      $\log c_0$ & Amplitude of the $\log c_{200}-\log M_{200}$ relation & [0, 1.3] & $\cZero^{+\cZeroup}_{-\cZerolow}$\\ \rule{0pt}{2.5ex}
      $f_{\rm off}$ & Fraction of miscentred clusters & $\mathcal{N}(0.3, 0.1)$ & ---\\ \rule{0pt}{2.5ex}
      $\sigma_{{\rm off}}$ & Miscentring scale (in $h^{-1}$Mpc) & [0, 0.5] & $\soffZero^{+\soffZeroup}_{-\soffZerolow}$\\ \rule{0pt}{2.5ex}
      $F_{\rm t}$ & Truncation factor of the BMO density profile & $\mathcal{N}(3, 0.5)$ & ---\\ \rule{0pt}{2.5ex}
      $(s,q)$ & Parameters entering the mass function correction factor & $\mathcal{N}(\mu_{\rm \scriptscriptstyle HMF}, C_{\rm \scriptscriptstyle HMF})$ & --- \\
    \end{tabular}
  \tablefoot{In the first and second columns we list the symbols and descriptions of the parameters, respectively. The third column reports the priors on the parameters, and in particular a range represents a uniform prior, while $\mathcal{N}(\mu,\sigma)$ stands for a Gaussian prior with mean $\mu$ and standard deviation $\sigma$. In the fourth column, we show the median values of the 1D marginalised posteriors, along with the 16th and 84th percentiles. The posterior is not reported in cases where it closely aligns with the prior.}
\end{table*}
The expected value of the stacked $g_+$ is expressed as follows:
\begin{align}\label{eq:DR4_WL_model}
\langle g_+(R,\Delta \lambda^*_{\rm ob},\Delta z_{\rm ob})\rangle = \;& (1 - f_{\rm off}) \langle g_{+,\rm cen}(R,\,\Delta \lambda^*_{\rm ob},\Delta z_{\rm ob})\rangle \,+ \nonumber\\
& + f_{\rm off} \langle g_{+,\rm off}(R,\Delta \lambda^*_{\rm ob},\Delta z_{\rm ob})\rangle\,,
\end{align}
where $f_{\rm off}$ is the fraction of haloes that belong to the miscentred population, while $\langle g_{+,\rm cen}\rangle$ and $\langle g_{+,\rm off}\rangle$ are the average $g_{+,\rm cen}$ and $g_{+,\rm off}$ of the sample, respectively. Specifically, $\langle g_{+,\rm cen}\rangle$ has the form
\begin{align}\label{eq:DR4_WL_model_centred}
\langle g_{+,\rm cen}(&R,\Delta\lambda^*_{\rm ob},\Delta z_{\rm ob})\rangle =
\frac{\mathcal{P}_{\rm clu}(\Delta \lambda^*_{\rm ob},\Delta z_{\rm ob})\,\langle\mathcal{P}_{\rm bkg}(\Delta z_{\rm ob})\rangle}{\langle n(\Delta \lambda^*_{\rm ob},\Delta z_{\rm ob})\rangle}\times\nonumber\\
&\times\int_0^\infty{\rm d}z_{\rm tr}\,\frac{{\rm d}^2 V}{{\rm d} z_{\rm tr}{\rm d}\Omega}\int_0^\infty{\rm d}M\,\frac{{\rm d} n(M,z_{\rm tr})}{{\rm d} M}\,\mathcal{B}_{\rm\scriptscriptstyle HMF}(M)\,\times\nonumber\\
&\times\,g_{+,\rm cen}(R^{\rm test},M,z_{\rm tr})\int_0^\infty{\rm d}\lambda^*_{\rm tr}\,\mathcal{C}_{\rm clu}(\lambda^*_{\rm tr},z_{\rm tr})\,P(\lambda^*_{\rm tr}|M,z_{\rm tr})\,\times\nonumber\\
&\times\int_{\Delta\lambda^*_{\rm ob}}{\rm d} \lambda^*_{\rm ob} \,\,P(\lambda^*_{\rm ob}|\lambda^*_{\rm tr},z_{\rm tr})\int_{\Delta z_{\rm ob}}{\rm d} z_{\rm ob} \,\,P(z_{\rm ob}|z_{\rm tr})\,,
\end{align}
while $\langle g_{+,\rm off}\rangle$ is obtained by replacing $g_{+,\rm cen}$ with $g_{+,\rm off}$ in Eq.\ \eqref{eq:DR4_WL_model_centred}. In Eq.\ \eqref{eq:DR4_WL_model_centred}, $n(\Delta\lambda^*_{\rm ob},\Delta z_{\rm ob})$ is the expected density of observed haloes not corrected for sample impurities, having the following expression:
\begin{align}
\langle n(\Delta \lambda^*_{\rm ob},\Delta &z_{\rm ob})\rangle =
\int_0^\infty{\rm d}z_{\rm tr}\,\frac{{\rm d}^2 V}{{\rm d} z_{\rm tr}{\rm d}\Omega}\int_0^\infty{\rm d}M\,\frac{{\rm d} n(M,z_{\rm tr})}{{\rm d} M}\,\mathcal{B}_{\rm\scriptscriptstyle HMF}(M)\times\nonumber\\
&\times\int_0^\infty{\rm d}\lambda^*_{\rm tr}\,\mathcal{C}_{\rm clu}(\lambda^*_{\rm tr},z_{\rm tr})\,P(\lambda^*_{\rm tr}|M,z_{\rm tr})\,\times\nonumber\\
&\times\int_{\Delta\lambda^*_{\rm ob}}{\rm d} \lambda^*_{\rm ob} \,\,P(\lambda^*_{\rm ob}|\lambda^*_{\rm tr},z_{\rm tr})\int_{\Delta z_{\rm ob}}{\rm d} z_{\rm ob} \,\,P(z_{\rm ob}|z_{\rm tr})\,.
\end{align}
The cluster sample purity, $\mathcal{P}_{\rm clu}$, in Eq.\ \eqref{eq:DR4_WL_model_centred} quantifies the suppression of the stacked weak-lensing signal due to false cluster detections, while $\langle\mathcal{P}_{\rm bkg}\rangle$ is the effective background sample purity, described in Sect.\ \ref{Sec_sys_2}. We remark that the contributions by halo triaxiality and projection effects were included in the covariance matrix, as discussed in Sect.\ \ref{sec:modelling:likelihood}. \\
\indent The measurements presented in Sect.\ \ref{sec:DR4_measure} were performed under the assumption of a fiducial cosmology. To address geometric distortions arising from this assumption, $g_{+,\rm cen}$ in Eq.\ \eqref{eq:DR4_WL_model_centred} is evaluated at the test projected radius $R^{\rm test}$, expressed as
\begin{equation}\label{eq:R_test}
R^{\rm test} = \theta D^{\rm test}_{\rm l}  = R^{\rm fid} \, \frac{D^{\rm test}_{\rm l}}{D^{\rm fid}_{\rm l}}\,,
\end{equation}
where $\theta$ is the angular separation from the cluster centre, $R^{\rm fid}$ is the projected radius in the fiducial cosmology, while $D_{\rm l}^{\rm fid}$ and $D_{\rm l}^{\rm test}$ represent the angular diameter distance of the cluster in the fiducial and test cosmologies, respectively. \\
\indent Lastly, we assumed a log-linear model for the $c_{200}-M_{200}$ relation, which is defined as
\begin{equation}\label{eq:cM_rel}
\log c_{200} = \log c_0 + c_M \log \frac{M}{M_{\rm piv}} + c_z \log \frac{1+z_{\rm tr}}{1+z_{\rm piv}}\,,
\end{equation}
where $M_{\rm piv}$ and $z_{\rm piv}$ are the same as those assumed in Eq.\ \ref{eq:scalingrelation}. As discussed in the following, $\log c_0$ is a free parameter in the analysis, while $c_M$ and $c_z$ are fixed to fiducial values. We point out that Eq.\ \eqref{eq:cM_rel} neglects the intrinsic scatter in the $\log c_{200}-\log M_{200}$ relation. In Sect.\ \ref{sec:DR4_results} we address the impact of this scatter on our results.

\subsection{Likelihood function and parameter priors}\label{sec:modelling:likelihood}
In this paper, we performed a joint Bayesian analysis of all the stacked weak-lensing and cluster count measurements, by means of an MCMC algorithm. The likelihood function is expressed as
\begin{equation}\label{eq:Likelihood}
\mathcal{L} = \mathcal{L}^{\rm\scriptscriptstyle WL}\mathcal{L}^{\rm\scriptscriptstyle C}\,,
\end{equation}
where $\mathcal{L}^{\rm\scriptscriptstyle WL}$ and $\mathcal{L}^{\rm\scriptscriptstyle C}$ are the cluster weak-lensing and count likelihoods, respectively. Notably, $\mathcal{L}^{\rm\scriptscriptstyle WL}$ takes the form
\begin{equation}
\mathcal{L}^{\rm\scriptscriptstyle WL} = \prod_{i=1}^{N^{\rm\scriptscriptstyle WL}_{\lambda^*}}\prod_{j=1}^{N^{\rm\scriptscriptstyle WL}_{z}} \mathcal{L}^{\rm\scriptscriptstyle WL}_{ij}\,,
\end{equation}
where $N^{\rm\scriptscriptstyle WL}_{\lambda^*}$ and $N^{\rm\scriptscriptstyle WL}_{z}$ are the numbers of cluster intrinsic richness and redshift bins used for the stacked measurements, respectively, while $\mathcal{L}^{\rm\scriptscriptstyle WL}_{ij}$ is a Gaussian likelihood function defined as
\begin{equation}
\mathcal{L}^{\rm\scriptscriptstyle WL}_{ij} \propto \exp(-\chi_{ij}^2/2)\,,
\end{equation}
with
\begin{align}
&\chi_{ij}^2=\sum_{k=1}^{N_R}\sum_{l=1}^{N_R}
  \left(g_{+,ijk}^{\rm ob}-g_{+,ijk}^{\rm mod} \right)\, C_{kl}^{-1}\, \left(
  g_{+,ijl}^{\rm ob}-g_{+,ijl}^{\rm mod}\right)\,.
\end{align}
Here, $g_{+}^{\rm ob}$ represents the observed stacked profile, given by Eq.\ \eqref{eq:DeltaSigma_measure_stack}, $g_{+}^{\rm mod}$ is the model, namely Eq.\ \eqref{eq:DR4_WL_model}, while the indices $k$ and $l$ run over the number of radial bins, $N_R$, and $C_{kl}^{-1}$ is the inverse of the covariance matrix. In particular, $C_{kl}$ is defined as\begin{equation}
C_{kl} = C_{kl}^{\rm BT} + C_{kl}^{\rm sys}\,,
\end{equation}
where $C_{kl}^{\rm BT}$ is estimated through a bootstrap resampling, as discussed in Sect.\ \ref{sec:WL_stacking_2}, while $C_{kl}^{\rm sys}$ accounts for residual uncertainties on systematic errors that are not included in the model (Eq.\ \ref{eq:DR4_WL_model_centred}). Specifically, $C_{kl}^{\rm sys}$ is written as
\begin{figure*}[t!]
\centering\includegraphics[width = \hsize-1.6cm, height = 6.8cm] {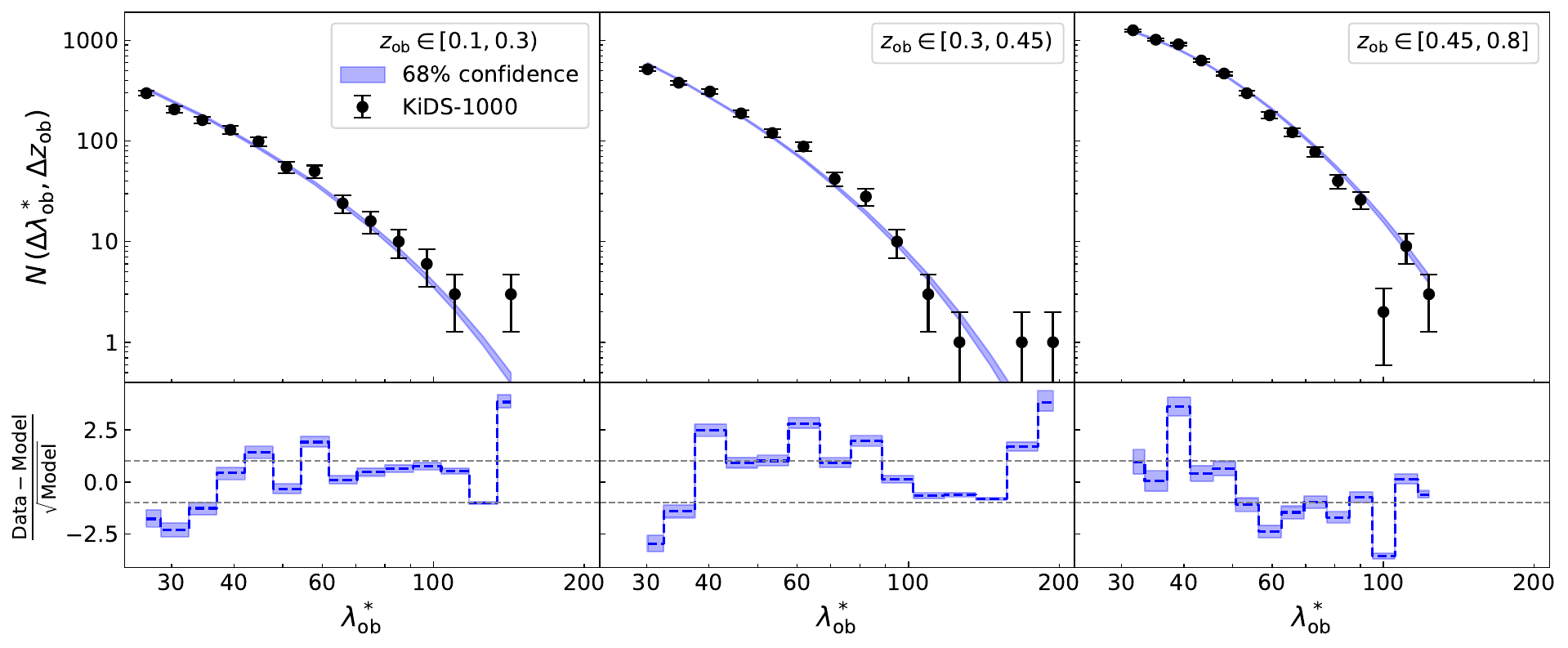}
\caption{Top panels: Counts of the AMICO KiDS-1000 galaxy clusters in the redshift bins adopted in the analysis (increasing from left to right). The black dots show the measures, with the error bars corresponding to Poisson uncertainties. The blue bands display the 68\% confidence levels of the model, derived from the posterior of all the free parameters considered in the joint analysis of counts and weak lensing. Bottom panels: Pearson residuals. The horizontal dashed grey lines show the interval between -1 and 1.}
\label{fig:counts}
\end{figure*}
\begin{equation}
C_{kl}^{\rm sys} = (\sigma_m^2 + \sigma_{\rm\scriptscriptstyle SOM}^2 + \sigma_{\rm\scriptscriptstyle OP}^2)\, g_{+,k}^{\rm ob}\,g_{+,l}^{\rm ob}\,,
\end{equation}
where $\sigma_m=0.02$ is the uncertainty on the multiplicative shear bias (Sect.\ \ref{sec:WL_stacking}), $\sigma_{\rm\scriptscriptstyle SOM}$ accounts for the uncertainty on the SOM-reconstructed background redshift distributions, amounting to $\sigma_{\rm\scriptscriptstyle SOM}=0.01$ for the first two cluster redshift bins and to $\sigma_{\rm\scriptscriptstyle SOM}=0.04$ for the last one (see Sect.\ \ref{Sec_sys_2}), while $\sigma_{\rm\scriptscriptstyle OP}=0.03$ is the residual uncertainty due to orientation and projection effects. Indeed, optical cluster finders preferentially select haloes with the major axis aligned with the line of sight, as these objects produce a larger density contrast with respect to the distribution of the field galaxies. In stacked weak-lensing analyses, this implies a boost in the weak-lensing signal \citep{Corless08,Dietrich14,Osato18,Zhang23,Giocoli23}. An opposite mass bias comes from projections of secondary haloes aligned with the detected clusters, which are in fact blended in a single detection. This may cause an overestimation of the cluster optical mass proxies, and a consequent bias in the mass-observable relation \citep{Myles21,Wu22}. For the AMICO clusters in KiDS-DR3, \citet{Bellagamba19} found that orientation effects counterbalance those caused by projections, leading to a negligible bias on the derived masses with a residual uncertainty of 3\%. Also \citet{Simet17} and \citet{Melchior17}, for galaxy clusters detected applying the red-sequence Matched-filter Probabilistic Percolation \citep[redMaPPer,][]{Rykoff14,Rykoff16} algorithm in the Dark Energy Survey \citep[DES,][]{DES05}, found that the combination of these bias sources is consistent with zero, deriving a mass corrective factor of $0.98\pm0.03$ \citep[see also][]{McClintock19}. Thus, we conservatively assumed a residual uncertainty on mass due to orientation and projections of 4\%. To relate this uncertainty on mass to that on weak-lensing profiles, we express the logarithmic dependence of $\Delta\Sigma_+$ on the mass $M$ as \citep{Melchior17}
\begin{equation}\label{eq:Melchior}
\Gamma = \frac{{\rm d}\ln \Delta\Sigma_+(M)}{{\rm d}\ln M}\,,
\end{equation}
with $\Gamma\simeq0.7$ being a good approximation for a broad range of cluster masses, redshifts, and cluster-centric distances \citep{Melchior17,Sereno17}. From Eq.\ \eqref{eq:Melchior}, we obtain
\begin{equation}\label{eq:WL_error_conversion}
\frac{\delta \Delta\Sigma_+}{\Delta\Sigma_+} = \Gamma \, \frac{\delta M}{M}\,.
\end{equation}
Based on Eq.\ \eqref{eq:WL_error_conversion}, assuming a $4$\% uncertainty on mass due to halo orientation and projection effects results in a relative uncertainty of about $3$\%, that is $\sigma_{\rm\scriptscriptstyle OP}=0.03$, on $\Delta\Sigma_+$ and, in turn, on $g_+$. \\
\indent Following \citet{Lima04} and \citet{Lacasa19}, the likelihood $\mathcal{L}^{\rm\scriptscriptstyle C}$ in Eq.\ \eqref{eq:Likelihood} is expressed as the convolution of Poisson and Gaussian distributions:
\begin{align}\label{eq:Likelihood_counts}
\mathcal{L}^{\rm\scriptscriptstyle C} = \int {\rm d}\boldsymbol{\delta}_{\rm b}^{N^{\rm\scriptscriptstyle C}_z}
&\left[\prod_{i=1}^{N^{\rm\scriptscriptstyle C}_{\lambda^*}}\prod_{j=1}^{N^{\rm\scriptscriptstyle C}_z}\text{Poiss}\left(N^{\rm ob}_{ij} \,\bigg|\, N^{\rm mod}_{ij} +
  \frac{{\rm d} N^{\rm mod}_{ij}}{{\rm d}\delta_{\rm b}}\delta_{{\rm b},j} \right)
  \right]\,\times\nonumber\\
  &\times\,\mathcal{N}(\boldsymbol{\delta}_{\rm b}|0,S)\,,
\end{align}
where $N^{\rm ob}$ is the observed number of clusters, $N^{\rm mod}$ corresponds to the cluster count model (Eq.\ \ref{eq:counts_model}), while ${\rm d} N^{\rm mod}/{\rm d}\delta_{\rm b}$ is the response of the counts to the variation $\boldsymbol{\delta}_{\rm b}$ of the background matter density. In particular, the cluster count response is analogous to Eq.\ \eqref{eq:counts_model}, where the integrand function is multiplied by the linear halo bias \citep[see e.g.][]{Lacasa19,Lesci22_counts}, for which we adopted the model by \citet{Tinker10}. In Eq.\ \eqref{eq:Likelihood_counts}, $\mathcal{N}(\boldsymbol{\delta}_{\rm b}|0,S)$ is a multivariate Gaussian probability density function describing the super-sample covariance (SSC) effects on cluster count measurements, which is a function of the background density fluctuation, $\boldsymbol{\delta}_{\rm b}$, is centred on zero, and has covariance matrix $S$. Notably, $\boldsymbol{\delta}_{\rm b}=\{\delta_{\text{b},1},...,\delta_{\text{b},N^{\rm\scriptscriptstyle C}_z}\}$, where $N^{\rm\scriptscriptstyle C}_z$ is the number of redshift bins used for cluster count measurements. Consequently, the $S$ matrix has dimension $N^{\rm\scriptscriptstyle C}_z\times N^{\rm\scriptscriptstyle C}_z$, and the matrix element $S_{pq}$ has the following functional form \citep{Lacasa19}:
\begin{equation}\label{eq:S_matrix}
S_{pq} = \frac{1}{8\pi^3 f_{\rm sky}^2} \sum_\ell (2\ell+1) \, \mathcal{C}_{\rm m}(\ell) \int {\rm d} k\,\, k^2 P(k) \frac{U_p(k,\ell)}{I_p} \frac{U_q(k,\ell)}{I_q}\,,
\end{equation}
where $f_{\rm sky}=\Omega/(4\pi)$ is the sky fraction covered by the survey, $\Omega$ is the survey area in steradians, $\mathcal{C}_{\rm m}(\ell)$ is the angular power spectrum of the footprint mask computed at the multipole $\ell$, accounting for the masked regions and the survey geometry, and $P(k)$ is the linear matter power spectrum computed at $z=0$. In addition, $U_p(k)$ in Eq.\ \eqref{eq:S_matrix} is expressed as
\begin{equation}
U_p(k,\ell) = \int \frac{{\rm d}V}{{\rm d}z}{\rm d}z \,\, W_p(z)\,G(z)\,j_\ell[kr(z)]\,,
\end{equation}
where $W_p(z)$ is a top-hat window function, $G(z)$ is the linear growth factor, $j_\ell$ is the spherical Bessel function, and $r(z)$ is the comoving distance, while $I_p$ in Eq.\ \eqref{eq:S_matrix} has the expression
\begin{equation}
I_p = \int \frac{{\rm d}V}{{\rm d}z}{\rm d}z \,\, W_p(z)\,.
\end{equation}
We note that $S$ and the cluster count response in Eq.\ \eqref{eq:Likelihood_counts} did not vary in our analysis. Indeed, these quantities were derived at the parameter posterior medians obtained from the MCMC analysis performed without including the SSC contribution. As the SSC was included in the likelihood, the MCMC analysis was repeated using the updated parameter median values until the posteriors converged. \\
\indent In Table \ref{tab:priors_and_posteriors} we list the parameter priors adopted in the analysis. Wide uniform priors were assumed for the cold dark matter density parameter at $z=0$, denoted as $\Omega_{\rm \scriptscriptstyle CDM}$, and for the amplitude of the primordial matter power spectrum, $A_{\rm s}$, such that $\Omega_{\rm m}$ and $\sigma_8$ are derived parameters in our analysis. We adopted Gaussian priors on the baryon density parameter, $\Omega_{\rm b}$, and on the primordial spectral index, $n_{\rm s}$, based on the results by \citet[][Table 2, TT, TE + EE + lowE + lensing, referred to as Planck18 hereafter]{Planck18}, assuming the same median values and imposing a standard deviation equal to 5$\sigma$. The prior on the normalised Hubble constant, $h$, is a Gaussian distribution centred on 0.7 with a standard deviation of 0.03. In addition, we assumed uniform priors on the $\log \lambda^*-\log M_{200}$ scaling relation parameters in Eq.\ \eqref{eq:scalingrelation}, namely $\alpha$, $\beta$, and $\gamma$, and on its intrinsic scatter, $\sigma_{\rm intr}$. A uniform prior was assumed also for the amplitude of the $\log c_{200}-\log M_{200}$ relation (Eq.\ \ref{eq:cM_rel}), namely $\log c_0$. Following \citet{Duffy08}, we fixed $c_M=-0.084$ and $c_z=-0.47$. Through these priors on $\sigma_{\rm intr}$ and $c_{200}$, we expect to account for the theoretical uncertainties on the contribution due to baryonic matter to the galaxy cluster profiles \citep{Schaller15,Henson17,Shirasaki18,Lee18,Beltz-Mohrmann21,Shao24}. Furthermore, we assumed a Gaussian prior with mean 0.3 and standard deviation 0.1 on the fraction of miscentred clusters, $f_{\rm off}$, imposing also $f_{\rm off}>0$. For the typical miscentring scale, we adopted the uniform prior $\sigma_{\rm off}\in[0,0.5]$ $h^{-1}$Mpc. These priors on the offset of galaxy-based cluster centres agree with the results from simulated \citep{Yan20,Sommer23} and observed \citep{Saro15,Zhang19,Seppi23,Ding24} galaxy, intra-cluster medium, and dark matter distributions. We considered a Gaussian prior on the truncation factor of the BMO profile, $F_{\rm t}$, with mean equal to 3 and standard deviation of 0.5, in agreement with \citet{OguriHamana11}. In addition, we imposed $F_{\rm t}>0$. \\
\indent In Eq.\ \eqref{eq:B_HMF} we introduced the $\mathcal{B}_{\rm \scriptscriptstyle HMF}$ correction factor for the \citet{Tinker08} halo mass function. Following \citet{Costanzi19}, we adopted a 2D Gaussian prior on the $s$ and $q$ parameters in Eq.\ \eqref{eq:B_HMF},\footnote{From a practical point of view, $s$ and $q$ are derived parameters in our analysis. Indeed, the 2D Gaussian prior is decomposed into two independent 1D priors via Cholesky decomposition, with this maintaining the correlation between $s$ and $q$.} having mean $\mu_{\rm \scriptscriptstyle HMF}=(0.037, 1.008)$ and covariance matrix $C_{\rm \scriptscriptstyle HMF}$ expressed as
\begin{equation}\label{covMtinker}
C_{\rm \scriptscriptstyle HMF} = 
\begin{pmatrix}
0.00019 & 0.00024 \\
0.00024 & 0.00038
\end{pmatrix}
\,.
\end{equation}
Despite the fact that this prescription does not account for the uncertainties due to baryonic physics, the latter are subdominant compared to the precision on $\mathcal{B}_{\rm \scriptscriptstyle HMF}$ \citep[see][and references therein]{Costanzi19}.

\section{Results}\label{sec:DR4_results}
\begin{figure}[t!]
\centering\includegraphics[width = \hsize, height = 6.3cm] {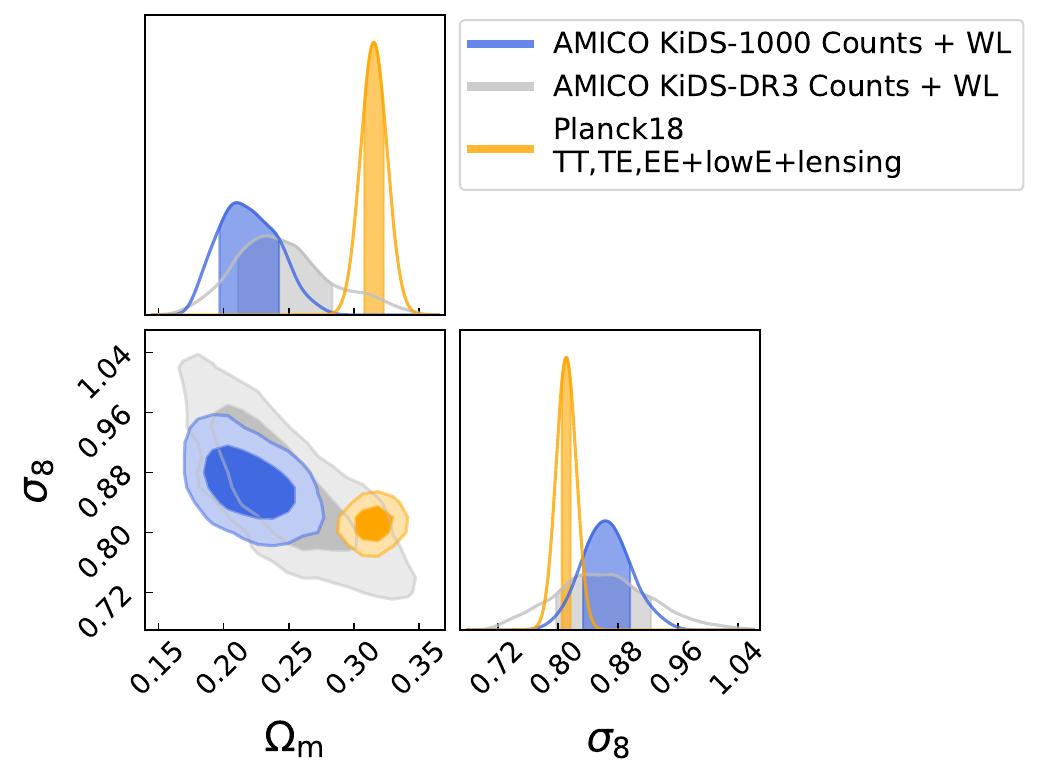}
\caption{Constraints on $\Omega_{\rm m}$ and $\sigma_8$, obtained from the joint modelling of cluster weak-lensing and count measurements described in this work (blue), by \citet{Lesci22_counts} (grey), and by \citetalias{Planck18} (orange). The confidence contours correspond to 68\% and 95\%, while the bands over the 1D marginalised posteriors represent the interval between 16th and 84th percentiles.}
\label{fig:posteriors_cosmology}
\end{figure}
In this section, we detail the constraints on cosmological parameters and on the $\log\lambda^*-\log M_{200}$ relation from the joint modelling of the stacked weak-lensing reduced shear profiles and counts of the AMICO KiDS-1000 clusters, carried out through a Bayesian analysis based on the likelihood function in Eq.\ \eqref{eq:Likelihood} and assuming the priors listed in Table \ref{tab:priors_and_posteriors}. We conducted a blind analysis by introducing biases in the completeness estimates of the cluster sample (see Sect. \ref{sec:selection_function}), with the results detailed in Sect. \ref{appendix:blinding}. While the cluster redshift cut at $z=0.8$ was driven by the low completeness and purity of the sample at higher redshifts (see Sect.\ \ref{sec:ClusterSample}), the $\lambda^*$ selection criteria were optimised to maintain robust model fits while maximising the cluster sample size. Compared to the weak-lensing analysis (Sect.\ \ref{sec:WL_stacking_2}), we adopted more stringent cuts in $\lambda^*$ for the counts, namely $\lambda^*>25$ for $z\in[0.1,0.3)$, $\lambda^*>28$ for $z\in[0.3,0.45)$, and $\lambda^*>30$ for $z\in[0.45,0.8]$. We verified that this sample selection has no significant impact on the final constraints and improves the quality of the cluster count fit. Differences between the lensing and count sample selections are expected, since low-$\lambda^*$ count measurements are very precise and more prone to unmodelled systematics. \\
\begin{figure*}[t!]
\centering\includegraphics[width = \hsize, height = 11.25cm] {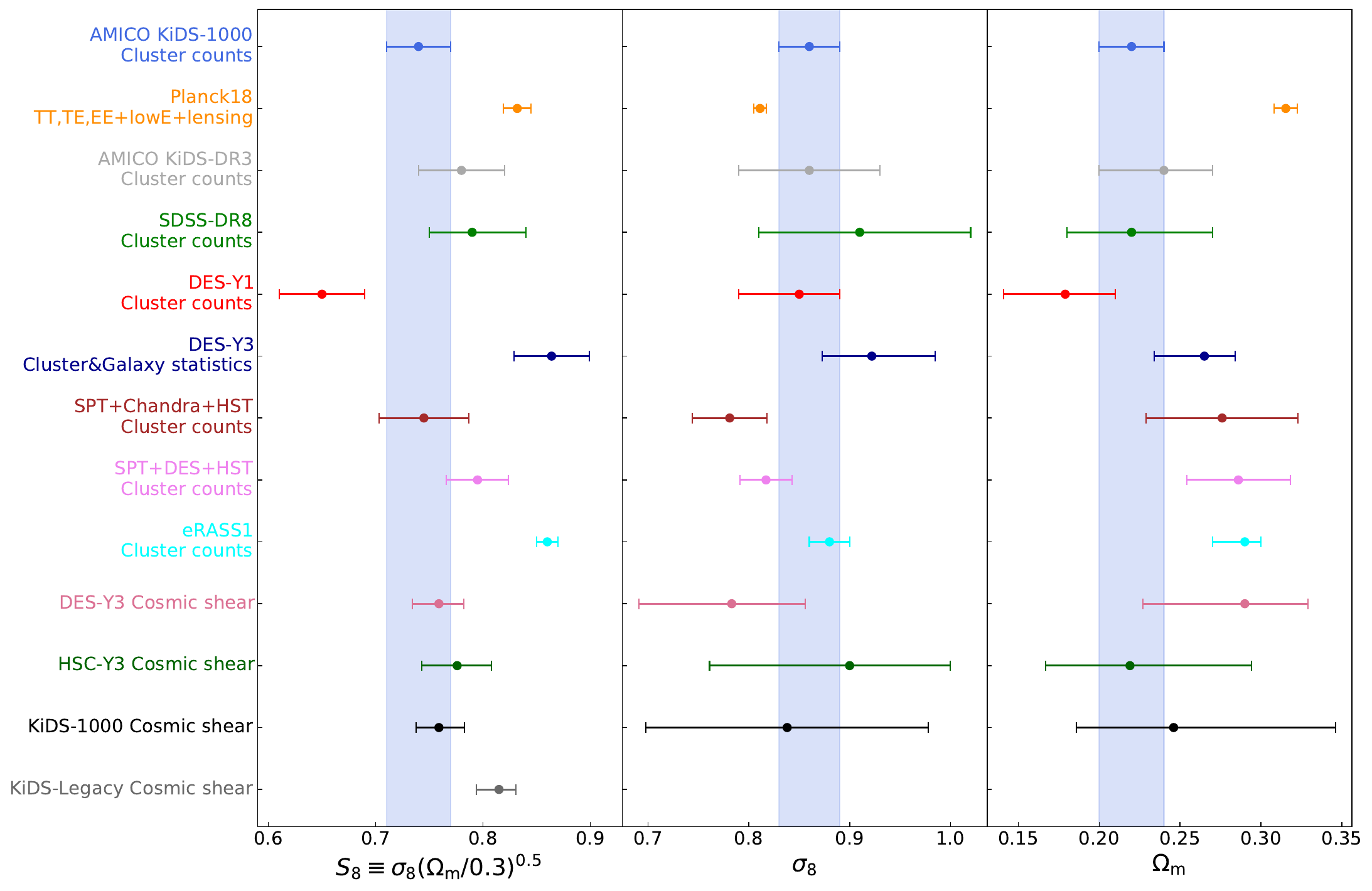}
\caption{Constraints on $S_8$ (left panel), $\sigma_8$ (middle panel), and $\Omega_{\rm m}$ (right panel) obtained, from top to bottom, in this work (blue), by \citetalias{Planck18} (orange), \citet{Lesci22_counts} (grey), \citet{Costanzi19} (green), \citet{Abbott20} (red), \citet{Abbott25} (dark blue), \citet{Bocquet19} (brown), \citet{Bocquet24} (violet), \citet{Ghirardini24} (cyan), \citet{Secco22}  (pink), \citet{Dalal23} (dark green), \citet{Asgari21} (black), and \citet{Wright_DR5} (dark grey). The median as well as the 16th and 84th percentiles are shown.}
\label{fig:comparison_cosmology}
\end{figure*}
\indent As mentioned in Sect.\ \ref{sec:modelling:likelihood}, we first run the pipeline without including the SSC. The median parameter values obtained from this initial run were then used to compute the $S$ matrix and the cluster count response entering Eq.\ \eqref{eq:Likelihood_counts}. This process was repeated iteratively until the posteriors converge. The resulting model values are overplotted to the stacked weak-lensing data in Fig.\ \ref{fig:g_t} and to the cluster count measurements in Fig.\ \ref{fig:counts}. By considering the weak-lensing measurements only, we obtained a reduced $\chi^2$ of $\chi_{\rm red}^2=1.1$. Here, in the computation of the degrees of freedom, we neglected the parameters entering the mass function correction factor, along with $\Omega_{\rm b}$, $n_{\rm s}$, and $h$, as their priors have no impact on the weak-lensing modelling. We found a good quality of the fit also for the counts, as shown by the Pearson residuals in the bottom panels of Fig.\ \ref{fig:counts}. \\
\indent In Sect.\ \ref{sec:DR4_results:baseline} we discuss the constraints on cosmological parameters and their comparison with the results from literature analyses, while Sect.\ \ref{sec:DR4_results:baseline_WL} focuses on the $\log\lambda^*-\log M_{200}$ relation. Finally, in Section \ref{sec:DR4_results:sys}, we assess the robustness of our findings against alternative modelling choices.

\subsection{Cosmological constraints}\label{sec:DR4_results:baseline}
From the initial MCMC run that does not include the SSC contribution, we obtained $\Omega_{\rm m}=0.21^{+0.02}_{-0.01}$, $\sigma_8=0.86^{+0.03}_{-0.03}$, and $S_8 \equiv \sigma_8(\Omega_{\rm m}/0.3)^{0.5}=0.72^{+0.02}_{-0.02}$. The inclusion of the SSC led to the following constraints:
\begin{align}
&\Omega_{\rm m} = \Om^{+\OmUp}_{-\OmLow}\,, \nonumber\\
&\sigma_8 = \seight^{+\seightUp}_{-\seightLow}\,, \nonumber\\
&S_8 = \Sotto^{+\SottoUp}_{-\SottoLow}\,.
\end{align}
We verified that the SSC iterative process converges after one iteration. These constraints, along with the ones on the other free model parameters considered in the analysis, are reported in Table \ref{tab:priors_and_posteriors}. We find a $1\sigma$ agreement with the results by \citet{Lesci22_counts}, who obtained $\Omega_{\rm m}=0.24^{+0.03}_{-0.04}$, $\sigma_8=0.86^{+0.07}_{-0.07}$, and $S_8=0.78^{+0.04}_{-0.04}$ by jointly modelling the cluster weak-lensing and count measurements of the AMICO clusters in KiDS-DR3. As we can see, thanks to the larger survey area and the extension in the covered cluster redshift range in our analysis (see Sect.\ \ref{sec:ClusterSample}), the statistical uncertainties on $\Omega_{\rm m}$ and $\sigma_8$ are halved, while the one on $S_8$ is reduced by 25\%. Defining the figure of merit (FoM) as
\begin{equation}
{\rm FoM} = \frac{1}{\sqrt{{\rm det}(C(\Omega_{\rm m},\sigma_8))}}\,,
\end{equation}
where $C(\Omega_{\rm m},\sigma_8)$ is the $\Omega_{\rm m}-\sigma_8$ covariance matrix, we obtain ${\rm FoM}=1680$. Compared to ${\rm FoM}=701$ from the constraints by \citet{Lesci22_counts}, the FoM more than doubles in KiDS-1000.\\
\indent Figure \ref{fig:posteriors_cosmology} displays the comparison with \citet{Lesci22_counts} results in the $\Omega_{\rm m}-\sigma_8$ parameter space, along with the constraints by \citetalias{Planck18} based on observations of the CMB power spectrum. The difference with \citetalias{Planck18} derived by \citet{Lesci22_counts} was of $2\sigma$ on $\Omega_{\rm m}$ and of about $1\sigma$ on $\sigma_8$ and $S_8$.\footnote{Throughout this paper, we express the differences between cosmological constraints as $x\sigma$, where $x=|M_1-M_2|\,(\sigma_1^2+\sigma_2^2)^{-0.5}$. Here, $M_1$ and $M_2$ are the median values of the parameter estimates 1 and 2, while $\sigma_1$ and $\sigma_2$ are the corresponding standard deviations.} Due to the lower median value and uncertainty, in this work we find a $4.5\sigma$ tension with \citetalias{Planck18} on $\Omega_{\rm m}$, while $\sigma_8$ agrees within $1.6\sigma$ with \citetalias{Planck18}. The lower $\Omega_{\rm m}$ compared to that obtained by \citet{Lesci22_counts} leads to a $2.8\sigma$ tension on $S_8$ with \citetalias{Planck18}. A tension on $\Omega_{\rm m}$, of about $3\sigma$, is found also with the results from the latest baryon acoustic oscillation (BAO) measurements based on the Dark Energy Spectroscopic Instrument \citep[DESI,][]{Adame25}. Indeed, assuming a flat $\Lambda$CDM model, they found $\Omega_{\rm m}=0.295\pm0.015$. \\
\indent Our results are compared with other literature analyses in Fig.\ \ref{fig:comparison_cosmology}. The $S_8$ constraint agrees within $1\sigma$ with the cluster count analysis by \citet{Costanzi19}, based on the photometric data from the SDSS Data
Release 8 \citep[SDSS-DR8,][]{Aihara11}, as well as with the counts of South Pole Telescope SZ survey \citep[SPT-SZ,][]{deHaan16} clusters by \citet{Bocquet19}, based on mass estimates from Chandra X-ray observations and weak-lensing measurements from the Hubble Space Telescope (HST) and the Magellan Telescope. We find agreement on $S_8$ also with the analysis by \citet{Bocquet24}, based on SZ-selected clusters in SPT and weak-lensing data from DES and HST. The result by \citet{Abbott20}, derived by modelling the abundance of DES-Y1 \citep{Drlica-Wagner18} optically selected clusters, agrees within $2\sigma$. This DES-Y1 result was confirmed by \citet{Aguena23}, who employed a different modelling of projection effects and a different MCMC sampler. Our $S_8$ result shows a 2.7$\sigma$ difference against the latest DES-Y3 result by \citet{Abbott25}, which combines cluster and galaxy statistics. Here, a key methodological distinction is our inclusion of small-scale cluster lensing, which \citet{Abbott25} exclude from their analysis. A large disagreement on $S_8$, of about $3.8\sigma$, is found with the counts of X-ray-selected clusters in the first eROSITA All-Sky Survey \citep[eRASS1,][]{Bulbul24} by \citet{Ghirardini24}, due to the $2.8\sigma$ difference in the $\Omega_{\rm m}$ constraint. Indeed, our $\sigma_8$ constraint agrees within $1\sigma$ with the result from \citet{Ghirardini24}, as well as with the findings from the other cluster count analyses mentioned earlier. The latter also show an agreement of about $1\sigma$ with our result on $\Omega_{\rm m}$. \\
\indent Figure \ref{fig:comparison_cosmology} shows that analyses based on optically selected clusters tend to yield lower values of $\Omega_{\rm m}$ compared to \citetalias{Planck18}. Following the DES-Y1 cosmological results by \citet{Abbott20}, extensive work was conducted to evaluate the impact of optical projection effects on cluster mass estimates \citep{Sunayama20,Wu22,Sunayama23,Park23,Zhou23} and, in turn, on cosmological constraints. As discussed in Sect.\ \ref{sec:DR4_results:sys}, we are confident that our results are not subject to such biases. \\
\indent Furthermore, as shown in Fig.\ \ref{fig:comparison_cosmology}, the $S_8$ result derived in this work is in excellent agreement with the cosmic shear analyses by \citet{Asgari21}, \citet{Secco22}, and \citet{Dalal23}, based on KiDS-1000 \citep{Giblin21}, DES-Y3 \citep{Gatti21}, and Hyper Suprime-Cam Year 3 \citep[HSC-Y3,][]{Li22} galaxy shape catalogues, respectively. We find a $2\sigma$ difference with the $S_8$ constraint by \citet{Wright_DR5}, who analysed the cosmic shear measurements from the latest KiDS-Legacy data \citep{Li23,Wright24}. The same level of discrepancy on $S_8$ was found by \citet{Wright_DR5} against the KiDS-1000 result by \citet{Asgari21}.

\subsection{Mass calibration}\label{sec:DR4_results:baseline_WL}
\begin{figure}[t!]
\centering\includegraphics[width = \hsize, height = \hsize] {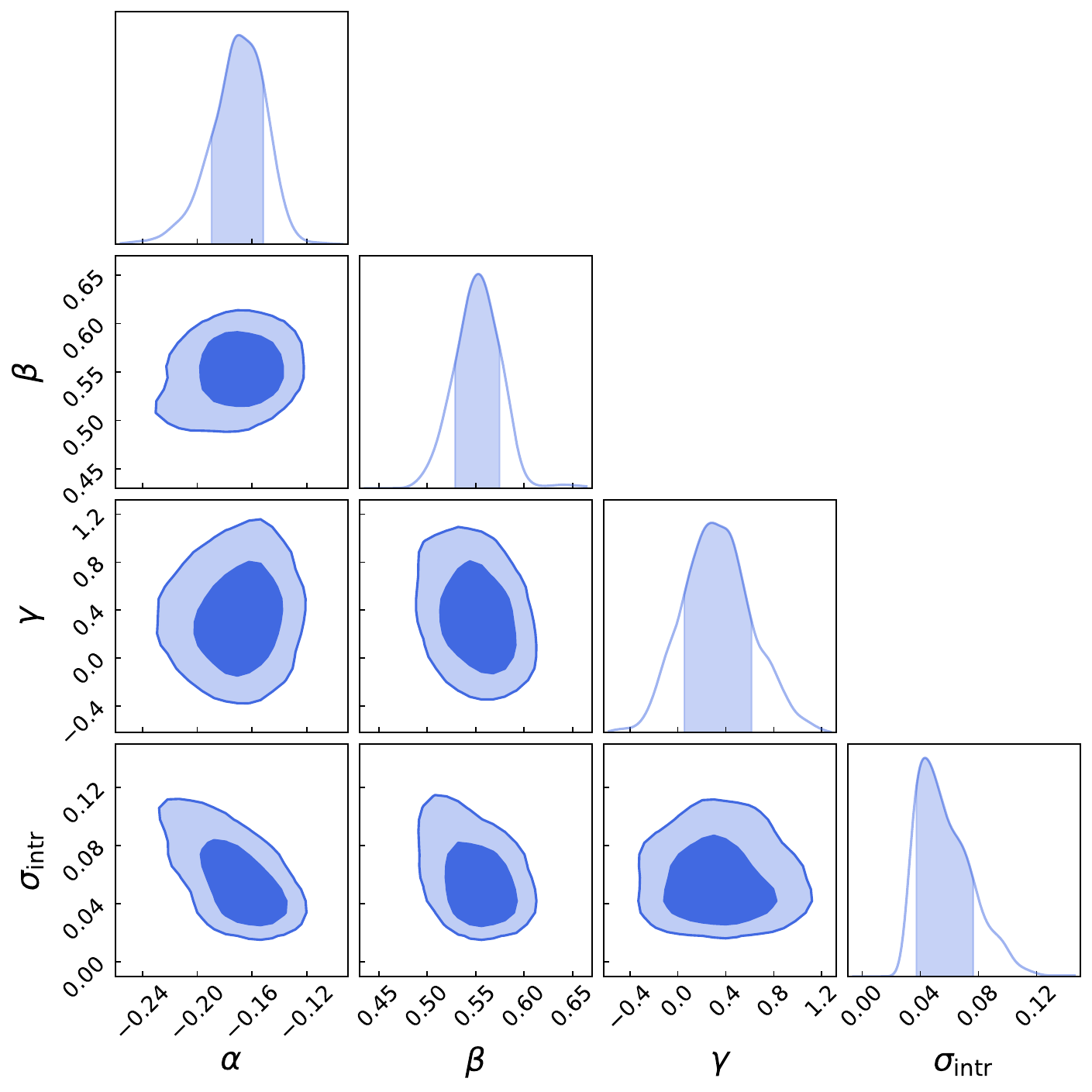}
\caption{Constraints on the $\log\lambda^*-\log M_{200}$ relation parameters, obtained from the joint modelling of cluster weak-lensing and count measurements. The confidence contours correspond to 68\% and 95\%, while the bands over the 1D marginalised posteriors represent the interval between the 16th and 84th percentiles.}
\label{fig:posteriors}
\end{figure}

\indent Regarding the parameters of the $\log\lambda^*-\log M_{200}$ relation, we derive $\alpha=\alphaScaling^{+\alphaScalingup}_{-\alphaScalinglow}$, $\beta=\betaScaling^{+\betaScalingup}_{-\betaScalinglow}$, $\gamma=\gammaScaling^{+\gammaScalingup}_{-\gammaScalinglow}$, and $\sigma_{\rm intr}=\scatterZero^{+\scatterZeroup}_{-\scatterZerolow}$ (see also Table \ref{tab:priors_and_posteriors} and Fig.\ \ref{fig:posteriors}). The result on $\sigma_{\rm intr}$ confirms the goodness of $\lambda^*$ as a mass proxy, as already found for AMICO KiDS-DR3 clusters by \citet{Sereno20} and \citet{Lesci22_counts}. In addition, this constraint agrees with literature results on the relation between cluster richness and weak-lensing mass based on real data \citep[see e.g.][]{Bleem20,Ghirardini24} and simulations \citep{Giocoli25}. Given the constraints on the $\log\lambda^*-\log M_{200}$ relation and on cosmological parameters, the expected mass value for a single galaxy cluster can be obtained from the formula
\begin{align}\label{eq:masses}
\langle M(&\lambda^*_{\rm ob}, z_{\rm ob})\rangle =
\frac{1}{\langle n(\lambda^*_{\rm ob},z_{\rm ob})\rangle}\times\nonumber\\
&\times\int_0^\infty{\rm d}z_{\rm tr}\,\frac{{\rm d}^2 V}{{\rm d} z_{\rm tr}{\rm d}\Omega}\int_0^\infty{\rm d}M\,\frac{{\rm d} n(M,z_{\rm tr})}{{\rm d} M}\,\mathcal{B}_{\rm\scriptscriptstyle HMF}(M)\,M\,\times\nonumber\\
&\times\,\int_0^\infty{\rm d}\lambda^*_{\rm tr}\,\mathcal{C}_{\rm clu}(\lambda^*_{\rm tr},z_{\rm tr})\,P(\lambda^*_{\rm tr}|M,z_{\rm tr})\,P(\lambda^*_{\rm ob}|\lambda^*_{\rm tr},z_{\rm tr})P(z_{\rm ob}|z_{\rm tr})\,,
\end{align}
where 
\begin{align}
\langle n&(\lambda^*_{\rm ob},z_{\rm ob})\rangle =
\int_0^\infty{\rm d}z_{\rm tr}\,\frac{{\rm d}^2 V}{{\rm d} z_{\rm tr}{\rm d}\Omega}\int_0^\infty{\rm d}M\,\frac{{\rm d} n(M,z_{\rm tr})}{{\rm d} M}\,\mathcal{B}_{\rm\scriptscriptstyle HMF}(M)\,\times\nonumber\\
&\times\,\int_0^\infty{\rm d}\lambda^*_{\rm tr}\,\mathcal{C}_{\rm clu}(\lambda^*_{\rm tr},z_{\rm tr})\,P(\lambda^*_{\rm tr}|M,z_{\rm tr})\,P(\lambda^*_{\rm ob}|\lambda^*_{\rm tr},z_{\rm tr})P(z_{\rm ob}|z_{\rm tr})\,.
\end{align}
By computing Eq.\ \eqref{eq:masses} at each MCMC step for each cluster in the sample, considering the whole parameter space defined in Table \ref{tab:priors_and_posteriors}, we obtained mean mass estimates and the corresponding standard deviations. The minimum cluster mass derived in this way is $M_{200}^{\rm min}=3\times10^{13}$ $h^{-1}$M$_\odot$ for $z\in[0.1,0.3)$, $M_{200}^{\rm min}=4\times10^{13}$ $h^{-1}$M$_\odot$ for $z\in[0.3,0.45)$, and $M_{200}^{\rm min}=5\times10^{13}$ $h^{-1}$M$_\odot$ for $z\in[0.45,0.8]$. Furthermore, the median mass precision corresponds to about 8\%. This precision shows a three percentage points enhancement compared to what derived by \citetalias{Bellagamba19} from the AMICO KiDS-DR3 sample \citep{Maturi19}. Thus, despite the fact that the galaxy weighted density in KiDS-1000, which amounts to 6.17 arcmin$^{-2}$ \citep{Giblin21}, is lower compared to the one associated with the KiDS-DR3 sample, namely 8.53 arcmin$^{-2}$ \citep{Hildebrandt17}, due to more stringent redshift and SOM-gold selections \citep{Wright20,Giblin21}, the larger survey area in KiDS-1000 implies a better precision on the mass scaling relation. We also remark that the analysis carried out in this work is not fully comparable with that performed by \citetalias{Bellagamba19}. The modelling procedure is different, as we derive the $\log\lambda^*-\log M_{200}$ relation directly from the stacked profiles, while \citetalias{Bellagamba19} first derived the average masses of the stacks and then modelled their relation with cluster observables. The weak-lensing modelling carried out in this work has the advantage of allowing for a proper marginalisation of the cosmological posteriors over the local cluster properties. Moreover, the set of modelling parameters differs from that assumed by \citetalias{Bellagamba19}, since, for example, we marginalise the $\log\lambda^*-\log M_{200}$ relation over the halo truncation factor, $F_{\rm t}$, and $\sigma_{\rm intr}$. A thorough comparison between AMICO KiDS-1000 and KiDS-DR3 mass estimates is detailed in Appendix \ref{appendix:mass_comparison}. In addition, in Appendix \ref{appendix:Delta} we discuss the mass calibration results in the case of alternative spherical overdensity definitions. \\
\indent We find that $F_{\rm t}$ and the fraction of miscentred clusters, $f_{\rm off}$, are not constrained, while we obtain $\sigma_{\rm off}=\soffZero^{+\soffZeroup}_{-\soffZerolow}$ $h^{-1}$Mpc. This result on the miscentring scale agrees with literature results from both simulations and observations \citep{Saro15,Zhang19,Yan20,Sommer23,Seppi23}. Furthermore, we verified that the constraint $\log c_0=\cZero^{+\cZeroup}_{-\cZerolow}$ yields a $\log c_{200}-\log M_{200}$ relation which is consistent within 1$\sigma$ with the one by \citet{Duffy08}. Compared to the results by \citet{Dutton14}, \citet{Meneghetti14}, \citet{Child18}, \citet{Diemer19}, and \citet{Ishiyama21}, our $\log c_{200}$ values are lower but remain consistent within 2$\sigma$. 

\subsection{Robustness of the results}\label{sec:DR4_results:sys}
Throughout this paper, we based our assumptions on simulations and prior observational constraints for parameters that cannot be fully determined by our data, as detailed in Sect.\ \ref{sec:modelling:likelihood}. In addition, the choice of sample selections is motivated solely by the goodness of the fit. For example, we excluded clusters at redshift $z>0.8$ because, if included, they would degrade the quality of our fit. In this section we test how much these assumptions affect the final results. In the tests presented in this section, we did not include the SSC contribution, as it is not expected to affect our conclusions. Therefore, the comparisons below are made against the posteriors obtained from the baseline analysis where SSC is also excluded. \\
\indent Firstly, to assess the impact of our assumptions on $c_{200}$, we set the mass evolution parameter of the concentration, namely $c_M$ in Eq.\ \eqref{eq:cM_rel}, as free to vary. By assuming the large flat prior $c_M\in[-1.5,1.5]$, we obtain $c_M=0.07^{+0.23}_{-0.19}$, that is well consistent with the value assumed for the baseline analysis, namely $c_M=-0.084$. The statistical uncertainty on $\log c_0$ is increased by 20\% this case, and the constraint on $\log c_0$ is consistent within $1\sigma$ with the one reported in Table \ref{tab:priors_and_posteriors}. The $\log\lambda^*-\log M_{200}$ relation and the cosmological posteriors remain unchanged. Furthermore, as discussed in Sect.\ \ref{sec:modelling:stacked_g}, we neglected the effect of the intrinsic scatter in the $\log c_{200}-\log M_{200}$ relation. If assumed to be log-normal, the intrinsic scatter on the natural logarithm of the concentration given a mass amounts to about 35\% \citep[see e.g.][]{Duffy08,Bhattacharya13,Diemer15,Umetsu20}. Applying this relative uncertainty to the median posterior value of $\log c_0$ reported in Table \ref{tab:priors_and_posteriors}, we obtain an intrinsic scatter of $\sigma_{\rm intr}^{\log c_0}=0.06$, that is less than half the statistical uncertainty on $\log c_0$, namely $\sigma_{\rm stat}^{\log c_0}=0.13$. Treating $\sigma_{\rm intr}^{\log c_0}$ and $\sigma_{\rm stat}^{\log c_0}$ as uncorrelated, the total uncertainty on $\log c_0$ amounts to $\sigma_{\rm tot}^{\log c_0}=0.14$, which is very close to $\sigma_{\rm stat}^{\log c_0}$. Thus, we do not expect $\sigma_{\rm intr}^{\log c_0}$ to have a significant role in our analysis. \\
\indent To investigate the impact of the Gaussian prior on the fraction of miscentred clusters, we assumed $f_{\rm off}=1$. Although this leads to the near halving of both the relative uncertainty and the median of $\sigma_{\rm off}$, the constraint on $\sigma_{\rm off}$ remains consistent within $1\sigma$ with the baseline. The lowering of the $\sigma_{\rm off}$ uncertainty is expected, thanks to the reduction of model parameters describing the population of miscentred clusters. In addition, the median $\sigma_{\rm off}$ decreases to compensate for the assumption that all clusters are miscentred. In this case, the posteriors on the $\log\lambda^*-\log M_{200}$ scaling relation parameters undergo negligible changes, and the $S_8$ variation is $\Delta S_8 \equiv S_8^{\rm test}-S_8^{\rm base} = 0.01$, where $S_8^{\rm test}$ and $S_8^{\rm base}$ are derived in this test and in the baseline analysis, respectively. We remark that, as discussed in Sect.\ \ref{sec:DR4_results:baseline}, the $S_8$ statistical uncertainty is 0.03. Overall, the minimal impact of miscentring assumptions on the final results was expected, given that we excluded an extended region around the cluster centres from the weak-lensing analysis (see Sect.\ \ref{sec:WL_stacking_2}) and that miscentring effects are mild in stacked cluster profiles \citep{Currie24}. \\
\indent Another useful choice of the radial range employed for weak-lensing measurements is the upper limit of 3.5 $h^{-1}$Mpc, as discussed in Sect.\ \ref{sec:WL_stacking_2}. Indeed, this ensures that the impact of anisotropic boosts, affecting the signal in the two-halo region of optically selected clusters \citep{Sunayama20,Wu22,Sunayama23,Park23,Zhou23}, is mitigated. To qualitatively but effectively reproduce the $g_+$ bias retrieved from simulations, we multiplied the $\Delta\Sigma_+$ two-halo term, derived from Eq.\ \eqref{eq:2halo}, by 1.2. This does not lead to any variations in the posteriors. We want to emphasise that the radial upper limit adopted in our analysis, similar to that employed by \citetalias{Bellagamba19} in KiDS-DR3, is a key difference with the analysis of DES cluster counts by \citet{Abbott20}, who found an $\Omega_{\rm m}$ constraint in significant tension with literature cosmic shear and CMB analyses. The study by \citet{Abbott20}, indeed, was based on the mass calibration by \citet{McClintock19}, who modelled the cluster weak-lensing signal up to 30 Mpc. \\
\indent Additionally, we tested the impact of excluding the stacked profile and count measurements in the redshift bin $z\in[0.45,0.8]$, where sample impurity and incompleteness are more significant. In this case, the median values of $\alpha$ and $\gamma$ do not vary, while their statistical uncertainties are doubled. The constraint on $\beta$ remains consistent within 1$\sigma$ but undergoes remarkable changes, as we derive a 1.7$\sigma$ increase of the median, while its uncertainty increases by 50\%. The constraint on $S_8$ changes accordingly, as we derive $\Delta S_8=0.02$ and a 50\% increase of its statistical uncertainty. All in all, the exclusion of the high-redshift bin from the analysis yields results consistent with those presented in the previous sections and, as expected, implies an increase in the width of the posteriors. Consequently, we conclude that any possible biases in the corrections for the low background selection purity at high $z$, discussed in Sect.\ \ref{Sec_sys_2}, and low cluster sample completeness, are not significant.\\
\indent Since the intrinsic scatter of the $\log\lambda^*-\log M_{200}$ relation is expected to depend on mass, we express $\sigma_{\rm intr}$ as
\begin{equation}
\sigma_{\rm intr} = \sigma_{\rm intr,0}+\sigma_{{\rm intr},M}\log(M/M_{\rm piv})\,,
\end{equation}
where we use the flat prior $\sigma_{{\rm intr},M}\in[-0.1,0.1]$, while for $\sigma_{\rm intr,0}$ we assume the same $\sigma_{\rm intr}$ prior in Table \ref{tab:priors_and_posteriors}. We find $\sigma_{{\rm intr},M}=-0.043\pm0.03$, $\sigma_{{\rm intr},0}=0.07\pm0.03$, and not significant changes in the scaling relation parameters, which yield $\Delta S_8=0.01$ and a 20\% increase in its statistical uncertainty. Thus, we conclude that the mass evolution of $\sigma_{\rm intr}$ has a negligible impact in our analysis.\\
\indent We do not expect intrinsic alignments to be important in our analysis, as they produce a sub-percent systematic error on cluster profiles \citep{Chisari14,Sereno18}. Other works validate the mass modelling through simulations designed to reproduce the data, testing the theoretical assumptions adopted in the analysis. The outcome of this approach is an estimate of the so-called weak-lensing mass bias \citep[see e.g.][]{Grandis21,Bocquet24_3}. Our analysis takes a different path, as we explicitly accounted for systematic uncertainties in the measurements and avoided simplified assumptions about halo properties, which can artificially increase both the mass bias and its scatter. For instance, we adopted a BMO profile with a free truncation parameter \citep[which performs better than a simple NFW; see][]{OguriHamana11}, and imposed conservative priors on the halo concentration, which is commonly fixed in mass bias studies. Despite these methodological differences, we expect our results to remain consistent with those derived from mass bias analyses.

\section{Conclusions}\label{sec:DR4_conclusions}
In this work, we jointly modelled the weak-lensing and count measurements of the galaxy clusters in the AMICO catalogue by \citetalias{AMICOKiDS-1000}, based on KiDS-1000 data \citep{Kuijken19} and covering an effective area of 839 deg$^{2}$. The weak-lensing measurements were based on 8730 clusters, while 7789 clusters contributed to the counts, covering the redshift range $z\in[0.1,0.8]$ and with a ${\rm S/N}>3.5$. With this joint modelling, we provided the first mass calibration of AMICO KiDS-1000 galaxy clusters, as well as the first cosmological constraints based on this sample.\\
\indent Based on the galaxy shape catalogue by \citet{Giblin21}, we measured the stacked cluster weak-lensing profiles in bins of $z$ and $\lambda^*$, relying on the combination of photo-$z$ and colour selections to define the background galaxy samples. We used a SOM analysis, based on the spectroscopic galaxy sample developed by \citet{vanDenBusch22} and \citet{Wright24}, to reconstruct the true galaxy redshift distribution of the background sample and assess the purity of background selection. We also measured the cluster abundance as a function of $z$ and $\lambda^*$. In the joint modelling of our probes, we accounted for the systematic uncertainties due to impure cluster detections and contaminants in the weak-lensing signal, halo orientation and miscentring, projection effects, uncertainties affecting the cluster $z$ and $\lambda^*$, truncation of cluster halo mass distributions, matter correlated with cluster haloes, multiplicative shear bias, geometric distortions, uncertainties in the halo mass function, and super-sample covariance. The theoretical uncertainties related to the baryonic impact on the cluster profiles are propagated into the final results through the assumption of not informative priors on $\sigma_{\rm intr}$ and on the halo concentration. We also employed a blinding of the analysis, briefly described in Appendix \ref{appendix:blinding} and detailed in \citetalias{AMICOKiDS-1000}, based on perturbing the cluster sample completeness. \\
\indent We obtained $\Omega_{\rm m} = \Om^{+\OmUp}_{-\OmLow}$, $\sigma_8 = \seight^{+\seightUp}_{-\seightLow}$, and $S_8 = \Sotto^{+\SottoUp}_{-\SottoLow}$. Compared to the joint modelling of cluster weak-lensing and counts of AMICO KiDS-DR3 clusters by \citet{Lesci22_counts}, the uncertainties on $\Omega_{\rm m}$ and $\sigma_8$ are halved and the constraints agree within $1\sigma$. We found excellent agreement on $S_8$ with recent cosmic shear and cluster count analyses. However, with respect to the $S_8$ result from the KiDS-Legacy cosmic shear analysis \citep{Wright_DR5}, we find the same level of discrepancy they derived against the KiDS-1000 cosmic shear result \citep{Asgari21}, amounting to $2\sigma$. As discussed in \citet{Stolzner_DR5} and \citet{Wright_DR5}, this significant shift in $S_8$ can be ascribed to improvements in data calibration and reduced statistical noise, compared to KiDS-1000. Changes in the ellipticity estimates and in the calibration of galaxy redshift distributions may have an impact on a future cluster mass calibration based on KiDS-Legacy data and, consequently, on the cosmological posteriors.\\
\indent We obtained a $2.8\sigma$ $S_8$ tension with \citetalias{Planck18}, driven by a $4.5\sigma$ tension on $\Omega_{\rm m}$. A $3\sigma$ tension on $\Omega_{\rm m}$ is also found with the DESI BAO results by \citet{Adame25}. On the other hand, our $\Omega_{\rm m}$ constraint agrees with the results from recent cluster count analyses \citep{Costanzi19,Bocquet19,Abbott20,Bocquet24}. As discussed in Sect.\ \ref{sec:DR4_results:sys}, our results reliably account for the known biases that affect optical data and, most importantly, are not affected by the selection effects that impact the two-halo term of cluster profiles, which were the main sources of systematic uncertainties affecting the constraints from DES-Y1 cluster counts \citep{Abbott20,Sunayama20,Costanzi21,Wu22,Zhou23}. All in all, the results on cosmological parameters presented in this paper are competitive, in terms of uncertainties, with recent cosmological analyses from the literature and confirm the $S_8$ tension between late-Universe and \citetalias{Planck18} CMB observations. \\
\indent Together with the cosmological parameters, we constrained the amplitude, $\alpha=\alphaScaling^{+\alphaScalingup}_{-\alphaScalinglow}$, the $\lambda^*$ slope, $\beta=\betaScaling^{+\betaScalingup}_{-\betaScalinglow}$, the redshift evolution parameter, $\gamma=\gammaScaling^{+\gammaScalingup}_{-\gammaScalinglow}$, and the intrinsic scatter, $\sigma_{\rm intr}=\scatterZero^{+\scatterZeroup}_{-\scatterZerolow}$, of the $\log\lambda^*-\log M_{200}$ relation. These constraints, along with those on cosmological parameters, are robust with respect to our modelling choices. Indeed, alternative assumptions on the halo miscentring and concentration models do not lead to significant variations in the final results. Furthermore, the exclusion of the high-redshift data from the analysis leads to negligible variations in our constraints. This implies that our estimates of cluster sample purity and foreground contamination are robust, given the uncertainties associated with our stacked weak-lensing measurements. The average mass precision amounts to 8\%, which is an improvement of three percentage points compared to the mass precision derived for the AMICO KiDS-DR3 clusters by \citetalias{Bellagamba19}, mainly driven by the larger survey area in KiDS-1000. Compared to \citetalias{Bellagamba19}, we included $\sigma_{\rm intr}$ in the modelling and marginalised the results over the halo truncation factor, $F_{\rm t}$. Further improvements compared to \citetalias{Bellagamba19} include the estimation of the true background redshift distributions and background sample purity as a function of lens redshift, derived through a SOM analysis. In addition, in this paper we calibrated the $\log\lambda^*-\log M_{200}$ relation by directly modelling the stacked weak-lensing profiles. This represents a significant improvement over the cluster count analysis by \citet{Lesci22_counts}, which was based on the mass calibration by \citetalias{Bellagamba19}, where the scaling relation relied on the mean masses assigned to each stack. In fact, the modelling presented in this work allows for a more effective propagation of the cluster profile uncertainties into the cosmological posteriors. \\
\indent For future analyses based on KiDS data, we plan to improve the cosmological constraints by combining the modelling presented in this paper with the cluster clustering analysis by Romanello et al.\ (in prep.), which is based on the same dataset. In addition, a significant enhancement of the cosmological constraints is anticipated from KiDS-Legacy \citep{Wright24}, which expands the survey area by 34\% and doubles the $i$-band exposure time compared to KiDS-1000 \citep{Kuijken19}. The calibration of the background redshift distributions, detailed in Sect.\ \ref{Sec_sys_2}, can be further refined by combining SOMs with spatial cross-correlations between spectroscopic samples and KiDS sources \citep[following][]{Wright25}. Another key advancement will be the inclusion of the KiDS cosmic shear likelihood in the analysis, which will enable constraints on neutrino masses and on the parameters of the dark energy equation of state \citep[see e.g.][]{Bocquet24_2,To25,Abbott25}. Tight constraints on the Hubble constant will be attained through the combination of baryon acoustic oscillation data and cluster counts \citep{Costanzi19}. Furthermore, cluster counts can be employed to test General Relativity and the $\Lambda$CDM model \citep[see e.g.][]{Artis24}. Alongside this, we will assess the impact of different richness-mass relation models \citep{Chen24,Grandis25} and test alternative lensing profile estimators to maximise the ${\rm S/N}$ \citep{ShirasakiTakada18}. Furthermore, the combination of weak-lensing observations with X-ray and SZ measurements can improve cluster mass precision and accuracy \citep{Costanzi21,Singh24}. AMICO-tailored simulations will enable more precise estimates of the offset between the cluster galaxy and dark matter distributions. Compared to the conservative priors on miscentring parameters discussed in Sect.\ \ref{sec:modelling:likelihood}, this will improve the precision of the $\log\lambda^*-\log M$ relation and, in turn, of cosmological parameter constraints.

\section*{Data availability}
With this paper, we release the $M_{200}$ estimates obtained from the baseline analysis, along with $M_{500}$ and $M_{200\rm m}$, whose derivation is detailed in Appendix \ref{appendix:Delta}. Mean and standard deviation of these masses are derived by following the methods described in Sect.\ \ref{sec:DR4_results:baseline_WL}, based on Eq.\ \eqref{eq:masses}.

\begin{acknowledgements}
Based on observations made with ESO Telescopes at the La Silla Paranal Observatory under programme IDs 177.A-3016, 177.A-3017, 177.A-3018 and 179.A-2004, and on data products produced by the KiDS consortium. The KiDS production team acknowledges support from: Deutsche Forschungsgemeinschaft, ERC, NOVA and NWO-M grants; Target; the University of Padova, and the University Federico II (Naples). We acknowledge the financial contribution from the grant PRIN-MUR 2022 20227RNLY3 “The concordance cosmological model: stress-tests with galaxy clusters” supported by Next Generation EU and from the grant ASI n.\ 2024-10-HH.0 “Attività scientifiche per la missione Euclid – fase E”. MS acknowledges financial contributions from contract ASI-INAF n.2017-14-H.0, contract INAF mainstream project 1.05.01.86.10, INAF Theory Grant 2023: Gravitational lensing detection of matter distribution at galaxy cluster boundaries and beyond (1.05.23.06.17), and contract Prin-MUR 2022 supported by Next Generation EU (n.20227RNLY3, The concordance cosmological model: stress-tests with galaxy clusters). GC acknowledges the support from the Next Generation EU funds within the National Recovery and Resilience Plan (PNRR), Mission 4 - Education and Research, Component 2 - From Research to Business (M4C2), Investment Line 3.1 - Strengthening and creation of Research Infrastructures, Project IR0000012 – “CTA+ - Cherenkov Telescope Array Plus”. MB is supported by the Polish National Science Center through grants no. 2020/38/E/ST9/00395 and 2020/39/B/ST9/03494. HH is supported by a DFG Heisenberg grant (Hi 1495/5-1), the DFG Collaborative Research Center SFB1491, an ERC Consolidator Grant (No. 770935), and the DLR project 50QE2305. SJ acknowledges the Ramón y Cajal Fellowship (RYC2022-036431-I) from the Spanish Ministry of Science and the Dennis Sciama Fellowship at the University of Portsmouth.
\end{acknowledgements}

\begingroup
\renewcommand{\baselinestretch}{1.0}
\small
\bibliographystyle{aa}
\bibliography{aanda}

\begin{thebibliography}{215}
\expandafter\ifx\csname natexlab\endcsname\relax\def\natexlab#1{#1}\fi

\bibitem[{{Abbott} {et~al.}(2020){Abbott}, {Aguena}, {Alarcon}, {Allam},
  {Allen}, {Annis}, {Avila}, {Bacon}, {Bechtol}, {Bermeo}, {Bernstein},
  {Bertin}, {Bhargava}, {Bocquet}, {Brooks}, {Brout}, {Buckley-Geer}, {Burke},
  {Carnero Rosell}, {Carrasco Kind}, {Carretero}, {Castander}, {Cawthon},
  {Chang}, {Chen}, {Choi}, {Costanzi}, {Crocce}, {da Costa}, {Davis}, {De
  Vicente}, {DeRose}, {Desai}, {Diehl}, {Dietrich}, {Dodelson}, {Doel},
  {Drlica-Wagner}, {Eckert}, {Eifler}, {Elvin-Poole}, {Estrada}, {Everett},
  {Evrard}, {Farahi}, {Ferrero}, {Flaugher}, {Fosalba}, {Frieman},
  {Garc{\'\i}a-Bellido}, {Gatti}, {Gaztanaga}, {Gerdes}, {Giannantonio},
  {Giles}, {Grandis}, {Gruen}, {Gruendl}, {Gschwend}, {Gutierrez}, {Hartley},
  {Hinton}, {Hollowood}, {Honscheid}, {Hoyle}, {Huterer}, {James}, {Jarvis},
  {Jeltema}, {Johnson}, {Johnson}, {Kent}, {Krause}, {Kron}, {Kuehn},
  {Kuropatkin}, {Lahav}, {Li}, {Lidman}, {Lima}, {Lin}, {MacCrann}, {Maia},
  {Mantz}, {Marshall}, {Martini}, {Mayers}, {Melchior}, {Mena-Fern{\'a}ndez},
  {Menanteau}, {Miquel}, {Mohr}, {Nichol}, {Nord}, {Ogando}, {Palmese},
  {Paz-Chinch{\'o}n}, {Plazas}, {Prat}, {Rau}, {Romer}, {Roodman}, {Rooney},
  {Rozo}, {Rykoff}, {Sako}, {Samuroff}, {S{\'a}nchez}, {Sanchez}, {Saro},
  {Scarpine}, {Schubnell}, {Scolnic}, {Serrano}, {Sevilla-Noarbe}, {Sheldon},
  {Smith}, {Smith}, {Suchyta}, {Swanson}, {Tarle}, {Thomas}, {To}, {Troxel},
  {Tucker}, {Varga}, {von der Linden}, {Walker}, {Wechsler}, {Weller},
  {Wilkinson}, {Wu}, {Yanny}, {Zhang}, {Zhang}, {Zuntz}, \& {DES
  Collaboration}}]{Abbott20}
{Abbott}, T.~M.~C., {Aguena}, M., {Alarcon}, A., {et~al.} 2020,
  \href{http://dx.doi.org/10.1103/PhysRevD.102.023509}{\color{blue}\prd},
  \href{https://ui.adsabs.harvard.edu/abs/2020PhRvD.102b3509A}{102, 023509}

\bibitem[{{Abbott} {et~al.}(2025){Abbott}, {Aguena}, {Alarcon}, {Anbajagane},
  {Andrade-Oliveira}, {Avila}, {Bacon}, {Becker}, {Bhargava}, {Blazek},
  {Bocquet}, {Brooks}, {Carnero Rosell}, {Carretero}, {Castander}, {Chang},
  {Choi}, {Conselice}, {Costanzi}, {Crocce}, {da Costa}, {Pereira}, {Davis},
  {Desai}, {Diehl}, {Dodelson}, {Doel}, {Elvin-Poole}, {Esteves}, {Everett},
  {Farahi}, {Fert{\'e}}, {Flaugher}, {Garc{\'\i}a-Bellido}, {Gatti},
  {Giannini}, {Giles}, {Grandis}, {Gruen}, {Gruendl}, {Gutierrez}, {Harrison},
  {Hinton}, {Hollowood}, {Honscheid}, {Jeffrey}, {Jeltema}, {Krause}, {Lahav},
  {Lee}, {Lidman}, {Lima}, {Lin}, {Mohr}, {Marshall}, {McCullough},
  {Mena-Fern}, {Miquel}, {Muir}, {Myles}, {Ogando}, {Palmese}, {Paterno},
  {Plazas Malag{\'o}n}, {Porredon}, {Prat}, {Romer}, {Roodman}, {Rozo},
  {Rykoff}, {Sanchez}, {Sanchez Cid}, {Sevilla-Noarbe}, {Smith}, {Suchyta},
  {Tarle}, {Thomas}, {To}, {Troxel}, {Vikram}, {Walker}, {Weinberg},
  {Weaverdyck}, {Wechsler}, {Weller}, {Wu}, {Yamamoto}, {Yanny}, {Zhang}, \&
  {Zhou}}]{Abbott25}
{Abbott}, T.~M.~C., {Aguena}, M., {Alarcon}, A., {et~al.} 2025,
  \href{https://ui.adsabs.harvard.edu/abs/2025arXiv250313632D}{\href{http://dx.doi.org/10.48550/arXiv.2503.13632}{\color{blue}arXiv
  e-prints}, arXiv:2503.13632}

\bibitem[{{Adame} {et~al.}(2025){Adame}, {Aguilar}, {Ahlen}, {Alam},
  {Alexander}, {Alvarez}, {Alves}, {Anand}, {Andrade}, {Armengaud}, {Avila},
  {Aviles}, {Awan}, {Bahr-Kalus}, {Bailey}, {Baltay}, {Bault}, {Behera},
  {BenZvi}, {Bera}, {Beutler}, {Bianchi}, {Blake}, {Blum}, {Brieden},
  {Brodzeller}, {Brooks}, {Buckley-Geer}, {Burtin}, {Calderon}, {Canning},
  {Carnero Rosell}, {Cereskaite}, {Cervantes-Cota}, {Chabanier}, {Chaussidon},
  {Chaves-Montero}, {Chen}, {Chen}, {Claybaugh}, {Cole}, {Cuceu}, {Davis},
  {Dawson}, {de la Macorra}, {de Mattia}, {Deiosso}, {Dey}, {Dey}, {Ding},
  {Doel}, {Edelstein}, {Eftekharzadeh}, {Eisenstein}, {Elliott}, {Fagrelius},
  {Fanning}, {Ferraro}, {Ereza}, {Findlay}, {Flaugher}, {Font-Ribera},
  {Forero-S{\'a}nchez}, {Forero-Romero}, {Frenk}, {Garcia-Quintero},
  {Gazta{\~n}aga}, {Gil-Mar{\'\i}n}, {Gontcho a Gontcho}, {Gonzalez-Morales},
  {Gonzalez-Perez}, {Gordon}, {Green}, {Gruen}, {Gsponer}, {Gutierrez}, {Guy},
  {Hadzhiyska}, {Hahn}, {Hanif}, {Herrera-Alcantar}, {Honscheid}, {Howlett},
  {Huterer}, {Ir{\v{s}}i{\v{c}}}, {Ishak}, {Juneau}, {Kara{\c{c}}ayl{\i}},
  {Kehoe}, {Kent}, {Kirkby}, {Kremin}, {Krolewski}, {Lai}, {Lan}, {Landriau},
  {Lang}, {Lasker}, {Le Goff}, {Le Guillou}, {Leauthaud}, {Levi}, {Li},
  {Linder}, {Lodha}, {Magneville}, {Manera}, {Margala}, {Martini}, {Maus},
  {McDonald}, {Medina-Varela}, {Meisner}, {Mena-Fern{\'a}ndez}, {Miquel},
  {Moon}, {Moore}, {Moustakas}, {Mueller}, {Mu{\~n}oz-Guti{\'e}rrez}, {Myers},
  {Nadathur}, {Napolitano}, {Neveux}, {Newman}, {Nguyen}, {Nie}, {Niz},
  {Noriega}, {Padmanabhan}, {Paillas}, {Palanque-Delabrouille}, {Pan},
  {Penmetsa}, {Percival}, {Pieri}, {Pinon}, {Poppett}, {Porredon}, {Prada},
  {P{\'e}rez-Fern{\'a}ndez}, {P{\'e}rez-R{\`a}fols}, {Rabinowitz}, {Raichoor},
  {Ram{\'\i}rez-P{\'e}rez}, {Ramirez-Solano}, {Rashkovetskyi}, {Ravoux},
  {Rezaie}, {Rich}, {Rocher}, {Rockosi}, {Roe}, {Rosado-Marin}, {Ross},
  {Rossi}, {Ruggeri}, {Ruhlmann-Kleider}, {Samushia}, {Sanchez}, {Saulder},
  {Schlafly}, {Schlegel}, {Schubnell}, {Seo}, {Shafieloo}, {Sharples},
  {Silber}, {Slosar}, {Smith}, {Sprayberry}, {Tan}, {Tarl{\'e}}, {Taylor},
  {Trusov}, {Ure{\~n}a-L{\'o}pez}, {Vaisakh}, {Valcin}, {Valdes},
  {Vargas-Maga{\~n}a}, {Verde}, {Walther}, {Wang}, {Wang}, {Weaver},
  {Weaverdyck}, {Wechsler}, {Weinberg}, {White}, {Yu}, {Yu}, {Yuan},
  {Y{\`e}che}, {Zaborowski}, {Zarrouk}, {Zhang}, {Zhao}, {Zhao}, {Zhou}, \&
  {Zhuang}}]{Adame25}
{Adame}, A.~G., {Aguilar}, J., {Ahlen}, S., {et~al.} 2025,
  \href{http://dx.doi.org/10.1088/1475-7516/2025/02/021}{\color{blue}\jcap},
  \href{https://ui.adsabs.harvard.edu/abs/2025JCAP...02..021A}{2025, 021}

\bibitem[{{Aguena} {et~al.}(2023){Aguena}, {Alves}, {Annis}, {Bacon},
  {Bocquet}, {Brooks}, {Carnero Rosell}, {Chang}, {Costanzi}, {Coviello}, {da
  Costa}, {Davis}, {De Vicente}, {Diehl}, {Doel}, {Esteves}, {Everett},
  {Ferrero}, {Fert{\'e}}, {Friedel}, {Frieman}, {Gatti}, {Giannini}, {Gruen},
  {Gruendl}, {Gutierrez}, {Herner}, {Hinton}, {Hollowood}, {Honscheid},
  {James}, {Jeltema}, {Kirby}, {Kuehn}, {Lahav}, {Li}, {Marshall},
  {McClintock}, {Mellor}, {Mena-Fern{\'a}ndez}, {Miquel}, {O'Donnell},
  {Palmese}, {Paterno}, {Pereira}, {Pieres}, {Plazas Malag{\'o}n},
  {Rodriguez-Monroy}, {Romer}, {Roodman}, {Sanchez}, {Schubnell},
  {Sevilla-Noarbe}, {Shin}, {Smith}, {Suchyta}, {Swanson}, {Tarle}, {Weller},
  {Wiseman}, {Wu}, {Zhang}, \& {Zhou}}]{Aguena23}
{Aguena}, M., {Alves}, O., {Annis}, J., {et~al.} 2023,
  \href{https://ui.adsabs.harvard.edu/abs/2023arXiv230906593A}{\href{http://dx.doi.org/10.48550/arXiv.2309.06593}{\color{blue}arXiv
  e-prints}, arXiv:2309.06593}

\bibitem[{{Aihara} {et~al.}(2011){Aihara}, {Allende Prieto}, {An}, {Anderson},
  {Aubourg}, {Balbinot}, {Beers}, {Berlind}, {Bickerton}, {Bizyaev}, {Blanton},
  {Bochanski}, {Bolton}, {Bovy}, {Brandt}, {Brinkmann}, {Brown}, {Brownstein},
  {Busca}, {Campbell}, {Carr}, {Chen}, {Chiappini}, {Comparat}, {Connolly},
  {Cortes}, {Croft}, {Cuesta}, {da Costa}, {Davenport}, {Dawson}, {Dhital},
  {Ealet}, {Ebelke}, {Edmondson}, {Eisenstein}, {Escoffier}, {Esposito},
  {Evans}, {Fan}, {Femen{\'\i}a Castell{\'a}}, {Font-Ribera}, {Frinchaboy},
  {Ge}, {Gillespie}, {Gilmore}, {Gonz{\'a}lez Hern{\'a}ndez}, {Gott}, {Gould},
  {Grebel}, {Gunn}, {Hamilton}, {Harding}, {Harris}, {Hawley}, {Hearty}, {Ho},
  {Hogg}, {Holtzman}, {Honscheid}, {Inada}, {Ivans}, {Jiang}, {Johnson},
  {Jordan}, {Jordan}, {Kazin}, {Kirkby}, {Klaene}, {Knapp}, {Kneib},
  {Kochanek}, {Koesterke}, {Kollmeier}, {Kron}, {Lampeitl}, {Lang}, {Le Goff},
  {Lee}, {Lin}, {Long}, {Loomis}, {Lucatello}, {Lundgren}, {Lupton}, {Ma},
  {MacDonald}, {Mahadevan}, {Maia}, {Makler}, {Malanushenko}, {Malanushenko},
  {Mandelbaum}, {Maraston}, {Margala}, {Masters}, {McBride}, {McGehee},
  {McGreer}, {M{\'e}nard}, {Miralda-Escud{\'e}}, {Morrison}, {Mullally},
  {Muna}, {Munn}, {Murayama}, {Myers}, {Naugle}, {Neto}, {Nguyen}, {Nichol},
  {O'Connell}, {Ogando}, {Olmstead}, {Oravetz}, {Padmanabhan},
  {Palanque-Delabrouille}, {Pan}, {Pandey}, {P{\^a}ris}, {Percival},
  {Petitjean}, {Pfaffenberger}, {Pforr}, {Phleps}, {Pichon}, {Pieri}, {Prada},
  {Price-Whelan}, {Raddick}, {Ramos}, {Reyl{\'e}}, {Rich}, {Richards}, {Rix},
  {Robin}, {Rocha-Pinto}, {Rockosi}, {Roe}, {Rollinde}, {Ross}, {Ross},
  {Rossetto}, {S{\'a}nchez}, {Sayres}, {Schlegel}, {Schlesinger}, {Schmidt},
  {Schneider}, {Sheldon}, {Shu}, {Simmerer}, {Simmons}, {Sivarani}, {Snedden},
  {Sobeck}, {Steinmetz}, {Strauss}, {Szalay}, {Tanaka}, {Thakar}, {Thomas},
  {Tinker}, {Tofflemire}, {Tojeiro}, {Tremonti}, {Vandenberg}, {Vargas
  Maga{\~n}a}, {Verde}, {Vogt}, {Wake}, {Wang}, {Weaver}, {Weinberg}, {White},
  {White}, {Yanny}, {Yasuda}, {Yeche}, \& {Zehavi}}]{Aihara11}
{Aihara}, H., {Allende Prieto}, C., {An}, D., {et~al.} 2011,
  \href{http://dx.doi.org/10.1088/0067-0049/193/2/29}{\color{blue}\apjs},
  \href{https://ui.adsabs.harvard.edu/abs/2011ApJS..193...29A}{193, 29}

\bibitem[{{Alam} {et~al.}(2015){Alam}, {Albareti}, {Allende Prieto}, {Anders},
  {Anderson}, {Anderton}, {Andrews}, {Armengaud}, {Aubourg}, {Bailey}, {Basu},
  {Bautista}, {Beaton}, {Beers}, {Bender}, {Berlind}, {Beutler}, {Bhardwaj},
  {Bird}, {Bizyaev}, {Blake}, {Blanton}, {Blomqvist}, {Bochanski}, {Bolton},
  {Bovy}, {Shelden Bradley}, {Brandt}, {Brauer}, {Brinkmann}, {Brown},
  {Brownstein}, {Burden}, {Burtin}, {Busca}, {Cai}, {Capozzi}, {Carnero
  Rosell}, {Carr}, {Carrera}, {Chambers}, {Chaplin}, {Chen}, {Chiappini},
  {Chojnowski}, {Chuang}, {Clerc}, {Comparat}, {Covey}, {Croft}, {Cuesta},
  {Cunha}, {da Costa}, {Da Rio}, {Davenport}, {Dawson}, {De Lee}, {Delubac},
  {Deshpande}, {Dhital}, {Dutra-Ferreira}, {Dwelly}, {Ealet}, {Ebelke},
  {Edmondson}, {Eisenstein}, {Ellsworth}, {Elsworth}, {Epstein}, {Eracleous},
  {Escoffier}, {Esposito}, {Evans}, {Fan}, {Fern{\'a}ndez-Alvar}, {Feuillet},
  {Filiz Ak}, {Finley}, {Finoguenov}, {Flaherty}, {Fleming}, {Font-Ribera},
  {Foster}, {Frinchaboy}, {Galbraith-Frew}, {Garc{\'\i}a},
  {Garc{\'\i}a-Hern{\'a}ndez}, {Garc{\'\i}a P{\'e}rez}, {Gaulme}, {Ge},
  {G{\'e}nova-Santos}, {Georgakakis}, {Ghezzi}, {Gillespie}, {Girardi},
  {Goddard}, {Gontcho}, {Gonz{\'a}lez Hern{\'a}ndez}, {Grebel}, {Green},
  {Grieb}, {Grieves}, {Gunn}, {Guo}, {Harding}, {Hasselquist}, {Hawley},
  {Hayden}, {Hearty}, {Hekker}, {Ho}, {Hogg}, {Holley-Bockelmann}, {Holtzman},
  {Honscheid}, {Huber}, {Huehnerhoff}, {Ivans}, {Jiang}, {Johnson},
  {Kinemuchi}, {Kirkby}, {Kitaura}, {Klaene}, {Knapp}, {Kneib}, {Koenig},
  {Lam}, {Lan}, {Lang}, {Laurent}, {Le Goff}, {Leauthaud}, {Lee}, {Lee},
  {Licquia}, {Liu}, {Long}, {L{\'o}pez-Corredoira}, {Lorenzo-Oliveira},
  {Lucatello}, {Lundgren}, {Lupton}, {Mack}, {Mahadevan}, {Maia}, {Majewski},
  {Malanushenko}, {Malanushenko}, {Manchado}, {Manera}, {Mao}, {Maraston},
  {Marchwinski}, {Margala}, {Martell}, {Martig}, {Masters}, {Mathur},
  {McBride}, {McGehee}, {McGreer}, {McMahon}, {M{\'e}nard}, {Menzel},
  {Merloni}, {M{\'e}sz{\'a}ros}, {Miller}, {Miralda-Escud{\'e}}, {Miyatake},
  {Montero-Dorta}, {More}, {Morganson}, {Morice-Atkinson}, {Morrison},
  {Mosser}, {Muna}, {Myers}, {Nandra}, {Newman}, {Neyrinck}, {Nguyen},
  {Nichol}, {Nidever}, {Noterdaeme}, {Nuza}, {O'Connell}, {O'Connell},
  {O'Connell}, {Ogando}, {Olmstead}, {Oravetz}, {Oravetz}, {Osumi}, {Owen},
  {Padgett}, {Padmanabhan}, {Paegert}, {Palanque-Delabrouille}, \&
  {Pan}}]{Alam15}
{Alam}, S., {Albareti}, F.~D., {Allende Prieto}, C., {et~al.} 2015,
  \href{http://dx.doi.org/10.1088/0067-0049/219/1/12}{\color{blue}\apjs},
  \href{https://ui.adsabs.harvard.edu/abs/2015ApJS..219...12A}{219, 12}

\bibitem[{{Angulo} {et~al.}(2012){Angulo}, {Springel}, {White}, {Jenkins},
  {Baugh}, \& {Frenk}}]{Angulo12}
{Angulo}, R.~E., {Springel}, V., {White}, S.~D.~M., {et~al.} 2012,
  \href{http://dx.doi.org/10.1111/j.1365-2966.2012.21830.x}{\color{blue}\mnras},
  \href{https://ui.adsabs.harvard.edu/abs/2012MNRAS.426.2046A}{426, 2046}

\bibitem[{{Arnaud} {et~al.}(2010){Arnaud}, {Pratt}, {Piffaretti},
  {B{\"o}hringer}, {Croston}, \& {Pointecouteau}}]{Arnaud10}
{Arnaud}, M., {Pratt}, G.~W., {Piffaretti}, R., {et~al.} 2010,
  \href{http://dx.doi.org/10.1051/0004-6361/200913416}{\color{blue}\aap},
  \href{https://ui.adsabs.harvard.edu/abs/2010A&A...517A..92A}{517, A92}

\bibitem[{{Artis} {et~al.}(2025){Artis}, {Bulbul}, {Grandis}, {Ghirardini},
  {Clerc}, {Seppi}, {Comparat}, {Cataneo}, {von der Linden}, {Bahar}, {Balzer},
  {Chiu}, {Gruen}, {Kleinebreil}, {Kluge}, {Krippendorf}, {Li}, {Liu},
  {Malavasi}, {Merloni}, {Miyatake}, {Miyazaki}, {Nandra}, {Okabe}, {Pacaud},
  {Predehl}, {Ramos-Ceja}, {Reiprich}, {Sanders}, {Schrabback}, {Zelmer}, \&
  {Zhang}}]{Artis24}
{Artis}, E., {Bulbul}, E., {Grandis}, S., {et~al.} 2025,
  \href{http://dx.doi.org/10.1051/0004-6361/202452584}{\color{blue}\aap},
  \href{https://ui.adsabs.harvard.edu/abs/2025A&A...696A...5A}{696, A5}

\bibitem[{{Asgari} {et~al.}(2021){Asgari}, {Lin}, {Joachimi}, {Giblin},
  {Heymans}, {Hildebrandt}, {Kannawadi}, {St{\"o}lzner}, {Tr{\"o}ster}, {van
  den Busch}, {Wright}, {Bilicki}, {Blake}, {de Jong}, {Dvornik}, {Erben},
  {Getman}, {Hoekstra}, {K{\"o}hlinger}, {Kuijken}, {Miller}, {Radovich},
  {Schneider}, {Shan}, \& {Valentijn}}]{Asgari21}
{Asgari}, M., {Lin}, C.-A., {Joachimi}, B., {et~al.} 2021,
  \href{http://dx.doi.org/10.1051/0004-6361/202039070}{\color{blue}\aap},
  \href{https://ui.adsabs.harvard.edu/abs/2021A&A...645A.104A}{645, A104}

\bibitem[{{Baltz} {et~al.}(2009){Baltz}, {Marshall}, \& {Oguri}}]{BMO}
{Baltz}, E.~A., {Marshall}, P., \& {Oguri}, M. 2009,
  \href{http://dx.doi.org/10.1088/1475-7516/2009/01/015}{\color{blue}\jcap},
  \href{https://ui.adsabs.harvard.edu/abs/2009JCAP...01..015B}{2009, 015}

\bibitem[{{Bardeau} {et~al.}(2007){Bardeau}, {Soucail}, {Kneib}, {Czoske},
  {Ebeling}, {Hudelot}, {Smail}, \& {Smith}}]{Bardeau07}
{Bardeau}, S., {Soucail}, G., {Kneib}, J.~P., {et~al.} 2007,
  \href{http://dx.doi.org/10.1051/0004-6361:20077443}{\color{blue}\aap},
  \href{https://ui.adsabs.harvard.edu/abs/2007A&A...470..449B}{470, 449}

\bibitem[{{Bartelmann} \& {Schneider}(2001)}]{Bartelmann01}
{Bartelmann}, M. \& {Schneider}, P. 2001,
  \href{http://dx.doi.org/10.1016/S0370-1573(00)00082-X}{\color{blue}\physrep},
  \href{https://ui.adsabs.harvard.edu/abs/2001PhR...340..291B}{340, 291}

\bibitem[{{Bellagamba} {et~al.}(2018){Bellagamba}, {Roncarelli}, {Maturi}, \&
  {Moscardini}}]{Bellagamba18}
{Bellagamba}, F., {Roncarelli}, M., {Maturi}, M., \& {Moscardini}, L. 2018,
  \href{http://dx.doi.org/10.1093/mnras/stx2701}{\color{blue}\mnras},
  \href{https://ui.adsabs.harvard.edu/abs/2018MNRAS.473.5221B}{473, 5221}

\bibitem[{{Bellagamba} {et~al.}(2019){Bellagamba}, {Sereno}, {Roncarelli},
  {Maturi}, {Radovich}, {Bardelli}, {Puddu}, {Moscardini}, {Getman},
  {Hildebrandt}, \& {Napolitano}}]{Bellagamba19}
{Bellagamba}, F., {Sereno}, M., {Roncarelli}, M., {et~al.} 2019,
  \href{http://dx.doi.org/10.1093/mnras/stz090}{\color{blue}\mnras},
  \href{https://ui.adsabs.harvard.edu/abs/2019MNRAS.484.1598B}{484, 1598}

\bibitem[{{Beltz-Mohrmann} \& {Berlind}(2021)}]{Beltz-Mohrmann21}
{Beltz-Mohrmann}, G.~D. \& {Berlind}, A.~A. 2021,
  \href{http://dx.doi.org/10.3847/1538-4357/ac1e27}{\color{blue}\apj},
  \href{https://ui.adsabs.harvard.edu/abs/2021ApJ...921..112B}{921, 112}

\bibitem[{{Ben{\'\i}tez}(2000)}]{Benitez00}
{Ben{\'\i}tez}, N. 2000,
  \href{http://dx.doi.org/10.1086/308947}{\color{blue}\apj},
  \href{https://ui.adsabs.harvard.edu/abs/2000ApJ...536..571B}{536, 571}

\bibitem[{{Bhattacharya} {et~al.}(2013){Bhattacharya}, {Habib}, {Heitmann}, \&
  {Vikhlinin}}]{Bhattacharya13}
{Bhattacharya}, S., {Habib}, S., {Heitmann}, K., \& {Vikhlinin}, A. 2013,
  \href{http://dx.doi.org/10.1088/0004-637X/766/1/32}{\color{blue}\apj},
  \href{https://ui.adsabs.harvard.edu/abs/2013ApJ...766...32B}{766, 32}

\bibitem[{{Biviano} \& {Mamon}(2023)}]{Biviano23}
{Biviano}, A. \& {Mamon}, G.~A. 2023,
  \href{http://dx.doi.org/10.1051/0004-6361/202244626}{\color{blue}\aap},
  \href{https://ui.adsabs.harvard.edu/abs/2023A&A...670A..17B}{670, A17}

\bibitem[{{Blake} {et~al.}(2016){Blake}, {Amon}, {Childress}, {Erben},
  {Glazebrook}, {Harnois-Deraps}, {Heymans}, {Hildebrandt}, {Hinton},
  {Janssens}, {Johnson}, {Joudaki}, {Klaes}, {Kuijken}, {Lidman}, {Marin},
  {Parkinson}, {Poole}, \& {Wolf}}]{Blake16}
{Blake}, C., {Amon}, A., {Childress}, M., {et~al.} 2016,
  \href{http://dx.doi.org/10.1093/mnras/stw1990}{\color{blue}\mnras},
  \href{https://ui.adsabs.harvard.edu/abs/2016MNRAS.462.4240B}{462, 4240}

\bibitem[{{Bleem} {et~al.}(2020){Bleem}, {Bocquet}, {Stalder}, {Gladders},
  {Ade}, {Allen}, {Anderson}, {Annis}, {Ashby}, {Austermann}, {Avila}, {Avva},
  {Bayliss}, {Beall}, {Bechtol}, {Bender}, {Benson}, {Bertin}, {Bianchini},
  {Blake}, {Brodwin}, {Brooks}, {Buckley-Geer}, {Burke}, {Carlstrom}, {Rosell},
  {Carrasco Kind}, {Carretero}, {Chang}, {Chiang}, {Citron}, {Moran},
  {Costanzi}, {Crawford}, {Crites}, {da Costa}, {de Haan}, {De Vicente},
  {Desai}, {Diehl}, {Dietrich}, {Dobbs}, {Eifler}, {Everett}, {Flaugher},
  {Floyd}, {Frieman}, {Gallicchio}, {Garc{\'\i}a-Bellido}, {George}, {Gerdes},
  {Gilbert}, {Gruen}, {Gruendl}, {Gschwend}, {Gupta}, {Gutierrez}, {Halverson},
  {Harrington}, {Henning}, {Heymans}, {Holder}, {Hollowood}, {Holzapfel},
  {Honscheid}, {Hrubes}, {Huang}, {Hubmayr}, {Irwin}, {James}, {Jeltema},
  {Joudaki}, {Khullar}, {Klein}, {Knox}, {Kuropatkin}, {Lee}, {Li}, {Lidman},
  {Lowitz}, {MacCrann}, {Mahler}, {Maia}, {Marshall}, {McDonald}, {McMahon},
  {Melchior}, {Menanteau}, {Meyer}, {Miquel}, {Mocanu}, {Mohr}, {Montgomery},
  {Nadolski}, {Natoli}, {Nibarger}, {Noble}, {Novosad}, {Padin}, {Palmese},
  {Parkinson}, {Patil}, {Paz-Chinch{\'o}n}, {Plazas}, {Pryke}, {Ramachandra},
  {Reichardt}, {Remolina Gonz{\'a}lez}, {Romer}, {Roodman}, {Ruhl}, {Rykoff},
  {Saliwanchik}, {Sanchez}, {Saro}, {Sayre}, {Schaffer}, {Schrabback},
  {Serrano}, {Sharon}, {Sievers}, {Smecher}, {Smith}, {Soares-Santos}, {Stark},
  {Story}, {Suchyta}, {Tarle}, {Tucker}, {Vanderlinde}, {Veach}, {Vieira},
  {Wang}, {Weller}, {Whitehorn}, {Wu}, {Yefremenko}, \& {Zhang}}]{Bleem20}
{Bleem}, L.~E., {Bocquet}, S., {Stalder}, B., {et~al.} 2020,
  \href{http://dx.doi.org/10.3847/1538-4365/ab6993}{\color{blue}\apjs},
  \href{https://ui.adsabs.harvard.edu/abs/2020ApJS..247...25B}{247, 25}

\bibitem[{{Bocquet} {et~al.}(2019){Bocquet}, {Dietrich}, {Schrabback}, {Bleem},
  {Klein}, {Allen}, {Applegate}, {Ashby}, {Bautz}, {Bayliss}, {Benson},
  {Brodwin}, {Bulbul}, {Canning}, {Capasso}, {Carlstrom}, {Chang}, {Chiu},
  {Cho}, {Clocchiatti}, {Crawford}, {Crites}, {de Haan}, {Desai}, {Dobbs},
  {Foley}, {Forman}, {Garmire}, {George}, {Gladders}, {Gonzalez}, {Grandis},
  {Gupta}, {Halverson}, {Hlavacek-Larrondo}, {Hoekstra}, {Holder}, {Holzapfel},
  {Hou}, {Hrubes}, {Huang}, {Jones}, {Khullar}, {Knox}, {Kraft}, {Lee}, {von
  der Linden}, {Luong-Van}, {Mantz}, {Marrone}, {McDonald}, {McMahon}, {Meyer},
  {Mocanu}, {Mohr}, {Morris}, {Padin}, {Patil}, {Pryke}, {Rapetti},
  {Reichardt}, {Rest}, {Ruhl}, {Saliwanchik}, {Saro}, {Sayre}, {Schaffer},
  {Shirokoff}, {Stalder}, {Stanford}, {Staniszewski}, {Stark}, {Story},
  {Strazzullo}, {Stubbs}, {Vanderlinde}, {Vieira}, {Vikhlinin}, {Williamson},
  \& {Zenteno}}]{Bocquet19}
{Bocquet}, S., {Dietrich}, J.~P., {Schrabback}, T., {et~al.} 2019,
  \href{http://dx.doi.org/10.3847/1538-4357/ab1f10}{\color{blue}\apj},
  \href{https://ui.adsabs.harvard.edu/abs/2019ApJ...878...55B}{878, 55}

\bibitem[{{Bocquet} {et~al.}(2024{\natexlab{a}}){Bocquet}, {Grandis}, {Bleem},
  {Klein}, {Mohr}, {Aguena}, {Alarcon}, {Allam}, {Allen}, {Alves}, {Amon},
  {Ansarinejad}, {Bacon}, {Bayliss}, {Bechtol}, {Becker}, {Benson},
  {Bernstein}, {Brodwin}, {Brooks}, {Campos}, {Canning}, {Carlstrom}, {Carnero
  Rosell}, {Carrasco Kind}, {Carretero}, {Cawthon}, {Chang}, {Chen}, {Choi},
  {Cordero}, {Costanzi}, {da Costa}, {Pereira}, {Davis}, {DeRose}, {Desai}, {de
  Haan}, {De Vicente}, {Diehl}, {Dodelson}, {Doel}, {Doux}, {Drlica-Wagner},
  {Eckert}, {Elvin-Poole}, {Everett}, {Ferrero}, {Fert{\'e}}, {Flores},
  {Frieman}, {Garc{\'\i}a-Bellido}, {Gatti}, {Giannini}, {Gladders}, {Gruen},
  {Gruendl}, {Harrison}, {Hartley}, {Herner}, {Hinton}, {Hollowood},
  {Holzapfel}, {Honscheid}, {Huang}, {Huff}, {James}, {Jarvis}, {Khullar},
  {Kim}, {Kraft}, {Kuehn}, {Kuropatkin}, {K{\'e}ruzor{\'e}}, {Lee}, {Leget},
  {MacCrann}, {Mahler}, {Mantz}, {Marshall}, {McCullough}, {McDonald},
  {Mena-Fern{\'a}ndez}, {Miquel}, {Myles}, {Navarro-Alsina}, {Ogando},
  {Palmese}, {Pandey}, {Pieres}, {Plazas Malag{\'o}n}, {Prat}, {Raveri},
  {Reichardt}, {Roberson}, {Rollins}, {Romer}, {Romero}, {Roodman}, {Ross},
  {Rykoff}, {Salvati}, {S{\'a}nchez}, {Sanchez}, {Sanchez Cid}, {Saro},
  {Schrabback}, {Schubnell}, {Secco}, {Sevilla-Noarbe}, {Sharon}, {Sheldon},
  {Shin}, {Smith}, {Somboonpanyakul}, {Stalder}, {Stark}, {Strazzullo},
  {Suchyta}, {Swanson}, {Tarle}, {To}, {Troxel}, {Tutusaus}, {Varga}, {von der
  Linden}, {Weaverdyck}, {Weller}, {Wiseman}, {Yanny}, {Yin}, {Young}, {Zhang},
  {Zuntz}, {(The DES}, \& {SPT Collaborations)}}]{Bocquet24_3}
{Bocquet}, S., {Grandis}, S., {Bleem}, L.~E., {et~al.} 2024{\natexlab{a}},
  \href{http://dx.doi.org/10.1103/PhysRevD.110.083509}{\color{blue}\prd},
  \href{https://ui.adsabs.harvard.edu/abs/2024PhRvD.110h3509B}{110, 083509}

\bibitem[{{Bocquet} {et~al.}(2024{\natexlab{b}}){Bocquet}, {Grandis}, {Bleem},
  {Klein}, {Mohr}, {Schrabback}, {Abbott}, {Ade}, {Aguena}, {Alarcon}, {Allam},
  {Allen}, {Alves}, {Amon}, {Anderson}, {Annis}, {Ansarinejad}, {Austermann},
  {Avila}, {Bacon}, {Bayliss}, {Beall}, {Bechtol}, {Becker}, {Bender},
  {Benson}, {Bernstein}, {Bhargava}, {Bianchini}, {Brodwin}, {Brooks},
  {Bryant}, {Campos}, {Canning}, {Carlstrom}, {Carnero Rosell}, {Carrasco
  Kind}, {Carretero}, {Castander}, {Cawthon}, {Chang}, {Chang}, {Chaubal},
  {Chen}, {Chiang}, {Choi}, {Chou}, {Citron}, {Corbett Moran}, {Cordero},
  {Costanzi}, {Crawford}, {Crites}, {da Costa}, {Pereira}, {Davis}, {Davis},
  {DeRose}, {Desai}, {de Haan}, {Diehl}, {Dobbs}, {Dodelson}, {Doux},
  {Drlica-Wagner}, {Eckert}, {Elvin-Poole}, {Everett}, {Everett}, {Ferrero},
  {Fert{\'e}}, {Flores}, {Frieman}, {Gallicchio}, {Garc{\'\i}a-Bellido},
  {Gatti}, {George}, {Giannini}, {Gladders}, {Gruen}, {Gruendl}, {Gupta},
  {Gutierrez}, {Halverson}, {Harrison}, {Hartley}, {Herner}, {Hinton},
  {Holder}, {Hollowood}, {Holzapfel}, {Honscheid}, {Hrubes}, {Huang},
  {Hubmayr}, {Huff}, {Huterer}, {Irwin}, {James}, {Jarvis}, {Khullar}, {Kim},
  {Knox}, {Kraft}, {Krause}, {Kuehn}, {Kuropatkin}, {K{\'e}ruzor{\'e}},
  {Lahav}, {Lee}, {Leget}, {Li}, {Lin}, {Lowitz}, {MacCrann}, {Mahler},
  {Mantz}, {Marshall}, {McCullough}, {McDonald}, {McMahon},
  {Mena-Fern{\'a}ndez}, {Menanteau}, {Meyer}, {Miquel}, {Montgomery}, {Myles},
  {Natoli}, {Navarro-Alsina}, {Nibarger}, {Noble}, {Novosad}, {Ogando},
  {Omori}, {Padin}, {Pandey}, {Paschos}, {Patil}, {Pieres}, {Plazas
  Malag{\'o}n}, {Porredon}, {Prat}, {Pryke}, {Raveri}, {Reichardt}, {Roberson},
  {Rollins}, {Romero}, {Roodman}, {Ruhl}, {Rykoff}, {Saliwanchik}, {Salvati},
  {S{\'a}nchez}, {Sanchez}, {Sanchez Cid}, {Saro}, {Schaffer}, {Secco},
  {Sevilla-Noarbe}, {Sharon}, {Sheldon}, {Shin}, {Sievers}, {Smecher}, {Smith},
  {Somboonpanyakul}, {Sommer}, {Stalder}, {Stark}, {Stephen}, {Strazzullo},
  {Suchyta}, {Tarle}, {To}, {Troxel}, {Tucker}, {Tutusaus}, {Varga}, {Veach},
  {Vieira}, {Vikhlinin}, {von der Linden}, {Wang}, {Weaverdyck}, {Weller},
  {Whitehorn}, {Wu}, {Yanny}, {Yefremenko}, {Yin}, {Young}, {Zebrowski},
  {Zhang}, {Zohren}, {Zuntz}, {(SPT}, \& {DES Collaborations)}}]{Bocquet24}
{Bocquet}, S., {Grandis}, S., {Bleem}, L.~E., {et~al.} 2024{\natexlab{b}},
  \href{http://dx.doi.org/10.1103/PhysRevD.110.083510}{\color{blue}\prd},
  \href{https://ui.adsabs.harvard.edu/abs/2024PhRvD.110h3510B}{110, 083510}

\bibitem[{{Bocquet} {et~al.}(2025){Bocquet}, {Grandis}, {Krause}, {To},
  {Bleem}, {Klein}, {Mohr}, {Schrabback}, {Alarcon}, {Alves}, {Amon},
  {Andrade-Oliveira}, {Baxter}, {Bechtol}, {Becker}, {Bernstein}, {Blazek},
  {Camacho}, {Campos}, {Carnero Rosell}, {Carrasco Kind}, {Cawthon}, {Chang},
  {Chen}, {Choi}, {Cordero}, {Crocce}, {Davis}, {DeRose}, {Diehl}, {Dodelson},
  {Doux}, {Drlica-Wagner}, {Eckert}, {Eifler}, {Elsner}, {Elvin-Poole},
  {Everett}, {Fang}, {Fert{\'e}}, {Fosalba}, {Friedrich}, {Frieman}, {Gatti},
  {Giannini}, {Gruen}, {Gruendl}, {Harrison}, {Hartley}, {Herner}, {Huang},
  {Huff}, {Huterer}, {Jarvis}, {Kuropatkin}, {Leget}, {Lemos}, {Liddle},
  {MacCrann}, {McCullough}, {Muir}, {Myles}, {Navarro-Alsina}, {Pandey},
  {Park}, {Porredon}, {Prat}, {Raveri}, {Rollins}, {Roodman}, {Rosenfeld},
  {Rykoff}, {S{\'a}nchez}, {Sanchez}, {Secco}, {Sevilla-Noarbe}, {Sheldon},
  {Shin}, {Troxel}, {Tutusaus}, {Varga}, {Weaverdyck}, {Wechsler}, {Wu},
  {Yanny}, {Yin}, {Zhang}, {Zuntz}, {Abbott}, {Ade}, {Aguena}, {Allam},
  {Allen}, {Anderson}, {Ansarinejad}, {Austermann}, {Bayliss}, {Beall},
  {Bender}, {Benson}, {Bianchini}, {Brodwin}, {Brooks}, {Bryant}, {Burke},
  {Canning}, {Carlstrom}, {Carretero}, {Castander}, {Chang}, {Chaubal},
  {Chiang}, {Chou}, {Citron}, {Corbett Moran}, {Costanzi}, {Crawford},
  {Crites}, {da Costa}, {Pereira}, {Davis}, {de Haan}, {Dobbs}, {Doel},
  {Everett}, {Farahi}, {Flaugher}, {Flores}, {Floyd}, {Gallicchio},
  {Gaztanaga}, {George}, {Gladders}, {Gupta}, {Gutierrez}, {Halverson},
  {Hinton}, {Hlavacek-Larrondo}, {Holder}, {Hollowood}, {Holzapfel}, {Hrubes},
  {Huang}, {Hubmayr}, {Irwin}, {James}, {K{\'e}ruzor{\'e}}, {Khullar}, {Kim},
  {Knox}, {Kraft}, {Kuehn}, {Lahav}, {Lee}, {Lee}, {Li}, {Lidman}, {Lima},
  {Lowitz}, {Mahler}, {Mantz}, {Marshall}, {McDonald}, {McMahon},
  {Mena-Fern{\'a}ndez}, {Meyer}, {Miquel}, {Montgomery}, {Natoli}, {Nibarger},
  {Noble}, {Novosad}, {Ogando}, {Padin}, {Paschos}, {Patil}, {Plazas
  Malag{\'o}n}, {Pryke}, {Reichardt}, {Roberson}, {Romer}, {Romero}, {Ruhl},
  {Saliwanchik}, {Salvati}, {Samuroff}, {Sanchez}, {Santiago}, {Sarkar},
  {Saro}, {Schaffer}, {Sharon}, {Sievers}, {Smecher}, {Smith},
  {Somboonpanyakul}, {Sommer}, {Stalder}, {Stark}, \& {Stephen}}]{Bocquet24_2}
{Bocquet}, S., {Grandis}, S., {Krause}, E., {et~al.} 2025,
  \href{http://dx.doi.org/10.1103/PhysRevD.111.063533}{\color{blue}\prd},
  \href{https://ui.adsabs.harvard.edu/abs/2025PhRvD.111f3533B}{111, 063533}

\bibitem[{{Bocquet} {et~al.}(2016){Bocquet}, {Saro}, {Dolag}, \&
  {Mohr}}]{Bocquet16}
{Bocquet}, S., {Saro}, A., {Dolag}, K., \& {Mohr}, J.~J. 2016,
  \href{http://dx.doi.org/10.1093/mnras/stv2657}{\color{blue}\mnras},
  \href{https://ui.adsabs.harvard.edu/abs/2016MNRAS.456.2361B}{456, 2361}

\bibitem[{{Borgani} \& {Kravtsov}(2011)}]{Borgani11}
{Borgani}, S. \& {Kravtsov}, A. 2011,
  \href{http://dx.doi.org/10.1166/asl.2011.1209}{\color{blue}Advanced Science
  Letters}, \href{https://ui.adsabs.harvard.edu/abs/2011ASL.....4..204B}{4,
  204}

\bibitem[{{Broadhurst} {et~al.}(2005){Broadhurst}, {Takada}, {Umetsu}, {Kong},
  {Arimoto}, {Chiba}, \& {Futamase}}]{Broadhurst05}
{Broadhurst}, T., {Takada}, M., {Umetsu}, K., {et~al.} 2005,
  \href{http://dx.doi.org/10.1086/428122}{\color{blue}\apjl},
  \href{https://ui.adsabs.harvard.edu/abs/2005ApJ...619L.143B}{619, L143}

\bibitem[{{Bulbul} {et~al.}(2024){Bulbul}, {Liu}, {Kluge}, {Zhang}, {Sanders},
  {Bahar}, {Ghirardini}, {Artis}, {Seppi}, {Garrel}, {Ramos-Ceja}, {Comparat},
  {Balzer}, {B{\"o}ckmann}, {Br{\"u}ggen}, {Clerc}, {Dennerl}, {Dolag},
  {Freyberg}, {Grandis}, {Gruen}, {Kleinebreil}, {Krippendorf}, {Lamer},
  {Merloni}, {Migkas}, {Nandra}, {Pacaud}, {Predehl}, {Reiprich}, {Schrabback},
  {Veronica}, {Weller}, \& {Zelmer}}]{Bulbul24}
{Bulbul}, E., {Liu}, A., {Kluge}, M., {et~al.} 2024,
  \href{http://dx.doi.org/10.1051/0004-6361/202348264}{\color{blue}\aap},
  \href{https://ui.adsabs.harvard.edu/abs/2024A&A...685A.106B}{685, A106}

\bibitem[{{Busillo} {et~al.}(2023){Busillo}, {Covone}, {Sereno}, {Ingoglia},
  {Radovich}, {Bardelli}, {Castignani}, {Giocoli}, {Lesci}, {Marulli},
  {Maturi}, {Moscardini}, {Puddu}, \& {Roncarelli}}]{Busillo23}
{Busillo}, V., {Covone}, G., {Sereno}, M., {et~al.} 2023,
  \href{http://dx.doi.org/10.1093/mnras/stad2190}{\color{blue}\mnras},
  \href{https://ui.adsabs.harvard.edu/abs/2023MNRAS.524.5050B}{524, 5050}

\bibitem[{{Cao} {et~al.}(2025){Cao}, {Wu}, {Costanzi}, {Farahi}, {Grandis},
  {Weinberg}, {Evrard}, {Rozo}, {Salcedo}, {To}, {Yang}, {Zhou}, \& {DES
  Collaboration}}]{Cao25}
{Cao}, S., {Wu}, H.-Y., {Costanzi}, M., {et~al.} 2025,
  \href{http://dx.doi.org/10.1103/r7tt-bzs7}{\color{blue}\prd},
  \href{https://ui.adsabs.harvard.edu/abs/2025PhRvD.112d3517C}{112, 043517}

\bibitem[{{Capaccioli} \& {Schipani}(2011)}]{VST}
{Capaccioli}, M. \& {Schipani}, P. 2011, The Messenger,
  \href{https://ui.adsabs.harvard.edu/abs/2011Msngr.146....2C}{146, 2}

\bibitem[{{Castignani} {et~al.}(2022){Castignani}, {Radovich}, {Combes},
  {Salom{\'e}}, {Maturi}, {Moscardini}, {Bardelli}, {Giocoli}, {Lesci},
  {Marulli}, {Puddu}, \& {Sereno}}]{Castignani22}
{Castignani}, G., {Radovich}, M., {Combes}, F., {et~al.} 2022,
  \href{http://dx.doi.org/10.1051/0004-6361/202243689}{\color{blue}\aap},
  \href{https://ui.adsabs.harvard.edu/abs/2022A&A...667A..52C}{667, A52}

\bibitem[{{Castignani} {et~al.}(2023){Castignani}, {Radovich}, {Combes},
  {Salom{\'e}}, {Moscardini}, {Bardelli}, {Giocoli}, {Lesci}, {Marulli},
  {Maturi}, {Puddu}, {Sereno}, \& {Tramonte}}]{Castignani23}
{Castignani}, G., {Radovich}, M., {Combes}, F., {et~al.} 2023,
  \href{http://dx.doi.org/10.1051/0004-6361/202245380}{\color{blue}\aap},
  \href{https://ui.adsabs.harvard.edu/abs/2023A&A...672A.139C}{672, A139}

\bibitem[{{Castro} {et~al.}(2021){Castro}, {Borgani}, {Dolag}, {Marra},
  {Quartin}, {Saro}, \& {Sefusatti}}]{Castro21}
{Castro}, T., {Borgani}, S., {Dolag}, K., {et~al.} 2021,
  \href{http://dx.doi.org/10.1093/mnras/staa3473}{\color{blue}\mnras},
  \href{https://ui.adsabs.harvard.edu/abs/2021MNRAS.500.2316C}{500, 2316}

\bibitem[{{Chen} {et~al.}(2024){Chen}, {Cui}, {Fang}, \& {Wen}}]{Chen24}
{Chen}, M., {Cui}, W., {Fang}, W., \& {Wen}, Z. 2024,
  \href{http://dx.doi.org/10.3847/1538-4357/ad3931}{\color{blue}\apj},
  \href{https://ui.adsabs.harvard.edu/abs/2024ApJ...966..227C}{966, 227}

\bibitem[{{CHEX-MATE Collaboration}(2021)}]{CHEX-MATE}
{CHEX-MATE Collaboration}. 2021,
  \href{http://dx.doi.org/10.1051/0004-6361/202039632}{\color{blue}\aap},
  \href{https://ui.adsabs.harvard.edu/abs/2021A&A...650A.104C}{650, A104}

\bibitem[{{Child} {et~al.}(2018){Child}, {Habib}, {Heitmann}, {Frontiere},
  {Finkel}, {Pope}, \& {Morozov}}]{Child18}
{Child}, H.~L., {Habib}, S., {Heitmann}, K., {et~al.} 2018,
  \href{http://dx.doi.org/10.3847/1538-4357/aabf95}{\color{blue}\apj},
  \href{https://ui.adsabs.harvard.edu/abs/2018ApJ...859...55C}{859, 55}

\bibitem[{{Chisari} {et~al.}(2014){Chisari}, {Mandelbaum}, {Strauss}, {Huff},
  \& {Bahcall}}]{Chisari14}
{Chisari}, N.~E., {Mandelbaum}, R., {Strauss}, M.~A., {Huff}, E.~M., \&
  {Bahcall}, N.~A. 2014,
  \href{http://dx.doi.org/10.1093/mnras/stu1786}{\color{blue}\mnras},
  \href{https://ui.adsabs.harvard.edu/abs/2014MNRAS.445..726C}{445, 726}

\bibitem[{{Corasaniti} {et~al.}(2021){Corasaniti}, {Sereno}, \&
  {Ettori}}]{Corasaniti21}
{Corasaniti}, P.-S., {Sereno}, M., \& {Ettori}, S. 2021,
  \href{http://dx.doi.org/10.3847/1538-4357/abe9a4}{\color{blue}\apj},
  \href{https://ui.adsabs.harvard.edu/abs/2021ApJ...911...82C}{911, 82}

\bibitem[{{Corless} \& {King}(2008)}]{Corless08}
{Corless}, V.~L. \& {King}, L.~J. 2008,
  \href{http://dx.doi.org/10.1111/j.1365-2966.2008.13744.x}{\color{blue}\mnras},
  \href{https://ui.adsabs.harvard.edu/abs/2008MNRAS.390..997C}{390, 997}

\bibitem[{{Costanzi} {et~al.}(2019){Costanzi}, {Rozo}, {Simet}, {Zhang},
  {Evrard}, {Mantz}, {Rykoff}, {Jeltema}, {Gruen}, {Allen}, {McClintock},
  {Romer}, {von der Linden}, {Farahi}, {DeRose}, {Varga}, {Weller}, {Giles},
  {Hollowood}, {Bhargava}, {Bermeo-Hernandez}, {Chen}, {Abbott}, {Abdalla},
  {Avila}, {Bechtol}, {Brooks}, {Buckley-Geer}, {Burke}, {Rosell}, {Kind},
  {Carretero}, {Crocce}, {Cunha}, {da Costa}, {Davis}, {De Vicente}, {Diehl},
  {Dietrich}, {Doel}, {Eifler}, {Estrada}, {Flaugher}, {Fosalba}, {Frieman},
  {Garc{\'\i}a-Bellido}, {Gaztanaga}, {Gerdes}, {Giannantonio}, {Gruendl},
  {Gschwend}, {Gutierrez}, {Hartley}, {Honscheid}, {Hoyle}, {James}, {Krause},
  {Kuehn}, {Kuropatkin}, {Lima}, {Lin}, {Maia}, {March}, {Marshall}, {Martini},
  {Menanteau}, {Miller}, {Miquel}, {Mohr}, {Ogando}, {Plazas}, {Roodman},
  {Sanchez}, {Scarpine}, {Schindler}, {Schubnell}, {Serrano}, {Sevilla-Noarbe},
  {Sheldon}, {Smith}, {Soares-Santos}, {Sobreira}, {Suchyta}, {Swanson},
  {Tarle}, {Thomas}, \& {Wechsler}}]{Costanzi19}
{Costanzi}, M., {Rozo}, E., {Simet}, M., {et~al.} 2019,
  \href{http://dx.doi.org/10.1093/mnras/stz1949}{\color{blue}\mnras},
  \href{https://ui.adsabs.harvard.edu/abs/2019MNRAS.488.4779C}{488, 4779}

\bibitem[{{Costanzi} {et~al.}(2021){Costanzi}, {Saro}, {Bocquet}, {Abbott},
  {Aguena}, {Allam}, {Amara}, {Annis}, {Avila}, {Bacon}, {Benson}, {Bhargava},
  {Brooks}, {Buckley-Geer}, {Burke}, {Carnero Rosell}, {Carrasco Kind},
  {Carretero}, {Choi}, {da Costa}, {Pereira}, {De Vicente}, {Desai}, {Diehl},
  {Dietrich}, {Doel}, {Eifler}, {Everett}, {Ferrero}, {Fert{\'e}}, {Flaugher},
  {Fosalba}, {Frieman}, {Garc{\'\i}a-Bellido}, {Gaztanaga}, {Gerdes},
  {Giannantonio}, {Giles}, {Grandis}, {Gruen}, {Gruendl}, {Gupta}, {Gutierrez},
  {Hartley}, {Hinton}, {Hollowood}, {Honscheid}, {James}, {Jeltema}, {Krause},
  {Kuehn}, {Kuropatkin}, {Lahav}, {Lima}, {MacCrann}, {Maia}, {Marshall},
  {Menanteau}, {Miquel}, {Mohr}, {Morgan}, {Myles}, {Ogando}, {Palmese},
  {Paz-Chinch{\'o}n}, {Plazas}, {Rapetti}, {Reichardt}, {Romer}, {Roodman},
  {Ruppin}, {Salvati}, {Samuroff}, {Sanchez}, {Scarpine}, {Serrano},
  {Sevilla-Noarbe}, {Singh}, {Smith}, {Soares-Santos}, {Stark}, {Suchyta},
  {Swanson}, {Tarle}, {Thomas}, {To}, {Tucker}, {Varga}, {Wechsler}, {Zhang},
  {DES}, \& {SPT Collaborations}}]{Costanzi21}
{Costanzi}, M., {Saro}, A., {Bocquet}, S., {et~al.} 2021,
  \href{http://dx.doi.org/10.1103/PhysRevD.103.043522}{\color{blue}\prd},
  \href{https://ui.adsabs.harvard.edu/abs/2021PhRvD.103d3522C}{103, 043522}

\bibitem[{{Cui} {et~al.}(2012){Cui}, {Borgani}, {Dolag}, {Murante}, \&
  {Tornatore}}]{Cui12}
{Cui}, W., {Borgani}, S., {Dolag}, K., {Murante}, G., \& {Tornatore}, L. 2012,
  \href{http://dx.doi.org/10.1111/j.1365-2966.2012.21037.x}{\color{blue}\mnras},
  \href{https://ui.adsabs.harvard.edu/abs/2012MNRAS.423.2279C}{423, 2279}

\bibitem[{{Currie} {et~al.}(2025){Currie}, {Miller}, {Shin}, {Baxter}, \&
  {Jain}}]{Currie24}
{Currie}, M., {Miller}, K., {Shin}, T.-h., {Baxter}, E., \& {Jain}, B. 2025,
  \href{http://dx.doi.org/10.1088/1475-7516/2025/07/072}{\color{blue}\jcap},
  \href{https://ui.adsabs.harvard.edu/abs/2025JCAP...07..072C}{2025, 072}

\bibitem[{{Dalal} {et~al.}(2023){Dalal}, {Li}, {Nicola}, {Zuntz}, {Strauss},
  {Sugiyama}, {Zhang}, {Rau}, {Mandelbaum}, {Takada}, {More}, {Miyatake},
  {Kannawadi}, {Shirasaki}, {Taniguchi}, {Takahashi}, {Osato}, {Hamana},
  {Oguri}, {Nishizawa}, {Malag{\'o}n}, {Sunayama}, {Alonso}, {Slosar}, {Luo},
  {Armstrong}, {Bosch}, {Hsieh}, {Komiyama}, {Lupton}, {Lust}, {MacArthur},
  {Miyazaki}, {Murayama}, {Nishimichi}, {Okura}, {Price}, {Tait}, {Tanaka}, \&
  {Wang}}]{Dalal23}
{Dalal}, R., {Li}, X., {Nicola}, A., {et~al.} 2023,
  \href{http://dx.doi.org/10.1103/PhysRevD.108.123519}{\color{blue}\prd},
  \href{https://ui.adsabs.harvard.edu/abs/2023PhRvD.108l3519D}{108, 123519}

\bibitem[{{de Haan} {et~al.}(2016){de Haan}, {Benson}, {Bleem}, {Allen},
  {Applegate}, {Ashby}, {Bautz}, {Bayliss}, {Bocquet}, {Brodwin}, {Carlstrom},
  {Chang}, {Chiu}, {Cho}, {Clocchiatti}, {Crawford}, {Crites}, {Desai},
  {Dietrich}, {Dobbs}, {Doucouliagos}, {Foley}, {Forman}, {Garmire}, {George},
  {Gladders}, {Gonzalez}, {Gupta}, {Halverson}, {Hlavacek-Larrondo},
  {Hoekstra}, {Holder}, {Holzapfel}, {Hou}, {Hrubes}, {Huang}, {Jones},
  {Keisler}, {Knox}, {Lee}, {Leitch}, {von der Linden}, {Luong-Van}, {Mantz},
  {Marrone}, {McDonald}, {McMahon}, {Meyer}, {Mocanu}, {Mohr}, {Murray},
  {Padin}, {Pryke}, {Rapetti}, {Reichardt}, {Rest}, {Ruel}, {Ruhl},
  {Saliwanchik}, {Saro}, {Sayre}, {Schaffer}, {Schrabback}, {Shirokoff},
  {Song}, {Spieler}, {Stalder}, {Stanford}, {Staniszewski}, {Stark}, {Story},
  {Stubbs}, {Vanderlinde}, {Vieira}, {Vikhlinin}, {Williamson}, \&
  {Zenteno}}]{deHaan16}
{de Haan}, T., {Benson}, B.~A., {Bleem}, L.~E., {et~al.} 2016,
  \href{http://dx.doi.org/10.3847/0004-637X/832/1/95}{\color{blue}\apj},
  \href{https://ui.adsabs.harvard.edu/abs/2016ApJ...832...95D}{832, 95}

\bibitem[{{de Jong} {et~al.}(2017){de Jong}, {Verdoes Kleijn}, {Erben},
  {Hildebrandt}, {Kuijken}, {Sikkema}, {Brescia}, {Bilicki}, {Napolitano},
  {Amaro}, {Begeman}, {Boxhoorn}, {Buddelmeijer}, {Cavuoti}, {Getman}, {Grado},
  {Helmich}, {Huang}, {Irisarri}, {La Barbera}, {Longo}, {McFarland},
  {Nakajima}, {Paolillo}, {Puddu}, {Radovich}, {Rifatto}, {Tortora},
  {Valentijn}, {Vellucci}, {Vriend}, {Amon}, {Blake}, {Choi}, {Conti}, {Gwyn},
  {Herbonnet}, {Heymans}, {Hoekstra}, {Klaes}, {Merten}, {Miller}, {Schneider},
  \& {Viola}}]{deJong17}
{de Jong}, J. T.~A., {Verdoes Kleijn}, G.~A., {Erben}, T., {et~al.} 2017,
  \href{http://dx.doi.org/10.1051/0004-6361/201730747}{\color{blue}\aap},
  \href{https://ui.adsabs.harvard.edu/abs/2017A&A...604A.134D}{604, A134}

\bibitem[{{Despali} {et~al.}(2016){Despali}, {Giocoli}, {Angulo}, {Tormen},
  {Sheth}, {Baso}, \& {Moscardini}}]{Despali16}
{Despali}, G., {Giocoli}, C., {Angulo}, R.~E., {et~al.} 2016,
  \href{http://dx.doi.org/10.1093/mnras/stv2842}{\color{blue}\mnras},
  \href{https://ui.adsabs.harvard.edu/abs/2016MNRAS.456.2486D}{456, 2486}

\bibitem[{{Diemer} \& {Joyce}(2019)}]{Diemer19}
{Diemer}, B. \& {Joyce}, M. 2019,
  \href{http://dx.doi.org/10.3847/1538-4357/aafad6}{\color{blue}\apj},
  \href{https://ui.adsabs.harvard.edu/abs/2019ApJ...871..168D}{871, 168}

\bibitem[{{Diemer} \& {Kravtsov}(2014)}]{DK14}
{Diemer}, B. \& {Kravtsov}, A.~V. 2014,
  \href{http://dx.doi.org/10.1088/0004-637X/789/1/1}{\color{blue}\apj},
  \href{https://ui.adsabs.harvard.edu/abs/2014ApJ...789....1D}{789, 1}

\bibitem[{{Diemer} \& {Kravtsov}(2015)}]{Diemer15}
{Diemer}, B. \& {Kravtsov}, A.~V. 2015,
  \href{http://dx.doi.org/10.1088/0004-637X/799/1/108}{\color{blue}\apj},
  \href{https://ui.adsabs.harvard.edu/abs/2015ApJ...799..108D}{799, 108}

\bibitem[{{Dietrich} {et~al.}(2019){Dietrich}, {Bocquet}, {Schrabback},
  {Applegate}, {Hoekstra}, {Grandis}, {Mohr}, {Allen}, {Bayliss}, {Benson},
  {Bleem}, {Brodwin}, {Bulbul}, {Capasso}, {Chiu}, {Crawford}, {Gonzalez}, {de
  Haan}, {Klein}, {von der Linden}, {Mantz}, {Marrone}, {McDonald},
  {Raghunathan}, {Rapetti}, {Reichardt}, {Saro}, {Stalder}, {Stark}, {Stern},
  \& {Stubbs}}]{Dietrich19}
{Dietrich}, J.~P., {Bocquet}, S., {Schrabback}, T., {et~al.} 2019,
  \href{http://dx.doi.org/10.1093/mnras/sty3088}{\color{blue}\mnras},
  \href{https://ui.adsabs.harvard.edu/abs/2019MNRAS.483.2871D}{483, 2871}

\bibitem[{{Dietrich} {et~al.}(2014){Dietrich}, {Zhang}, {Song}, {Davis},
  {McKay}, {Baruah}, {Becker}, {Benoist}, {Busha}, {da Costa}, {Hao}, {Maia},
  {Miller}, {Ogando}, {Romer}, {Rozo}, {Rykoff}, \& {Wechsler}}]{Dietrich14}
{Dietrich}, J.~P., {Zhang}, Y., {Song}, J., {et~al.} 2014,
  \href{http://dx.doi.org/10.1093/mnras/stu1282}{\color{blue}\mnras},
  \href{https://ui.adsabs.harvard.edu/abs/2014MNRAS.443.1713D}{443, 1713}

\bibitem[{{Ding} {et~al.}(2025){Ding}, {Dalal}, {Sunayama}, {Strauss}, {Oguri},
  {Okabe}, {Hilton}, {Monteiro-Oliveira}, {Sif{\'o}n}, \& {Staggs}}]{Ding24}
{Ding}, J., {Dalal}, R., {Sunayama}, T., {et~al.} 2025,
  \href{http://dx.doi.org/10.1093/mnras/stae2601}{\color{blue}\mnras},
  \href{https://ui.adsabs.harvard.edu/abs/2025MNRAS.536..572D}{536, 572}

\bibitem[{{Doubrawa} {et~al.}(2023){Doubrawa}, {Cypriano}, {Finoguenov},
  {Lopes}, {Maturi}, {Gonzalez}, \& {Dupke}}]{Doubrawa23}
{Doubrawa}, L., {Cypriano}, E.~S., {Finoguenov}, A., {et~al.} 2023,
  \href{http://dx.doi.org/10.1093/mnras/stad3024}{\color{blue}\mnras},
  \href{https://ui.adsabs.harvard.edu/abs/2023MNRAS.526.4285D}{526, 4285}

\bibitem[{{Driver} {et~al.}(2022){Driver}, {Bellstedt}, {Robotham}, {Baldry},
  {Davies}, {Liske}, {Obreschkow}, {Taylor}, {Wright}, {Alpaslan}, {Bamford},
  {Bauer}, {Bland-Hawthorn}, {Bilicki}, {Bravo}, {Brough}, {Casura}, {Cluver},
  {Colless}, {Conselice}, {Croom}, {de Jong}, {D'Eugenio}, {De Propris},
  {Dogruel}, {Drinkwater}, {Dvornik}, {Farrow}, {Frenk}, {Giblin}, {Graham},
  {Grootes}, {Gunawardhana}, {Hashemizadeh}, {H{\"a}u{\ss}ler}, {Heymans},
  {Hildebrandt}, {Holwerda}, {Hopkins}, {Jarrett}, {Heath Jones}, {Kelvin},
  {Koushan}, {Kuijken}, {Lara-L{\'o}pez}, {Lange}, {L{\'o}pez-S{\'a}nchez},
  {Loveday}, {Mahajan}, {Meyer}, {Moffett}, {Napolitano}, {Norberg}, {Owers},
  {Radovich}, {Raouf}, {Peacock}, {Phillipps}, {Pimbblet}, {Popescu}, {Said},
  {Sansom}, {Seibert}, {Sutherland}, {Thorne}, {Tuffs}, {Turner}, {van der
  Wel}, {van Kampen}, \& {Wilkins}}]{Driver22}
{Driver}, S.~P., {Bellstedt}, S., {Robotham}, A. S.~G., {et~al.} 2022,
  \href{http://dx.doi.org/10.1093/mnras/stac472}{\color{blue}\mnras},
  \href{https://ui.adsabs.harvard.edu/abs/2022MNRAS.513..439D}{513, 439}

\bibitem[{{Driver} {et~al.}(2011){Driver}, {Hill}, {Kelvin}, {Robotham},
  {Liske}, {Norberg}, {Baldry}, {Bamford}, {Hopkins}, {Loveday}, {Peacock},
  {Andrae}, {Bland-Hawthorn}, {Brough}, {Brown}, {Cameron}, {Ching}, {Colless},
  {Conselice}, {Croom}, {Cross}, {de Propris}, {Dye}, {Drinkwater}, {Ellis},
  {Graham}, {Grootes}, {Gunawardhana}, {Jones}, {van Kampen}, {Maraston},
  {Nichol}, {Parkinson}, {Phillipps}, {Pimbblet}, {Popescu}, {Prescott},
  {Roseboom}, {Sadler}, {Sansom}, {Sharp}, {Smith}, {Taylor}, {Thomas},
  {Tuffs}, {Wijesinghe}, {Dunne}, {Frenk}, {Jarvis}, {Madore}, {Meyer},
  {Seibert}, {Staveley-Smith}, {Sutherland}, \& {Warren}}]{Driver11}
{Driver}, S.~P., {Hill}, D.~T., {Kelvin}, L.~S., {et~al.} 2011,
  \href{http://dx.doi.org/10.1111/j.1365-2966.2010.18188.x}{\color{blue}\mnras},
  \href{https://ui.adsabs.harvard.edu/abs/2011MNRAS.413..971D}{413, 971}

\bibitem[{{Drlica-Wagner} {et~al.}(2018){Drlica-Wagner}, {Sevilla-Noarbe},
  {Rykoff}, {Gruendl}, {Yanny}, {Tucker}, {Hoyle}, {Carnero Rosell},
  {Bernstein}, {Bechtol}, {Becker}, {Benoit-L{\'e}vy}, {Bertin}, {Carrasco
  Kind}, {Davis}, {de Vicente}, {Diehl}, {Gruen}, {Hartley}, {Leistedt}, {Li},
  {Marshall}, {Neilsen}, {Rau}, {Sheldon}, {Smith}, {Troxel}, {Wyatt}, {Zhang},
  {Abbott}, {Abdalla}, {Allam}, {Banerji}, {Brooks}, {Buckley-Geer}, {Burke},
  {Capozzi}, {Carretero}, {Cunha}, {D'Andrea}, {da Costa}, {DePoy}, {Desai},
  {Dietrich}, {Doel}, {Evrard}, {Fausti Neto}, {Flaugher}, {Fosalba},
  {Frieman}, {Garc{\'\i}a-Bellido}, {Gerdes}, {Giannantonio}, {Gschwend},
  {Gutierrez}, {Honscheid}, {James}, {Jeltema}, {Kuehn}, {Kuhlmann},
  {Kuropatkin}, {Lahav}, {Lima}, {Lin}, {Maia}, {Martini}, {McMahon},
  {Melchior}, {Menanteau}, {Miquel}, {Nichol}, {Ogando}, {Plazas}, {Romer},
  {Roodman}, {Sanchez}, {Scarpine}, {Schindler}, {Schubnell}, {Smith}, {Smith},
  {Soares-Santos}, {Sobreira}, {Suchyta}, {Tarle}, {Vikram}, {Walker},
  {Wechsler}, {Zuntz}, \& {DES Collaboration}}]{Drlica-Wagner18}
{Drlica-Wagner}, A., {Sevilla-Noarbe}, I., {Rykoff}, E.~S., {et~al.} 2018,
  \href{http://dx.doi.org/10.3847/1538-4365/aab4f5}{\color{blue}\apjs},
  \href{https://ui.adsabs.harvard.edu/abs/2018ApJS..235...33D}{235, 33}

\bibitem[{{Duffy} {et~al.}(2008){Duffy}, {Schaye}, {Kay}, \& {Dalla
  Vecchia}}]{Duffy08}
{Duffy}, A.~R., {Schaye}, J., {Kay}, S.~T., \& {Dalla Vecchia}, C. 2008,
  \href{http://dx.doi.org/10.1111/j.1745-3933.2008.00537.x}{\color{blue}\mnras},
  \href{https://ui.adsabs.harvard.edu/abs/2008MNRAS.390L..64D}{390, L64}

\bibitem[{{Dutton} \& {Macci{\`o}}(2014)}]{Dutton14}
{Dutton}, A.~A. \& {Macci{\`o}}, A.~V. 2014,
  \href{http://dx.doi.org/10.1093/mnras/stu742}{\color{blue}\mnras},
  \href{https://ui.adsabs.harvard.edu/abs/2014MNRAS.441.3359D}{441, 3359}

\bibitem[{{Dvornik} {et~al.}(2017){Dvornik}, {Cacciato}, {Kuijken}, {Viola},
  {Hoekstra}, {Nakajima}, {van Uitert}, {Brouwer}, {Choi}, {Erben}, {Fenech
  Conti}, {Farrow}, {Herbonnet}, {Heymans}, {Hildebrandt}, {Hopkins},
  {McFarland}, {Norberg}, {Schneider}, {Sif{\'o}n}, {Valentijn}, \&
  {Wang}}]{Dvornik17}
{Dvornik}, A., {Cacciato}, M., {Kuijken}, K., {et~al.} 2017,
  \href{http://dx.doi.org/10.1093/mnras/stx705}{\color{blue}\mnras},
  \href{https://ui.adsabs.harvard.edu/abs/2017MNRAS.468.3251D}{468, 3251}

\bibitem[{{Eckert} {et~al.}(2020){Eckert}, {Finoguenov}, {Ghirardini},
  {Grandis}, {Kaefer}, {Sanders}, \& {Ramos-Ceja}}]{Eckert20}
{Eckert}, D., {Finoguenov}, A., {Ghirardini}, V., {et~al.} 2020,
  \href{http://dx.doi.org/10.21105/astro.2009.13944}{\color{blue}The Open
  Journal of Astrophysics},
  \href{https://ui.adsabs.harvard.edu/abs/2020OJAp....3E..12E}{3, 12}

\bibitem[{{Edge} {et~al.}(2013){Edge}, {Sutherland}, {Kuijken}, {Driver},
  {McMahon}, {Eales}, \& {Emerson}}]{VIKINGS}
{Edge}, A., {Sutherland}, W., {Kuijken}, K., {et~al.} 2013, The Messenger,
  \href{https://ui.adsabs.harvard.edu/abs/2013Msngr.154...32E}{154, 32}

\bibitem[{{Ettori} {et~al.}(2013){Ettori}, {Donnarumma}, {Pointecouteau},
  {Reiprich}, {Giodini}, {Lovisari}, \& {Schmidt}}]{Ettori13}
{Ettori}, S., {Donnarumma}, A., {Pointecouteau}, E., {et~al.} 2013,
  \href{http://dx.doi.org/10.1007/s11214-013-9976-7}{\color{blue}\ssr},
  \href{https://ui.adsabs.harvard.edu/abs/2013SSRv..177..119E}{177, 119}

\bibitem[{{Euclid Collaboration: Adam} {et~al.}(2019){Euclid Collaboration:
  Adam}, {Vannier}, {Maurogordato}, {Biviano}, {Adami}, {Ascaso}, {Bellagamba},
  {Benoist}, {Cappi}, {D{\'\i}az-S{\'a}nchez}, {Durret}, {Farrens}, {Gonzalez},
  {Iovino}, {Licitra}, {Maturi}, {Mei}, {Merson}, {Munari}, {Pell{\'o}},
  {Ricci}, {Rocci}, {Roncarelli}, {Sarron}, {Amoura}, {Andreon}, {Apostolakos},
  {Arnaud}, {Bardelli}, {Bartlett}, {Baugh}, {Borgani}, {Brodwin}, {Castander},
  {Castignani}, {Cucciati}, {De Lucia}, {Dubath}, {Fosalba}, {Giocoli},
  {Hoekstra}, {Mamon}, {Melin}, {Moscardini}, {Paltani}, {Radovich},
  {Sartoris}, {Schultheis}, {Sereno}, {Weller}, {Burigana}, {Carvalho},
  {Corcione}, {Kurki-Suonio}, {Lilje}, {Sirri}, {Toledo-Moreo}, \&
  {Zamorani}}]{Adam19}
{Euclid Collaboration: Adam}, R., {Vannier}, M., {Maurogordato}, S., {et~al.}
  2019, \href{http://dx.doi.org/10.1051/0004-6361/201935088}{\color{blue}\aap},
  \href{https://ui.adsabs.harvard.edu/abs/2019A&A...627A..23E}{627, A23}

\bibitem[{{Euclid Collaboration: Castro} {et~al.}(2024{\natexlab{a}}){Euclid
  Collaboration: Castro}, {Borgani}, {Costanzi}, {Dakin}, {Dolag}, {Fumagalli},
  {Ragagnin}, {Saro}, {Le Brun}, {Aghanim}, {Amara}, {Andreon}, {Auricchio},
  {Baldi}, {Bardelli}, {Bodendorf}, {Bonino}, {Branchini}, {Brescia},
  {Brinchmann}, {Camera}, {Capobianco}, {Carbone}, {Carretero}, {Casas},
  {Castellano}, {Cavuoti}, {Cimatti}, {Congedo}, {Conselice}, {Conversi},
  {Copin}, {Corcione}, {Courbin}, {Courtois}, {Cropper}, {Da Silva},
  {Degaudenzi}, {Di Giorgio}, {Dinis}, {Dubath}, {Duncan}, {Dupac}, {Farina},
  {Farrens}, {Ferriol}, {Frailis}, {Franceschi}, {Fumana}, {Galeotta},
  {Gillis}, {Giocoli}, {Grazian}, {Grupp}, {Haugan}, {Holmes}, {Hormuth},
  {Hornstrup}, {Jahnke}, {Keih{\"a}nen}, {Kermiche}, {Kiessling}, {Kilbinger},
  {Kubik}, {Kunz}, {Kurki-Suonio}, {Ligori}, {Lilje}, {Lindholm}, {Lloro},
  {Maiorano}, {Mansutti}, {Marggraf}, {Markovic}, {Martinet}, {Marulli},
  {Massey}, {Maurogordato}, {Medinaceli}, {Meneghetti}, {Merlin}, {Meylan},
  {Moresco}, {Moscardini}, {Munari}, {Niemi}, {Padilla}, {Paltani}, {Pasian},
  {Pettorino}, {Pires}, {Polenta}, {Poncet}, {Popa}, {Pozzetti}, {Raison},
  {Rebolo}, {Renzi}, {Rhodes}, {Riccio}, {Romelli}, {Roncarelli}, {Saglia},
  {Sapone}, {Sartoris}, {Schneider}, {Schrabback}, {Secroun}, {Seidel},
  {Serrano}, {Sirignano}, {Sirri}, {Stanco}, {Starck}, {Tallada-Cresp{\'\i}},
  {Taylor}, {Tereno}, {Toledo-Moreo}, {Torradeflot}, {Tutusaus}, {Valentijn},
  {Valenziano}, {Vassallo}, {Veropalumbo}, {Wang}, {Weller}, {Zacchei},
  {Zamorani}, {Zoubian}, {Zucca}, {Biviano}, {Bozzo}, {Cerna}, {Colodro-Conde},
  {Di Ferdinando}, {Mauri}, {Neissner}, {Sakr}, {Scottez}, {Tenti}, {Viel},
  {Wiesmann}, {Akrami}, {Anselmi}, {Baccigalupi}, {Ballardini}, {Borlaff},
  {Bruton}, {Burigana}, {Cabanac}, {Cappi}, {Carvalho}, {Castignani},
  {Ca{\~n}as-Herrera}, {Chambers}, {Cooray}, {Coupon}, {Cucciati},
  {D{\'\i}az-S{\'a}nchez}, {Davini}, {de la Torre}, {De Lucia}, {Desprez}, {Di
  Domizio}, {Dole}, {Escoffier}, {Ferrero}, {Finelli}, {Gabarra}, {Ganga},
  {Garcia-Bellido}, {Giacomini}, {Gozaliasl}, {Hildebrandt}, {Ili{\'c}},
  {Jimanez Mun{\~n}oz}, {Kajava}, {Kansal}, {Kirkpatrick}, {Legrand},
  {Loureiro}, {Macias-Perez}, {Magliocchetti}, {Mainetti}, {Maoli},
  {Martinelli}, {Martins}, {Matthew}, {Maturi}, {Maurin}, {Metcalf},
  {Migliaccio}, {Monaco}, {Morgante}, {Nadathur}, {Patrizii}, {Pezzotta},
  {Popa}, \& {Porciani}}]{Castro23}
{Euclid Collaboration: Castro}, T., {Borgani}, S., {Costanzi}, M., {et~al.}
  2024{\natexlab{a}},
  \href{http://dx.doi.org/10.1051/0004-6361/202348388}{\color{blue}\aap},
  \href{https://ui.adsabs.harvard.edu/abs/2024A&A...685A.109E}{685, A109}

\bibitem[{{Euclid Collaboration: Castro} {et~al.}(2023){Euclid Collaboration:
  Castro}, {Fumagalli}, {Angulo}, {Bocquet}, {Borgani}, {Carbone}, {Dakin},
  {Dolag}, {Giocoli}, {Monaco}, {Ragagnin}, {Saro}, {Sefusatti}, {Costanzi},
  {Le Brun}, {Corasaniti}, {Amara}, {Amendola}, {Baldi}, {Bender}, {Bodendorf},
  {Branchini}, {Brescia}, {Camera}, {Capobianco}, {Carretero}, {Castellano},
  {Cavuoti}, {Cimatti}, {Cledassou}, {Congedo}, {Conversi}, {Copin},
  {Corcione}, {Courbin}, {Da Silva}, {Degaudenzi}, {Douspis}, {Dubath},
  {Duncan}, {Dupac}, {Farrens}, {Ferriol}, {Fosalba}, {Frailis}, {Franceschi},
  {Galeotta}, {Garilli}, {Gillis}, {Grazian}, {Grupp}, {Haugan}, {Hormuth},
  {Hornstrup}, {Hudelot}, {Jahnke}, {Kermiche}, {Kitching}, {Kunz},
  {Kurki-Suonio}, {Lilje}, {Lloro}, {Mansutti}, {Marggraf}, {Marulli},
  {Meneghetti}, {Merlin}, {Meylan}, {Moresco}, {Moscardini}, {Munari}, {Niemi},
  {Padilla}, {Paltani}, {Pasian}, {Pedersen}, {Pettorino}, {Pires}, {Polenta},
  {Poncet}, {Popa}, {Pozzetti}, {Raison}, {Rebolo}, {Renzi}, {Rhodes},
  {Riccio}, {Romelli}, {Saglia}, {Sapone}, {Sartoris}, {Schneider}, {Seidel},
  {Sirri}, {Stanco}, {Tallada Cresp{\'\i}}, {Taylor}, {Toledo-Moreo},
  {Torradeflot}, {Tutusaus}, {Valentijn}, {Valenziano}, {Vassallo}, {Wang},
  {Weller}, {Zacchei}, {Zamorani}, {Andreon}, {Bardelli}, {Bozzo},
  {Colodro-Conde}, {Di Ferdinando}, {Farina}, {Graci{\'a}-Carpio}, {Lindholm},
  {Neissner}, {Scottez}, {Tenti}, {Zucca}, {Baccigalupi},
  {Balaguera-Antol{\'\i}nez}, {Ballardini}, {Bernardeau}, {Biviano},
  {Blanchard}, {Borlaff}, {Burigana}, {Cabanac}, {Cappi}, {Carvalho}, {Casas},
  {Castignani}, {Cooray}, {Coupon}, {Courtois}, {Davini}, {De Lucia},
  {Desprez}, {Dole}, {Escartin}, {Escoffier}, {Finelli}, {Ganga},
  {Garcia-Bellido}, {George}, {Gozaliasl}, {Hildebrandt}, {Hook}, {Ili{\'c}},
  {Kansal}, {Keihanen}, {Kirkpatrick}, {Loureiro}, {Macias-Perez},
  {Magliocchetti}, {Maoli}, {Marcin}, {Martinelli}, {Martinet}, {Matthew},
  {Maturi}, {Metcalf}, {Morgante}, {Nadathur}, {Nucita}, {Patrizii}, {Peel},
  {Popa}, {Porciani}, {Potter}, {Pourtsidou}, {P{\"o}ntinen}, {S{\'a}nchez},
  {Sakr}, {Schirmer}, {Sereno}, {Spurio Mancini}, {Teyssier}, {Valiviita},
  {Veropalumbo}, \& {Viel}}]{Castro23_HMF}
{Euclid Collaboration: Castro}, T., {Fumagalli}, A., {Angulo}, R.~E., {et~al.}
  2023, \href{http://dx.doi.org/10.1051/0004-6361/202244674}{\color{blue}\aap},
  \href{https://ui.adsabs.harvard.edu/abs/2023A&A...671A.100E}{671, A100}

\bibitem[{{Euclid Collaboration: Castro} {et~al.}(2024{\natexlab{b}}){Euclid
  Collaboration: Castro}, {Fumagalli}, {Angulo}, {Bocquet}, {Borgani},
  {Costanzi}, {Dakin}, {Dolag}, {Monaco}, {Saro}, {Sefusatti}, {Aghanim},
  {Amendola}, {Andreon}, {Baccigalupi}, {Baldi}, {Bodendorf}, {Bonino},
  {Branchini}, {Brescia}, {Caillat}, {Camera}, {Capobianco}, {Carbone},
  {Carretero}, {Casas}, {Castellano}, {Castignani}, {Cavuoti}, {Cimatti},
  {Colodro-Conde}, {Congedo}, {Conselice}, {Conversi}, {Copin}, {Costille},
  {Courbin}, {Courtois}, {Da Silva}, {Degaudenzi}, {De Lucia}, {Di Giorgio},
  {Douspis}, {Dupac}, {Dusini}, {Farina}, {Farrens}, {Ferriol}, {Fosalba},
  {Frailis}, {Franceschi}, {Fumana}, {Galeotta}, {Gillis}, {Giocoli},
  {G{\'o}mez-Alvarez}, {Grazian}, {Grupp}, {Guzzo}, {Haugan}, {Holmes},
  {Hormuth}, {Hornstrup}, {Ili{\'c}}, {Jahnke}, {Jhabvala}, {Joachimi},
  {Keih{\"a}nen}, {Kermiche}, {Kiessling}, {Kilbinger}, {Kubik}, {Kunz},
  {Kurki-Suonio}, {Lilje}, {Lindholm}, {Lloro}, {Maiorano}, {Mansutti},
  {Marggraf}, {Markovic}, {Martinelli}, {Martinet}, {Marulli}, {Massey},
  {Maurogordato}, {Medinaceli}, {Melchior}, {Mellier}, {Meneghetti}, {Merlin},
  {Meylan}, {Moscardini}, {Munari}, {Niemi}, {Padilla}, {Paltani}, {Pasian},
  {Pedersen}, {Percival}, {Pettorino}, {Pires}, {Polenta}, {Poncet}, {Popa},
  {Pozzetti}, {Raison}, {Renzi}, {Riccio}, {Romelli}, {Roncarelli}, {Saglia},
  {Sakr}, {Salvignol}, {S{\'a}nchez}, {Sapone}, {Sartoris}, {Schirmer},
  {Secroun}, {Serrano}, {Sirignano}, {Sirri}, {Stanco}, {Steinwagner},
  {Tallada-Cresp{\'\i}}, {Taylor}, {Tereno}, {Toledo-Moreo}, {Torradeflot},
  {Tutusaus}, {Valenziano}, {Vassallo}, {Verdoes Kleijn}, {Wang}, {Weller},
  {Zacchei}, {Zamorani}, {Zucca}, {Biviano}, {Bolzonella}, {Bozzo}, {Burigana},
  {Calabrese}, {Di Ferdinando}, {Escartin Vigo}, {Finelli}, {Gracia-Carpio},
  {Matthew}, {Mauri}, {Pezzotta}, {P{\"o}ntinen}, {Porciani}, {Scottez},
  {Tenti}, {Viel}, {Wiesmann}, {Akrami}, {Allevato}, {Anselmi}, {Archidiacono},
  {Atrio-Barandela}, {Balaguera-Antolinez}, {Ballardini}, {Bertacca},
  {Bethermin}, {Blanchard}, {Blot}, {B{\"o}hringer}, {Bruton}, {Cabanac},
  {Calabro}, {Ca{\~n}as-Herrera}, {Cappi}, {Caro}, {Carvalho}, {Chambers},
  {Cooray}, {De Caro}, {de la Torre}, {Desprez}, {D{\'\i}az-S{\'a}nchez},
  {Diaz}, {Di Domizio}, {Dole}, {Escoffier}, {Ferrari}, {Ferreira}, {Ferrero},
  {Finoguenov}, {Fontana}, {Fornari}, {Gabarra}, {Ganga},
  {Garc{\'\i}a-Bellido}, {Gasparetto}, {Gautard}, {Gaztanaga}, {Giacomini}, \&
  {Gianotti}}]{Castro24}
{Euclid Collaboration: Castro}, T., {Fumagalli}, A., {Angulo}, R.~E., {et~al.}
  2024{\natexlab{b}},
  \href{http://dx.doi.org/10.1051/0004-6361/202451230}{\color{blue}\aap},
  \href{https://ui.adsabs.harvard.edu/abs/2024A&A...691A..62E}{691, A62}

\bibitem[{{Euclid Collaboration: Giocoli} {et~al.}(2024){Euclid Collaboration:
  Giocoli}, {Meneghetti}, {Rasia}, {Borgani}, {Despali}, {Lesci}, {Marulli},
  {Moscardini}, {Sereno}, {Cui}, {Knebe}, {Yepes}, {Castro}, {Corasaniti},
  {Pires}, {Castignani}, {Schrabback}, {Pratt}, {Le Brun}, {Aghanim},
  {Amendola}, {Auricchio}, {Baldi}, {Bodendorf}, {Bonino}, {Branchini},
  {Brescia}, {Brinchmann}, {Camera}, {Capobianco}, {Carbone}, {Carretero},
  {Castander}, {Castellano}, {Cavuoti}, {Cledassou}, {Congedo}, {Conselice},
  {Conversi}, {Copin}, {Corcione}, {Courbin}, {Cropper}, {Da Silva},
  {Degaudenzi}, {Dinis}, {Dubath}, {Dupac}, {Dusini}, {Farrens}, {Ferriol},
  {Fosalba}, {Frailis}, {Franceschi}, {Fumana}, {Galeotta}, {Garilli},
  {Gillis}, {Grazian}, {Grupp}, {Haugan}, {Holmes}, {Hornstrup}, {Jahnke},
  {K{\"u}mmel}, {Kermiche}, {Kilbinger}, {Kunz}, {Kurki-Suonio}, {Ligori},
  {Lilje}, {Lloro}, {Maiorano}, {Mansutti}, {Marggraf}, {Markovic}, {Massey},
  {Maurogordato}, {Mei}, {Merlin}, {Meylan}, {Moresco}, {Munari}, {Niemi},
  {Nightingale}, {Nutma}, {Padilla}, {Paltani}, {Pasian}, {Pedersen},
  {Pettorino}, {Polenta}, {Poncet}, {Popa}, {Raison}, {Renzi}, {Rhodes},
  {Riccio}, {Romelli}, {Roncarelli}, {Rossetti}, {Saglia}, {Sapone},
  {Sartoris}, {Schneider}, {Secroun}, {Serrano}, {Sirignano}, {Sirri},
  {Stanco}, {Starck}, {Tallada-Cresp{\'\i}}, {Taylor}, {Tereno},
  {Toledo-Moreo}, {Torradeflot}, {Tutusaus}, {Valentijn}, {Valenziano},
  {Vassallo}, {Wang}, {Weller}, {Zamorani}, {Zoubian}, {Andreon}, {Bardelli},
  {Boucaud}, {Bozzo}, {Colodro-Conde}, {Di Ferdinando}, {Fabbian}, {Farina},
  {Israel}, {Keih{\"a}nen}, {Lindholm}, {Mauri}, {Neissner}, {Schirmer},
  {Scottez}, {Tenti}, {Zucca}, {Akrami}, {Baccigalupi}, {Ballardini},
  {Bernardeau}, {Biviano}, {Borlaff}, {Burigana}, {Cabanac}, {Cappi},
  {Carvalho}, {Casas}, {Chambers}, {Cooray}, {Courtois}, {Davini}, {de la
  Torre}, {De Lucia}, {Desprez}, {Dole}, {Escartin}, {Escoffier}, {Ferrero},
  {Finelli}, {Gabarra}, {Ganga}, {Garcia-Bellido}, {George}, {Giacomini},
  {Gozaliasl}, {Hildebrandt}, {Hook}, {Jimenez Mu{\~n}oz}, {Joachimi},
  {Kajava}, {Kansal}, {Kirkpatrick}, {Legrand}, {Loureiro}, {Macias-Perez},
  {Magliocchetti}, {Mainetti}, {Maoli}, {Marcin}, {Martinelli}, {Martinet},
  {Martins}, {Matthew}, {Maurin}, {Metcalf}, {Monaco}, {Morgante}, {Nadathur},
  {Nucita}, {Patrizii}, {Peel}, {Pollack}, {Popa}, \& {Porciani}}]{Giocoli23}
{Euclid Collaboration: Giocoli}, C., {Meneghetti}, M., {Rasia}, E., {et~al.}
  2024, \href{http://dx.doi.org/10.1051/0004-6361/202346058}{\color{blue}\aap},
  \href{https://ui.adsabs.harvard.edu/abs/2024A&A...681A..67E}{681, A67}

\bibitem[{{Euclid Collaboration: Lesci} {et~al.}(2024){Euclid Collaboration:
  Lesci}, {Sereno}, {Radovich}, {Castignani}, {Bisigello}, {Marulli},
  {Moscardini}, {Baumont}, {Covone}, {Farrens}, {Giocoli}, {Ingoglia}, {Miranda
  La Hera}, {Vannier}, {Biviano}, {Maurogordato}, {Aghanim}, {Amara},
  {Andreon}, {Auricchio}, {Baldi}, {Bardelli}, {Bender}, {Bodendorf}, {Bonino},
  {Branchini}, {Brescia}, {Brinchmann}, {Camera}, {Capobianco}, {Carbone},
  {Carretero}, {Casas}, {Castander}, {Castellano}, {Cavuoti}, {Cimatti},
  {Congedo}, {Conselice}, {Conversi}, {Copin}, {Corcione}, {Courbin},
  {Courtois}, {Da Silva}, {Degaudenzi}, {Di Giorgio}, {Dinis}, {Dubath},
  {Duncan}, {Dupac}, {Dusini}, {Farina}, {Ferriol}, {Fosalba}, {Fotopoulou},
  {Frailis}, {Franceschi}, {Franzetti}, {Fumana}, {Galeotta}, {Garilli},
  {Gillis}, {Grazian}, {Grupp}, {Haugan}, {Hook}, {Hormuth}, {Hornstrup},
  {Hudelot}, {Jahnke}, {K{\"u}mmel}, {Kermiche}, {Kiessling}, {Kilbinger},
  {Kubik}, {Kunz}, {Kurki-Suonio}, {Ligori}, {Lilje}, {Lindholm}, {Lloro},
  {Maiorano}, {Mansutti}, {Marggraf}, {Markovic}, {Martinet}, {Massey},
  {Medinaceli}, {Melchior}, {Mellier}, {Meneghetti}, {Merlin}, {Meylan},
  {Moresco}, {Munari}, {Nakajima}, {Niemi}, {Padilla}, {Paltani}, {Pasian},
  {Pedersen}, {Pettorino}, {Pires}, {Polenta}, {Poncet}, {Popa}, {Pozzetti},
  {Raison}, {Rebolo}, {Renzi}, {Rhodes}, {Riccio}, {Romelli}, {Roncarelli},
  {Rossetti}, {Saglia}, {Sapone}, {Sartoris}, {Schirmer}, {Schneider},
  {Secroun}, {Seidel}, {Serrano}, {Sirignano}, {Sirri}, {Skottfelt}, {Stanco},
  {Starck}, {Tallada-Cresp{\'\i}}, {Taylor}, {Teplitz}, {Tereno},
  {Toledo-Moreo}, {Torradeflot}, {Tutusaus}, {Valentijn}, {Valenziano},
  {Vassallo}, {Veropalumbo}, {Wang}, {Weller}, {Zacchei}, {Zamorani},
  {Zoubian}, {Zucca}, {Bolzonella}, {Bozzo}, {Colodro-Conde}, {Di Ferdinando},
  {Graci{\'a}-Carpio}, {Marcin}, {Mauri}, {Neissner}, {Nucita}, {Sakr},
  {Scottez}, {Tenti}, {Viel}, {Wiesmann}, {Akrami}, {Anselmi}, {Baccigalupi},
  {Ballardini}, {Borgani}, {Borlaff}, {Bruton}, {Burigana}, {Cabanac},
  {Calabro}, {Cappi}, {Carvalho}, {Castro}, {Ca{\~n}as-Herrera}, {Chambers},
  {Cooray}, {Coupon}, {Cucciati}, {Davini}, {de la Torre}, {De Lucia},
  {Desprez}, {Di Domizio}, {Dole}, {D{\'\i}az-S{\'a}nchez}, {Escartin Vigo},
  {Escoffier}, {Ferrero}, {Finelli}, {Gabarra}, {Ganga}, {Garc{\'\i}a-Bellido},
  {Giacomini}, {Gozaliasl}, {Gwyn}, {Hildebrandt}, {Huertas-Company}, {Jimenez
  Mu{\~n}oz}, {Kajava}, {Kansal}, {Kirkpatrick}, {Legrand}, {Loureiro},
  {Macias-Perez}, {Magliocchetti}, {Mainetti}, {Maoli}, {Martinelli},
  {Martins}, {Matthew}, {Maturi}, {Maurin}, {Metcalf}, {Migliaccio}, {Monaco},
  {Morgante}, {Nadathur}, {Patrizii}, {Pezzotta}, {Porciani}, {Potter},
  {P{\"o}ntinen}, {Reimberg}, {Rocci}, {S{\'a}nchez}, {Schneider},
  {Schultheis}, {Sefusatti}, {Simon}, {Spurio Mancini}, {Stanford},
  {Steinwagner}, {Testera}, {Teyssier}, {Toft}, {Tosi}, {Troja}, {Tucci},
  {Valiviita}, \& {Vergani}}]{Lesci_colours}
{Euclid Collaboration: Lesci}, G.~F., {Sereno}, M., {Radovich}, M., {et~al.}
  2024, \href{http://dx.doi.org/10.1051/0004-6361/202348743}{\color{blue}\aap},
  \href{https://ui.adsabs.harvard.edu/abs/2024A&A...684A.139E}{684, A139}

\bibitem[{{Euclid Collaboration: Mellier} {et~al.}(2025){Euclid Collaboration:
  Mellier}, {Abdurro'uf}, {Acevedo Barroso}, {Ach{\'u}carro}, {Adamek}, {Adam},
  {Addison}, {Aghanim}, {Aguena}, {Ajani}, {Akrami}, {Al-Bahlawan}, {Alavi},
  {Albuquerque}, {Alestas}, {Alguero}, {Allaoui}, {Allen}, {Allevato},
  {Alonso-Tetilla}, {Altieri}, {Alvarez-Candal}, {Alvi}, {Amara}, {Amendola},
  {Amiaux}, {Andika}, {Andreon}, {Andrews}, {Angora}, {Angulo}, {Annibali},
  {Anselmi}, {Anselmi}, {Arcari}, {Archidiacono}, {Aric{\`o}}, {Arnaud},
  {Arnouts}, {Asgari}, {Asorey}, {Atayde}, {Atek}, {Atrio-Barandela}, {Aubert},
  {Aubourg}, {Auphan}, {Auricchio}, {Aussel}, {Aussel}, {Avelino},
  {Avgoustidis}, {Avila}, {Awan}, {Azzollini}, {Baccigalupi}, {Bachelet},
  {Bacon}, {Baes}, {Bagley}, {Bahr-Kalus}, {Balaguera-Antolinez}, {Balbinot},
  {Balcells}, {Baldi}, {Baldry}, {Balestra}, {Ballardini}, {Ballester},
  {Balogh}, {Ba{\~n}ados}, {Barbier}, {Bardelli}, {Baron}, {Barreiro},
  {Barrena}, {Barriere}, {Barros}, {Barthelemy}, {Bartolo}, {Basset},
  {Battaglia}, {Battisti}, {Baugh}, {Baumont}, {Bazzanini}, {Beaulieu},
  {Beckmann}, {Belikov}, {Bel}, {Bellagamba}, {Bella}, {Bellini}, {Benabed},
  {Bender}, {Benevento}, {Bennett}, {Benson}, {Bergamini}, {Bermejo-Climent},
  {Bernardeau}, {Bertacca}, {Berthe}, {Berthier}, {Bethermin}, {Beutler},
  {Bevillon}, {Bhargava}, {Bhatawdekar}, {Bianchi}, {Bisigello}, {Biviano},
  {Blake}, {Blanchard}, {Blazek}, {Blot}, {Bosco}, {Bodendorf}, {Boenke},
  {B{\"o}hringer}, {Boldrini}, {Bolzonella}, {Bonchi}, {Bonici}, {Bonino},
  {Bonino}, {Bonvin}, {Bon}, {Booth}, {Borgani}, {Borlaff}, {Borsato}, {Bose},
  {Botticella}, {Boucaud}, {Bouche}, {Boucher}, {Boutigny}, {Bouvard},
  {Bouwens}, {Bouy}, {Bowler}, {Bozza}, {Bozzo}, {Branchini}, {Brando},
  {Brau-Nogue}, {Brekke}, {Bremer}, {Brescia}, {Breton}, {Brinchmann},
  {Brinckmann}, {Brockley-Blatt}, {Brodwin}, {Brouard}, {Brown}, {Bruton},
  {Bucko}, {Buddelmeijer}, {Buenadicha}, {Buitrago}, {Burger}, {Burigana},
  {Busillo}, {Busonero}, {Cabanac}, {Cabayol-Garcia}, {Cagliari}, {Caillat},
  {Caillat}, {Calabrese}, {Calabro}, {Calderone}, {Calura}, {Camacho Quevedo},
  {Camera}, {Campos}, {Ca{\~n}as-Herrera}, {Candini}, {Cantiello},
  {Capobianco}, {Cappellaro}, {Cappelluti}, {Cappi}, {Caputi}, {Cara},
  {Carbone}, {Cardone}, {Carella}, {Carlberg}, {Carle}, {Carminati}, {Caro},
  {Carrasco}, {Carretero}, {Carrilho}, {Carron Duque}, \& {Carry}}]{Mellier24}
{Euclid Collaboration: Mellier}, Y., {Abdurro'uf}, {Acevedo Barroso}, J.~A.,
  {et~al.} 2025,
  \href{http://dx.doi.org/10.1051/0004-6361/202450810}{\color{blue}\aap},
  \href{https://ui.adsabs.harvard.edu/abs/2025A&A...697A...1E}{697, A1}

\bibitem[{{Euclid Collaboration: Sereno} {et~al.}(2024){Euclid Collaboration:
  Sereno}, {Farrens}, {Ingoglia}, {Lesci}, {Baumont}, {Covone}, {Giocoli},
  {Marulli}, {Miranda La Hera}, {Vannier}, {Biviano}, {Maurogordato},
  {Moscardini}, {Aghanim}, {Andreon}, {Auricchio}, {Baldi}, {Bardelli},
  {Bellagamba}, {Bodendorf}, {Bonino}, {Branchini}, {Brescia}, {Brinchmann},
  {Camera}, {Capobianco}, {Carbone}, {Cardone}, {Carretero}, {Casas},
  {Castellano}, {Cavuoti}, {Cimatti}, {Cledassou}, {Congedo}, {Conselice},
  {Conversi}, {Copin}, {Corcione}, {Courbin}, {Courtois}, {Cropper}, {Da
  Silva}, {Degaudenzi}, {Di Giorgio}, {Dinis}, {Dubath}, {Duncan}, {Dupac},
  {Dusini}, {Farina}, {Ferriol}, {Frailis}, {Franceschi}, {Fumana}, {Galeotta},
  {Garilli}, {Gillis}, {Grazian}, {Grupp}, {Haugan}, {Holmes}, {Hook},
  {Hormuth}, {Hornstrup}, {Hudelot}, {Jahnke}, {Joachimi}, {Keih{\"a}nen},
  {Kermiche}, {Kiessling}, {Kubik}, {Kunz}, {Kurki-Suonio}, {Ligori}, {Lilje},
  {Lindholm}, {Lloro}, {Maino}, {Maiorano}, {Mansutti}, {Marggraf}, {Markovic},
  {Martinet}, {Massey}, {Medinaceli}, {Mei}, {Mellier}, {Meneghetti}, {Merlin},
  {Meylan}, {Moresco}, {Munari}, {Niemi}, {Nutma}, {Padilla}, {Paltani},
  {Pasian}, {Pedersen}, {Pettorino}, {Pires}, {Polenta}, {Poncet}, {Popa},
  {Raison}, {Rebolo}, {Renzi}, {Rhodes}, {Riccio}, {Romelli}, {Roncarelli},
  {Rossetti}, {Saglia}, {Sapone}, {Sartoris}, {Schirmer}, {Schneider},
  {Schrabback}, {Secroun}, {Seidel}, {Serrano}, {Sirignano}, {Sirri}, {Stanco},
  {Starck}, {Tallada-Cresp{\'\i}}, {Taylor}, {Tereno}, {Toledo-Moreo},
  {Torradeflot}, {Tutusaus}, {Valentijn}, {Valenziano}, {Vassallo},
  {Veropalumbo}, {Wang}, {Weller}, {Zacchei}, {Zamorani}, {Zoubian}, {Zucca},
  {Boucaud}, {Bozzo}, {Cerna}, {Colodro-Conde}, {Di Ferdinando}, {Farinelli},
  {Israel}, {Mauri}, {Neissner}, {Scottez}, {Tenti}, {Wiesmann}, {Akrami},
  {Allevato}, {Baccigalupi}, {Ballardini}, {Benielli}, {Borgani}, {Borlaff},
  {Burigana}, {Cabanac}, {Cappi}, {Carvalho}, {Castignani}, {Castro},
  {Ca{\~n}as-Herrera}, {Chambers}, {Cooray}, {Coupon}, {Davini}, {De Lucia},
  {Desprez}, {Di Domizio}, {Dole}, {Escartin Vigo}, {Escoffier}, {Ferrero},
  {Gabarra}, {Gaztanaga}, {George}, {Giacomini}, {Gozaliasl}, {Hildebrandt},
  {Kajava}, {Kansal}, {Kirkpatrick}, {Legrand}, {Liebing}, {Loureiro},
  {Macias-Perez}, {Magliocchetti}, {Mainetti}, {Maoli}, {Martinelli},
  {Martins}, {Matthew}, {Maturi}, \& {Maurin}}]{Sereno24}
{Euclid Collaboration: Sereno}, M., {Farrens}, S., {Ingoglia}, L., {et~al.}
  2024, \href{http://dx.doi.org/10.1051/0004-6361/202348680}{\color{blue}\aap},
  \href{https://ui.adsabs.harvard.edu/abs/2024A&A...689A.252E}{689, A252}

\bibitem[{{Fenech Conti} {et~al.}(2017){Fenech Conti}, {Herbonnet}, {Hoekstra},
  {Merten}, {Miller}, \& {Viola}}]{Fenech17}
{Fenech Conti}, I., {Herbonnet}, R., {Hoekstra}, H., {et~al.} 2017,
  \href{http://dx.doi.org/10.1093/mnras/stx200}{\color{blue}\mnras},
  \href{https://ui.adsabs.harvard.edu/abs/2017MNRAS.467.1627F}{467, 1627}

\bibitem[{{Fumagalli} {et~al.}(2024){Fumagalli}, {Costanzi}, {Saro}, {Castro},
  \& {Borgani}}]{Fumagalli24}
{Fumagalli}, A., {Costanzi}, M., {Saro}, A., {Castro}, T., \& {Borgani}, S.
  2024, \href{http://dx.doi.org/10.1051/0004-6361/202348296}{\color{blue}\aap},
  \href{https://ui.adsabs.harvard.edu/abs/2024A&A...682A.148F}{682, A148}

\bibitem[{{Gatti} {et~al.}(2021){Gatti}, {Sheldon}, {Amon}, {Becker}, {Troxel},
  {Choi}, {Doux}, {MacCrann}, {Navarro-Alsina}, {Harrison}, {Gruen},
  {Bernstein}, {Jarvis}, {Secco}, {Fert{\'e}}, {Shin}, {McCullough}, {Rollins},
  {Chen}, {Chang}, {Pandey}, {Tutusaus}, {Prat}, {Elvin-Poole}, {Sanchez},
  {Plazas}, {Roodman}, {Zuntz}, {Abbott}, {Aguena}, {Allam}, {Annis}, {Avila},
  {Bacon}, {Bertin}, {Bhargava}, {Brooks}, {Burke}, {Carnero Rosell}, {Carrasco
  Kind}, {Carretero}, {Castander}, {Conselice}, {Costanzi}, {Crocce}, {da
  Costa}, {Davis}, {De Vicente}, {Desai}, {Diehl}, {Dietrich}, {Doel},
  {Drlica-Wagner}, {Eckert}, {Everett}, {Ferrero}, {Frieman},
  {Garc{\'\i}a-Bellido}, {Gerdes}, {Giannantonio}, {Gruendl}, {Gschwend},
  {Gutierrez}, {Hartley}, {Hinton}, {Hollowood}, {Honscheid}, {Hoyle}, {Huff},
  {Huterer}, {Jain}, {James}, {Jeltema}, {Krause}, {Kron}, {Kuropatkin},
  {Lima}, {Maia}, {Marshall}, {Miquel}, {Morgan}, {Myles}, {Palmese},
  {Paz-Chinch{\'o}n}, {Rykoff}, {Samuroff}, {Sanchez}, {Scarpine}, {Schubnell},
  {Serrano}, {Sevilla-Noarbe}, {Smith}, {Soares-Santos}, {Suchyta}, {Swanson},
  {Tarle}, {Thomas}, {To}, {Tucker}, {Varga}, {Wechsler}, {Weller}, {Wester},
  \& {Wilkinson}}]{Gatti21}
{Gatti}, M., {Sheldon}, E., {Amon}, A., {et~al.} 2021,
  \href{http://dx.doi.org/10.1093/mnras/stab918}{\color{blue}\mnras},
  \href{https://ui.adsabs.harvard.edu/abs/2021MNRAS.504.4312G}{504, 4312}

\bibitem[{{Ghirardini} {et~al.}(2024){Ghirardini}, {Bulbul}, {Artis}, {Clerc},
  {Garrel}, {Grandis}, {Kluge}, {Liu}, {Bahar}, {Balzer}, {Chiu}, {Comparat},
  {Gruen}, {Kleinebreil}, {Krippendorf}, {Merloni}, {Nandra}, {Okabe},
  {Pacaud}, {Predehl}, {Ramos-Ceja}, {Reiprich}, {Sanders}, {Schrabback},
  {Seppi}, {Zelmer}, {Zhang}, {Bornemann}, {Brunner}, {Burwitz}, {Coutinho},
  {Dennerl}, {Freyberg}, {Friedrich}, {Gaida}, {Gueguen}, {Haberl}, {Kink},
  {Lamer}, {Li}, {Liu}, {Maitra}, {Meidinger}, {Mueller}, {Miyatake},
  {Miyazaki}, {Robrade}, {Schwope}, \& {Stewart}}]{Ghirardini24}
{Ghirardini}, V., {Bulbul}, E., {Artis}, E., {et~al.} 2024,
  \href{http://dx.doi.org/10.1051/0004-6361/202348852}{\color{blue}\aap},
  \href{https://ui.adsabs.harvard.edu/abs/2024A&A...689A.298G}{689, A298}

\bibitem[{{Giblin} {et~al.}(2021){Giblin}, {Heymans}, {Asgari}, {Hildebrandt},
  {Hoekstra}, {Joachimi}, {Kannawadi}, {Kuijken}, {Lin}, {Miller},
  {Tr{\"o}ster}, {van den Busch}, {Wright}, {Bilicki}, {Blake}, {de Jong},
  {Dvornik}, {Erben}, {Getman}, {Napolitano}, {Schneider}, {Shan}, \&
  {Valentijn}}]{Giblin21}
{Giblin}, B., {Heymans}, C., {Asgari}, M., {et~al.} 2021,
  \href{http://dx.doi.org/10.1051/0004-6361/202038850}{\color{blue}\aap},
  \href{https://ui.adsabs.harvard.edu/abs/2021A&A...645A.105G}{645, A105}

\bibitem[{{Giocoli} {et~al.}(2025){Giocoli}, {Despali}, {Meneghetti}, {Rasia},
  {Moscardini}, {Borgani}, {Lesci}, {Marulli}, {Cui}, \& {Yepes}}]{Giocoli25}
{Giocoli}, C., {Despali}, G., {Meneghetti}, M., {et~al.} 2025,
  \href{http://dx.doi.org/10.1051/0004-6361/202553871}{\color{blue}\aap},
  \href{https://ui.adsabs.harvard.edu/abs/2025A&A...697A.184G}{697, A184}

\bibitem[{{Giocoli} {et~al.}(2021){Giocoli}, {Marulli}, {Moscardini}, {Sereno},
  {Veropalumbo}, {Gigante}, {Maturi}, {Radovich}, {Bellagamba}, {Roncarelli},
  {Bardelli}, {Contarini}, {Covone}, {Harnois-D{\'e}raps}, {Ingoglia}, {Lesci},
  {Nanni}, \& {Puddu}}]{Giocoli21}
{Giocoli}, C., {Marulli}, F., {Moscardini}, L., {et~al.} 2021,
  \href{http://dx.doi.org/10.1051/0004-6361/202140795}{\color{blue}\aap},
  \href{https://ui.adsabs.harvard.edu/abs/2021A&A...653A..19G}{653, A19}

\bibitem[{{Giocoli} {et~al.}(2024){Giocoli}, {Palmucci}, {Lesci}, {Moscardini},
  {Despali}, {Marulli}, {Maturi}, {Radovich}, {Sereno}, {Bardelli},
  {Castignani}, {Covone}, {Ingoglia}, {Romanello}, {Roncarelli}, \&
  {Puddu}}]{Giocoli24}
{Giocoli}, C., {Palmucci}, L., {Lesci}, G.~F., {et~al.} 2024,
  \href{http://dx.doi.org/10.1051/0004-6361/202449561}{\color{blue}\aap},
  \href{https://ui.adsabs.harvard.edu/abs/2024A&A...687A..79G}{687, A79}

\bibitem[{{Giocoli} {et~al.}(2012){Giocoli}, {Tormen}, \& {Sheth}}]{Giocoli12}
{Giocoli}, C., {Tormen}, G., \& {Sheth}, R.~K. 2012,
  \href{http://dx.doi.org/10.1111/j.1365-2966.2012.20594.x}{\color{blue}\mnras},
  \href{https://ui.adsabs.harvard.edu/abs/2012MNRAS.422..185G}{422, 185}

\bibitem[{{Grandis} {et~al.}(2021){Grandis}, {Bocquet}, {Mohr}, {Klein}, \&
  {Dolag}}]{Grandis21}
{Grandis}, S., {Bocquet}, S., {Mohr}, J.~J., {Klein}, M., \& {Dolag}, K. 2021,
  \href{http://dx.doi.org/10.1093/mnras/stab2414}{\color{blue}\mnras},
  \href{https://ui.adsabs.harvard.edu/abs/2021MNRAS.507.5671G}{507, 5671}

\bibitem[{{Grandis} {et~al.}(2025){Grandis}, {Costanzi}, {Mohr}, {Bleem}, {Wu},
  {Aguena}, {Allam}, {Andrade-Oliveira}, {Bocquet}, {Brooks}, {Carnero Rosell},
  {Carretero}, {da Costa}, {Pereira}, {Davis}, {Desai}, {Diehl}, {Doel},
  {Everett}, {Flaugher}, {Frieman}, {Garc{\'\i}a-Bellido}, {Gaztanaga},
  {Gruen}, {Gruendl}, {Gutierrez}, {Hinton}, {Hlacacek-Larrondo}, {Hollowood},
  {Honscheid}, {James}, {Klein}, {Marshall}, {Mena-Fern{\'a}ndez}, {Miquel},
  {Palmese}, {Plazas Malag{\'o}n}, {Reichardt}, {Romer}, {Samuroff}, {Sanchez
  Cid}, {Sanchez}, {Santiago}, {Saro}, {Sevilla-Noarbe}, {Smith},
  {Soares-Santos}, {Sommer}, {Suchyta}, {Tarle}, {To}, {Tucker}, {Weaverdyck},
  {Weller}, \& {Wiseman}}]{Grandis25}
{Grandis}, S., {Costanzi}, M., {Mohr}, J.~J., {et~al.} 2025,
  \href{http://dx.doi.org/10.1051/0004-6361/202554177}{\color{blue}\aap},
  \href{https://ui.adsabs.harvard.edu/abs/2025A&A...700A..15G}{700, A15}

\bibitem[{{Grandis} {et~al.}(2024){Grandis}, {Ghirardini}, {Bocquet}, {Garrel},
  {Mohr}, {Liu}, {Kluge}, {Kimmig}, {Reiprich}, {Alarcon}, {Amon}, {Artis},
  {Bahar}, {Balzer}, {Bechtol}, {Becker}, {Bernstein}, {Bulbul}, {Campos},
  {Carnero Rosell}, {Carrasco Kind}, {Cawthon}, {Chang}, {Chen}, {Chiu},
  {Choi}, {Clerc}, {Comparat}, {Cordero}, {Davis}, {Derose}, {Diehl},
  {Dodelson}, {Doux}, {Drlica-Wagner}, {Eckert}, {Elvin-Poole}, {Everett},
  {Ferte}, {Gatti}, {Giannini}, {Giles}, {Gruen}, {Gruendl}, {Harrison},
  {Hartley}, {Herner}, {Huff}, {Kleinebreil}, {Kuropatkin}, {Leget},
  {Maccrann}, {Mccullough}, {Merloni}, {Myles}, {Nandra}, {Navarro-Alsina},
  {Okabe}, {Pacaud}, {Pandey}, {Prat}, {Predehl}, {Ramos}, {Raveri}, {Rollins},
  {Roodman}, {Ross}, {Rykoff}, {Sanchez}, {Sanders}, {Schrabback}, {Secco},
  {Seppi}, {Sevilla-Noarbe}, {Sheldon}, {Shin}, {Troxel}, {Tutusaus}, {Varga},
  {Wu}, {Yanny}, {Yin}, {Zhang}, {Zhang}, {Alves}, {Bhargava}, {Brooks},
  {Burke}, {Carretero}, {Costanzi}, {da Costa}, {Pereira}, {De Vicente},
  {Desai}, {Doel}, {Ferrero}, {Flaugher}, {Friedel}, {Frieman},
  {Garc{\'\i}a-Bellido}, {Gutierrez}, {Hinton}, {Hollowood}, {Honscheid},
  {James}, {Jeffrey}, {Lahav}, {Lee}, {Marshall}, {Menanteau}, {Ogando},
  {Pieres}, {Plazas Malag{\'o}n}, {Romer}, {Sanchez}, {Schubnell}, {Smith},
  {Suchyta}, {Swanson}, {Tarle}, {Weaverdyck}, \& {Weller}}]{Grandis24}
{Grandis}, S., {Ghirardini}, V., {Bocquet}, S., {et~al.} 2024,
  \href{http://dx.doi.org/10.1051/0004-6361/202348615}{\color{blue}\aap},
  \href{https://ui.adsabs.harvard.edu/abs/2024A&A...687A.178G}{687, A178}

\bibitem[{{Hartlap} {et~al.}(2007){Hartlap}, {Simon}, \&
  {Schneider}}]{Hartlap07}
{Hartlap}, J., {Simon}, P., \& {Schneider}, P. 2007,
  \href{http://dx.doi.org/10.1051/0004-6361:20066170}{\color{blue}\aap},
  \href{https://ui.adsabs.harvard.edu/abs/2007A&A...464..399H}{464, 399}

\bibitem[{{Hennig} {et~al.}(2017){Hennig}, {Mohr}, {Zenteno}, {Desai},
  {Dietrich}, {Bocquet}, {Strazzullo}, {Saro}, {Abbott}, {Abdalla}, {Bayliss},
  {Benoit-L{\'e}vy}, {Bernstein}, {Bertin}, {Brooks}, {Capasso}, {Capozzi},
  {Carnero}, {Carrasco Kind}, {Carretero}, {Chiu}, {D'Andrea}, {daCosta},
  {Diehl}, {Doel}, {Eifler}, {Evrard}, {Fausti-Neto}, {Fosalba}, {Frieman},
  {Gangkofner}, {Gonzalez}, {Gruen}, {Gruendl}, {Gupta}, {Gutierrez},
  {Honscheid}, {Hlavacek-Larrondo}, {James}, {Kuehn}, {Kuropatkin}, {Lahav},
  {March}, {Marshall}, {Martini}, {McDonald}, {Melchior}, {Miller}, {Miquel},
  {Neilsen}, {Nord}, {Ogando}, {Plazas}, {Reichardt}, {Romer}, {Rozo},
  {Rykoff}, {Sanchez}, {Santiago}, {Schubnell}, {Sevilla-Noarbe}, {Smith},
  {Soares-Santos}, {Sobreira}, {Stalder}, {Stanford}, {Suchyta}, {Swanson},
  {Tarle}, {Thomas}, {Vikram}, {Walker}, \& {Zhang}}]{Hennig17}
{Hennig}, C., {Mohr}, J.~J., {Zenteno}, A., {et~al.} 2017,
  \href{http://dx.doi.org/10.1093/mnras/stx175}{\color{blue}\mnras},
  \href{https://ui.adsabs.harvard.edu/abs/2017MNRAS.467.4015H}{467, 4015}

\bibitem[{{Henson} {et~al.}(2017){Henson}, {Barnes}, {Kay}, {McCarthy}, \&
  {Schaye}}]{Henson17}
{Henson}, M.~A., {Barnes}, D.~J., {Kay}, S.~T., {McCarthy}, I.~G., \& {Schaye},
  J. 2017, \href{http://dx.doi.org/10.1093/mnras/stw2899}{\color{blue}\mnras},
  \href{https://ui.adsabs.harvard.edu/abs/2017MNRAS.465.3361H}{465, 3361}

\bibitem[{{Hildebrandt} {et~al.}(2016){Hildebrandt}, {Choi}, {Heymans},
  {Blake}, {Erben}, {Miller}, {Nakajima}, {van Waerbeke}, {Viola},
  {Buddendiek}, {Harnois-D{\'e}raps}, {Hojjati}, {Joachimi}, {Joudaki},
  {Kitching}, {Wolf}, {Gwyn}, {Johnson}, {Kuijken}, {Sheikhbahaee}, {Tudorica},
  \& {Yee}}]{Hildebrandt16}
{Hildebrandt}, H., {Choi}, A., {Heymans}, C., {et~al.} 2016,
  \href{http://dx.doi.org/10.1093/mnras/stw2013}{\color{blue}\mnras},
  \href{https://ui.adsabs.harvard.edu/abs/2016MNRAS.463..635H}{463, 635}

\bibitem[{{Hildebrandt} {et~al.}(2021){Hildebrandt}, {van den Busch}, {Wright},
  {Blake}, {Joachimi}, {Kuijken}, {Tr{\"o}ster}, {Asgari}, {Bilicki}, {de
  Jong}, {Dvornik}, {Erben}, {Getman}, {Giblin}, {Heymans}, {Kannawadi}, {Lin},
  \& {Shan}}]{Hildebrandt21}
{Hildebrandt}, H., {van den Busch}, J.~L., {Wright}, A.~H., {et~al.} 2021,
  \href{http://dx.doi.org/10.1051/0004-6361/202039018}{\color{blue}\aap},
  \href{https://ui.adsabs.harvard.edu/abs/2021A&A...647A.124H}{647, A124}

\bibitem[{{Hildebrandt} {et~al.}(2017){Hildebrandt}, {Viola}, {Heymans},
  {Joudaki}, {Kuijken}, {Blake}, {Erben}, {Joachimi}, {Klaes}, {Miller},
  {Morrison}, {Nakajima}, {Verdoes Kleijn}, {Amon}, {Choi}, {Covone}, {de
  Jong}, {Dvornik}, {Fenech Conti}, {Grado}, {Harnois-D{\'e}raps}, {Herbonnet},
  {Hoekstra}, {K{\"o}hlinger}, {McFarland}, {Mead}, {Merten}, {Napolitano},
  {Peacock}, {Radovich}, {Schneider}, {Simon}, {Valentijn}, {van den Busch},
  {van Uitert}, \& {Van Waerbeke}}]{Hildebrandt17}
{Hildebrandt}, H., {Viola}, M., {Heymans}, C., {et~al.} 2017,
  \href{http://dx.doi.org/10.1093/mnras/stw2805}{\color{blue}\mnras},
  \href{https://ui.adsabs.harvard.edu/abs/2017MNRAS.465.1454H}{465, 1454}

\bibitem[{{Hilton} {et~al.}(2021){Hilton}, {Sif{\'o}n}, {Naess},
  {Madhavacheril}, {Oguri}, {Rozo}, {Rykoff}, {Abbott}, {Adhikari}, {Aguena},
  {Aiola}, {Allam}, {Amodeo}, {Amon}, {Annis}, {Ansarinejad}, {Aros-Bunster},
  {Austermann}, {Avila}, {Bacon}, {Battaglia}, {Beall}, {Becker}, {Bernstein},
  {Bertin}, {Bhandarkar}, {Bhargava}, {Bond}, {Brooks}, {Burke}, {Calabrese},
  {Carrasco Kind}, {Carretero}, {Choi}, {Choi}, {Conselice}, {da Costa},
  {Costanzi}, {Crichton}, {Crowley}, {D{\"u}nner}, {Denison}, {Devlin},
  {Dicker}, {Diehl}, {Dietrich}, {Doel}, {Duff}, {Duivenvoorden}, {Dunkley},
  {Everett}, {Ferraro}, {Ferrero}, {Fert{\'e}}, {Flaugher}, {Frieman},
  {Gallardo}, {Garc{\'\i}a-Bellido}, {Gaztanaga}, {Gerdes}, {Giles}, {Golec},
  {Gralla}, {Grandis}, {Gruen}, {Gruendl}, {Gschwend}, {Gutierrez}, {Han},
  {Hartley}, {Hasselfield}, {Hill}, {Hilton}, {Hincks}, {Hinton}, {Ho},
  {Honscheid}, {Hoyle}, {Hubmayr}, {Huffenberger}, {Hughes}, {Jaelani}, {Jain},
  {James}, {Jeltema}, {Kent}, {Knowles}, {Koopman}, {Kuehn}, {Lahav}, {Lima},
  {Lin}, {Lokken}, {Loubser}, {MacCrann}, {Maia}, {Marriage}, {Martin},
  {McMahon}, {Melchior}, {Menanteau}, {Miquel}, {Miyatake}, {Moodley},
  {Morgan}, {Mroczkowski}, {Nati}, {Newburgh}, {Niemack}, {Nishizawa},
  {Ogando}, {Orlowski-Scherer}, {Page}, {Palmese}, {Partridge},
  {Paz-Chinch{\'o}n}, {Phakathi}, {Plazas}, {Robertson}, {Romer}, {Carnero
  Rosell}, {Salatino}, {Sanchez}, {Schaan}, {Schillaci}, {Sehgal}, {Serrano},
  {Shin}, {Simon}, {Smith}, {Soares-Santos}, {Spergel}, {Staggs}, {Storer},
  {Suchyta}, {Swanson}, {Tarle}, {Thomas}, {To}, {Trac}, {Ullom}, {Vale}, {Van
  Lanen}, {Vavagiakis}, {De Vicente}, {Wilkinson}, {Wollack}, {Xu}, \&
  {Zhang}}]{Hilton21}
{Hilton}, M., {Sif{\'o}n}, C., {Naess}, S., {et~al.} 2021,
  \href{http://dx.doi.org/10.3847/1538-4365/abd023}{\color{blue}\apjs},
  \href{https://ui.adsabs.harvard.edu/abs/2021ApJS..253....3H}{253, 3}

\bibitem[{{Ho} {et~al.}(2022){Ho}, {Ntampaka}, {Rau}, {Chen}, {Lansberry},
  {Ruehle}, \& {Trac}}]{Ho22}
{Ho}, M., {Ntampaka}, M., {Rau}, M.~M., {et~al.} 2022,
  \href{http://dx.doi.org/10.1038/s41550-022-01711-1}{\color{blue}Nature
  Astronomy}, \href{https://ui.adsabs.harvard.edu/abs/2022NatAs...6..936H}{6,
  936}

\bibitem[{{Hoekstra} {et~al.}(2012){Hoekstra}, {Mahdavi}, {Babul}, \&
  {Bildfell}}]{Hoekstra12}
{Hoekstra}, H., {Mahdavi}, A., {Babul}, A., \& {Bildfell}, C. 2012,
  \href{http://dx.doi.org/10.1111/j.1365-2966.2012.22072.x}{\color{blue}\mnras},
  \href{https://ui.adsabs.harvard.edu/abs/2012MNRAS.427.1298H}{427, 1298}

\bibitem[{{Hu} \& {Kravtsov}(2003)}]{Hu03}
{Hu}, W. \& {Kravtsov}, A.~V. 2003,
  \href{http://dx.doi.org/10.1086/345846}{\color{blue}\apj},
  \href{https://ui.adsabs.harvard.edu/abs/2003ApJ...584..702H}{584, 702}

\bibitem[{{Ingoglia} {et~al.}(2022){Ingoglia}, {Covone}, {Sereno}, {Giocoli},
  {Bardelli}, {Bellagamba}, {Castignani}, {Farrens}, {Hildebrandt}, {Joudaki},
  {Jullo}, {Lanzieri}, {Lesci}, {Marulli}, {Maturi}, {Moscardini}, {Nanni},
  {Puddu}, {Radovich}, {Roncarelli}, {Sapio}, \& {Schimd}}]{Ingoglia22}
{Ingoglia}, L., {Covone}, G., {Sereno}, M., {et~al.} 2022,
  \href{http://dx.doi.org/10.1093/mnras/stac046}{\color{blue}\mnras},
  \href{https://ui.adsabs.harvard.edu/abs/2022MNRAS.511.1484I}{511, 1484}

\bibitem[{Ishiyama {et~al.}(2021)Ishiyama, Prada, Klypin, Sinha, Metcalf,
  Jullo, Altieri, Cora, Croton, de~la Torre, Mill{\'{a}}n-Calero, Oogi, Ruedas,
  \& Vega-Mart{\'{\i}}nez}]{Ishiyama21}
Ishiyama, T., Prada, F., Klypin, A.~A., {et~al.} 2021,
  \href{http://dx.doi.org/10.1093/mnras/stab1755}{\color{blue}\mnras},
  \href{https://ui.adsabs.harvard.edu/abs/2021MNRAS.506.4210I/abstract}{506,
  4210}

\bibitem[{{Johnston} {et~al.}(2007){Johnston}, {Sheldon}, {Wechsler}, {Rozo},
  {Koester}, {Frieman}, {McKay}, {Evrard}, {Becker}, \& {Annis}}]{Johnston07}
{Johnston}, D.~E., {Sheldon}, E.~S., {Wechsler}, R.~H., {et~al.} 2007,
  \href{https://ui.adsabs.harvard.edu/abs/2007arXiv0709.1159J}{arXiv e-prints,
  arXiv:0709.1159}

\bibitem[{{Kleinebreil} {et~al.}(2025){Kleinebreil}, {Grandis}, {Schrabback},
  {Ghirardini}, {Chiu}, {Liu}, {Kluge}, {Reiprich}, {Artis}, {Bahar}, {Balzer},
  {Bulbul}, {Clerc}, {Comparat}, {Garrel}, {Gruen}, {Li}, {Miyatake},
  {Miyazaki}, {Ramos-Ceja}, {Sanders}, {Seppi}, {Okabe}, \&
  {Zhang}}]{Kleinebreil24}
{Kleinebreil}, F., {Grandis}, S., {Schrabback}, T., {et~al.} 2025,
  \href{http://dx.doi.org/10.1051/0004-6361/202449599}{\color{blue}\aap},
  \href{https://ui.adsabs.harvard.edu/abs/2025A&A...695A.216K}{695, A216}

\bibitem[{{Kodi Ramanah} {et~al.}(2020){Kodi Ramanah}, {Wojtak}, {Ansari},
  {Gall}, \& {Hjorth}}]{Kodi20}
{Kodi Ramanah}, D., {Wojtak}, R., {Ansari}, Z., {Gall}, C., \& {Hjorth}, J.
  2020, \href{http://dx.doi.org/10.1093/mnras/staa2886}{\color{blue}\mnras},
  \href{https://ui.adsabs.harvard.edu/abs/2020MNRAS.499.1985K}{499, 1985}

\bibitem[{Kohonen(1982)}]{Kohonen1982}
Kohonen, T. 1982,
  \href{http://dx.doi.org/10.1007/BF00337288}{\color{blue}Biological
  Cybernetics}, \href{https://doi.org/10.1007/BF00337288}{43, 59}

\bibitem[{{Kuijken}(2011)}]{OmegaCAM}
{Kuijken}, K. 2011, The Messenger,
  \href{https://ui.adsabs.harvard.edu/abs/2011Msngr.146....8K}{146, 8}

\bibitem[{{Kuijken} {et~al.}(2019){Kuijken}, {Heymans}, {Dvornik},
  {Hildebrandt}, {de Jong}, {Wright}, {Erben}, {Bilicki}, {Giblin}, {Shan},
  {Getman}, {Grado}, {Hoekstra}, {Miller}, {Napolitano}, {Paolilo}, {Radovich},
  {Schneider}, {Sutherland}, {Tewes}, {Tortora}, {Valentijn}, \& {Verdoes
  Kleijn}}]{Kuijken19}
{Kuijken}, K., {Heymans}, C., {Dvornik}, A., {et~al.} 2019,
  \href{http://dx.doi.org/10.1051/0004-6361/201834918}{\color{blue}\aap},
  \href{https://ui.adsabs.harvard.edu/abs/2019A&A...625A...2K}{625, A2}

\bibitem[{{Lacasa} \& {Grain}(2019)}]{Lacasa19}
{Lacasa}, F. \& {Grain}, J. 2019,
  \href{http://dx.doi.org/10.1051/0004-6361/201834343}{\color{blue}\aap},
  \href{https://ui.adsabs.harvard.edu/abs/2019A&A...624A..61L}{624, A61}

\bibitem[{{Laigle} {et~al.}(2016){Laigle}, {McCracken}, {Ilbert}, {Hsieh},
  {Davidzon}, {Capak}, {Hasinger}, {Silverman}, {Pichon}, {Coupon}, {Aussel},
  {Le Borgne}, {Caputi}, {Cassata}, {Chang}, {Civano}, {Dunlop}, {Fynbo},
  {Kartaltepe}, {Koekemoer}, {Le F{\`e}vre}, {Le Floc'h}, {Leauthaud}, {Lilly},
  {Lin}, {Marchesi}, {Milvang-Jensen}, {Salvato}, {Sanders}, {Scoville},
  {Smolcic}, {Stockmann}, {Taniguchi}, {Tasca}, {Toft}, {Vaccari}, \&
  {Zabl}}]{Laigle16}
{Laigle}, C., {McCracken}, H.~J., {Ilbert}, O., {et~al.} 2016,
  \href{http://dx.doi.org/10.3847/0067-0049/224/2/24}{\color{blue}\apjs},
  \href{https://ui.adsabs.harvard.edu/abs/2016ApJS..224...24L}{224, 24}

\bibitem[{{Lee} {et~al.}(2018){Lee}, {Le Brun}, {Haq}, {Deering}, {King},
  {Applegate}, \& {McCarthy}}]{Lee18}
{Lee}, B.~E., {Le Brun}, A.~M.~C., {Haq}, M.~E., {et~al.} 2018,
  \href{http://dx.doi.org/10.1093/mnras/sty1377}{\color{blue}\mnras},
  \href{https://ui.adsabs.harvard.edu/abs/2018MNRAS.479..890L}{479, 890}

\bibitem[{{Lesci} {et~al.}(2022{\natexlab{a}}){Lesci}, {Marulli}, {Moscardini},
  {Sereno}, {Veropalumbo}, {Maturi}, {Giocoli}, {Radovich}, {Bellagamba},
  {Roncarelli}, {Bardelli}, {Contarini}, {Covone}, {Ingoglia}, {Nanni}, \&
  {Puddu}}]{Lesci22_counts}
{Lesci}, G.~F., {Marulli}, F., {Moscardini}, L., {et~al.} 2022{\natexlab{a}},
  \href{http://dx.doi.org/10.1051/0004-6361/202040194}{\color{blue}\aap},
  \href{https://ui.adsabs.harvard.edu/abs/2022A&A...659A..88L}{659, A88}

\bibitem[{{Lesci} {et~al.}(2022{\natexlab{b}}){Lesci}, {Nanni}, {Marulli},
  {Moscardini}, {Veropalumbo}, {Maturi}, {Sereno}, {Radovich}, {Bellagamba},
  {Roncarelli}, {Bardelli}, {Castignani}, {Covone}, {Giocoli}, {Ingoglia}, \&
  {Puddu}}]{Lesci22b}
{Lesci}, G.~F., {Nanni}, L., {Marulli}, F., {et~al.} 2022{\natexlab{b}},
  \href{http://dx.doi.org/10.1051/0004-6361/202243538}{\color{blue}\aap},
  \href{https://ui.adsabs.harvard.edu/abs/2022A&A...665A.100L}{665, A100}

\bibitem[{{Lewis} \& {Challinor}(2011)}]{CAMB}
{Lewis}, A. \& {Challinor}, A.
  \href{https://ui.adsabs.harvard.edu/abs/2011ascl.soft02026L}{2011,
  ascl:1102.026}

\bibitem[{{Li} {et~al.}(2023){Li}, {Kuijken}, {Hoekstra}, {Miller}, {Heymans},
  {Hildebrandt}, {van den Busch}, {Wright}, {Yoon}, {Bilicki}, {Bravo}, \&
  {Lagos}}]{Li23}
{Li}, S.-S., {Kuijken}, K., {Hoekstra}, H., {et~al.} 2023,
  \href{http://dx.doi.org/10.1051/0004-6361/202245210}{\color{blue}\aap},
  \href{https://ui.adsabs.harvard.edu/abs/2023A&A...670A.100L}{670, A100}

\bibitem[{{Li} {et~al.}(2022){Li}, {Miyatake}, {Luo}, {More}, {Oguri},
  {Hamana}, {Mandelbaum}, {Shirasaki}, {Takada}, {Armstrong}, {Kannawadi},
  {Takita}, {Miyazaki}, {Nishizawa}, {Plazas Malagon}, {Strauss}, {Tanaka}, \&
  {Yoshida}}]{Li22}
{Li}, X., {Miyatake}, H., {Luo}, W., {et~al.} 2022,
  \href{http://dx.doi.org/10.1093/pasj/psac006}{\color{blue}\pasj},
  \href{https://ui.adsabs.harvard.edu/abs/2022PASJ...74..421L}{74, 421}

\bibitem[{{Lima} \& {Hu}(2004)}]{Lima04}
{Lima}, M. \& {Hu}, W. 2004,
  \href{http://dx.doi.org/10.1103/PhysRevD.70.043504}{\color{blue}\prd},
  \href{https://ui.adsabs.harvard.edu/abs/2004PhRvD..70d3504L}{70, 043504}

\bibitem[{{Liske} {et~al.}(2015){Liske}, {Baldry}, {Driver}, {Tuffs},
  {Alpaslan}, {Andrae}, {Brough}, {Cluver}, {Grootes}, {Gunawardhana},
  {Kelvin}, {Loveday}, {Robotham}, {Taylor}, {Bamford}, {Bland-Hawthorn},
  {Brown}, {Drinkwater}, {Hopkins}, {Meyer}, {Norberg}, {Peacock}, {Agius},
  {Andrews}, {Bauer}, {Ching}, {Colless}, {Conselice}, {Croom}, {Davies}, {De
  Propris}, {Dunne}, {Eardley}, {Ellis}, {Foster}, {Frenk}, {H{\"a}u{\ss}ler},
  {Holwerda}, {Howlett}, {Ibarra}, {Jarvis}, {Jones}, {Kafle}, {Lacey},
  {Lange}, {Lara-L{\'o}pez}, {L{\'o}pez-S{\'a}nchez}, {Maddox}, {Madore},
  {McNaught-Roberts}, {Moffett}, {Nichol}, {Owers}, {Palamara}, {Penny},
  {Phillipps}, {Pimbblet}, {Popescu}, {Prescott}, {Proctor}, {Sadler},
  {Sansom}, {Seibert}, {Sharp}, {Sutherland}, {V{\'a}zquez-Mata}, {van Kampen},
  {Wilkins}, {Williams}, \& {Wright}}]{Liske15}
{Liske}, J., {Baldry}, I.~K., {Driver}, S.~P., {et~al.} 2015,
  \href{http://dx.doi.org/10.1093/mnras/stv1436}{\color{blue}\mnras},
  \href{https://ui.adsabs.harvard.edu/abs/2015MNRAS.452.2087L}{452, 2087}

\bibitem[{{Mantz} {et~al.}(2015){Mantz}, {von der Linden}, {Allen},
  {Applegate}, {Kelly}, {Morris}, {Rapetti}, {Schmidt}, {Adhikari}, {Allen},
  {Burchat}, {Burke}, {Cataneo}, {Donovan}, {Ebeling}, {Shandera}, \&
  {Wright}}]{Mantz15}
{Mantz}, A.~B., {von der Linden}, A., {Allen}, S.~W., {et~al.} 2015,
  \href{http://dx.doi.org/10.1093/mnras/stu2096}{\color{blue}\mnras},
  \href{https://ui.adsabs.harvard.edu/abs/2015MNRAS.446.2205M}{446, 2205}

\bibitem[{{Marulli} {et~al.}(2021){Marulli}, {Veropalumbo},
  {Garc{\'\i}a-Farieta}, {Moresco}, {Moscardini}, \& {Cimatti}}]{Marulli21}
{Marulli}, F., {Veropalumbo}, A., {Garc{\'\i}a-Farieta}, J.~E., {et~al.} 2021,
  \href{http://dx.doi.org/10.3847/1538-4357/ac0e8c}{\color{blue}\apj},
  \href{https://ui.adsabs.harvard.edu/abs/2021ApJ...920...13M}{920, 13}

\bibitem[{{Marulli} {et~al.}(2016){Marulli}, {Veropalumbo}, \& {Moresco}}]{cbl}
{Marulli}, F., {Veropalumbo}, A., \& {Moresco}, M. 2016,
  \href{http://dx.doi.org/10.1016/j.ascom.2016.01.005}{\color{blue}Astronomy
  and Computing},
  \href{https://ui.adsabs.harvard.edu/abs/2016A&C....14...35M}{14, 35}

\bibitem[{{Maturi} {et~al.}(2019){Maturi}, {Bellagamba}, {Radovich},
  {Roncarelli}, {Sereno}, {Moscardini}, {Bardelli}, \& {Puddu}}]{Maturi19}
{Maturi}, M., {Bellagamba}, F., {Radovich}, M., {et~al.} 2019,
  \href{http://dx.doi.org/10.1093/mnras/stz294}{\color{blue}\mnras},
  \href{https://ui.adsabs.harvard.edu/abs/2019MNRAS.485..498M}{485, 498}

\bibitem[{{Maturi} {et~al.}(2023){Maturi}, {Finoguenov}, {Lopes}, {Gonz{\'a}lez
  Delgado}, {Dupke}, {Cypriano}, {Carrasco}, {Diego}, {Penna-Lima}, {Doubrawa},
  {V{\'\i}lchez}, {Moscardini}, {Marra}, {Bonoli},
  {Rodr{\'\i}guez-Mart{\'\i}n}, {Zitrin}, {M{\'a}rquez},
  {Hern{\'a}n-Caballero}, {Jim{\'e}nez-Teja}, {Abramo}, {Alcaniz}, {Benitez},
  {Carneiro}, {Cenarro}, {Crist{\'o}bal-Hornillos}, {Ederoclite},
  {L{\'o}pez-Sanjuan}, {Mar{\'\i}n-Franch}, {Mendes de Oliveira}, {Moles},
  {Sodr{\'e}}, {Taylor}, {Varela}, {V{\'a}zquez Rami{\'o}}, \&
  {Fern{\'a}ndez-Ontiveros}}]{Maturi23}
{Maturi}, M., {Finoguenov}, A., {Lopes}, P.~A.~A., {et~al.} 2023,
  \href{http://dx.doi.org/10.1051/0004-6361/202245323}{\color{blue}\aap},
  \href{https://ui.adsabs.harvard.edu/abs/2023A&A...678A.145M}{678, A145}

\bibitem[{{Maturi} {et~al.}(2025){Maturi}, {Radovich}, {Moscardini}, {Lesci},
  {Castignani}, {Marulli}, {Puddu}, {Romanello}, {Sereno}, {Giocoli},
  {Ingoglia}, {Bardelli}, {Giblin}, {Hildebrandt}, \&
  {Joudaki}}]{AMICOKiDS-1000}
{Maturi}, M., {Radovich}, M., {Moscardini}, L., {et~al.} 2025,
  \href{https://ui.adsabs.harvard.edu/abs/2025arXiv250714338M}{arXiv e-prints,
  arXiv:2507.14338}

\bibitem[{{McClintock} {et~al.}(2019){McClintock}, {Varga}, {Gruen}, {Rozo},
  {Rykoff}, {Shin}, {Melchior}, {DeRose}, {Seitz}, {Dietrich}, {Sheldon},
  {Zhang}, {von der Linden}, {Jeltema}, {Mantz}, {Romer}, {Allen}, {Becker},
  {Bermeo}, {Bhargava}, {Costanzi}, {Everett}, {Farahi}, {Hamaus}, {Hartley},
  {Hollowood}, {Hoyle}, {Israel}, {Li}, {MacCrann}, {Morris}, {Palmese},
  {Plazas}, {Pollina}, {Rau}, {Simet}, {Soares-Santos}, {Troxel}, {Vergara
  Cervantes}, {Wechsler}, {Zuntz}, {Abbott}, {Abdalla}, {Allam}, {Annis},
  {Avila}, {Bridle}, {Brooks}, {Burke}, {Carnero Rosell}, {Carrasco Kind},
  {Carretero}, {Castander}, {Crocce}, {Cunha}, {D'Andrea}, {da Costa}, {Davis},
  {De Vicente}, {Diehl}, {Doel}, {Drlica-Wagner}, {Evrard}, {Flaugher},
  {Fosalba}, {Frieman}, {Garc{\'\i}a-Bellido}, {Gaztanaga}, {Gerdes},
  {Giannantonio}, {Gruendl}, {Gutierrez}, {Honscheid}, {James}, {Kirk},
  {Krause}, {Kuehn}, {Lahav}, {Li}, {Lima}, {March}, {Marshall}, {Menanteau},
  {Miquel}, {Mohr}, {Nord}, {Ogando}, {Roodman}, {Sanchez}, {Scarpine},
  {Schindler}, {Sevilla-Noarbe}, {Smith}, {Smith}, {Sobreira}, {Suchyta},
  {Swanson}, {Tarle}, {Tucker}, {Vikram}, {Walker}, {Weller}, \& {DES
  Collaboration}}]{McClintock19}
{McClintock}, T., {Varga}, T.~N., {Gruen}, D., {et~al.} 2019,
  \href{http://dx.doi.org/10.1093/mnras/sty2711}{\color{blue}\mnras},
  \href{https://ui.adsabs.harvard.edu/abs/2019MNRAS.482.1352M}{482, 1352}

\bibitem[{{Medezinski} {et~al.}(2018{\natexlab{a}}){Medezinski}, {Battaglia},
  {Umetsu}, {Oguri}, {Miyatake}, {Nishizawa}, {Sif{\'o}n}, {Spergel}, {Chiu},
  {Lin}, {Bahcall}, \& {Komiyama}}]{Medezinski18}
{Medezinski}, E., {Battaglia}, N., {Umetsu}, K., {et~al.} 2018{\natexlab{a}},
  \href{http://dx.doi.org/10.1093/pasj/psx128}{\color{blue}\pasj},
  \href{https://ui.adsabs.harvard.edu/abs/2018PASJ...70S..28M}{70, S28}

\bibitem[{{Medezinski} {et~al.}(2007){Medezinski}, {Broadhurst}, {Umetsu},
  {Coe}, {Ben{\'\i}tez}, {Ford}, {Rephaeli}, {Arimoto}, \&
  {Kong}}]{Medezinski07}
{Medezinski}, E., {Broadhurst}, T., {Umetsu}, K., {et~al.} 2007,
  \href{http://dx.doi.org/10.1086/518638}{\color{blue}\apj},
  \href{https://ui.adsabs.harvard.edu/abs/2007ApJ...663..717M}{663, 717}

\bibitem[{{Medezinski} {et~al.}(2018{\natexlab{b}}){Medezinski}, {Oguri},
  {Nishizawa}, {Speagle}, {Miyatake}, {Umetsu}, {Leauthaud}, {Murata},
  {Mandelbaum}, {Sif{\'o}n}, {Strauss}, {Huang}, {Simet}, {Okabe}, {Tanaka}, \&
  {Komiyama}}]{hsc_med+al18b}
{Medezinski}, E., {Oguri}, M., {Nishizawa}, A.~J., {et~al.} 2018{\natexlab{b}},
  \href{http://dx.doi.org/10.1093/pasj/psy009}{\color{blue}\pasj},
  \href{https://ui.adsabs.harvard.edu/abs/2018PASJ...70...30M}{70, 30}

\bibitem[{{Melchior} {et~al.}(2017){Melchior}, {Gruen}, {McClintock}, {Varga},
  {Sheldon}, {Rozo}, {Amara}, {Becker}, {Benson}, {Bermeo}, {Bridle},
  {Clampitt}, {Dietrich}, {Hartley}, {Hollowood}, {Jain}, {Jarvis}, {Jeltema},
  {Kacprzak}, {MacCrann}, {Rykoff}, {Saro}, {Suchyta}, {Troxel}, {Zuntz},
  {Bonnett}, {Plazas}, {Abbott}, {Abdalla}, {Annis}, {Benoit-L{\'e}vy},
  {Bernstein}, {Bertin}, {Brooks}, {Buckley-Geer}, {Carnero Rosell}, {Carrasco
  Kind}, {Carretero}, {Cunha}, {D'Andrea}, {da Costa}, {Desai}, {Eifler},
  {Flaugher}, {Fosalba}, {Garc{\'\i}a-Bellido}, {Gaztanaga}, {Gerdes},
  {Gruendl}, {Gschwend}, {Gutierrez}, {Honscheid}, {James}, {Kirk}, {Krause},
  {Kuehn}, {Kuropatkin}, {Lahav}, {Lima}, {Maia}, {March}, {Martini},
  {Menanteau}, {Miller}, {Miquel}, {Mohr}, {Nichol}, {Ogando}, {Romer},
  {Sanchez}, {Scarpine}, {Sevilla-Noarbe}, {Smith}, {Soares-Santos},
  {Sobreira}, {Swanson}, {Tarle}, {Thomas}, {Walker}, {Weller}, \&
  {Zhang}}]{Melchior17}
{Melchior}, P., {Gruen}, D., {McClintock}, T., {et~al.} 2017,
  \href{http://dx.doi.org/10.1093/mnras/stx1053}{\color{blue}\mnras},
  \href{https://ui.adsabs.harvard.edu/abs/2017MNRAS.469.4899M}{469, 4899}

\bibitem[{{Melchior} {et~al.}(2015){Melchior}, {Suchyta}, {Huff}, {Hirsch},
  {Kacprzak}, {Rykoff}, {Gruen}, {Armstrong}, {Bacon}, {Bechtol}, {Bernstein},
  {Bridle}, {Clampitt}, {Honscheid}, {Jain}, {Jouvel}, {Krause}, {Lin},
  {MacCrann}, {Patton}, {Plazas}, {Rowe}, {Vikram}, {Wilcox}, {Young}, {Zuntz},
  {Abbott}, {Abdalla}, {Allam}, {Banerji}, {Bernstein}, {Bernstein}, {Bertin},
  {Buckley-Geer}, {Burke}, {Castander}, {da Costa}, {Cunha}, {Depoy}, {Desai},
  {Diehl}, {Doel}, {Estrada}, {Evrard}, {Neto}, {Fernandez}, {Finley},
  {Flaugher}, {Frieman}, {Gaztanaga}, {Gerdes}, {Gruendl}, {Gutierrez},
  {Jarvis}, {Karliner}, {Kent}, {Kuehn}, {Kuropatkin}, {Lahav}, {Maia},
  {Makler}, {Marriner}, {Marshall}, {Merritt}, {Miller}, {Miquel}, {Mohr},
  {Neilsen}, {Nichol}, {Nord}, {Reil}, {Roe}, {Roodman}, {Sako}, {Sanchez},
  {Santiago}, {Schindler}, {Schubnell}, {Sevilla-Noarbe}, {Sheldon}, {Smith},
  {Soares-Santos}, {Swanson}, {Sypniewski}, {Tarle}, {Thaler}, {Thomas},
  {Tucker}, {Walker}, {Wechsler}, {Weller}, \& {Wester}}]{Melchior15}
{Melchior}, P., {Suchyta}, E., {Huff}, E., {et~al.} 2015,
  \href{http://dx.doi.org/10.1093/mnras/stv398}{\color{blue}\mnras},
  \href{https://ui.adsabs.harvard.edu/abs/2015MNRAS.449.2219M}{449, 2219}

\bibitem[{{Meneghetti} {et~al.}(2014){Meneghetti}, {Rasia}, {Vega}, {Merten},
  {Postman}, {Yepes}, {Sembolini}, {Donahue}, {Ettori}, {Umetsu}, {Balestra},
  {Bartelmann}, {Ben{\'\i}tez}, {Biviano}, {Bouwens}, {Bradley}, {Broadhurst},
  {Coe}, {Czakon}, {De Petris}, {Ford}, {Giocoli}, {Gottl{\"o}ber}, {Grillo},
  {Infante}, {Jouvel}, {Kelson}, {Koekemoer}, {Lahav}, {Lemze}, {Medezinski},
  {Melchior}, {Mercurio}, {Molino}, {Moscardini}, {Monna}, {Moustakas},
  {Moustakas}, {Nonino}, {Rhodes}, {Rosati}, {Sayers}, {Seitz}, {Zheng}, \&
  {Zitrin}}]{Meneghetti14}
{Meneghetti}, M., {Rasia}, E., {Vega}, J., {et~al.} 2014,
  \href{http://dx.doi.org/10.1088/0004-637X/797/1/34}{\color{blue}\apj},
  \href{https://ui.adsabs.harvard.edu/abs/2014ApJ...797...34M}{797, 34}

\bibitem[{{Miller} {et~al.}(2013){Miller}, {Heymans}, {Kitching}, {van
  Waerbeke}, {Erben}, {Hildebrandt}, {Hoekstra}, {Mellier}, {Rowe}, {Coupon},
  {Dietrich}, {Fu}, {Harnois-D{\'e}raps}, {Hudson}, {Kilbinger}, {Kuijken},
  {Schrabback}, {Semboloni}, {Vafaei}, \& {Velander}}]{Miller13}
{Miller}, L., {Heymans}, C., {Kitching}, T.~D., {et~al.} 2013,
  \href{http://dx.doi.org/10.1093/mnras/sts454}{\color{blue}\mnras},
  \href{https://ui.adsabs.harvard.edu/abs/2013MNRAS.429.2858M}{429, 2858}

\bibitem[{{Miller} {et~al.}(2007){Miller}, {Kitching}, {Heymans}, {Heavens}, \&
  {van Waerbeke}}]{Miller07}
{Miller}, L., {Kitching}, T.~D., {Heymans}, C., {Heavens}, A.~F., \& {van
  Waerbeke}, L. 2007,
  \href{http://dx.doi.org/10.1111/j.1365-2966.2007.12363.x}{\color{blue}\mnras},
  \href{https://ui.adsabs.harvard.edu/abs/2007MNRAS.382..315M}{382, 315}

\bibitem[{{Myles} {et~al.}(2021){Myles}, {Gruen}, {Mantz}, {Allen}, {Morris},
  {Rykoff}, {Costanzi}, {To}, {DeRose}, {Wechsler}, {Rozo}, {Jeltema},
  {Carrasco}, {Kremin}, \& {Kron}}]{Myles21}
{Myles}, J., {Gruen}, D., {Mantz}, A.~B., {et~al.} 2021,
  \href{http://dx.doi.org/10.1093/mnras/stab1243}{\color{blue}\mnras},
  \href{https://ui.adsabs.harvard.edu/abs/2021MNRAS.505...33M}{505, 33}

\bibitem[{{Navarro} {et~al.}(1997){Navarro}, {Frenk}, \& {White}}]{NFW}
{Navarro}, J.~F., {Frenk}, C.~S., \& {White}, S. D.~M. 1997,
  \href{http://dx.doi.org/10.1086/304888}{\color{blue}\apj},
  \href{https://ui.adsabs.harvard.edu/abs/1997ApJ...490..493N}{490, 493}

\bibitem[{{Oguri} {et~al.}(2012){Oguri}, {Bayliss}, {Dahle}, {Sharon},
  {Gladders}, {Natarajan}, {Hennawi}, \& {Koester}}]{Oguri12}
{Oguri}, M., {Bayliss}, M.~B., {Dahle}, H., {et~al.} 2012,
  \href{http://dx.doi.org/10.1111/j.1365-2966.2011.20248.x}{\color{blue}\mnras},
  \href{https://ui.adsabs.harvard.edu/abs/2012MNRAS.420.3213O}{420, 3213}

\bibitem[{{Oguri} \& {Hamana}(2011)}]{OguriHamana11}
{Oguri}, M. \& {Hamana}, T. 2011,
  \href{http://dx.doi.org/10.1111/j.1365-2966.2011.18481.x}{\color{blue}\mnras},
  \href{https://ui.adsabs.harvard.edu/abs/2011MNRAS.414.1851O}{414, 1851}

\bibitem[{{Oguri} \& {Takada}(2011)}]{Oguri11}
{Oguri}, M. \& {Takada}, M. 2011,
  \href{http://dx.doi.org/10.1103/PhysRevD.83.023008}{\color{blue}\prd},
  \href{https://ui.adsabs.harvard.edu/abs/2011PhRvD..83b3008O}{83, 023008}

\bibitem[{{Okabe} {et~al.}(2010){Okabe}, {Zhang}, {Finoguenov}, {Takada},
  {Smith}, {Umetsu}, \& {Futamase}}]{Okabe10}
{Okabe}, N., {Zhang}, Y.~Y., {Finoguenov}, A., {et~al.} 2010,
  \href{http://dx.doi.org/10.1088/0004-637X/721/1/875}{\color{blue}\apj},
  \href{https://ui.adsabs.harvard.edu/abs/2010ApJ...721..875O}{721, 875}

\bibitem[{{Osato} {et~al.}(2018){Osato}, {Nishimichi}, {Oguri}, {Takada}, \&
  {Okumura}}]{Osato18}
{Osato}, K., {Nishimichi}, T., {Oguri}, M., {Takada}, M., \& {Okumura}, T.
  2018, \href{http://dx.doi.org/10.1093/mnras/sty762}{\color{blue}\mnras},
  \href{https://ui.adsabs.harvard.edu/abs/2018MNRAS.477.2141O}{477, 2141}

\bibitem[{{Palmese} {et~al.}(2020){Palmese}, {Annis}, {Burgad}, {Farahi},
  {Soares-Santos}, {Welch}, {da Silva Pereira}, {Lin}, {Bhargava}, {Hollowood},
  {Wilkinson}, {Giles}, {Jeltema}, {Romer}, {Evrard}, {Hilton}, {Vergara
  Cervantes}, {Bermeo}, {Mayers}, {DeRose}, {Gruen}, {Hartley}, {Lahav},
  {Leistedt}, {McClintock}, {Rozo}, {Rykoff}, {Varga}, {Wechsler}, {Zhang},
  {Avila}, {Brooks}, {Buckley-Geer}, {Burke}, {Carnero Rosell}, {Carrasco
  Kind}, {Carretero}, {Castander}, {Collins}, {da Costa}, {Desai}, {De
  Vicente}, {Diehl}, {Dietrich}, {Doel}, {Flaugher}, {Fosalba}, {Frieman},
  {Garc{\'\i}a-Bellido}, {Gerdes}, {Gruendl}, {Gschwend}, {Gutierrez},
  {Honscheid}, {James}, {Krause}, {Kuehn}, {Kuropatkin}, {Liddle}, {Lima},
  {Maia}, {Mann}, {Marshall}, {Menanteau}, {Miquel}, {Ogando}, {Plazas},
  {Roodman}, {Rooney}, {Sahlen}, {Sanchez}, {Scarpine}, {Schubnell}, {Serrano},
  {Sevilla-Noarbe}, {Sobreira}, {Stott}, {Suchyta}, {Swanson}, {Tarle},
  {Thomas}, {Tucker}, {Viana}, {Vikram}, {Walker}, \& {DES
  Collaboration}}]{Palmese20}
{Palmese}, A., {Annis}, J., {Burgad}, J., {et~al.} 2020,
  \href{http://dx.doi.org/10.1093/mnras/staa526}{\color{blue}\mnras},
  \href{https://ui.adsabs.harvard.edu/abs/2020MNRAS.493.4591P}{493, 4591}

\bibitem[{{Park} {et~al.}(2023){Park}, {Sunayama}, {Takada}, {Kobayashi},
  {Miyatake}, {More}, {Nishimichi}, \& {Sugiyama}}]{Park23}
{Park}, Y., {Sunayama}, T., {Takada}, M., {et~al.} 2023,
  \href{http://dx.doi.org/10.1093/mnras/stac3410}{\color{blue}\mnras},
  \href{https://ui.adsabs.harvard.edu/abs/2023MNRAS.518.5171P}{518, 5171}

\bibitem[{{Payerne} {et~al.}(2025){Payerne}, {Zhang}, {Aguena}, {Combet},
  {Guillemin}, {Ricci}, {Amouroux}, {Avestruz}, {Barroso}, {Farahi}, {Kovacs},
  {Murray}, {Rau}, {Rykoff}, {Schmidt}, \& {LSST Dark Energy Science
  Collaboration}}]{Payerne25}
{Payerne}, C., {Zhang}, Z., {Aguena}, M., {et~al.} 2025,
  \href{http://dx.doi.org/10.1051/0004-6361/202554107}{\color{blue}\aap},
  \href{https://ui.adsabs.harvard.edu/abs/2025A&A...700A..34P}{700, A34}

\bibitem[{{Pereira} {et~al.}(2020){Pereira}, {Palmese}, {Varga}, {McClintock},
  {Soares-Santos}, {Burgad}, {Annis}, {Farahi}, {Lin}, {Choi}, {DeRose},
  {Esteves}, {Gatti}, {Gruen}, {Hartley}, {Hoyle}, {Jeltema}, {MacCrann},
  {Roodman}, {S{\'a}nchez}, {Shin}, {von der Linden}, {Zuntz}, {Abbott},
  {Aguena}, {Avila}, {Bertin}, {Bhargava}, {Bridle}, {Brooks}, {Burke},
  {Carnero Rosell}, {Carrasco Kind}, {Carretero}, {Costanzi}, {da Costa},
  {Desai}, {Diehl}, {Dietrich}, {Doel}, {Estrada}, {Everett}, {Flaugher},
  {Fosalba}, {Frieman}, {Garc{\'\i}a-Bellido}, {Gaztanaga}, {Gerdes},
  {Gruendl}, {Gschwend}, {Gutierrez}, {Hinton}, {Hollowood}, {Honscheid},
  {James}, {Kuehn}, {Kuropatkin}, {Lahav}, {Lima}, {Maia}, {March}, {Marshall},
  {Melchior}, {Menanteau}, {Miquel}, {Ogando}, {Paz-Chinch{\'o}n}, {Plazas},
  {Romer}, {Sanchez}, {Scarpine}, {Schubnell}, {Serrano}, {Sevilla-Noarbe},
  {Smith}, {Suchyta}, {Swanson}, {Tarle}, {Wechsler}, {Weller}, {Zhang},
  {Zhang}, \& {DES Collaboration}}]{Pereira20}
{Pereira}, M.~E.~S., {Palmese}, A., {Varga}, T.~N., {et~al.} 2020,
  \href{http://dx.doi.org/10.1093/mnras/staa2687}{\color{blue}\mnras},
  \href{https://ui.adsabs.harvard.edu/abs/2020MNRAS.498.5450P}{498, 5450}

\bibitem[{{Planck Collaboration} {et~al.}(2020){Planck Collaboration},
  {Aghanim}, {Akrami}, {Ashdown}, {Aumont}, {Baccigalupi}, {Ballardini},
  {Banday}, {Barreiro}, {Bartolo}, {Basak}, {Battye}, {Benabed}, {Bernard},
  {Bersanelli}, {Bielewicz}, {Bock}, {Bond}, {Borrill}, {Bouchet}, {Boulanger},
  {Bucher}, {Burigana}, {Butler}, {Calabrese}, {Cardoso}, {Carron},
  {Challinor}, {Chiang}, {Chluba}, {Colombo}, {Combet}, {Contreras}, {Crill},
  {Cuttaia}, {de Bernardis}, {de Zotti}, {Delabrouille}, {Delouis}, {Di
  Valentino}, {Diego}, {Dor{\'e}}, {Douspis}, {Ducout}, {Dupac}, {Dusini},
  {Efstathiou}, {Elsner}, {En{\ss}lin}, {Eriksen}, {Fantaye}, {Farhang},
  {Fergusson}, {Fernandez-Cobos}, {Finelli}, {Forastieri}, {Frailis},
  {Fraisse}, {Franceschi}, {Frolov}, {Galeotta}, {Galli}, {Ganga},
  {G{\'e}nova-Santos}, {Gerbino}, {Ghosh}, {Gonz{\'a}lez-Nuevo}, {G{\'o}rski},
  {Gratton}, {Gruppuso}, {Gudmundsson}, {Hamann}, {Handley}, {Hansen},
  {Herranz}, {Hildebrandt}, {Hivon}, {Huang}, {Jaffe}, {Jones}, {Karakci},
  {Keih{\"a}nen}, {Keskitalo}, {Kiiveri}, {Kim}, {Kisner}, {Knox},
  {Krachmalnicoff}, {Kunz}, {Kurki-Suonio}, {Lagache}, {Lamarre}, {Lasenby},
  {Lattanzi}, {Lawrence}, {Le Jeune}, {Lemos}, {Lesgourgues}, {Levrier},
  {Lewis}, {Liguori}, {Lilje}, {Lilley}, {Lindholm}, {L{\'o}pez-Caniego},
  {Lubin}, {Ma}, {Mac{\'\i}as-P{\'e}rez}, {Maggio}, {Maino}, {Mandolesi},
  {Mangilli}, {Marcos-Caballero}, {Maris}, {Martin}, {Martinelli},
  {Mart{\'\i}nez-Gonz{\'a}lez}, {Matarrese}, {Mauri}, {McEwen}, {Meinhold},
  {Melchiorri}, {Mennella}, {Migliaccio}, {Millea}, {Mitra},
  {Miville-Desch{\^e}nes}, {Molinari}, {Montier}, {Morgante}, {Moss}, {Natoli},
  {N{\o}rgaard-Nielsen}, {Pagano}, {Paoletti}, {Partridge}, {Patanchon},
  {Peiris}, {Perrotta}, {Pettorino}, {Piacentini}, {Polastri}, {Polenta},
  {Puget}, {Rachen}, {Reinecke}, {Remazeilles}, {Renzi}, {Rocha}, {Rosset},
  {Roudier}, {Rubi{\~n}o-Mart{\'\i}n}, {Ruiz-Granados}, {Salvati}, {Sandri},
  {Savelainen}, {Scott}, {Shellard}, {Sirignano}, {Sirri}, {Spencer},
  {Sunyaev}, {Suur-Uski}, {Tauber}, {Tavagnacco}, {Tenti}, {Toffolatti},
  {Tomasi}, {Trombetti}, {Valenziano}, {Valiviita}, {Van Tent}, {Vibert},
  {Vielva}, {Villa}, {Vittorio}, {Wandelt}, {Wehus}, {White}, {White},
  {Zacchei}, \& {Zonca}}]{Planck18}
{Planck Collaboration}, {Aghanim}, N., {Akrami}, Y., {et~al.} 2020,
  \href{http://dx.doi.org/10.1051/0004-6361/201833910}{\color{blue}\aap},
  \href{https://ui.adsabs.harvard.edu/abs/2020A&A...641A...6P}{641, A6}

\bibitem[{{Planck Collaboration XX}(2014)}]{PlanckXX2013}
{Planck Collaboration XX}. 2014,
  \href{http://dx.doi.org/10.1051/0004-6361/201321521}{\color{blue}\aap},
  \href{https://ui.adsabs.harvard.edu/abs/2014A&A...571A..20P}{571, A20}

\bibitem[{{Planck Collaboration XXIV}(2016)}]{Planck_counts}
{Planck Collaboration XXIV}. 2016,
  \href{http://dx.doi.org/10.1051/0004-6361/201525833}{\color{blue}\aap},
  \href{https://ui.adsabs.harvard.edu/abs/2016A&A...594A..24P}{594, A24}

\bibitem[{{Pratt} {et~al.}(2019){Pratt}, {Arnaud}, {Biviano}, {Eckert},
  {Ettori}, {Nagai}, {Okabe}, \& {Reiprich}}]{Pratt19}
{Pratt}, G.~W., {Arnaud}, M., {Biviano}, A., {et~al.} 2019,
  \href{http://dx.doi.org/10.1007/s11214-019-0591-0}{\color{blue}\ssr},
  \href{https://ui.adsabs.harvard.edu/abs/2019SSRv..215...25P}{215, 25}

\bibitem[{{Puddu} {et~al.}(2021){Puddu}, {Radovich}, {Sereno}, {Bardelli},
  {Maturi}, {Moscardini}, {Bellagamba}, {Giocoli}, {Marulli}, \&
  {Roncarelli}}]{Puddu21}
{Puddu}, E., {Radovich}, M., {Sereno}, M., {et~al.} 2021,
  \href{http://dx.doi.org/10.1051/0004-6361/202039259}{\color{blue}\aap},
  \href{https://ui.adsabs.harvard.edu/abs/2021A&A...645A...9P}{645, A9}

\bibitem[{{Radovich} {et~al.}(2020){Radovich}, {Tortora}, {Bellagamba},
  {Maturi}, {Moscardini}, {Puddu}, {Roncarelli}, {Roy}, {Bardelli}, {Marulli},
  {Sereno}, {Getman}, \& {Napolitano}}]{Radovich20}
{Radovich}, M., {Tortora}, C., {Bellagamba}, F., {et~al.} 2020,
  \href{http://dx.doi.org/10.1093/mnras/staa2705}{\color{blue}\mnras},
  \href{https://ui.adsabs.harvard.edu/abs/2020MNRAS.498.4303R}{498, 4303}

\bibitem[{{Ragagnin} {et~al.}(2021){Ragagnin}, {Saro}, {Singh}, \&
  {Dolag}}]{Ragagnin21}
{Ragagnin}, A., {Saro}, A., {Singh}, P., \& {Dolag}, K. 2021,
  \href{http://dx.doi.org/10.1093/mnras/staa3523}{\color{blue}\mnras},
  \href{https://ui.adsabs.harvard.edu/abs/2021MNRAS.500.5056R}{500, 5056}

\bibitem[{{Rau} {et~al.}(2024){Rau}, {K{\'e}ruzor{\'e}}, {Ramachandra}, \&
  {Bleem}}]{Rau24}
{Rau}, M.~M., {K{\'e}ruzor{\'e}}, F., {Ramachandra}, N., \& {Bleem}, L. 2024,
  \href{https://ui.adsabs.harvard.edu/abs/2024arXiv240611950R}{arXiv e-prints,
  arXiv:2406.11950}

\bibitem[{{Romanello} {et~al.}(2024){Romanello}, {Marulli}, {Moscardini},
  {Lesci}, {Sartoris}, {Contarini}, {Giocoli}, {Bardelli}, {Busillo},
  {Castignani}, {Covone}, {Ingoglia}, {Maturi}, {Puddu}, {Radovich},
  {Roncarelli}, \& {Sereno}}]{Romanello24}
{Romanello}, M., {Marulli}, F., {Moscardini}, L., {et~al.} 2024,
  \href{http://dx.doi.org/10.1051/0004-6361/202348305}{\color{blue}\aap},
  \href{https://ui.adsabs.harvard.edu/abs/2024A&A...682A..72R}{682, A72}

\bibitem[{{Rykoff} {et~al.}(2014){Rykoff}, {Rozo}, {Busha}, {Cunha},
  {Finoguenov}, {Evrard}, {Hao}, {Koester}, {Leauthaud}, {Nord}, {Pierre},
  {Reddick}, {Sadibekova}, {Sheldon}, \& {Wechsler}}]{Rykoff14}
{Rykoff}, E.~S., {Rozo}, E., {Busha}, M.~T., {et~al.} 2014,
  \href{http://dx.doi.org/10.1088/0004-637X/785/2/104}{\color{blue}\apj},
  \href{https://ui.adsabs.harvard.edu/abs/2014ApJ...785..104R}{785, 104}

\bibitem[{{Rykoff} {et~al.}(2016){Rykoff}, {Rozo}, {Hollowood},
  {Bermeo-Hernandez}, {Jeltema}, {Mayers}, {Romer}, {Rooney}, {Saro}, {Vergara
  Cervantes}, {Wechsler}, {Wilcox}, {Abbott}, {Abdalla}, {Allam}, {Annis},
  {Benoit-L{\'e}vy}, {Bernstein}, {Bertin}, {Brooks}, {Burke}, {Capozzi},
  {Carnero Rosell}, {Carrasco Kind}, {Castander}, {Childress}, {Collins},
  {Cunha}, {D'Andrea}, {da Costa}, {Davis}, {Desai}, {Diehl}, {Dietrich},
  {Doel}, {Evrard}, {Finley}, {Flaugher}, {Fosalba}, {Frieman}, {Glazebrook},
  {Goldstein}, {Gruen}, {Gruendl}, {Gutierrez}, {Hilton}, {Honscheid}, {Hoyle},
  {James}, {Kay}, {Kuehn}, {Kuropatkin}, {Lahav}, {Lewis}, {Lidman}, {Lima},
  {Maia}, {Mann}, {Marshall}, {Martini}, {Melchior}, {Miller}, {Miquel},
  {Mohr}, {Nichol}, {Nord}, {Ogando}, {Plazas}, {Reil}, {Sahl{\'e}n},
  {Sanchez}, {Santiago}, {Scarpine}, {Schubnell}, {Sevilla-Noarbe}, {Smith},
  {Soares-Santos}, {Sobreira}, {Stott}, {Suchyta}, {Swanson}, {Tarle},
  {Thomas}, {Tucker}, {Uddin}, {Viana}, {Vikram}, {Walker}, {Zhang}, \& {DES
  Collaboration}}]{Rykoff16}
{Rykoff}, E.~S., {Rozo}, E., {Hollowood}, D., {et~al.} 2016,
  \href{http://dx.doi.org/10.3847/0067-0049/224/1/1}{\color{blue}\apjs},
  \href{https://ui.adsabs.harvard.edu/abs/2016ApJS..224....1R}{224, 1}

\bibitem[{{Saro} {et~al.}(2015){Saro}, {Bocquet}, {Rozo}, {Benson}, {Mohr},
  {Rykoff}, {Soares-Santos}, {Bleem}, {Dodelson}, {Melchior}, {Sobreira},
  {Upadhyay}, {Weller}, {Abbott}, {Abdalla}, {Allam}, {Armstrong}, {Banerji},
  {Bauer}, {Bayliss}, {Benoit-L{\'e}vy}, {Bernstein}, {Bertin}, {Brodwin},
  {Brooks}, {Buckley-Geer}, {Burke}, {Carlstrom}, {Capasso}, {Capozzi},
  {Carnero Rosell}, {Carrasco Kind}, {Chiu}, {Covarrubias}, {Crawford},
  {Crocce}, {D'Andrea}, {da Costa}, {DePoy}, {Desai}, {de Haan}, {Diehl},
  {Dietrich}, {Doel}, {Cunha}, {Eifler}, {Evrard}, {Fausti Neto}, {Fernandez},
  {Flaugher}, {Fosalba}, {Frieman}, {Gangkofner}, {Gaztanaga}, {Gerdes},
  {Gruen}, {Gruendl}, {Gupta}, {Hennig}, {Holzapfel}, {Honscheid}, {Jain},
  {James}, {Kuehn}, {Kuropatkin}, {Lahav}, {Li}, {Lin}, {Maia}, {March},
  {Marshall}, {Martini}, {McDonald}, {Miller}, {Miquel}, {Nord}, {Ogando},
  {Plazas}, {Reichardt}, {Romer}, {Roodman}, {Sako}, {Sanchez}, {Schubnell},
  {Sevilla}, {Smith}, {Stalder}, {Stark}, {Strazzullo}, {Suchyta}, {Swanson},
  {Tarle}, {Thaler}, {Thomas}, {Tucker}, {Vikram}, {von der Linden}, {Walker},
  {Wechsler}, {Wester}, {Zenteno}, \& {Ziegler}}]{Saro15}
{Saro}, A., {Bocquet}, S., {Rozo}, E., {et~al.} 2015,
  \href{http://dx.doi.org/10.1093/mnras/stv2141}{\color{blue}\mnras},
  \href{https://ui.adsabs.harvard.edu/abs/2015MNRAS.454.2305S}{454, 2305}

\bibitem[{{Sartoris} {et~al.}(2016){Sartoris}, {Biviano}, {Fedeli}, {Bartlett},
  {Borgani}, {Costanzi}, {Giocoli}, {Moscardini}, {Weller}, {Ascaso},
  {Bardelli}, {Maurogordato}, \& {Viana}}]{Sartoris16}
{Sartoris}, B., {Biviano}, A., {Fedeli}, C., {et~al.} 2016,
  \href{http://dx.doi.org/10.1093/mnras/stw630}{\color{blue}\mnras},
  \href{https://ui.adsabs.harvard.edu/abs/2016MNRAS.459.1764S}{459, 1764}

\bibitem[{{Schaller} {et~al.}(2015){Schaller}, {Frenk}, {Bower}, {Theuns},
  {Jenkins}, {Schaye}, {Crain}, {Furlong}, {Dalla Vecchia}, \&
  {McCarthy}}]{Schaller15}
{Schaller}, M., {Frenk}, C.~S., {Bower}, R.~G., {et~al.} 2015,
  \href{http://dx.doi.org/10.1093/mnras/stv1067}{\color{blue}\mnras},
  \href{https://ui.adsabs.harvard.edu/abs/2015MNRAS.451.1247S}{451, 1247}

\bibitem[{{Schechter}(1976)}]{Schechter1976}
{Schechter}, P. 1976,
  \href{http://dx.doi.org/10.1086/154079}{\color{blue}\apj},
  \href{https://ui.adsabs.harvard.edu/abs/1976ApJ...203..297S}{203, 297}

\bibitem[{{Scheck} {et~al.}(2023){Scheck}, {Sanders}, {Biffi}, {Dolag},
  {Bulbul}, \& {Liu}}]{Scheck23}
{Scheck}, D., {Sanders}, J.~S., {Biffi}, V., {et~al.} 2023,
  \href{http://dx.doi.org/10.1051/0004-6361/202244582}{\color{blue}\aap},
  \href{https://ui.adsabs.harvard.edu/abs/2023A&A...670A..33S}{670, A33}

\bibitem[{{Schneider} \& {Seitz}(1995)}]{Schneider1995}
{Schneider}, P. \& {Seitz}, C. 1995,
  \href{http://dx.doi.org/10.48550/arXiv.astro-ph/9407032}{\color{blue}\aap},
  \href{https://ui.adsabs.harvard.edu/abs/1995A&A...294..411S}{294, 411}

\bibitem[{{Schrabback} {et~al.}(2018){Schrabback}, {Applegate}, {Dietrich},
  {Hoekstra}, {Bocquet}, {Gonzalez}, {von der Linden}, {McDonald}, {Morrison},
  {Raihan}, {Allen}, {Bayliss}, {Benson}, {Bleem}, {Chiu}, {Desai}, {Foley},
  {de Haan}, {High}, {Hilbert}, {Mantz}, {Massey}, {Mohr}, {Reichardt}, {Saro},
  {Simon}, {Stern}, {Stubbs}, \& {Zenteno}}]{Schrabback18}
{Schrabback}, T., {Applegate}, D., {Dietrich}, J.~P., {et~al.} 2018,
  \href{http://dx.doi.org/10.1093/mnras/stx2666}{\color{blue}\mnras},
  \href{https://ui.adsabs.harvard.edu/abs/2018MNRAS.474.2635S}{474, 2635}

\bibitem[{{Schrabback} {et~al.}(2021){Schrabback}, {Bocquet}, {Sommer},
  {Zohren}, {van den Busch}, {Hern{\'a}ndez-Mart{\'\i}n}, {Hoekstra}, {Raihan},
  {Schirmer}, {Applegate}, {Bayliss}, {Benson}, {Bleem}, {Dietrich}, {Floyd},
  {Hilbert}, {Hlavacek-Larrondo}, {McDonald}, {Saro}, {Stark}, \&
  {Weissgerber}}]{Schrabback21}
{Schrabback}, T., {Bocquet}, S., {Sommer}, M., {et~al.} 2021,
  \href{http://dx.doi.org/10.1093/mnras/stab1386}{\color{blue}\mnras},
  \href{https://ui.adsabs.harvard.edu/abs/2021MNRAS.505.3923S}{505, 3923}

\bibitem[{{Secco} {et~al.}(2022){Secco}, {Samuroff}, {Krause}, {Jain},
  {Blazek}, {Raveri}, {Campos}, {Amon}, {Chen}, {Doux}, {Choi}, {Gruen},
  {Bernstein}, {Chang}, {DeRose}, {Myles}, {Fert{\'e}}, {Lemos}, {Huterer},
  {Prat}, {Troxel}, {MacCrann}, {Liddle}, {Kacprzak}, {Fang}, {S{\'a}nchez},
  {Pandey}, {Dodelson}, {Chintalapati}, {Hoffmann}, {Alarcon}, {Alves},
  {Andrade-Oliveira}, {Baxter}, {Bechtol}, {Becker}, {Brandao-Souza},
  {Camacho}, {Carnero Rosell}, {Carrasco Kind}, {Cawthon}, {Cordero}, {Crocce},
  {Davis}, {Di Valentino}, {Drlica-Wagner}, {Eckert}, {Eifler}, {Elidaiana},
  {Elsner}, {Elvin-Poole}, {Everett}, {Fosalba}, {Friedrich}, {Gatti},
  {Giannini}, {Gruendl}, {Harrison}, {Hartley}, {Herner}, {Huang}, {Huff},
  {Jarvis}, {Jeffrey}, {Kuropatkin}, {Leget}, {Muir}, {Mccullough}, {Navarro
  Alsina}, {Omori}, {Park}, {Porredon}, {Rollins}, {Roodman}, {Rosenfeld},
  {Ross}, {Rykoff}, {Sanchez}, {Sevilla-Noarbe}, {Sheldon}, {Shin}, {Troja},
  {Tutusaus}, {Varga}, {Weaverdyck}, {Wechsler}, {Yanny}, {Yin}, {Zhang},
  {Zuntz}, {Abbott}, {Aguena}, {Allam}, {Annis}, {Bacon}, {Bertin}, {Bhargava},
  {Bridle}, {Brooks}, {Buckley-Geer}, {Burke}, {Carretero}, {Costanzi}, {da
  Costa}, {De Vicente}, {Diehl}, {Dietrich}, {Doel}, {Ferrero}, {Flaugher},
  {Frieman}, {Garc{\'\i}a-Bellido}, {Gaztanaga}, {Gerdes}, {Giannantonio},
  {Gschwend}, {Gutierrez}, {Hinton}, {Hollowood}, {Honscheid}, {Hoyle},
  {James}, {Jeltema}, {Kuehn}, {Lahav}, {Lima}, {Lin}, {Maia}, {Marshall},
  {Martini}, {Melchior}, {Menanteau}, {Miquel}, {Mohr}, {Morgan}, {Ogando},
  {Palmese}, {Paz-Chinch{\'o}n}, {Petravick}, {Pieres}, {Plazas Malag{\'o}n},
  {Rodriguez-Monroy}, {Romer}, {Sanchez}, {Scarpine}, {Schubnell}, {Scolnic},
  {Serrano}, {Smith}, {Soares-Santos}, {Suchyta}, {Swanson}, {Tarle}, {Thomas},
  {To}, \& {DES Collaboration}}]{Secco22}
{Secco}, L.~F., {Samuroff}, S., {Krause}, E., {et~al.} 2022,
  \href{http://dx.doi.org/10.1103/PhysRevD.105.023515}{\color{blue}\prd},
  \href{https://ui.adsabs.harvard.edu/abs/2022PhRvD.105b3515S}{105, 023515}

\bibitem[{{Seitz} \& {Schneider}(1997)}]{Seitz97}
{Seitz}, C. \& {Schneider}, P. 1997,
  \href{http://dx.doi.org/10.48550/arXiv.astro-ph/9601079}{\color{blue}\aap},
  \href{https://ui.adsabs.harvard.edu/abs/1997A&A...318..687S}{318, 687}

\bibitem[{{Seppi} {et~al.}(2024){Seppi}, {Comparat}, {Ghirardini}, {Garrel},
  {Artis}, {S{\'a}nchez}, {Liu}, {Clerc}, {Bulbul}, {Grandis}, {Kluge},
  {Reiprich}, {Merloni}, {Zhang}, {Bahar}, {Shreeram}, {Sanders}, {Ramos-Ceja},
  \& {Krumpe}}]{Seppi24}
{Seppi}, R., {Comparat}, J., {Ghirardini}, V., {et~al.} 2024,
  \href{http://dx.doi.org/10.1051/0004-6361/202348843}{\color{blue}\aap},
  \href{https://ui.adsabs.harvard.edu/abs/2024A&A...686A.196S}{686, A196}

\bibitem[{{Seppi} {et~al.}(2023){Seppi}, {Comparat}, {Nandra}, {Dolag},
  {Biffi}, {Bulbul}, {Liu}, {Ghirardini}, \& {Ider-Chitham}}]{Seppi23}
{Seppi}, R., {Comparat}, J., {Nandra}, K., {et~al.} 2023,
  \href{http://dx.doi.org/10.1051/0004-6361/202245138}{\color{blue}\aap},
  \href{https://ui.adsabs.harvard.edu/abs/2023A&A...671A..57S}{671, A57}

\bibitem[{{Sereno}(2015)}]{Sereno15_Comalit3}
{Sereno}, M. 2015,
  \href{http://dx.doi.org/10.1093/mnras/stu2505}{\color{blue}\mnras},
  \href{https://ui.adsabs.harvard.edu/abs/2015MNRAS.450.3665S}{450, 3665}

\bibitem[{{Sereno} {et~al.}(2017){Sereno}, {Covone}, {Izzo}, {Ettori},
  {Coupon}, \& {Lieu}}]{Sereno17}
{Sereno}, M., {Covone}, G., {Izzo}, L., {et~al.} 2017,
  \href{http://dx.doi.org/10.1093/mnras/stx2085}{\color{blue}\mnras},
  \href{https://ui.adsabs.harvard.edu/abs/2017MNRAS.472.1946S}{472, 1946}

\bibitem[{{Sereno} \& {Ettori}(2015)}]{Sereno15}
{Sereno}, M. \& {Ettori}, M. 2015,
  \href{http://dx.doi.org/10.1093/mnras/stv814}{\color{blue}\mnras},
  \href{https://ui.adsabs.harvard.edu/abs/2015MNRAS.450.3675S}{450, 3675}

\bibitem[{{Sereno} {et~al.}(2020){Sereno}, {Ettori}, {Lesci}, {Marulli},
  {Maturi}, {Moscardini}, {Radovich}, {Bellagamba}, \& {Roncarelli}}]{Sereno20}
{Sereno}, M., {Ettori}, S., {Lesci}, G.~F., {et~al.} 2020,
  \href{http://dx.doi.org/10.1093/mnras/staa1902}{\color{blue}\mnras},
  \href{https://ui.adsabs.harvard.edu/abs/2020MNRAS.497..894S}{497, 894}

\bibitem[{{Sereno} {et~al.}(2018){Sereno}, {Giocoli}, {Izzo}, {Marulli},
  {Veropalumbo}, {Ettori}, {Moscardini}, {Covone}, {Ferragamo}, {Barrena}, \&
  {Streblyanska}}]{Sereno18}
{Sereno}, M., {Giocoli}, C., {Izzo}, L., {et~al.} 2018,
  \href{http://dx.doi.org/10.1038/s41550-018-0508-y}{\color{blue}Nature
  Astronomy}, \href{https://ui.adsabs.harvard.edu/abs/2018NatAs...2..744S}{2,
  744}

\bibitem[{{Sereno} {et~al.}(2021){Sereno}, {Lovisari}, {Cui}, \&
  {Schellenberger}}]{Sereno21}
{Sereno}, M., {Lovisari}, L., {Cui}, W., \& {Schellenberger}, G. 2021,
  \href{http://dx.doi.org/10.1093/mnras/stab2435}{\color{blue}\mnras},
  \href{https://ui.adsabs.harvard.edu/abs/2021MNRAS.507.5214S}{507, 5214}

\bibitem[{{Sereno} {et~al.}(2025){Sereno}, {Maurogordato}, {Cappi}, {Barrena},
  {Benoist}, {Haines}, {Radovich}, {Nonino}, {Ettori}, {Ferragamo}, {Gavazzi},
  {Huot}, {Pizzuti}, {Pratt}, {Streblyanska}, {Zarattini}, {Castignani},
  {Eckert}, {Gastaldello}, {Kay}, {Lovisari}, {Maughan}, {Pointecouteau},
  {Rasia}, {Rossetti}, \& {Sayers}}]{Sereno24_2}
{Sereno}, M., {Maurogordato}, S., {Cappi}, A., {et~al.} 2025,
  \href{http://dx.doi.org/10.1051/0004-6361/202451610}{\color{blue}\aap},
  \href{https://ui.adsabs.harvard.edu/abs/2025A&A...693A...2S}{693, A2}

\bibitem[{{Shao} \& {Anbajagane}(2024)}]{Shao24}
{Shao}, M. \& {Anbajagane}, D. 2024,
  \href{http://dx.doi.org/10.33232/001c.116969}{\color{blue}The Open Journal of
  Astrophysics},
  \href{https://ui.adsabs.harvard.edu/abs/2024OJAp....7E..29S}{7, 29}

\bibitem[{{Sheldon} {et~al.}(2004){Sheldon}, {Johnston}, {Frieman}, {Scranton},
  {McKay}, {Connolly}, {Budav{\'a}ri}, {Zehavi}, {Bahcall}, {Brinkmann}, \&
  {Fukugita}}]{Sheldon04}
{Sheldon}, E.~S., {Johnston}, D.~E., {Frieman}, J.~A., {et~al.} 2004,
  \href{http://dx.doi.org/10.1086/383293}{\color{blue}\aj},
  \href{https://ui.adsabs.harvard.edu/abs/2004AJ....127.2544S}{127, 2544}

\bibitem[{{Sheth} {et~al.}(2001){Sheth}, {Mo}, \& {Tormen}}]{Sheth01}
{Sheth}, R.~K., {Mo}, H.~J., \& {Tormen}, G. 2001,
  \href{http://dx.doi.org/10.1046/j.1365-8711.2001.04006.x}{\color{blue}\mnras},
  \href{https://ui.adsabs.harvard.edu/abs/2001MNRAS.323....1S}{323, 1}

\bibitem[{{Sheth} \& {Tormen}(1999)}]{Sheth1999}
{Sheth}, R.~K. \& {Tormen}, G. 1999,
  \href{http://dx.doi.org/10.1046/j.1365-8711.1999.02692.x}{\color{blue}\mnras},
  \href{https://ui.adsabs.harvard.edu/abs/1999MNRAS.308..119S}{308, 119}

\bibitem[{{Shirasaki} {et~al.}(2018){Shirasaki}, {Lau}, \&
  {Nagai}}]{Shirasaki18}
{Shirasaki}, M., {Lau}, E.~T., \& {Nagai}, D. 2018,
  \href{http://dx.doi.org/10.1093/mnras/sty763}{\color{blue}\mnras},
  \href{https://ui.adsabs.harvard.edu/abs/2018MNRAS.477.2804S}{477, 2804}

\bibitem[{{Shirasaki} \& {Takada}(2018)}]{ShirasakiTakada18}
{Shirasaki}, M. \& {Takada}, M. 2018,
  \href{http://dx.doi.org/10.1093/mnras/sty1327}{\color{blue}\mnras},
  \href{https://ui.adsabs.harvard.edu/abs/2018MNRAS.478.4277S}{478, 4277}

\bibitem[{{Simet} {et~al.}(2017){Simet}, {McClintock}, {Mandelbaum}, {Rozo},
  {Rykoff}, {Sheldon}, \& {Wechsler}}]{Simet17}
{Simet}, M., {McClintock}, T., {Mandelbaum}, R., {et~al.} 2017,
  \href{http://dx.doi.org/10.1093/mnras/stw3250}{\color{blue}\mnras},
  \href{https://ui.adsabs.harvard.edu/abs/2017MNRAS.466.3103S}{466, 3103}

\bibitem[{{Singh} {et~al.}(2025){Singh}, {Mohr}, {Davies}, {Bocquet},
  {Grandis}, {Klein}, {Marshall}, {Aguena}, {Allam}, {Alves},
  {Andrade-Oliveira}, {Bacon}, {Bhargava}, {Brooks}, {Carnero Rosell},
  {Carretero}, {Costanzi}, {da Costa}, {Pereira}, {Desai}, {Diehl}, {Doel},
  {Everett}, {Flaugher}, {Frieman}, {Garc{\'\i}a-Bellido}, {Gaztanaga},
  {Gruendl}, {Gutierrez}, {Hollowood}, {Honscheid}, {James}, {Kuehn}, {Lima},
  {Mena-Fern{\'a}ndez}, {Menanteau}, {Miquel}, {Myles}, {Pieres}, {Romer},
  {Samuroff}, {Sanchez}, {Sanchez Cid}, {Sevilla-Noarbe}, {Smith}, {Suchyta},
  {Swanson}, {Tarle}, {To}, {Tucker}, {Vikram}, {Weaverdyck}, \&
  {Wiseman}}]{Singh24}
{Singh}, A., {Mohr}, J.~J., {Davies}, C.~T., {et~al.} 2025,
  \href{http://dx.doi.org/10.1051/0004-6361/202451516}{\color{blue}\aap},
  \href{https://ui.adsabs.harvard.edu/abs/2025A&A...695A..49S}{695, A49}

\bibitem[{{Singh} {et~al.}(2017){Singh}, {Mandelbaum}, {Seljak}, {Slosar}, \&
  {Vazquez Gonzalez}}]{Singh17}
{Singh}, S., {Mandelbaum}, R., {Seljak}, U., {Slosar}, A., \& {Vazquez
  Gonzalez}, J. 2017,
  \href{http://dx.doi.org/10.1093/mnras/stx1828}{\color{blue}\mnras},
  \href{https://ui.adsabs.harvard.edu/abs/2017MNRAS.471.3827S}{471, 3827}

\bibitem[{{Sommer} {et~al.}(2024){Sommer}, {Schrabback}, {Ragagnin}, \&
  {Rockenfeller}}]{Sommer23}
{Sommer}, M.~W., {Schrabback}, T., {Ragagnin}, A., \& {Rockenfeller}, R. 2024,
  \href{http://dx.doi.org/10.1093/mnras/stae1580}{\color{blue}\mnras},
  \href{https://ui.adsabs.harvard.edu/abs/2024MNRAS.532.3359S}{532, 3359}

\bibitem[{{Stern} {et~al.}(2019){Stern}, {Dietrich}, {Bocquet}, {Applegate},
  {Mohr}, {Bridle}, {Carrasco Kind}, {Gruen}, {Jarvis}, {Kacprzak}, {Saro},
  {Sheldon}, {Troxel}, {Zuntz}, {Benson}, {Capasso}, {Chiu}, {Desai},
  {Rapetti}, {Reichardt}, {Saliwanchik}, {Schrabback}, {Gupta}, {Abbott},
  {Abdalla}, {Avila}, {Bertin}, {Brooks}, {Burke}, {Carnero Rosell},
  {Carretero}, {Castander}, {D'Andrea}, {da Costa}, {Davis}, {De Vicente},
  {Diehl}, {Doel}, {Estrada}, {Evrard}, {Flaugher}, {Fosalba}, {Frieman},
  {Garc{\'\i}a-Bellido}, {Gaztanaga}, {Gruendl}, {Gschwend}, {Gutierrez},
  {Hollowood}, {Jeltema}, {Kirk}, {Kuehn}, {Kuropatkin}, {Lahav}, {Lima},
  {Maia}, {March}, {Melchior}, {Menanteau}, {Miquel}, {Plazas}, {Romer},
  {Sanchez}, {Schindler}, {Schubnell}, {Sevilla-Noarbe}, {Smith}, {Smith},
  {Sobreira}, {Suchyta}, {Swanson}, {Tarle}, {Walker}, {DES Collaboration}, \&
  {SPT Collaboration}}]{Stern19}
{Stern}, C., {Dietrich}, J.~P., {Bocquet}, S., {et~al.} 2019,
  \href{http://dx.doi.org/10.1093/mnras/stz234}{\color{blue}\mnras},
  \href{https://ui.adsabs.harvard.edu/abs/2019MNRAS.485...69S}{485, 69}

\bibitem[{{St\"olzner} {et~al.}(2025){St\"olzner}, {Wright}, {Asgari},
  {Heymans}, {Hildebrandt}, {Hoekstra}, {Joachimi}, {Kuijken}, {Li}, {Mahony},
  {Reischke}, {Yoon}, {Bilicki}, {Burger}, {Chisari}, {Dvornik}, {Georgiou},
  {Giblin}, {Harnois-D\textbackslash'eraps}, {Jalan}, {William}, {Joudaki},
  {Lesci}, {Linke}, {Loureiro}, {Maturi}, {Moscardini}, {Napolitano}, {Porth},
  {Radovich}, {Tr\textbackslash''oster}, {von Wietersheim-Kramsta}, {Wittje},
  {Yan}, \& {Zhang}}]{Stolzner_DR5}
{St\"olzner}, B., {Wright}, A.~H., {Asgari}, M., {et~al.} 2025,
  \href{https://ui.adsabs.harvard.edu/abs/2025arXiv250319442S}{arXiv e-prints,
  arXiv:2503.19442}

\bibitem[{{Sunayama}(2023)}]{Sunayama23}
{Sunayama}, T. 2023,
  \href{http://dx.doi.org/10.1093/mnras/stad786}{\color{blue}\mnras},
  \href{https://ui.adsabs.harvard.edu/abs/2023MNRAS.521.5064S}{521, 5064}

\bibitem[{{Sunayama} {et~al.}(2020){Sunayama}, {Park}, {Takada}, {Kobayashi},
  {Nishimichi}, {Kurita}, {More}, {Oguri}, \& {Osato}}]{Sunayama20}
{Sunayama}, T., {Park}, Y., {Takada}, M., {et~al.} 2020,
  \href{http://dx.doi.org/10.1093/mnras/staa1646}{\color{blue}\mnras},
  \href{https://ui.adsabs.harvard.edu/abs/2020MNRAS.496.4468S}{496, 4468}

\bibitem[{{Sutherland} {et~al.}(2015){Sutherland}, {Emerson}, {Dalton},
  {Atad-Ettedgui}, {Beard}, {Bennett}, {Bezawada}, {Born}, {Caldwell}, {Clark},
  {Craig}, {Henry}, {Jeffers}, {Little}, {McPherson}, {Murray}, {Stewart},
  {Stobie}, {Terrett}, {Ward}, {Whalley}, \& {Woodhouse}}]{VISTA}
{Sutherland}, W., {Emerson}, J., {Dalton}, G., {et~al.} 2015,
  \href{http://dx.doi.org/10.1051/0004-6361/201424973}{\color{blue}\aap},
  \href{https://ui.adsabs.harvard.edu/abs/2015A&A...575A..25S}{575, A25}

\bibitem[{{Teyssier} {et~al.}(2011){Teyssier}, {Moore}, {Martizzi}, {Dubois},
  \& {Mayer}}]{Teyssier11}
{Teyssier}, R., {Moore}, B., {Martizzi}, D., {Dubois}, Y., \& {Mayer}, L. 2011,
  \href{http://dx.doi.org/10.1111/j.1365-2966.2011.18399.x}{\color{blue}\mnras},
  \href{https://ui.adsabs.harvard.edu/abs/2011MNRAS.414..195T}{414, 195}

\bibitem[{{The Dark Energy Survey Collaboration}(2005)}]{DES05}
{The Dark Energy Survey Collaboration}. 2005,
  \href{https://ui.adsabs.harvard.edu/abs/2005astro.ph.10346T}{arXiv e-prints,
  astro}

\bibitem[{{Tinker} {et~al.}(2008){Tinker}, {Kravtsov}, {Klypin}, {Abazajian},
  {Warren}, {Yepes}, {Gottl{\"o}ber}, \& {Holz}}]{Tinker08}
{Tinker}, J., {Kravtsov}, A.~V., {Klypin}, A., {et~al.} 2008,
  \href{http://dx.doi.org/10.1086/591439}{\color{blue}\apj},
  \href{https://ui.adsabs.harvard.edu/abs/2008ApJ...688..709T}{688, 709}

\bibitem[{{Tinker} {et~al.}(2010){Tinker}, {Robertson}, {Kravtsov}, {Klypin},
  {Warren}, {Yepes}, \& {Gottl{\"o}ber}}]{Tinker10}
{Tinker}, J.~L., {Robertson}, B.~E., {Kravtsov}, A.~V., {et~al.} 2010,
  \href{http://dx.doi.org/10.1088/0004-637X/724/2/878}{\color{blue}\apj},
  \href{https://ui.adsabs.harvard.edu/abs/2010ApJ...724..878T}{724, 878}

\bibitem[{{To} {et~al.}(2025){To}, {Krause}, {Chang}, {Wu}, {Wechsler}, {Rozo},
  {Weinberg}, {Anbajagane}, {Avila}, {Blazek}, {Bocquet}, {Costanzi}, {De
  Vicente}, {Elvin-Poole}, {Fert{\'e}}, {Grandis}, {Muir}, {Porredon},
  {Samuroff}, {Sanchez}, {Sanchez Cid}, {Sevilla-Noarbe}, {Weaverdyck},
  {Abbott}, {Aguena}, {Andrade-Oliveira}, {Bacon}, {Becker}, {Brooks}, {Carnero
  Rosell}, {Carretero}, {Choi}, {da Costa}, {Pereira}, {Davis}, {Desai},
  {Doel}, {Everett}, {Frieman}, {Garc{\'\i}a-Bellido}, {Gatti}, {Gaztanaga},
  {Giannini}, {Gruen}, {Gutierrez}, {Hinton}, {Hollowood}, {Honscheid},
  {Jeltema}, {Kuehn}, {Lee}, {Marshall}, {Mena-Fern}, {Miquel}, {Mohr},
  {Myles}, {Palmese}, {Plazas Malag{\'o}n}, {Romer}, {Shin}, {Smith},
  {Suchyta}, {Tarle}, {Vikram}, {Walker}, \& {Weller}}]{To25}
{To}, C.-H., {Krause}, E., {Chang}, C., {et~al.} 2025,
  \href{https://ui.adsabs.harvard.edu/abs/2025arXiv250313631T}{\href{http://dx.doi.org/10.48550/arXiv.2503.13631}{\color{blue}arXiv
  e-prints}, arXiv:2503.13631}

\bibitem[{{Toni} {et~al.}(2025){Toni}, {Gozaliasl}, {Maturi}, {Moscardini},
  {Finoguenov}, {Castignani}, {Gentile}, {Virolainen}, {Casey}, {Kartaltepe},
  {Akins}, {Allen}, {Arango-Toro}, {Babul}, {Brinch}, {Drakos}, {Faisst},
  {Franco}, {Griffiths}, {Harish}, {Hasinger}, {Ilbert}, {Jin}, {Khostovan},
  {Koekemoer}, {Korpi-Lagg}, {Larson}, {Lertprasertpong}, {Liu}, {Magdis},
  {Massey}, {McCracken}, {McKinney}, {Paquereau}, {Rhodes}, {Robertson},
  {Sargent}, {Shuntov}, {Tanaka}, {Taamoli}, {Tempel}, {Toft}, {Vardoulaki}, \&
  {Yang}}]{Toni25}
{Toni}, G., {Gozaliasl}, G., {Maturi}, M., {et~al.} 2025,
  \href{http://dx.doi.org/10.1051/0004-6361/202553759}{\color{blue}\aap},
  \href{https://ui.adsabs.harvard.edu/abs/2025A&A...697A.197T}{697, A197}

\bibitem[{{Toni} {et~al.}(2024){Toni}, {Maturi}, {Finoguenov}, {Moscardini}, \&
  {Castignani}}]{Toni24}
{Toni}, G., {Maturi}, M., {Finoguenov}, A., {Moscardini}, L., \& {Castignani},
  G. 2024,
  \href{http://dx.doi.org/10.1051/0004-6361/202348832}{\color{blue}\aap},
  \href{https://ui.adsabs.harvard.edu/abs/2024A&A...687A..56T}{687, A56}

\bibitem[{{Tramonte} {et~al.}(2023){Tramonte}, {Ma}, {Yan}, {Maturi},
  {Castignani}, {Sereno}, {Bardelli}, {Giocoli}, {Marulli}, {Moscardini},
  {Puddu}, {Radovich}, {Van Waerbeke}, \& {Wright}}]{Tramonte23}
{Tramonte}, D., {Ma}, Y.-Z., {Yan}, Z., {et~al.} 2023,
  \href{http://dx.doi.org/10.3847/1538-4365/acbcca}{\color{blue}\apjs},
  \href{https://ui.adsabs.harvard.edu/abs/2023ApJS..265...55T}{265, 55}

\bibitem[{{Umetsu}(2020)}]{Umetsu20}
{Umetsu}, K. 2020,
  \href{http://dx.doi.org/10.1007/s00159-020-00129-w}{\color{blue}\aapr},
  \href{https://ui.adsabs.harvard.edu/abs/2020A&ARv..28....7U}{28, 7}

\bibitem[{{Umetsu} {et~al.}(2014){Umetsu}, {Medezinski}, {Nonino}, {Merten},
  {Postman}, {Meneghetti}, {Donahue}, {Czakon}, {Molino}, {Seitz}, {Gruen},
  {Lemze}, {Balestra}, {Ben{\'\i}tez}, {Biviano}, {Broadhurst}, {Ford},
  {Grillo}, {Koekemoer}, {Melchior}, {Mercurio}, {Moustakas}, {Rosati}, \&
  {Zitrin}}]{Umetsu14}
{Umetsu}, K., {Medezinski}, E., {Nonino}, M., {et~al.} 2014,
  \href{http://dx.doi.org/10.1088/0004-637X/795/2/163}{\color{blue}\apj},
  \href{https://ui.adsabs.harvard.edu/abs/2014ApJ...795..163U}{795, 163}

\bibitem[{{Umetsu} {et~al.}(2020){Umetsu}, {Sereno}, {Lieu}, {Miyatake},
  {Medezinski}, {Nishizawa}, {Giles}, {Gastaldello}, {McCarthy}, {Kilbinger},
  {Birkinshaw}, {Ettori}, {Okabe}, {Chiu}, {Coupon}, {Eckert}, {Fujita},
  {Higuchi}, {Koulouridis}, {Maughan}, {Miyazaki}, {Oguri}, {Pacaud}, {Pierre},
  {Rapetti}, \& {Smith}}]{Umetsu20b}
{Umetsu}, K., {Sereno}, M., {Lieu}, M., {et~al.} 2020,
  \href{http://dx.doi.org/10.3847/1538-4357/ab6bca}{\color{blue}\apj},
  \href{https://ui.adsabs.harvard.edu/abs/2020ApJ...890..148U}{890, 148}

\bibitem[{{van den Busch} {et~al.}(2020){van den Busch}, {Hildebrandt},
  {Wright}, {Morrison}, {Blake}, {Joachimi}, {Erben}, {Heymans}, {Kuijken}, \&
  {Taylor}}]{vandenBusch20}
{van den Busch}, J.~L., {Hildebrandt}, H., {Wright}, A.~H., {et~al.} 2020,
  \href{http://dx.doi.org/10.1051/0004-6361/202038835}{\color{blue}\aap},
  \href{https://ui.adsabs.harvard.edu/abs/2020A&A...642A.200V}{642, A200}

\bibitem[{{van den Busch} {et~al.}(2022){van den Busch}, {Wright},
  {Hildebrandt}, {Bilicki}, {Asgari}, {Joudaki}, {Blake}, {Heymans},
  {Kannawadi}, {Shan}, \& {Tr{\"o}ster}}]{vanDenBusch22}
{van den Busch}, J.~L., {Wright}, A.~H., {Hildebrandt}, H., {et~al.} 2022,
  \href{http://dx.doi.org/10.1051/0004-6361/202142083}{\color{blue}\aap},
  \href{https://ui.adsabs.harvard.edu/abs/2022A&A...664A.170V}{664, A170}

\bibitem[{{Vazza} {et~al.}(2017){Vazza}, {Jones}, {Br{\"u}ggen}, {Brunetti},
  {Gheller}, {Porter}, \& {Ryu}}]{Vazza17}
{Vazza}, F., {Jones}, T.~W., {Br{\"u}ggen}, M., {et~al.} 2017,
  \href{http://dx.doi.org/10.1093/mnras/stw2351}{\color{blue}\mnras},
  \href{https://ui.adsabs.harvard.edu/abs/2017MNRAS.464..210V}{464, 210}

\bibitem[{{Velliscig} {et~al.}(2014){Velliscig}, {van Daalen}, {Schaye},
  {McCarthy}, {Cacciato}, {Le Brun}, \& {Dalla Vecchia}}]{Velliscig14}
{Velliscig}, M., {van Daalen}, M.~P., {Schaye}, J., {et~al.} 2014,
  \href{http://dx.doi.org/10.1093/mnras/stu1044}{\color{blue}\mnras},
  \href{https://ui.adsabs.harvard.edu/abs/2014MNRAS.442.2641V}{442, 2641}

\bibitem[{{Viola} {et~al.}(2015){Viola}, {Cacciato}, {Brouwer}, {Kuijken},
  {Hoekstra}, {Norberg}, {Robotham}, {van Uitert}, {Alpaslan}, {Baldry},
  {Choi}, {de Jong}, {Driver}, {Erben}, {Grado}, {Graham}, {Heymans},
  {Hildebrandt}, {Hopkins}, {Irisarri}, {Joachimi}, {Loveday}, {Miller},
  {Nakajima}, {Schneider}, {Sif{\'o}n}, \& {Verdoes Kleijn}}]{Viola15}
{Viola}, M., {Cacciato}, M., {Brouwer}, M., {et~al.} 2015,
  \href{http://dx.doi.org/10.1093/mnras/stv1447}{\color{blue}\mnras},
  \href{https://ui.adsabs.harvard.edu/abs/2015MNRAS.452.3529V}{452, 3529}

\bibitem[{{Weaver} {et~al.}(2022){Weaver}, {Kauffmann}, {Ilbert}, {McCracken},
  {Moneti}, {Toft}, {Brammer}, {Shuntov}, {Davidzon}, {Hsieh}, {Laigle},
  {Anastasiou}, {Jespersen}, {Vinther}, {Capak}, {Casey}, {McPartland},
  {Milvang-Jensen}, {Mobasher}, {Sanders}, {Zalesky}, {Arnouts}, {Aussel},
  {Dunlop}, {Faisst}, {Franx}, {Furtak}, {Fynbo}, {Gould}, {Greve}, {Gwyn},
  {Kartaltepe}, {Kashino}, {Koekemoer}, {Kokorev}, {Le F{\`e}vre}, {Lilly},
  {Masters}, {Magdis}, {Mehta}, {Peng}, {Riechers}, {Salvato}, {Sawicki},
  {Scarlata}, {Scoville}, {Shirley}, {Silverman}, {Sneppen}, {Smolc̆i{\'c}},
  {Steinhardt}, {Stern}, {Tanaka}, {Taniguchi}, {Teplitz}, {Vaccari}, {Wang},
  \& {Zamorani}}]{Weaver22}
{Weaver}, J.~R., {Kauffmann}, O.~B., {Ilbert}, O., {et~al.} 2022,
  \href{http://dx.doi.org/10.3847/1538-4365/ac3078}{\color{blue}\apjs},
  \href{https://ui.adsabs.harvard.edu/abs/2022ApJS..258...11W}{258, 11}

\bibitem[{Wehrens \& Buydens(2007)}]{Wehrens07}
Wehrens, R. \& Buydens, L. M.~C. 2007,
  \href{http://dx.doi.org/10.18637/jss.v021.i05}{\color{blue}Journal of
  Statistical Software},
  \href{https://www.jstatsoft.org/article/view/v021i05}{21, 1}

\bibitem[{Wehrens \& Kruisselbrink(2018)}]{Wehrens18}
Wehrens, R. \& Kruisselbrink, J. 2018,
  \href{http://dx.doi.org/10.18637/jss.v087.i07}{\color{blue}Journal of
  Statistical Software},
  \href{https://www.jstatsoft.org/article/view/v087i07}{87, 1}

\bibitem[{{Wen} \& {Han}(2024)}]{Wen24}
{Wen}, Z.~L. \& {Han}, J.~L. 2024,
  \href{http://dx.doi.org/10.3847/1538-4365/ad409d}{\color{blue}\apjs},
  \href{https://ui.adsabs.harvard.edu/abs/2024ApJS..272...39W}{272, 39}

\bibitem[{{Wright} {et~al.}(2025{\natexlab{a}}){Wright}, {Hildebrandt}, {van
  den Busch}, {Bilicki}, {Heymans}, {Joachimi}, {Mahony}, {Reischke},
  {St\textbackslash''olzner}, {Wittje}, {Asgari}, {Chisari}, {Dvornik},
  {Georgiou}, {Giblin}, {Hoekstra}, {Jalan}, {William}, {Joudaki}, {Kuijken},
  {Lesci}, {Li}, {Linke}, {Loureiro}, {Maturi}, {Moscardin}, {Porth},
  {Radovich}, {Tr\textbackslash''oster}, {von Wietersheim-Kramsta}, {Yan},
  {Yoon}, \& {Zhang}}]{Wright25}
{Wright}, A.~H., {Hildebrandt}, H., {van den Busch}, J.~L., {et~al.}
  2025{\natexlab{a}},
  \href{https://ui.adsabs.harvard.edu/abs/2025arXiv250319440W}{arXiv e-prints,
  arXiv:2503.19440}

\bibitem[{{Wright} {et~al.}(2020){Wright}, {Hildebrandt}, {van den Busch}, \&
  {Heymans}}]{Wright20}
{Wright}, A.~H., {Hildebrandt}, H., {van den Busch}, J.~L., \& {Heymans}, C.
  2020, \href{http://dx.doi.org/10.1051/0004-6361/201936782}{\color{blue}\aap},
  \href{https://ui.adsabs.harvard.edu/abs/2020A&A...637A.100W}{637, A100}

\bibitem[{{Wright} {et~al.}(2024){Wright}, {Kuijken}, {Hildebrandt},
  {Radovich}, {Bilicki}, {Dvornik}, {Getman}, {Heymans}, {Hoekstra}, {Li},
  {Miller}, {Napolitano}, {Xia}, {Asgari}, {Brescia}, {Buddelmeijer}, {Burger},
  {Castignani}, {Cavuoti}, {de Jong}, {Edge}, {Giblin}, {Giocoli},
  {Harnois-D{\'e}raps}, {Jalan}, {Joachimi}, {John William}, {Joudaki},
  {Kannawadi}, {Kaur}, {La Barbera}, {Linke}, {Mahony}, {Maturi}, {Moscardini},
  {Nakoneczny}, {Paolillo}, {Porth}, {Puddu}, {Reischke}, {Schneider},
  {Sereno}, {Shan}, {Sif{\'o}n}, {St{\"o}lzner}, {Tr{\"o}ster}, {Valentijn},
  {van den Busch}, {Verdoes Kleijn}, {Wittje}, {Yan}, {Yao}, {Yoon}, \&
  {Zhang}}]{Wright24}
{Wright}, A.~H., {Kuijken}, K., {Hildebrandt}, H., {et~al.} 2024,
  \href{http://dx.doi.org/10.1051/0004-6361/202346730}{\color{blue}\aap},
  \href{https://ui.adsabs.harvard.edu/abs/2024A&A...686A.170W}{686, A170}

\bibitem[{{Wright} {et~al.}(2025{\natexlab{b}}){Wright}, {St\"olzner},
  {Asgari}, {Bilicki}, {Giblin}, {Heymans}, {Hildebrandt}, {Hoekstra},
  {Joachimi}, {Kuijken}, {Li}, {Reischke}, {von Wietersheim-Kramsta}, {Yoon},
  {Burger}, {Chisari}, {de Jong}, {Dvornik}, {Georgiou},
  {Harnois-D\textbackslash'eraps}, {Jalan}, {William}, {Joudaki}, {Lesci},
  {Linke}, {Loureiro}, {Mahony}, {Maturi}, {Miller}, {Moscardini},
  {Napolitano}, {Porth}, {Radovich}, {Schneider}, {Tr\textbackslash''oster},
  {Wittje}, {Yan}, \& {Zhang}}]{Wright_DR5}
{Wright}, A.~H., {St\"olzner}, B., {Asgari}, M., {et~al.} 2025{\natexlab{b}},
  \href{https://ui.adsabs.harvard.edu/abs/2025arXiv250319441W}{arXiv e-prints,
  arXiv:2503.19441}

\bibitem[{{Wu} {et~al.}(2022){Wu}, {Costanzi}, {To}, {Salcedo}, {Weinberg},
  {Annis}, {Bocquet}, {da Silva Pereira}, {DeRose}, {Esteves}, {Farahi},
  {Grandis}, {Rozo}, {Rykoff}, {Varga}, {Wechsler}, {Zeng}, {Zhang}, {Zhang},
  \& {DES Collaboration}}]{Wu22}
{Wu}, H.-Y., {Costanzi}, M., {To}, C.-H., {et~al.} 2022,
  \href{http://dx.doi.org/10.1093/mnras/stac2048}{\color{blue}\mnras},
  \href{https://ui.adsabs.harvard.edu/abs/2022MNRAS.515.4471W}{515, 4471}

\bibitem[{{Yan} {et~al.}(2020){Yan}, {Raza}, {Van Waerbeke}, {Mead},
  {McCarthy}, {Tr{\"o}ster}, \& {Hinshaw}}]{Yan20}
{Yan}, Z., {Raza}, N., {Van Waerbeke}, L., {et~al.} 2020,
  \href{http://dx.doi.org/10.1093/mnras/staa295}{\color{blue}\mnras},
  \href{https://ui.adsabs.harvard.edu/abs/2020MNRAS.493.1120Y}{493, 1120}

\bibitem[{{Yang} {et~al.}(2006){Yang}, {Mo}, {van den Bosch}, {Jing},
  {Weinmann}, \& {Meneghetti}}]{Yang06}
{Yang}, X., {Mo}, H.~J., {van den Bosch}, F.~C., {et~al.} 2006,
  \href{http://dx.doi.org/10.1111/j.1365-2966.2006.11091.x}{\color{blue}\mnras},
  \href{https://ui.adsabs.harvard.edu/abs/2006MNRAS.373.1159Y}{373, 1159}

\bibitem[{{Zhang} {et~al.}(2019){Zhang}, {Jeltema}, {Hollowood}, {Everett},
  {Rozo}, {Farahi}, {Bermeo}, {Bhargava}, {Giles}, {Romer}, {Wilkinson},
  {Rykoff}, {Mantz}, {Diehl}, {Evrard}, {Stern}, {Gruen}, {von der Linden},
  {Splettstoesser}, {Chen}, {Costanzi}, {Allen}, {Collins}, {Hilton}, {Klein},
  {Mann}, {Manolopoulou}, {Morris}, {Mayers}, {Sahlen}, {Stott}, {Vergara
  Cervantes}, {Viana}, {Wechsler}, {Allam}, {Avila}, {Bechtol}, {Bertin},
  {Brooks}, {Burke}, {Carnero Rosell}, {Carrasco Kind}, {Carretero},
  {Castander}, {da Costa}, {De Vicente}, {Desai}, {Dietrich}, {Doel},
  {Flaugher}, {Fosalba}, {Frieman}, {Garc{\'\i}a-Bellido}, {Gaztanaga},
  {Gruendl}, {Gschwend}, {Gutierrez}, {Hartley}, {Honscheid}, {Hoyle},
  {Krause}, {Kuehn}, {Kuropatkin}, {Lima}, {Maia}, {Marshall}, {Melchior},
  {Menanteau}, {Miller}, {Miquel}, {Ogando}, {Plazas}, {Sanchez}, {Scarpine},
  {Schindler}, {Serrano}, {Sevilla-Noarbe}, {Smith}, {Soares-Santos},
  {Suchyta}, {Swanson}, {Tarle}, {Thomas}, {Tucker}, {Vikram}, {Wester}, \&
  {DES Collaboration}}]{Zhang19}
{Zhang}, Y., {Jeltema}, T., {Hollowood}, D.~L., {et~al.} 2019,
  \href{http://dx.doi.org/10.1093/mnras/stz1361}{\color{blue}\mnras},
  \href{https://ui.adsabs.harvard.edu/abs/2019MNRAS.487.2578Z}{487, 2578}

\bibitem[{{Zhang} {et~al.}(2023){Zhang}, {Wu}, {Zhang}, {Frieman}, {To},
  {DeRose}, {Costanzi}, {Wechsler}, {Adhikari}, {Rykoff}, {Jeltema}, {Evrard},
  \& {Rozo}}]{Zhang23}
{Zhang}, Z., {Wu}, H.-Y., {Zhang}, Y., {et~al.} 2023,
  \href{http://dx.doi.org/10.1093/mnras/stad1404}{\color{blue}\mnras},
  \href{https://ui.adsabs.harvard.edu/abs/2023MNRAS.523.1994Z}{523, 1994}

\bibitem[{{Zhou} {et~al.}(2024){Zhou}, {Wu}, {Salcedo}, {Grandis}, {Jeltema},
  {Leauthaud}, {Costanzi}, {Sunayama}, {Weinberg}, {Zhang}, {Rozo}, {To},
  {Bocquet}, {Varga}, \& {Kwiecien}}]{Zhou23}
{Zhou}, C., {Wu}, H.-Y., {Salcedo}, A.~N., {et~al.} 2024,
  \href{http://dx.doi.org/10.1103/PhysRevD.110.103508}{\color{blue}\prd},
  \href{https://ui.adsabs.harvard.edu/abs/2024PhRvD.110j3508Z}{110, 103508}

\bibitem[{{Zhu} {et~al.}(2021){Zhu}, {Xu}, {Hu}, {Shan}, {Zhu}, {Fan}, {Zhao},
  {Gu}, \& {Wu}}]{Zhu21}
{Zhu}, Z., {Xu}, H., {Hu}, D., {et~al.} 2021,
  \href{http://dx.doi.org/10.3847/1538-4357/abd327}{\color{blue}\apj},
  \href{https://ui.adsabs.harvard.edu/abs/2021ApJ...908...17Z}{908, 17}

\end{thebibliography}
\endgroup

\appendix

\section{Weak-lensing null tests}\label{appendix:null_tests}
\begin{figure}[ht]
\centering\includegraphics[width = 8.5cm, height = 7.2cm] {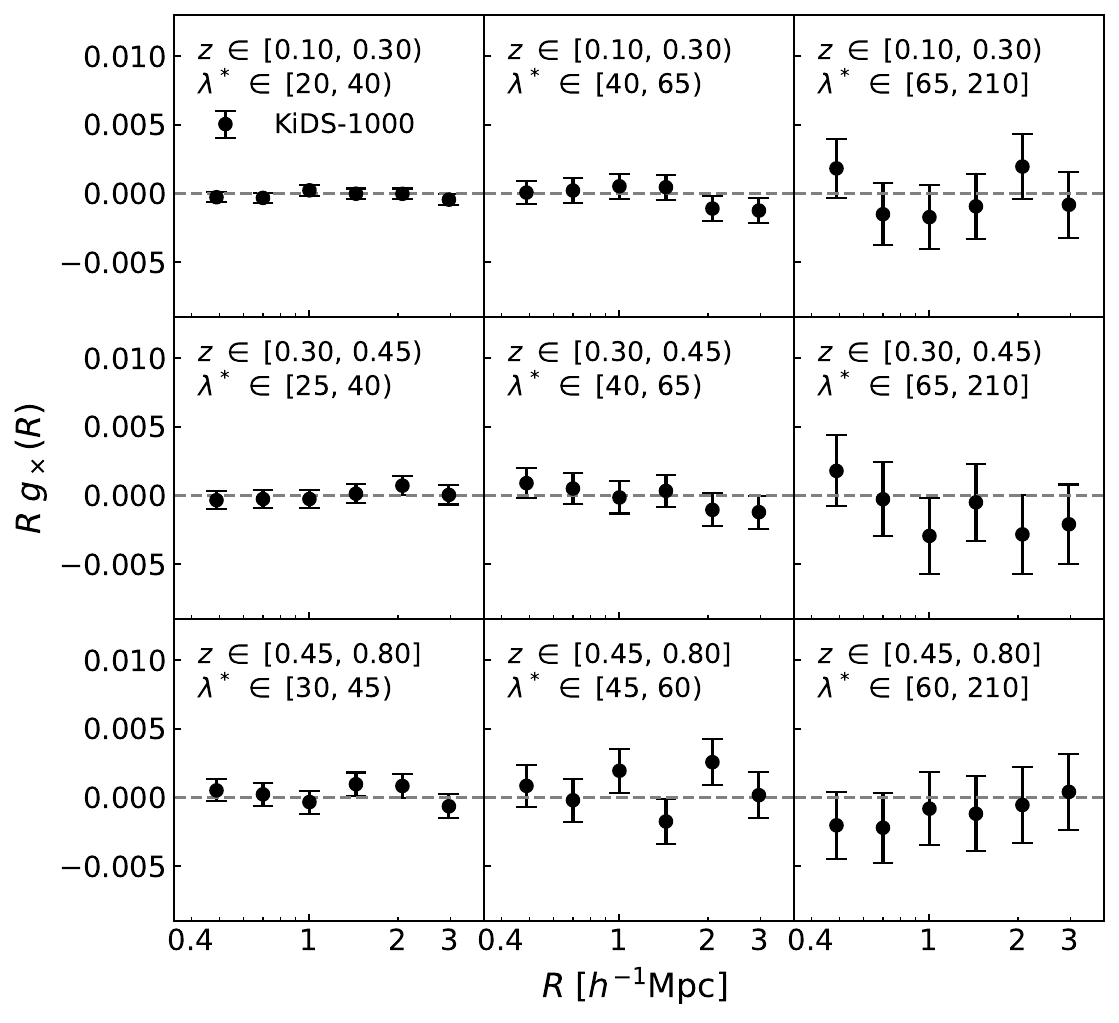}
\caption{Stacked $R\,g_\times(R)$ profiles of the AMICO KiDS-1000 galaxy clusters in bins of $z$ (increasing from top to bottom) and $\lambda^*$ (increasing from left to right). The error bars are derived through bootstrap resampling.}
\label{fig:g_x}
\end{figure}
\begin{figure}[ht]
\centering\includegraphics[width = 9cm, height = 3.8cm] {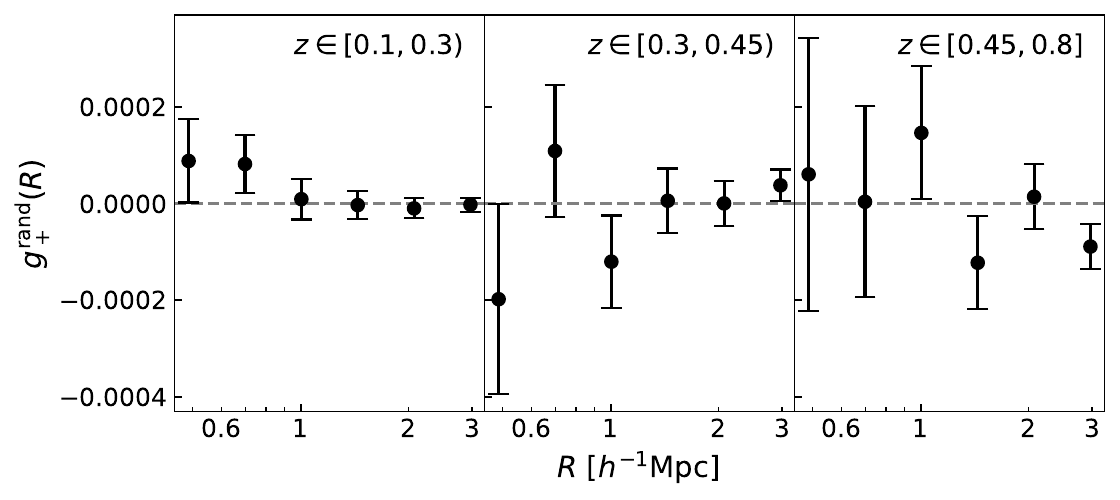}
\caption{Stacked random reduced tangential shear, $g_+^{\rm rand}(R)$, in bins of $z$ (increasing from the left to the right panel) and averaged over $\lambda^*$. The error bars are derived through bootstrap resampling.}
\label{fig:g_random}
\end{figure}
\noindent In the absence of residual systematic uncertainties, the cross-component of the reduced shear, namely $g_\times$, should be consistent with zero. Figure \ref{fig:g_x} shows that $g_\times$ does not exhibit substantial deviation from zero across the entire radial range examined in the analysis. We derived a reduced $\chi^2$ of $\chi^2_{\rm red}=0.6$, in agreement with what derived by \citetalias{Bellagamba19} from KiDS-DR3 data. In addition, to further assess the presence of possible additive biases, caused for example by\ PSF effects and spurious objects in the shear sample \citep[see e.g.][]{Dvornik17}, we measured the reduced tangential shear around random positions, namely $g_+^{\rm rand}$. This test is based on a random catalogue 50 times larger than the observed cluster sample, built up accounting for the angular mask used for the AMICO detection pipeline. Random redshifts are sampled from those in the observed catalogue. In Fig.\ \ref{fig:g_random} we show that $g_+^{\rm rand}$, stacked in the same redshift bins used in the analysis, is consistent with zero. This agrees with the results by \citet{Singh17} and \citet{Giocoli21} for example, who found deviations from zero only on scales larger than those considered in this work. As our $g_+^{\rm rand}$ estimates are consistent with zero, we decide not to subtract them from the stacked signal presented in Sect.\ \ref{sec:DR4_measure}.

\section{Blinding}\label{appendix:blinding}
\begin{table}[t]
\caption{\label{tab:unblinding}Parameter constraints using the original and blind versions of the sample completeness.}
  \centering
    \begin{tabular}{l c c c r} 
      Parameter & Original & Blind 1 & Blind 2 \\ 
      \hline
      \rule{0pt}{4ex}
      $\alpha$ & $-0.17^{+0.02}_{-0.02}$ & $-0.17^{+0.02}_{-0.02}$ & $-0.16^{+0.02}_{-0.02}$\\ \rule{0pt}{2.5ex}
      $\beta$ & $0.54^{+0.02}_{-0.02}$ & $0.57^{+0.02}_{-0.02}$ & $0.54^{+0.02}_{-0.02}$\\ \rule{0pt}{2.5ex}
      $\gamma$ & $0.38^{+0.11}_{-0.11}$ & $0.24^{+0.12}_{-0.12}$ & $0.48^{+0.10}_{-0.11}$\\ \rule{0pt}{2.5ex}
      $\sigma_{\rm intr}$ & $0.05^{+0.02}_{-0.02}$ & $0.05^{+0.02}_{-0.02}$ & $0.05^{+0.02}_{-0.02}$\\ \rule{0pt}{2.5ex}
      $\Omega_{\rm m}$ & $0.21^{+0.02}_{-0.01}$ & $0.21^{+0.02}_{-0.01}$ & $0.22^{+0.02}_{-0.02}$\\ \rule{0pt}{2.5ex}
      $\sigma_8$ & $0.86^{+0.03}_{-0.03}$ & $0.87^{+0.03}_{-0.03}$ & $0.86^{+0.03}_{-0.03}$\\ \rule{0pt}{2.5ex}
      $S_8$ & $0.72^{+0.02}_{-0.02}$ & $0.73^{+0.02}_{-0.02}$ & $0.74^{+0.02}_{-0.02}$
      \end{tabular}
  \tablefoot{In the first column we list the symbols of the parameters. The constraints derived using the original, Blind 1, and Blind 2 completeness estimates are listed in the second, third, and fourth column, respectively. Median, 16th and 84th percentiles are reported. The SSC contribution was not included in these tests.}
\end{table}
\begin{figure*}[t!]
\centering\includegraphics[width = \hsize-1cm, height = \hsize-1cm] {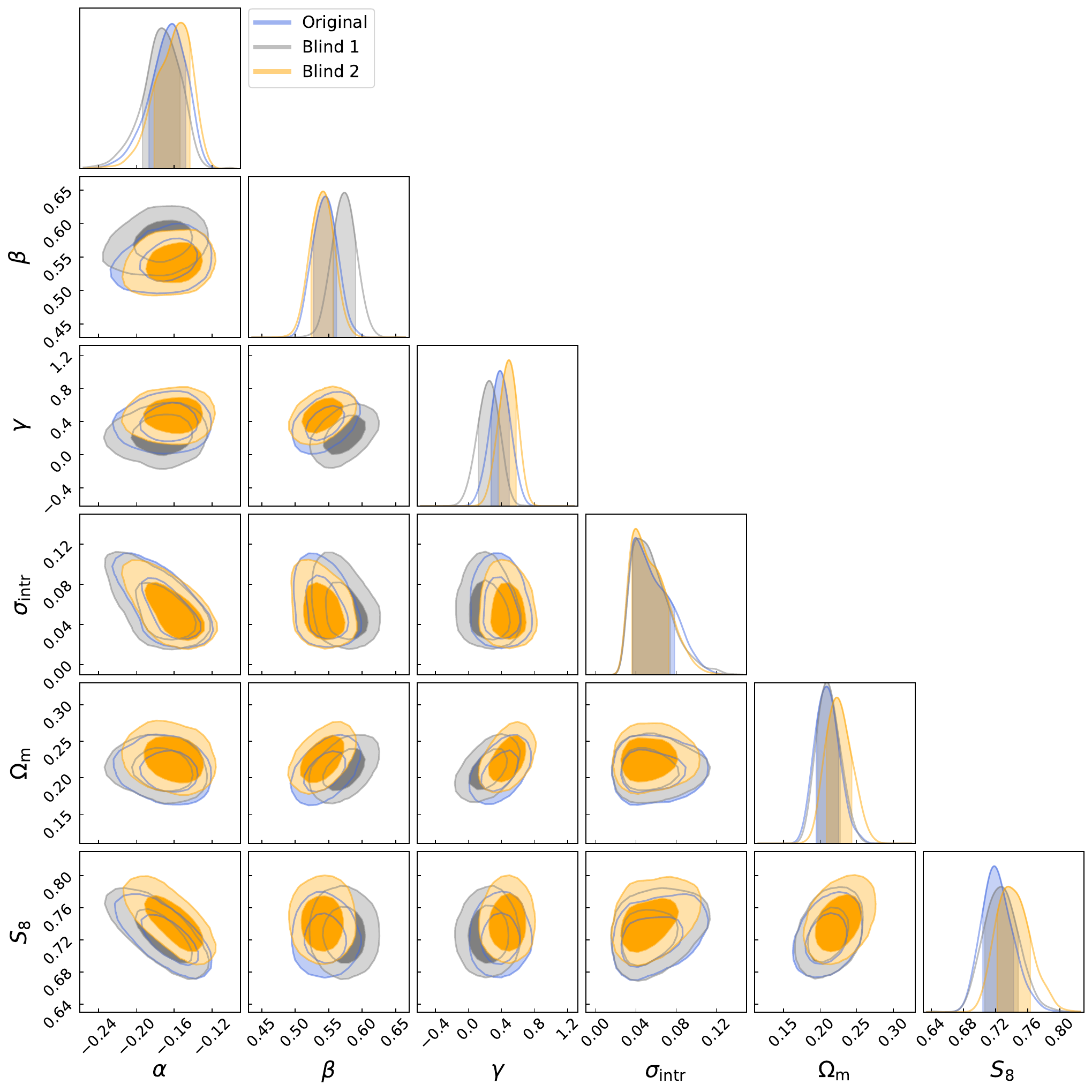}
\caption{Parameter constraints obtained by using the original (blue), Blind 1 (grey), and Blind 2 (orange) versions of the sample completeness. The confidence contours correspond to 68\% and 95\%, while the bands over the 1D marginalised posteriors represent the interval between 16th and 84th percentiles.}
\label{fig:unblinding}
\end{figure*}
\begin{figure}[t]
\centering\includegraphics[width = \hsize, height = \hsize-1cm] {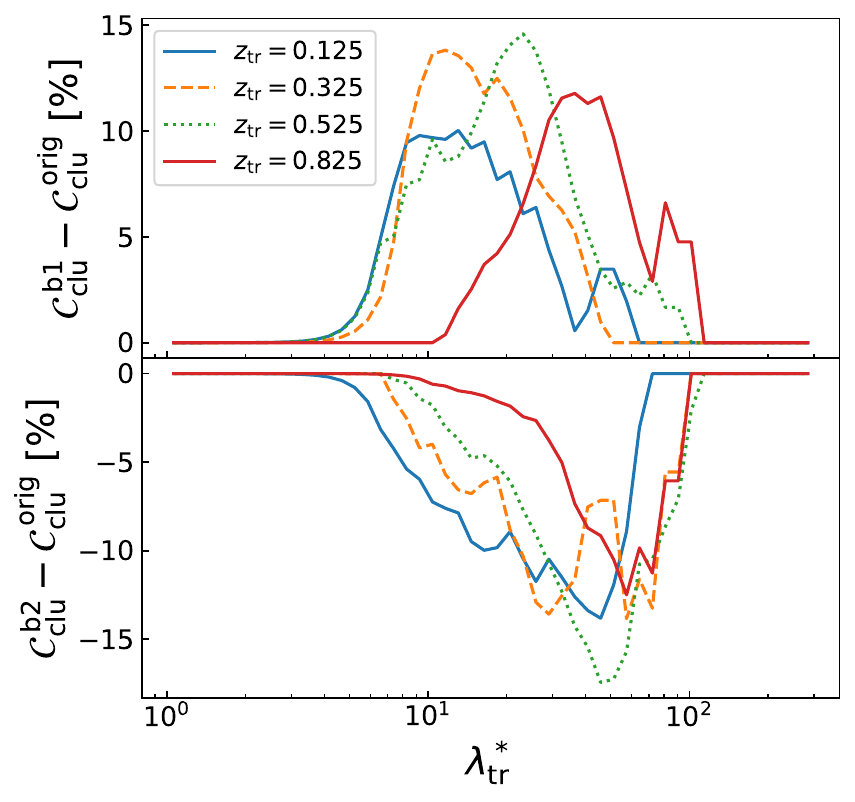}
\caption{Differences between the original cluster completeness and those in Blind 1 (top panel) and Blind 2 (bottom panel) cases, for different true redshift values (see the legend) and as a function of the true intrinsic richness.}
\label{fig:completeness_blinding}
\end{figure}
In this work, we carried out the joint modelling of cluster weak lensing and counts by blinding the cluster sample completeness. The blinding strategy consists in computing the ratio of halo counts between a reference cosmology and a set of perturbed cosmologies, assuming a theoretical mass function. This ratio is then mapped to the mass proxy using the scaling relation by \citetalias{Bellagamba19}, based on KiDS-DR3 data. Then, the completeness of detected clusters with low ${\rm S/N}$ is artificially perturbed based on this ratio, while high ${\rm S/N}$ detections remain unchanged. Finally, three completeness estimates are provided, one of which corresponds to the unperturbed original. GFL performed all tests and developed the full analysis pipeline, including the sample selection criteria, by processing only one of the blinded datasets. This choice was motivated in part by computational cost, but more importantly, to ensure that GFL had no comparison with the other completeness estimates. Only after all authors judged the pipeline to be robust and stable, we proceeded to freeze it and analyse the other blinded datasets. Specifically, likelihood models, covariance matrices, parameter priors, and the length of the MCMC were kept the same across all blinded cases, and were defined based solely on our initial blinded dataset. On the other hand, the $\lambda^*_{\rm ob}$ and $z_{\rm ob}$ selections can in principle vary across different blinded dataset versions. In fact, our selection criteria were designed to ensure cosmological results remain insensitive to these choices, while maximising the number of selected clusters. Only MM, who constructed the mock and observed catalogues, was aware of the identity of the true sample completeness prior to unblinding, and he kept it secret as an external person would have done. Unblinding was conducted only after the sample selection was fixed and MCMC runs were completed for all blinded datasets. For more details on the blinding strategy, we refer the reader to \citetalias{AMICOKiDS-1000}. \\
\indent We found that, across all three blinded dataset versions, the same sample selection criteria yield the most efficient balance between sample size and robustness of the results. Due to the limited computational resources, we performed the three analyses without including the SSC contribution. The results are listed in Table \ref{tab:unblinding}, where we refer to the two perturbed versions of the completeness as `Blind 1' and `Blind 2'. As displayed also in Fig.\ \ref{fig:unblinding}, $\alpha$ and $\sigma_{\rm intr}$ are not affected by the blinding, while the median $\gamma$ shows shifts of $1\sigma$ with respect to the original. The parameter $\beta$ undergoes substantial changes only in the Blind 1 case. Nonetheless, $\gamma$ and $\beta$ remain consistent within $1\sigma$ in all cases. The same holds for $S_8$, as we find $0.5\sigma$ and $1\sigma$ shifts of the median, compared to the original, in Blind 1 and Blind 2 cases, respectively. These results show that the blinding strategy we proposed has the potential to effectively bias the cosmological constraints. However, the completeness perturbations we introduced were not sufficiently large to produce the desired effects, namely shifts in the median $S_8$ of 2 to 3$\sigma$. Notwithstanding this, the integrity of the blinding strategy remained intact, as the pipeline freezing based on the initial blinded dataset prevented any methodological bias. \\
\indent Cosmological simulations with the characteristics of the survey, such as galaxy spatial distribution, depth, and area, will be crucial to employ this blinding strategy in future studies. In addition, we point out that the magnitude of completeness perturbations is limited by the definition of completeness itself. This is evident from Fig.\ \ref{fig:completeness_blinding}, displaying the differences between the perturbed completeness estimates, referred to as $\mathcal{C}_{\rm clu}^{\,\rm b1}$ for Blind 1 and $\mathcal{C}_{\rm clu}^{\,\rm b2}$ for Blind 2, and the original one, $\mathcal{C}_{\rm clu}^{\,\rm orig}$. As we can see, at the largest $\lambda^*_{\rm tr}$ values and low $z_{\rm tr}$, $\mathcal{C}_{\rm clu}^{\,\rm b2}$ shows larger differences against $\mathcal{C}_{\rm clu}^{\,\rm orig}$ compared to $\mathcal{C}_{\rm clu}^{\,\rm b2}$, because $\mathcal{C}_{\rm clu}^{\,\rm orig}$ is close to 100\% at those scales. This also explains why $\mathcal{C}_{\rm clu}^{\,\rm b2}$ yields the largest difference in the $S_8$ constraint. We conclude that for future surveys, such as \textit{Euclid} \citep{Mellier24}, this blinding method could prove to be extremely effective, as small completeness perturbations that are negligible in KiDS could become significant for larger cluster samples. 

\section{Comparison with AMICO KiDS-DR3 mass estimates}\label{appendix:mass_comparison}
\begin{figure*}[t!]
\centering\includegraphics[width = \hsize-1cm, height = 7.3cm] {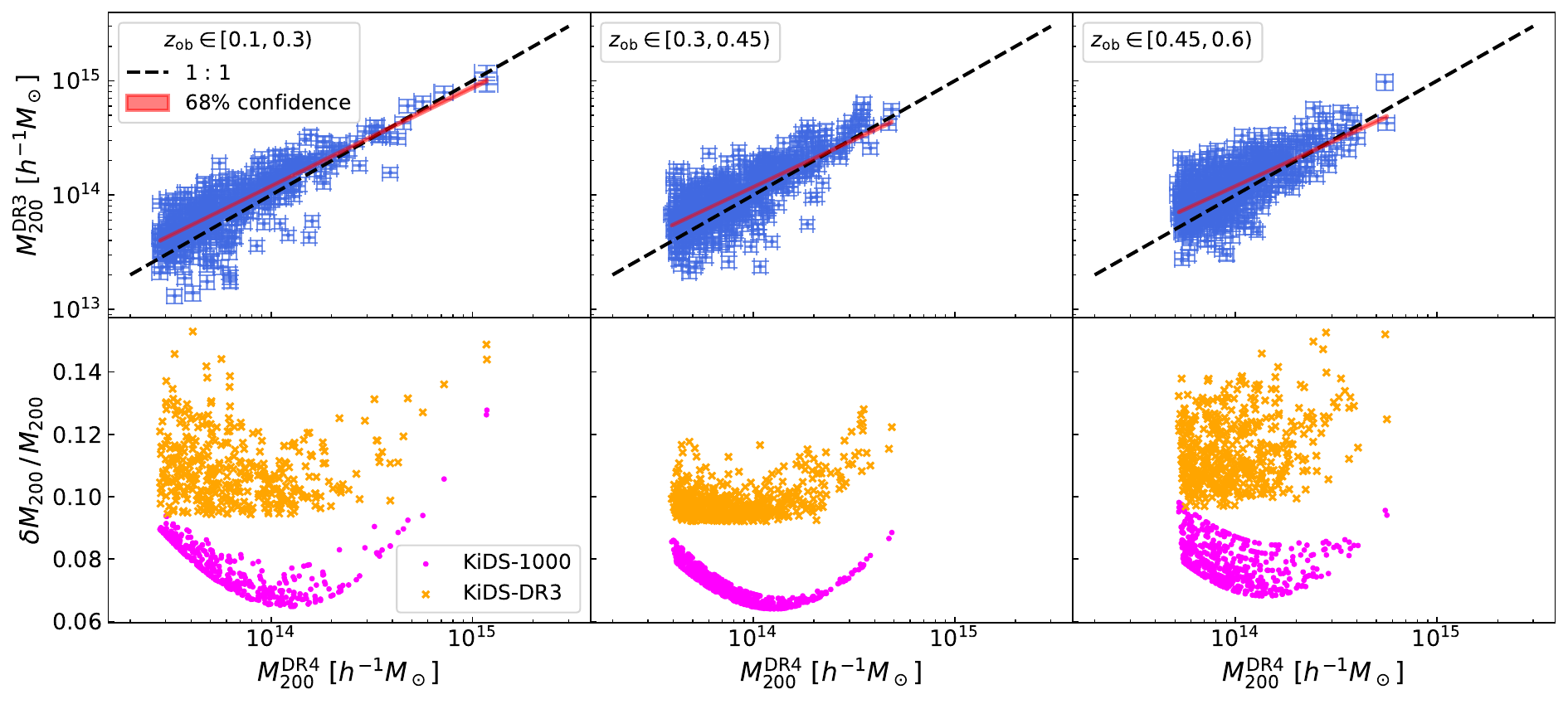}
\caption{Comparison between $M_{200}^{\rm DR3}$ and $M_{200}^{\rm DR4}$ for $z_{\rm ob}\in[0.1,0.3)$ (left panels), $z_{\rm ob}\in[0.3,0.45)$ (middle panels), and $z_{\rm ob}\in[0.45,0.6)$ (right panels), applying the $\lambda^*_{\rm ob}$ cuts listed in Table \ref{tab:sample} to the AMICO KiDS-1000 cluster sample. Top panels: Mean $M_{200}^{\rm DR3}$ against the mean $M_{200}^{\rm DR4}$. The error bars are derived by marginalising the mass estimates over all the free model parameter posteriors. The dashed black lines represent the 1:1 relation, while the red bands represent the 68\% confidence of the linear model used for the fit. Bottom panels: Mass precision in KiDS-DR4 (magenta dots) and in KiDS-DR3 (orange crosses).}
\label{fig:mass_comparison_DR3}
\end{figure*}
In this section, we discuss the comparison between our mass calibration results and those derived by \citetalias{Bellagamba19}, based on AMICO KiDS-DR3 data. As detailed in Sect.\ \ref{sec:DR4_results:baseline_WL}, our analysis improves the mass precision of AMICO clusters by three percentage points compared to KiDS-DR3 (see also the bottom panels in Fig.\ \ref{fig:mass_comparison_DR3}). To further compare these mass estimates, we matched the AMICO KiDS-1000 clusters considered in this work, with ${\rm S/N}>3.5$ and satisfying the selections listed in Table \ref{tab:sample}, with the AMICO KiDS-DR3 clusters used in the weak-lensing mass calibration by \citetalias{Bellagamba19}, having ${\rm S/N}>3.5$ and $z\in[0.1,0.6]$. Specifically, we assumed a maximum separation between cluster centres of 60 arcsec and differences in the mean photometric redshifts within $1\sigma$, obtaining about 1400 matches. Then we used the methods described in Sect.\ \ref{sec:DR4_results:baseline_WL} to derive the AMICO KiDS-1000 masses, namely $M_{200}^{\rm DR4}$.
To derive AMICO KiDS-DR3 masses, referred to as $M_{200}^{\rm DR3}$, we assumed the DR3 estimates of $\lambda^*_{\rm ob}$ and $z_{\rm ob}$ and adopt the $\log M_{200}-\log\lambda^*$ relation by \citetalias{Bellagamba19}. The relation between $M_{200}^{\rm DR3}$ and $M_{200}^{\rm DR4}$ is shown in the top panels of Fig.\ \ref{fig:mass_comparison_DR3}, based on the redshift bins adopted in \citetalias{Bellagamba19}, namely $z\in[0.1,0.3)$, $z\in[0.3,0.45)$, and $z\in[0.45,0.6)$. To quantify the agreement between $M_{200}^{\rm DR3}$ and $M_{200}^{\rm DR4}$, we fit the following relation
\begin{equation}
\log \left(\frac{M_{200}^{\rm DR3}}{10^{14}h^{-1}M_\odot}\right) = A \log \left(\frac{M_{200}^{\rm DR4}}{10^{14}h^{-1}M_\odot}\right) + B\,,
\end{equation}
through a Bayesian MCMC analysis, assuming a Gaussian likelihood and uniform priors on $A$ and $B$. We find $A=0.864\pm0.009$ and $B=0.078\pm0.003$ for $z\in[0.1,0.3)$, $A=0.830\pm0.009$ and $B=0.070\pm0.002$ for $z\in[0.3,0.45)$, $A=0.809\pm0.011$ and $B=0.078\pm0.002$ for $z\in[0.45,0.6)$. As displayed in Fig.\ \eqref{fig:mass_comparison_DR3}, we find an excellent agreement between $M_{200}^{\rm DR3}$ and $M_{200}^{\rm DR4}$ at intermediate and high masses, while $M_{200}^{\rm DR3}$ is systematically larger than $M_{200}^{\rm DR4}$ at the lowest mass values. \\
\indent We verified that KiDS-DR3 and KiDS-1000 galaxy shape measurements lead to stacked cluster weak-lensing signals which are consistent within 1$\sigma$, given the same measurement procedure and the same cluster sample. No biases are derived from the comparison of these measurements. Consequently, the discrepancies between $M_{200}^{\rm DR3}$ and $M_{200}^{\rm DR4}$ are tied to differences in the modelling of the cluster weak-lensing profiles. For example, \citetalias{Bellagamba19} imposed $\sigma_{\rm intr}=0$, while in our analysis $\sigma_{\rm intr}$ is a free parameter and this leads to lower mass estimates. Another difference with \citetalias{Bellagamba19} is the SOM analysis described in Sect.\ \ref{Sec_sys_2}. We reconstructed the true background redshift distribution and defined the background sample purity as a function of lens redshift, including these calibrations in the halo profile model (Sect.\ \ref{sec_modelling}). On the other hand, based on reference photometric and spectroscopic galaxy catalogues, \citetalias{Bellagamba19} accounted for the bias in the mean of the background redshift distribution, neglecting its full shape. This bias and the one due to foreground contamination, both averaged over the whole cluster redshift range considered, were treated as additional statistical errors included in the covariance matrix. This conservative approach, driven by the lower quality of the data compared to our present analysis, may have contributed to a possible bias in the mass estimates by \citetalias{Bellagamba19}. Lastly, \citetalias{Bellagamba19} did not account for the impact of systematic and statistical uncertainties associated with cluster redshifts and mass proxies, whereas we have explicitly incorporated these uncertainties into our likelihood. All in all, the offset between $M_{200}^{\rm DR3}$ and $M_{200}^{\rm DR4}$ can explain the lower $S_8$ value retrieved in this work compared to that obtained by \citet{Lesci22_counts}. Specifically, $M_{200}^{\rm DR4}<M_{200}^{\rm DR3}$ at low masses motivates the lower $\Omega_{\rm m}$ value in KiDS-1000 discussed in Sect.\ \ref{sec:DR4_results:baseline}.

\section{Alternative overdensity definitions}\label{appendix:Delta}
\begin{figure*}[t!]
\centering\includegraphics[width = \hsize-1cm, height = 6.7cm] {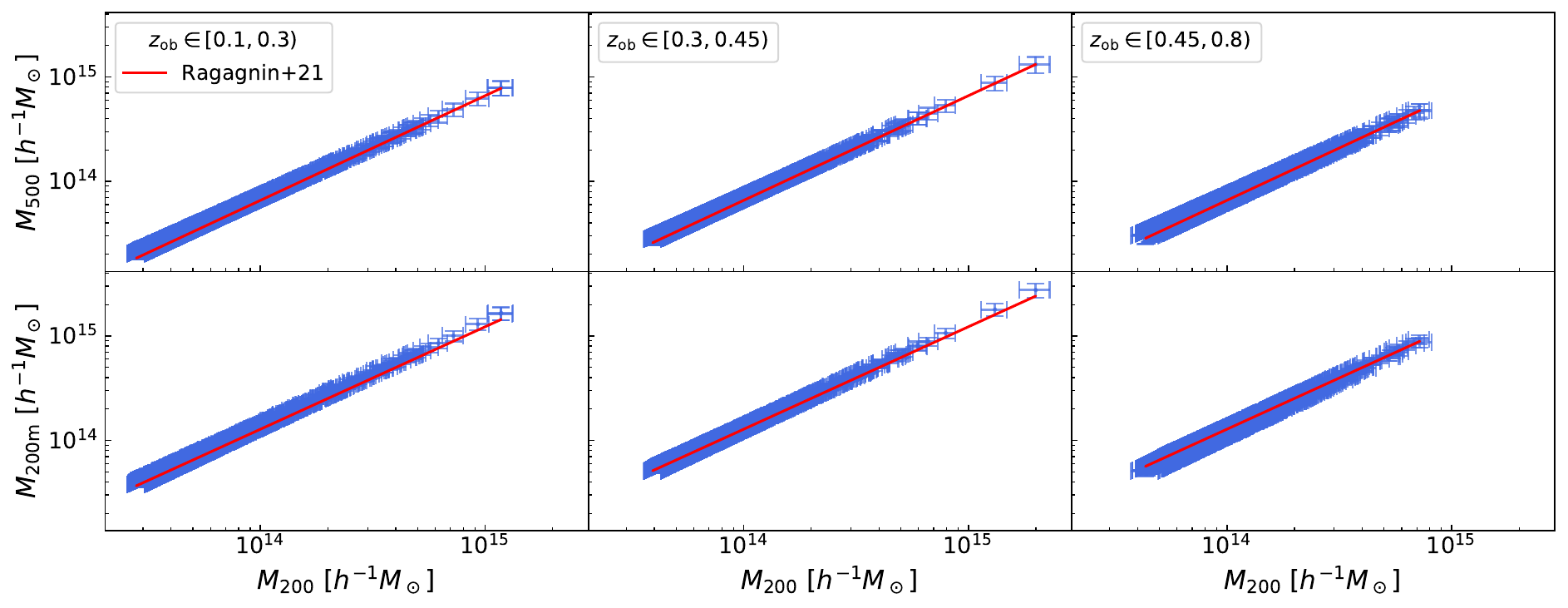}
\caption{$M_{500}$ (top panels) and $M_{200\rm m}$ (bottom panels) compared against $M_{200}$, for $z_{\rm ob}\in[0.1,0.3)$ (left panels), $z_{\rm ob}\in[0.3,0.45)$ (middle panels), and $z_{\rm ob}\in[0.45,0.8)$ (right panels), applying the $\lambda^*_{\rm ob}$ cuts listed in Table \ref{tab:sample}. The blue dots represent the estimates derived in this work, obtained by running the pipeline under the assumption of alternative overdensity definitions. The red lines represent the mass conversion from $M_{200}$ to a different overdensity based on the theoretical relation by \citet{Ragagnin21}.}
\label{fig:mass_comparison_Delta}
\end{figure*}
\begin{table}[t]
\caption{\label{tab:priors_and_posteriors:2}Results for alternative overdensity definitions.}
  \centering
    \begin{tabular}{l c c c r} 
      Parameter & $M_{200}$ & $M_{500}$ & $M_{200\rm m}$ \\ 
      \hline
      \rule{0pt}{4ex}
      $\alpha$ & $\alphaScaling^{+\alphaScalingup}_{-\alphaScalinglow}$ & $-0.08^{+0.03}_{-0.03}$ & $-0.23^{+0.01}_{-0.01}$\\ \rule{0pt}{2.5ex}
      $\beta$ & $\betaScaling^{+\betaScalingup}_{-\betaScalinglow}$ & $0.56^{+0.03}_{-0.03}$ & $0.55^{+0.02}_{-0.02}$\\ \rule{0pt}{2.5ex}
      $\gamma$ & $\gammaScaling^{+\gammaScalingup}_{-\gammaScalinglow}$ & $0.36^{+0.32}_{-0.22}$ & $0.61^{+0.28}_{-0.26}$\\ \rule{0pt}{2.5ex}
      $\sigma_{\rm intr}$ & $\scatterZero^{+\scatterZeroup}_{-\scatterZerolow}$ & $0.05^{+0.02}_{-0.01}$ & $0.05^{+0.02}_{-0.01}$\\ \rule{0pt}{2.5ex}
      $\log c_0$ & $\cZero^{+\cZeroup}_{-\cZerolow}$ & $0.23^{+0.12}_{-0.12}$ & $0.52^{+0.13}_{-0.13}$\\ \rule{0pt}{2.5ex}
      $\sigma_{{\rm off}}$ & $\soffZero^{+\soffZeroup}_{-\soffZerolow}$ & $0.27^{+0.12}_{-0.10}$ & $0.23^{+0.15}_{-0.13}$\\
      \end{tabular}
  \tablefoot{In the first column we list the symbols of the parameters. The results for $M_{500}$ and $M_{200\rm m}$ are listed in the second and third columns, respectively. Median, 16th and 84th percentiles are reported.}
\end{table}
To facilitate the use of the mass calibration presented here in future studies, we provide the scaling relation between $\lambda^*$ and mass based on alternative overdensity definitions. We focus on $M_{500}$, that is the mass enclosed within the critical radius $R_{500}$, which is widely used in SZ and X-ray studies of clusters \citep[see e.g.][]{Hilton21,Scheck23}, and on $M_{200\rm m}$, namely the mass enclosed within a sphere whose average density is 200 times the mean cosmic matter density at the cluster redshift, which is commonly used in the case of photometric cluster samples \citep[see e.g.][]{McClintock19} as an alternative to $M_{200}$. In the case of $M_{500}$, we assumed the priors listed in Table \ref{tab:priors_and_posteriors} except for the halo truncation factor, $F_{\rm t}$, in Eq.\ \eqref{rho_BMO}. Indeed, as $R_{500}\simeq 0.7\times R_{200}$ \citep{Hu03,Sereno15_Comalit3}, we divided the prior distribution parameters in Table \ref{tab:priors_and_posteriors} by 0.7, obtaining a Gaussian prior on $F_{\rm t}$ having mean $\mu=4.3$ and standard deviation $\sigma=0.7$. Since the mass and redshift evolution parameters of the concentration-mass relation are consistent in the cases of $M_{200}$ and $M_{500}$ \citep{Ragagnin21}, we assumed the same priors on $c_M$ and $c_z$ as those adopted in the baseline analysis, namely $c_M=-0.084$ and $c_z=-0.47$ \citep[corresponding to the results by][]{Duffy08}. Analogous to what done for $M_{500}$, in the case of $M_{200\rm m}$ we assumed a prior on $F_{\rm t}$ with $\mu=1.7$ and $\sigma=0.3$, while we imposed $c_M=-0.081$ and $c_z=-1.01$ following \citet{Duffy08}. \\
\indent We found no remarkable variations in the cosmological parameter posteriors and in the quality of the fit by assuming alternative overdensity definitions. The parameter constraints are listed in Table \ref{tab:priors_and_posteriors:2}, where we also report the case of $M_{200}$ to ease the comparison. The $\sigma_{\rm intr}$ and $\sigma_{\rm off}$ posteriors are consistent within $1\sigma$ across all cases. In line with the results obtained for $M_{200}$, described in Sect.\ \ref{sec:DR4_results:baseline_WL}, $F_{\rm t}$ and $f_{\rm off}$ are not constrained. The constraints on $\beta$ and $\gamma$ agree within $1\sigma$, while $\alpha$ ($\log c_0$) is larger (lower) for $M_{500}$ and lower (larger) for $M_{\rm 200m}$, respectively, in agreement with the definitions of $M_{500}$ and $M_{200\rm m}$. As a sanity check, Fig.\ \ref{fig:mass_comparison_Delta} compares our estimates of $M_{500}$ and $M_{200\rm m}$, obtained using the methods detailed in Sect.\ \ref{sec:DR4_results:baseline_WL}, with the predictions from the \citet{Ragagnin21} relation based on hydrodynamical simulations. The latter is used to convert our $M_{200}$ estimates into $M_{500}$ and $M_{200\rm m}$. For this conversion, we assumed the median redshift of the samples defined by the redshift bins shown in Fig.\ \ref{fig:mass_comparison_Delta} and the median of the cosmological posteriors discussed in Sect.\ \ref{sec:DR4_results:baseline}. As displayed in the figure, we find an excellent agreement with simulations. 

\end{document}